\documentclass[twocolumn]{aastex631}

\usepackage[whole]{bxcjkjatype} 

\usepackage{color}
\usepackage{amsmath, amssymb}
\usepackage[normalem]{ulem} 
\usepackage{multirow}
\usepackage{hyperref}

\usepackage{stackengine}
\usepackage{scalerel}
\newcommand{\ellipse}{{\raisebox{0pt}{\scalebox{3}[1.5]{$\circ$}}}}

\newcommand{\ellipsev}{{\raisebox{0pt}{\scalebox{1.5}[3]{$\circ$}}}}

\newcommand\ellipsesd{{\ThisStyle{\ensurestackMath{%
  \raisebox{-1.5\LMpt}{\stackinset{c}{}{c}{0.2\LMpt}%
  {\SavedStyle\scaleobj{.5}{\ellipse}}{\SavedStyle\ellipse}}}}}}

\newcommand\ellipsesdv{{\ThisStyle{\ensurestackMath{%
  \raisebox{-5\LMpt}{\stackinset{c}{}{c}{0.43\LMpt}%
  {\SavedStyle\scaleobj{.5}{\ellipsev}}{\SavedStyle\ellipsev}}}}}}

\newcommand\ellipsesc{{\ThisStyle{\ensurestackMath{%
  \raisebox{-5\LMpt}{\stackinset{c}{}{c}{0.43\LMpt}%
  {\SavedStyle\scaleobj{1.0}{\ellipse}}{\SavedStyle\ellipsev}}}}}}

\newcommand{\rin}{{r_\mathrm{in}}}
\newcommand{\rout}{{r_\mathrm{out}}}

\newcommand{\bb}{{\mathbf{b}}}
\newcommand{\vv}{{\mathbf{v}}}
\newcommand{\Pv}{{\mathbf{P}}}
\newcommand{\Bv}{{\mathbf{B}}}
\newcommand{\jv}{{\mathbf{j}}}

\newcommand{\ev}{{\mathbf{e}}}
\newcommand{\gv}{{\mathbf{g}}}
\newcommand{\Cfunc}{{\mathcal{C}}}
\newcommand{\Lfunc}{{\mathcal{L}}}
\newcommand{\omv}{{\text{\boldmath$\omega$\unboldmath}}}
\newcommand{\fv}{{\mathbf{f}}}
\newcommand{\vhat}[1]{{\text{\boldmath$\hat{#1}$\unboldmath}}}
\newcommand{\lfrac}[2]{{{#1}/{#2}}}

\newcommand{\cm}{{\mathrm{\,cm}}}

\newcommand{\second}{{\mathrm{\,s}}}

\newcommand{\km}{{\mathrm{\,km}}}

\newcommand{\kms}{{\km\second^{-1}}}

\newcommand{\au}{{\mathrm{\,au}}}

\newcommand{\yr}{{\mathrm{\,yr}}}

\newcommand{\gm}{{\mathrm{\,g}}}

\newcommand{\gram}{{\gm}}

\newcommand{\msunNSP}{{M_\sun}}
\newcommand{\msun}{{\,\msunNSP}}

\newcommand{\solarmass}{\msun}
\newcommand{\solarmassyr}{{\solarmass\yr^{-1}}}

\newcommand{\stellarmass}{{M_\ast}}
\newcommand{\K}{{\mathrm{\,K}}}

\newcommand{\Micron}{{\mathrm{\,\ensuremath{\mu}m}}}

\newcommand{\MA}{{M_\mathrm{A}}}
\newcommand{\alphab}{{\alpha_b}}

\newcommand{\shell}{_s}
\newcommand{\rsh}{r\shell}
\newcommand{\vsh}{v\shell}
\newcommand{\wind}{_\mathrm{w}}
\newcommand{\rhow}{\rho\wind}
\newcommand{\barrhow}{\bar\rho\wind}
\newcommand{\vw}{v\wind}
\newcommand{\aw}{a\wind}
\newcommand{\amb}{_\mathrm{a}}
\newcommand{\rhoa}{\rho\amb}
\newcommand{\barrhoa}{\bar\rho\amb}
\newcommand{\aamb}{a\amb}
\newcommand{\MC}{_\mathrm{MC}}
\newcommand{\vMC}{v\MC}
\newcommand{\rMC}{r\MC}

\defcitealias{shu1991}{SRLL}
\defcitealias{shang2020}{Paper I}
\defcitealias{shang_PII}{Paper II}
\defcitealias{shang_PIII}{Paper III}
\newcommand{\Synline}{{\textbf{\textsl{Synline}}}}

\newcommand{\MOS}{{MS}}
\newcommand{\vMOS}{v_\mathrm{\MOS{}}}
\newcommand{\rMOS}{r_\mathrm{\MOS{}}}

\accepted{November 28, 2022}
\shorttitle{A Unified Model for Bipolar Outflows from Young Stars: Kinematic Signatures}
\shortauthors{Shang et al.}

\begin{document}
\title{A Unified Model for Bipolar Outflows from Young Stars:\\ Kinematic Signatures of Jets, Winds, and Their Magnetic Interplay with the Ambient Toroids}

\author[0000-0001-8385-9838]{Hsien Shang （尚賢）}
\affiliation{Institute of Astronomy and Astrophysics, Academia Sinica,  Taipei 10617, Taiwan}

\author[0000-0002-1624-6545]{Chun-Fan Liu（劉君帆）}
\affiliation{Institute of Astronomy and Astrophysics, Academia Sinica, Taipei 10617, Taiwan}

\author[0000-0001-5557-5387]{Ruben Krasnopolsky}
\affiliation{Institute of Astronomy and Astrophysics, Academia Sinica,  Taipei 10617, Taiwan}

\author[0000-0001-7800-0025]{Liang-Yao Wang（王亮堯）}
\affiliation{Institute of Astronomy and Astrophysics, Academia Sinica, Taipei 10617, Taiwan}

\correspondingauthor{Hsien Shang}
\email{shang@asiaa.sinica.edu.tw}


\begin{abstract}
Kinematic signatures of the jet, winds, multicavities, and episodic shells arising in the unified model of bipolar outflows developed in Shang et al.\ (2020), in which an outflow forms by radially directed, wide-angle toroidally magnetized winds interacting with magnetized isothermal toroids, are extracted in the form of position--velocity diagrams. Elongated outflow lobes, driven by magnetized winds and their interplay with the environment, are dominated by extended bubble structures with mixing layers beyond the conventional thin-shell models. The axial cylindrically stratified density jet carries a broad profile near the base, across the projected velocity of the wide-angle wind, and narrows down along the axis with the collimated flow. The reverse shock encloses the magnetized free wind, forms an innermost cavity, and deflects the flow pattern. Shear, Kelvin--Helmholtz instabilities, and pseudopulses add fine and distinctive features between the jet--shell components, and the fluctuating jet velocities. The broad webbed velocity features connect the extremely high and the low velocities across the multicavities, mimicking nested outflowing slower-wind components. Rings and ovals in the perpendicular cuts trace multicavities at different heights, and the compressed ambient gap regions enrich the low-velocity features with protruding spikes. Our kinematic signatures capture the observed systematics of the high-, intermediate-, and low-velocity components from Class 0 to II jet--outflow systems in molecular and atomic lines. The nested shells observed in HH 212, HH 30, and DG Tau B are naturally explained. Outflows as bubbles are ubiquitous and form an inevitable integrative outcome of the interaction between wind and ambient media.

\end{abstract}

\keywords{\href{http://astrothesaurus.org/uat/1607}{Stellar jets (1607)}; \href{http://astrothesaurus.org/uat/1636}{Stellar winds (1636)}; \href{http://astrothesaurus.org/uat/1635}{Stellar wind bubbles (1635)}; \href{http://astrothesaurus.org/uat/267}{Collapsing clouds (267)}; \href{http://astrothesaurus.org/uat/1834}{Young stellar objects (1834)}; \href{http://astrothesaurus.org/uat/1302}{Protostars (1302)}; \href{http://astrothesaurus.org/uat/1569}{Star formation (1569)}; \href{http://astrothesaurus.org/uat/1966}{Magnetohydrodynamical simulations (1966)}; \href{http://astrothesaurus.org/uat/844}{Interstellar line emission (844)}}

\section{INTRODUCTION}
\label{sec:intro}

The study of jets and outflows as signposts of star formation started with the serendipitous discoveries of the Herbig--Haro (HH) objects \citep{herbig1951,haro1952}. These early HH objects are often accompanied by strong emission lines such as [\ion{S}{2}] and [\ion{O}{1}], and high-excitation lines [\ion{O}{3}] and [\ion{Ne}{3}] with supersonic velocities \citep{schwartz1975}. These properties led \citet{schwartz1978} to hypothesize a stellar source driving a wind (a stellar wind) of $100\kms$ and a mass-loss rate of $\sim 1\times 10^{-5}$ to $1\times 10^{-6}\,M_\odot\yr^{-1}$, at least for a brief period of time to excite HH 1 and 2. Subsequent proper-motion measurements of HH 28 and 29 by \citet{cudworth1979} gave a tangential velocity of $145\kms$, and those of HH 1 and 2 \citep{HJ1981}, HH 39 knots \citep{JH1982}, showed very high tangential ``latitude-dependent'' velocities, away from their driving infrared sources. Broad CO spectra from the Class I source L1551 prompted \citet{snell1980} to hypothesize an origin in stellar wind sweeping into the dense surrounding cloud for the emission blobs coming from the outflowing lobe-like structures. These measurements revolutionized the understanding of the HH objects and related the outflow phenomena in molecular lines to the supersonic motions found in those angle-dependent velocity profiles.

Historically, a variety of models have been developed to describe the HH objects in terms of the interaction between a stellar supersonic wind and the ambient medium \citep{schwartz1978,Norman1979,canto_rodriguez1980,canto1980,canto_etal1980,smith1986,konigl1982}. The framework of \citet[][hereafter SRLL]{shu1991} connects the bipolar CO outflows with the momentum input from a (stellar) wind \citep{shu1988}, and the density distribution of the ambient cloud core that gives birth to the star. Conservation laws of mass and momentum give a simple conceptual solution of a self-similar thin shell of the outflows whose shapes are determined by an angle-dependent function. These wind-driven momentum-conserving thin-shell models descending from the early stellar wind hypotheses of off-axis HH objects establish the bases of the modern wide-angle wind paradigm of the molecular outflows \citep[e.g.,][]{shu1991,masson1992,li1996,matzner1999,lee2001}.

Nowadays, HH objects are identified with the jet knots on the axes, forming the basis of the jet-driven bow-shock models. The connection between the formation of HH objects and the collimated jets from young stars was established by the linear alignment of HH 46 and 47 \citep{bok1978,schwartz1977ApJ,schwartz1977ApJS}. The linear alignment of HH 47A, HH 46, and HH 47C indicated high collimation in the bipolar flow \citep{dopita1982}. \citet{MF1983} found emission features of the HH objects resembling those of the HH 46/47 jet, leading to the identification of HH objects as ``jets'' from young stars. The subsequent discovery of HH 34 as a collimated jet pointing toward a series of bow shocks \citep{reipurth1986} indicates the connection of a jet and aligned bow shocks. Many similar collimated jets associated with HH objects from young stars \citep[see the review by][]{RB2001} have bow shocks appearing in the highly pointed jet-like momentum output carried in the gas \citep[e.g.,][]{raga1993a,raga1993, masson1993,stone1993,stone1993b,lee2000,ostriker2001,offner2011, offner2014,offner2017}.

For the shocked shells formed when a spherical wind runs into an ambient medium, \citet{KM1992a,KM1992b} introduced the structure and evolution of a spherical wind-blown radiative bubble. Two shocks arise in the interaction of a wind (radiative, adiabatic, or in between) with the ambient medium: a forward (ambient) shock, and a reverse (wind) shock. The shocked wind and the shocked ambient medium are separated by a contact discontinuity. The nature of bubbles adds a new dimension onto the structures formed during the interaction of the media.

In the framework depicted above, \citet{shang2006} developed a unified model of bipolar outflows integrating the wind-driven and jet-driven paradigms. \citepalias[][hereafter Paper II]{shang2020} considered an extension including bubble structures. In \citet{shang2006}, the wind is dominated by the toroidal component of the magnetic field, as expected from the asymptotic regime of an X-wind demonstrated in \citet{shu1995} and \citet{shang1998}, and a magnetocentrifugal wind arises from the innermost regions of the disk
\citep{li1996,krasnopolsky2003,shang2007_PPV}. The outflows form by the interaction of these winds with the ambient medium.

The ambient environment is characterized by a series of magnetized, isothermal ``toroids'' representing molecular cloud cores before the onset of dynamical collapse \citep{li1996b,allen2003b,allen2003,galli2006}. 
Analytic formulations \citep{shu1991,shu1995,li1996b} can serve good purpose for HD thin-shell approximation \citep{matzner1999} using one toroid density profile \citep{lee2000}. 
Full 2D MHD simulations with series of toroid configurations were carried out in \citet{shang2006}, followed by
\citet{wang2015,wang2019}, and \citet{shang2020}, using the Zeus family of codes \citep{stone1992b,stone1992c,krasnopolsky2010}.
The morphology correlates well with the systematic evolution of the outflow shapes \citep{arce2006,velusamy2014} from the Class 0 to II stages of evolution.
 
The self-similar bubble nature of the outflow lobes in \citetalias{shang2020} can only be revealed when the numerical resolution is sufficiently high. The fine structures are bounded by reverse and forward shocks, separated by a tangential discontinuity (TD). The wind--toroid interface is unstable to the Kelvin--Helmholtz instability (KHI) due to substantial shear, leading to the formation of extended magnetized mixing layers within the outflow lobes. Magnetic forces in the compressed toroidally magnetized wind can generate vorticity, magnetic pseudopulses, and nonlinear patterns, forming a feedback loop to the growth of structures. These become key morphological features belonging to an elongated magnetized wind-blown bubble, which maintains self-similarity in time.

Molecular outflows carrying both the jet and wind characters in the earliest phases are natural testbeds of the unified theoretical framework. The youngest Class 0 sources with elongated molecular shells surrounding the highly collimated molecular SiO or CO jets are identified in the submillimeter to millimeter wavelengths (non-exhaustively including HH 211, HH 212, L1448C(N), IRAS 04166+2706, CARMA-7, L1157, Serpens SMM1, and Ser-emb 8(N); see, e.g., \citealt{jhan2021}, \citealt{lee2021_hh212}, \citealt{hirano2010}, \citealt{wang2014}, \citealt{plunkett2015_nat}, \citealt{podio2016}, \citealt{tychoniec2019}, and the references therein). These sources usually possess distinct extremely high velocity (EHV) associated with their highly collimated molecular jets, and the expected low-velocity (LV) cavity. In Class I phase, similar phenomena can be identified with bright optical or near-infrared jets giving the high-velocity component (HVC) surrounded by wide LV molecular cavity walls (e.g.\ HH 111, HH 46/47, L1551, DG Tau B, and HH 30; \citealt{lefloch2007}, \citealt{zhang2019}, \citealt{stojimirovic2006}, \citealt{zapata2015}, \citealt{louvet2018}). In T Tauri phase, high-velocity (HV) jets or microjets in bright atomic forbidden lines are mostly visible without obvious thick cavities, except for some flat-spectrum sources \citep{adams1988,yokogawa2001} such as FS Tau B, DG Tau A, and HL Tau \citep{liu2012,agra-amboage2014,takami2007}. Additionally, in some active T Tauri stars, additional mysterious peaks in LV atomic oxygen [\ion{O}{1}] have been known to exist \citep[see, e.g.,][]{HEG1995,simon2016,banzatti2019}.

High-resolution and high-sensitivity instruments from optical–infrared (such as the Very Large Telescope and Gemini) to submillimeter (such as the Atacama Large Millimeter/submillimeter
Array, ALMA) wavelengths have revealed additional finer structures in those very young sources with collimated jet phenomena from Class 0 to II sources. Accumulating evidence of multiple molecular shells organized in nested layers is found in these youngest sources in ALMA observations, which have been described as ``onion-like'' or as ``nested cones'', coincidentally, in evolutionary phases spread from Class 0 to II \citep[e.g.,][]{lee2021_hh212,devalon2020,harsono2021,guedel2018}. These nested-shell structures also come with accompanying peculiar velocity features distinct from the high-velocity jet or the conventional wide low-velocity cavity. Together with the previously unidentified low-velocity [\ion{O}{1}] peaks in T Tauri sources, some trends of systematics appear in the correlated phenomena. Yet a systematic explanation for these connected features has not existed in the literature until this work.

The purpose of the present paper is to document the rich kinematics displayed by the model developed in \citetalias{shang2020}, which goes beyond the basic picture expected out of the conventional simple high-velocity jet and low-velocity shell combinations (e.g.\ \citealt{bally2016}). 
The patterns of interplay generate their own kinematic signatures differentiable from the smooth bulk flows in the position--velocity diagrams. We document these kinematic signatures as evidence of outflows being elongated bubbles by figure panels demonstrating the corresponding patterns step by step, for the parameter space explored (e.g.\ in Secs.\ \ref{sec:PVD} and \ref{sec:perp_PV}).

Our work here offers a self-consistent systematic explanation to the unexplained nested-shell velocity structures as multicavities naturally arising in a magnetized bubble as a result of interplay as exemplified in our framework. Our framework also offers a natural foundation to the ubiquity of molecular outflows as magnetized bubbles and occurrence of their internal structures. These structures arise with their distinct kinematic features. We highlight such connections in Section \ref{subsec:velocity_components}.
We note that the newly released image of the Class 0 protostar L1527 from the James Webb Space Telescope (JWST) lends the most spectacular support to the multicavity structures expected out of a magnetized wide-angle wind interacting with its ambient medium.

The organization of the paper is as follows. Section \ref{sec:framework} illustrates the theoretical framework to investigate the kinematics of the proposed unified model.
Section \ref{sec:setup} gives the basic data processing setups adopted and utilized. Section \ref{sec:2Dinterior}
constructs the internal structures of the 2D thin hydrodynamic elongated momentum-conserving shell with locally oblique shocks formed by the wind and toroid 
density profiles.
Section \ref{sec:struc_hydromagnetic_bubbles} describes the internal structures of the nonspherical elongated hydromagnetic bubbles. Section \ref{sec:MCcurves} 
traces curves in position--velocity space representing the outer shape of hydrodynamic and hydromagnetic models of outflow lobes.
Sections \ref{sec:PVD} and \ref{sec:perp_PV} give results for the velocity profiles and flow patterns of the cases explored in \citetalias{shang2020} and their position--velocity diagrams of column densities (PVDCDs), respectively, parallel and perpendicular to the outflow axes.
In Section \ref{sec:discussions}, we discuss the physical and observational implications, and we summarize our findings in Section \ref{sec:summary}. 

\section{Theoretical Framework}
\label{sec:framework}

In this section we review the theoretical concepts underlying the unified jet and wide-angle wind framework, as well as the analytical and computational methodologies adopted to investigate the properties of the proposed model \citep{shang2006,wang2015,shang2020}.
A system of toroidally magnetized wide-angle winds with a stratified profile of density $\rho\propto \varpi^{-2}$ and magnetic field $b_\phi\propto \varpi^{-1}$ can be formulated in the absence of rotation, for an exactly radial cold flow of constant velocity, with a set of simplified equations (which correspond to a wind in equilibrium in the cold limit). This system of winds can interact with the ambient system of singular isothermal toroids, which may be poloidally magnetized. These toroids represent the end states of quasi-static hydromagnetic evolution followed by gravitational collapse \citep{li1996b,allen2003b,allen2003,galli2006}.

Through the magnetic interplay between the inner wind and the series of ambient toroids, large-scale outflows can form. We highlight the conceptual considerations of the classical momentum-conserving shell model of molecular outflow as presented in \citetalias{shu1991}, the inclusion of magnetic pressure developed in \citetalias{shang2020}, and the theoretical nature of the spherical wind-blown bubbles as presented in \citet{KM1992a,KM1992b}.
We develop an analytic bubble-within-shell model to help illustrate conceptually the internal structures of the thin hydrodynamic momentum-conserving shell. We generalize this conceptual approach to treat the shapes of the elongated hydromagnetic bubbles with thick interiors in the integrated unified framework.

The overall wind characters considered in this work apply in general to magnetocentrifugal winds (as X-winds and disk winds) launched from the innermost portion of the disks, where the launch regions are much smaller than the size scales before which the bubble establishes its self-similarity. However, the X-winds have natural angle-dependent asymptotic behaviors \citep{shu1995} and fanning streamlines underlying the strongest wide-angle wind at the base. They possess inner ``hollow cones'' occupied by coronal magnetic fields \citep{shu1997} from the inner boundary interacting with the stellar magnetosphere, which require very high-angular resolution beyond
that sufficient to resolve the simple jet features.

\subsection{Physics of the Magnetized Wind}
\label{subsubsec:windMHD}

The governing equations utilized in this work are those of time-dependent axisymmetric MHD\@. They can be simplified to study a steady-state magnetized wind through
a force balance similar to the asymptotics of a cold X-wind \citep{shu1995} at the mid-to-large horizontal scale ($\varpi$) far away from the launching region, leading to equations fully derived in \citetalias{shang2020} for a cold radially directed flow of constant superfast velocity, and briefly summarized in Appendix \ref{sec:appendix:setups}.

For a magnetocentrifugal wind coming out from the innermost region of a disk, only a small region of the disk or simply a very small rotating disk is required for the launching. Such a wind is dominated by a wound-up toroidal component well beyond a transition region. The toroidal velocity of the wind declines steadily at large distances due to angular momentum conservation so that the wind flow becomes approximately poloidal ($v_\phi=0$) with a purely toroidal field frozen inside ($\bb_p=0$), mass density $\rho=\rhow$, and gas pressure $p=\aw^2 \rho$. For a completely cold wind, $p=0$, the simplified steady-state Equations (\ref{eqn:wind:vtheta0:mass})--(\ref{eqn:wind:vtheta0:induction}) admit an exact solution with a radial wind of constant velocity $\vv=v_r \vhat{r}$, cylindrically stratified density $\rhow\propto \varpi^{-2}$ and force-free magnetic field $\bb=b_\phi \vhat{\phi}$.
This basic radial wind justifies boundary conditions that are lightweight options for use in numerical simulations of flow propagation and interaction with the ambient medium.
Based on this solution, the wind is set up on the inner radial boundary at $r=\rin$ as
\begin{equation}
  \label{eqn:initcon}
  \rhow(\varpi)=\frac{D_{0}}{\varpi^{2}}, \quad
  b_{\phi}(\varpi)=\frac{b_{0}}{\varpi}, \quad
  v_{r}=v_{0}, \quad
  v_\theta=0
\end{equation}
where $\varpi=r \sin\theta$ is the cylindrical radius from the axis. Alternatively, we define
\begin{equation}
  r^2\rhow = D_0 / \sin^2\theta\ , \label{eqn:rhow} \\
\end{equation}
or, $\barrhow\equiv r^2\rhow$, as part of the wind boundary condition setup, and also in free-wind regions.  

The boundary condition value $b_0$ gives the scale of the wind magnetic field. That value is parameterized using the Alfv\'en Mach number at the boundary condition, $\MA\equiv v_0/v_\mathrm{A}=v_0\sqrt{D_0}/{b_0}$, fixing the magnetic scale $b_0=v_0\sqrt{D_0}/\MA$.

\subsubsection{Magnetic Forces in the Wind}

The forces due to the $b_\phi$ component, dominant or exclusive within the wind, can be expressed under axisymmetry within a current function formulation, as follows.
We define a current function $\Cfunc\equiv-\varpi b_\phi$, proportional to the total current carried in one hemisphere of the wind. This function can be used to find the current density and force term due to an axisymmetric $b_\phi$.
The current density $\jv_p$ 
\begin{equation}
\label{eqn:jp_Cfunc}
    \begin{split}
        \jv_p&=\nabla\times(b_\phi\vhat\phi)
    =\frac{1}{\varpi}\vhat\phi \times \nabla \Cfunc\ ,
    \end{split}
\end{equation}
and the magnetic force can be computed using $\Cfunc$, leading to a force function $\Lfunc=\Cfunc^2/2$ behaving in a way similar to a pressure in that its gradient helps to obtain the force per unit mass $\fv_c$ due to $b_\phi$
\begin{equation}
\label{eqn:lfunc_gradient}
    \begin{split}
    \fv_c=
        \frac{\jv_p}{\rho} \times(b_\phi\vhat\phi)
     &=
        -\frac{\Cfunc}{\rho\varpi^2}\nabla \Cfunc 
        =
        -\frac{1}{\rho\varpi^2}\nabla\Lfunc
        \ .
    \end{split}
\end{equation}

At the inner boundary, $\Cfunc=-\Cfunc_0$, where $\Cfunc_0 = b_0=\sqrt{D_0}v_0/M_{\mathrm A}$. 
For a free wind, the value of $\Cfunc$ stays exactly or nearly exactly constant. In the free-wind regions, the relevant gradients vanish, and the wind is current-free and force-free.

Interactions of the wind with the ambient environment gives rise to nonzero gradients of $\Cfunc$ and $\Lfunc$, and induces currents and magnetic forces. 

The compressed wind region allows not only density but also magnetic field $b_\phi$ to be compressed, generating gradients of $\Lfunc=(b_\phi\varpi)^2/2$.
The respective magnetic force components $f_{r,c}=\left(-\frac{1}{\rho\varpi^2}\partial_r\Lfunc\right)$
and $f_{\theta,c}=\left(-\frac{1}{r\rho\varpi^2}\partial_\theta\Lfunc\right)$, zero in the free wind,
can be generated inside the bulk of the compressed wind, leading to corresponding variations in $v_r$ and $v_\theta$, and accelerations.
The variations of $\Cfunc$ and $\Lfunc$ distinguish the free wind from the compressed wind across the reverse shock boundary and compressed ambient material across the wind--ambient interface 
when the wind dynamically interacts with the ambient toroid.

The compressed $b_\phi$ can also generate vorticity $\omega_\phi$ according to the torque
\begin{equation}
\label{eqn:torque_bphi}
    \begin{split}
 \nabla\times\left(\frac{\jv\times\bb}{\rho}\right)
 &=
 \frac{\nabla\rho\times\nabla (b_\phi^2/2)}{\rho^2}
 -\ev_\phi\partial_z(b_\phi^2/\rho\varpi)\\
 &=-\nabla\left(\frac{1}{\rho\varpi^2}\right)
 \times\nabla\Lfunc\ .
    \end{split}
\end{equation}
Magnetic pseudopulses facilitate and sustain mixing across the boundary through the strong KHI feedback (see Section \ref{subsec:pseudopulses}).

\subsection{Initial Ambient Medium: Isothermal Toroids}
\label{subsubsec:toroids}
We base our initial ambient medium on the ALS toroids as in \citet[][ALS]{allen2003b}, with minor additions and scalings, explained in this subsection.
In the initial ambient region, 
$v_\phi=0$ and $b^\mathrm{T}_\phi=0$, and the
magnetic field of the toroids is purely poloidal, taking the same form as Equation (\ref{eqn:bp}), with an angular dependence given by a magnetic flux function $\phi(\theta)$ as below:
\begin{equation}
\label{eqn:toroid_scaling}
\rho^{\mathrm{T}}(r,\theta)=\frac{\aamb^{2}}{2\pi Gr^{2}}R(\theta),\quad \Phi^\mathrm{T}=\frac{\aamb^2r}{(4{\pi}G)^{1/2}}\phi(\theta)\ ,
\end{equation}
where $R(\theta)$ and $\phi(\theta)$ are dimensionless angular functions defined through the solution of an ordinary differential equation (ODE) given in \citet{li1996b}, and $\aamb$ is the isothermal sound speed of the ambient gas. 
The linear sequence given by $n\neq 0$ labels a separate value of the mass-to-flux ratio in dimensionless form, and also measures the degree of magnetic support against self-gravity.
The $n=0$ solution of the ODE $R_{n=0}(\theta)=1$ is a constant value representing a spherically symmetric ambient medium. For $n>0$, the ambient medium has an opening region featuring very low-density values, more and more open and less and less dense as $n$ increases. 

This function, which depends on the parameter $n\geq 0$, is subject to the normalization condition
$\int_0^{\pi/2} R(\theta) \sin\theta\,d\theta = 1 + n/4$, and a set of boundary conditions that includes the limiting property $R(\theta)\approx a_0(n)\sin^n(\theta)$ for sufficiently small $\theta\rightarrow 0$, making the toroids more open at the axis as $n$ increases.
This functional form plays a role for the hydrodynamic shell to be shown in Section \ref{subsec:chi_limit} and Appendix \ref{sec:appendix:chi_limit}.

The initial ambient medium is set up as minor additions and scalings given in Equations (\ref{eqn:alpharho})--(\ref{eqn:alphab}) of the ALS toroids defined in Equation (\ref{eqn:toroid_scaling}). The initial ambient density and magnetic field are given as
\begin{eqnarray}
\label{eqn:alpharho}
    \rhoa(t=0) &=& \alpha_\rho \rho^\mathrm{T}(r,\theta) + \rho^S(r)\\
\label{eqn:alphab}
    \bb(t=0) &=& \alpha_b \bb^\mathrm{T}(r,\theta)
\end{eqnarray}
where $\rho^S\equiv{}D^Sr^{-2}$, in which $D^S$ is a constant equal to either $2.25\times10^{11}\gram\cm^{-1}$, or zero (defining either of the ``tapered'' or ``untapered'' toroids).
The term in Equation (\ref{eqn:alphab}) defining the initial magnetic field is derived from Equation (\ref{eqn:toroid_scaling}), and 
the components of $\bb^\mathrm{T}$ are found using Equation (\ref{eqn:bp_components}) multiplied by a scaling constant $\alphab$, set to be $1$, $0.1$, or $0$.
Equation (\ref{eqn:alpharho}) is used alternatively throughout this work as
\begin{equation}
r^2 \rhoa = \alpha_\rho\frac{\aamb^2}{2\pi G}R(\theta) + D^S \label{eqn:rhoa}
\end{equation}
motivating the definition 
$\barrhoa\equiv r^2\rhoa$, particularly relevant to the initial conditions, and to unperturbed portions of the ambient medium. The value $D^S$ represents the addition of a spherical density (equal to a tenuous $n=0$ toroid) where $D^S$ is much smaller than
$\aamb^2/2\pi G$, the density scale defined for a regular toroid [without the angle factor $R(\theta)$]. Almost all of the runs set $\alpha_\rho=1$, but a few runs lack an ALS toroid in their initial setup and have $\alpha_\rho=0$ instead. 
We often refer to the initial and later ambient medium of our typical run as a toroid (in a general sense), because it is derived from the ALS toroid setup.

\paragraph{Equation of state}
The gas pressure $p=a^2\rho$ is calculated from the two-temperature equation of state \citep{wang2015} by a sound speed $a$ based on the wind mass fraction $f=\rhow/\rho$ and $a^2 = f \aw^2 + (1-f) \aamb^2$, where $\rhow$ and $\rho$ are wind and total densities. The sound speeds for the wind and the ambient medium,
$\aw$ and 
$\aamb$ are given in Table \ref{tab:quantities}.

\subsection{Formation of Outflows}
\label{subsec:outflow_formation}

\begin{figure*}
\plotone{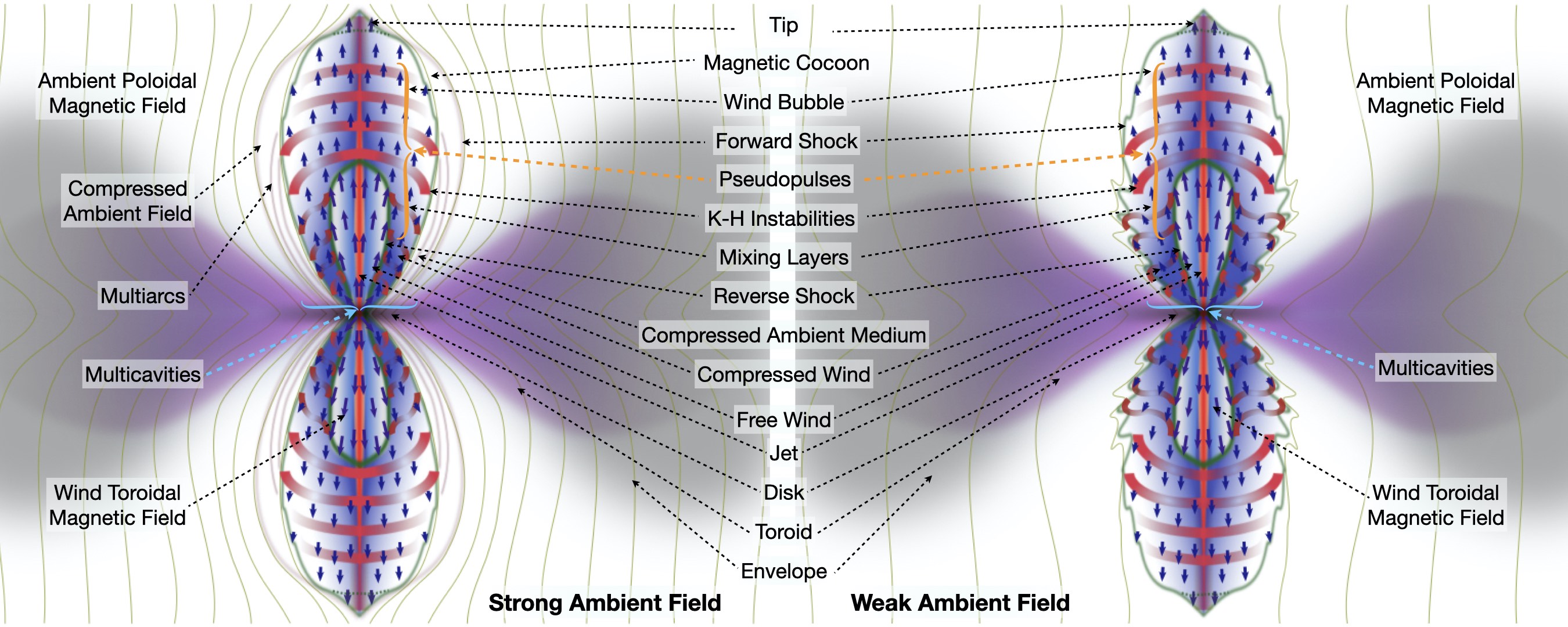}
\caption{Schematic illustrations of the outflow internal structures resulting from the interplay of a magnetized wide-angle wind and a toroid threaded by a strong (left) and weak (right) magnetic field.}
\label{fig:1}
\end{figure*}

\subsubsection{Shapes of Momentum-Conserving Thin Shells}
\label{subsubsec:thinshells}

\citetalias{shu1991} builds upon the assumption that a thin momentum-conserving (MC) swept-up shell of radius $\rMC\equiv\rsh$ can form by the input momentum of the wind, moving out with shell velocity $\vMC\equiv\vsh=d\rMC/dt$ exactly radially along each local piece of the shell where no distribution of matter happens along the $\theta$-direction. This can be parameterized by an angle-dependent ambient density distribution function $Q(\theta)$ and a momentum input function by the wind $P(\theta)$, defined in units such that the rate of change in mass $\mathcal{M}$ and momentum $\mathcal{M}\vMC$ per steradian are
$\frac{d\mathcal{M}}{dt}=\frac{\aamb^2}{2\pi G} Q(\theta) \vMC$ and
$\frac{d}{dt}(\mathcal{M}\vMC)=\frac{\dot{\mathcal{M}}_\mathrm{w}\vw}{4\pi} P(\theta)$, where $\vw$ is a radial wind velocity, $\vMC$ is the radial velocity of the momentum-conserving shell model, $\vsh$ is any generic radial shell velocity (numerical, observational, or analytical).
The equations as in \citetalias{shu1991} represent a hydrodynamic angle-dependent MC thin shell.

In real systems, the shell can acquire mass from both sweeping up the ambient medium and from collecting wind material:
\begin{equation}
\label{eqn:mass_rates}
 \begin{split}
  \frac{d\mathcal{M}}{dt} &=\frac{a^2}{2\pi G} Q(\theta) \vMC  + \frac{\dot{\mathcal{M}}_\mathrm{w}}{4\pi} P(\theta) \\
 &= r^2 \rhoa \vMC + r^2 \rhow (\vw -\vMC)\ .
 \end{split}
\end{equation}
The shell velocity
$\vMC$ is not strongly dependent on time, so the rate of momentum per steradian can be approximated as $r^2\rhow\vw (\vw-\vMC)=\frac{d\mathcal{M} \vMC}{dt}$ to $\approx \vMC\frac{d\mathcal{M}}{dt}$, which leads to the formula $r^2\rhow\vw (\vw-\vMC) \approx \vMC [r^2 \rhoa \vMC + r^2 \rhow (\vw - \vMC) ]$.
Collecting the terms in $\rhow$ on the left-hand side, and those in $\rhoa$ on the right-hand side, and canceling the radial factors allows us to rewrite the formula as
\begin{equation}
\label{eqn:shape:ram}
    \rhow (\vw-\vMC)^2 = \rhoa \vMC^2\ ,
\end{equation}
which can be used to find the shape of an MC thin shell in the hydrodynamic limit. 

The two sides of
Equation\ (\ref{eqn:shape:ram}) can be interpreted as pressure balance, respectively, representing the inner and outer ram pressures in a local frame comoving with the radial shell velocity  $\vsh=\vMC(\theta)$, with the hydrodynamic momentum-conserving value
\begin{equation}
\label{eqn:mc_eqn}
  \begin{split}
   \vMC(\theta)=\dot{r}\MC(\theta) &= {v_0}{\left[{1+\left(\frac{\rhow}{\rhoa}\right)^{-1/2}}
   \right]^{-1}}\ .
   \end{split}
\end{equation}
The consideration from Equations (\ref{eqn:mass_rates})--(\ref{eqn:shape:ram})
links the 
\citetalias{shu1991} MC formulation
to the expression adopted in \citet[][]{KM1992b} for a spherical bubble driven by a constant-velocity wind, at each individual angle $\theta$. This forms the basis of the integration of an elongated angle-dependent bubble inside the MC shell as realized in \citetalias{shang2020}.
The thin shells of the hydrodynamic MC model presented here
represent the shape of the outer contour of an outflow lobe. Outflow shapes of magnetized cases are examined in Section \ref{subsubsec:MHDshells} below.

\subsubsection{Shapes of Magnetized Outflow Bubbles}
\label{subsubsec:MHDshells}

Magnetized outflow bubbles can be explored using the methods of the previous subsection, once more with the objective of representing the shape of the outer contour of the outflow lobe through one-dimensional curves, analog to Equation (\ref{eqn:mc_eqn}), but now for a magnetized case.

We adopt here the notation $\vMOS$ instead of the specific $\vMC$ for generality.
For magnetized winds, the wind magnetic pressure $b_\phi^2/2$ is added to the inner ram pressure, modifying Equation (\ref{eqn:shape:ram}) in a way similar to \citet{draine1983}, and resulting in equation
\begin{equation}
\label{eqn:shape:vs_windbphi}
    \rhow (\vw-\vMOS)^2 + b_\phi^2/2= \rhoa \vMOS^2,
\end{equation}
although the ram pressures are not isotropic (nor do the magnetized winds always lead to thin shells).
With the presence of a poloidal magnetic field, a poloidal magnetic pressure $b_p^2/2$ term can be added to the right-hand side of Equation (\ref{eqn:shape:vs_windbphi}), as in \citet{konigl1982}:
\begin{equation}
\label{eqn:shape:vs_bp}
\rhow (\vw-\vMOS)^2 + b_\phi^2/2= \rhoa \vMOS^2
+ b_p^2/2.
\end{equation}

On the other hand, the poloidal magnetic field is strongly amplified in a region of compressed ambient field $b_p$ (Figure\ \ref{fig:1}, and also \citealt{konigl1982}), and that needs to be taken into account through an efficiency factor $e_p(\theta)$. An efficiency factor $\beta(\theta)$ is multiplied to the $\vw$ term, in order to treat velocity behaviors inside the magnetized cocoon
deviating from the strict thin-shell approximation, geometry effects, and shell thickness. Combining these terms and factors, the pressure balance equation can be written as
\begin{equation}
\label{eqn:shape:vs_windbphibp}
    \rhow (\beta\vw-\vMOS)^2 + b_\phi^2/2= \rhoa \vMOS^2 + e_p b_p^2/2\ .
\end{equation}
Solutions of this quadratic Equation (\ref{eqn:shape:vs_windbphibp}) allow estimates and fits of the shape function $\rMOS=\vMOS t$.

Inside the magnetized cocoon enclosing the outflow bubble, the magnetic field $b_p$ is amplified with respect to its initial value in the toroids due to the exclusion of poloidal field from the wind region. This amplification can be estimated by conservation of magnetic flux and considering cross-sectional areas characterizing features of the magnetized compression, and it is the main effect producing $e_p$. 
The balance of the total ram pressure with that of the poloidal field seems to be a good predictor of the outermost edge of the $b_p$-compressed ambient material for $n=1$ and $n=2$ cases of $\alpha_b=1$, and all $n$'s for $\alpha_b=0.1$. The updated shape curves are illustrated in Figure \ref{fig:MC_MA_i}.

Most of our models show strong self-similarity as demonstrated phenomenologically in \citetalias{shang2020}, and they follow the principle of momentum conservation and pressure balance. 
The updated formula (Equation (\ref{eqn:shape:vs_windbphibp})) using $\beta(\theta)$ and $e_p$ fits is used for tracing outflow shapes and production of the analytic PV-space curves for the magnetized outflow shape curves in Section \ref{subsec:MHD_Wind_curves}.
Details of the fits are presented in Appendix \ref{sec:appendix:ep_beta}.

\subsubsection{Internal Structures of Nonspherical Bubbles}
\label{subsubsec:hydromagnetic_bubbles}

In the context of a wind-blown bubble, \citet{KM1992a} considered the dynamics of a bubble blown by a constant-velocity wind, and their evolution in a uniform medium. \citet{KM1992b} generalized the formulation to a power-law density distribution $\rho_a(r)\propto r^{-k_\rho}$ driven by a constant-velocity wind, and the evolution of hydrodynamical momentum-conserving bubbles as thin shells in which both the shocked wind and ambient medium are radiative. A special case of $r^{-2}$ dependence of the ambient density profile, such as that in \citetalias{shu1991} for a singular isothermal sphere, is discussed in their Appendix A, in which the bubble as a thin shell expands with a constant velocity, as in Section \ref{subsubsec:thinshells}.
 
The outflow shapes in the hydrodynamic limit result from wind and ambient (toroid) density profiles as obtained in \citetalias{shang2020}. They are elongated (nearly momentum-conserving) bubbles driven by a wind in interaction with a toroid, both of them angle-dependent.
For a wind constant in time at each angle, self-similarity is achieved (see Appendix A and Equation [2.10] of \citealt{KM1992b}) when the system evolves beyond the influence of initial injection, as shown in Figures 10 and 14 in \citetalias{shang2020}.
Within a hydrodynamic bubble, a reverse (wind) shock (RS), a contact discontinuity (CD), and a forward (ambient) shock (FS) form from inside out when steady injection of mass and energy into the ambient medium occurs. The CD separates the shocked wind and shocked ambient medium in a 1D sphere, and the two shock fronts separate the physical space into four respective regions of distinct physical properties. This basic 1D spherical bubble structure is best illustrated in the grid of models in Appendix A of \citetalias{shang2020} with the relevant equations of state (two-temperature, and $\gamma=1.0001$).
In the situation where the shocks are oblique, shown in Appendix C of \citetalias{shang2020}, the contact discontinuity becomes a TD, and significant shear can develop and build up across the shocks.

\citetalias{shang2020} reveals the 2D structures of highly elongated, nonspherical hydromagnetic bubbles through the exploration of a large parameter space. Instead of the simple 1D RS--CD--FS structures present in the spherical bubble, the hydrodynamic nonspherical bubble formed with substantial shear across the now very oblique RS--FS surfaces along the TD\@.

``Multicavities'' form associated with the RS--FS in the nonspherical magnetized bubbles and are evidently delineated by the respective shock boundaries. The primary wind is confined by the innermost RS cavity near the base around the jet in the strongly magnetized wind. 
Complex structures are shown to develop inside the outflow lobes, contrary to the hydrodynamical thin-shell models. 

The interplay allows fairly complex structures to originate from simple setups. These interfaces and their crossing are demonstrated in the variations of the $\Cfunc$-function, wind fraction $f$, vorticity $\omv$, and the $\theta$-component of the velocity in \citetalias{shang2020}. The correlations with $f$ reveal the extended mixing regions confined by the RS on the wind side and the FS on the ambient side. 
Figures \ref{fig:2} and \ref{fig:3} show a glimpse of the structures formed based on simulations from the $6400\times 2016$ set for the $100\kms$ wind. The morphology of these hydromagnetic bubbles appears very elongated and fairly collimated, and shows some ``spindle''-like appearance. They appear to be ``prolate''-shaped.

In the following sections,
we demonstrate the conceptual development of forming hydrodynamic RS, CD, and FS layers from a nonspherical prolate-shaped bubble. We construct the structures of the hydrodynamic bubbles analytically by locally ``plane-parallel'' shocks in 1D and 2D, and implement laminar layers of shear. The concept is then generalized to physically thick and locally oblique structures for the magnetized mixing layers. We construct analytic properties in position--velocity space at incremental stages to form the baselines of signatures. These analytic models and kinematic signatures will be studied and juxtaposed together with those extracted from the MHD simulations. The traces of the underlying processes with their respective signatures can be thus identified.

\addtocounter{figure}{1}
\begin{figure*}
\figurenum{\arabic{figure}}
\centering
\epsscale{0.95}
\plotone{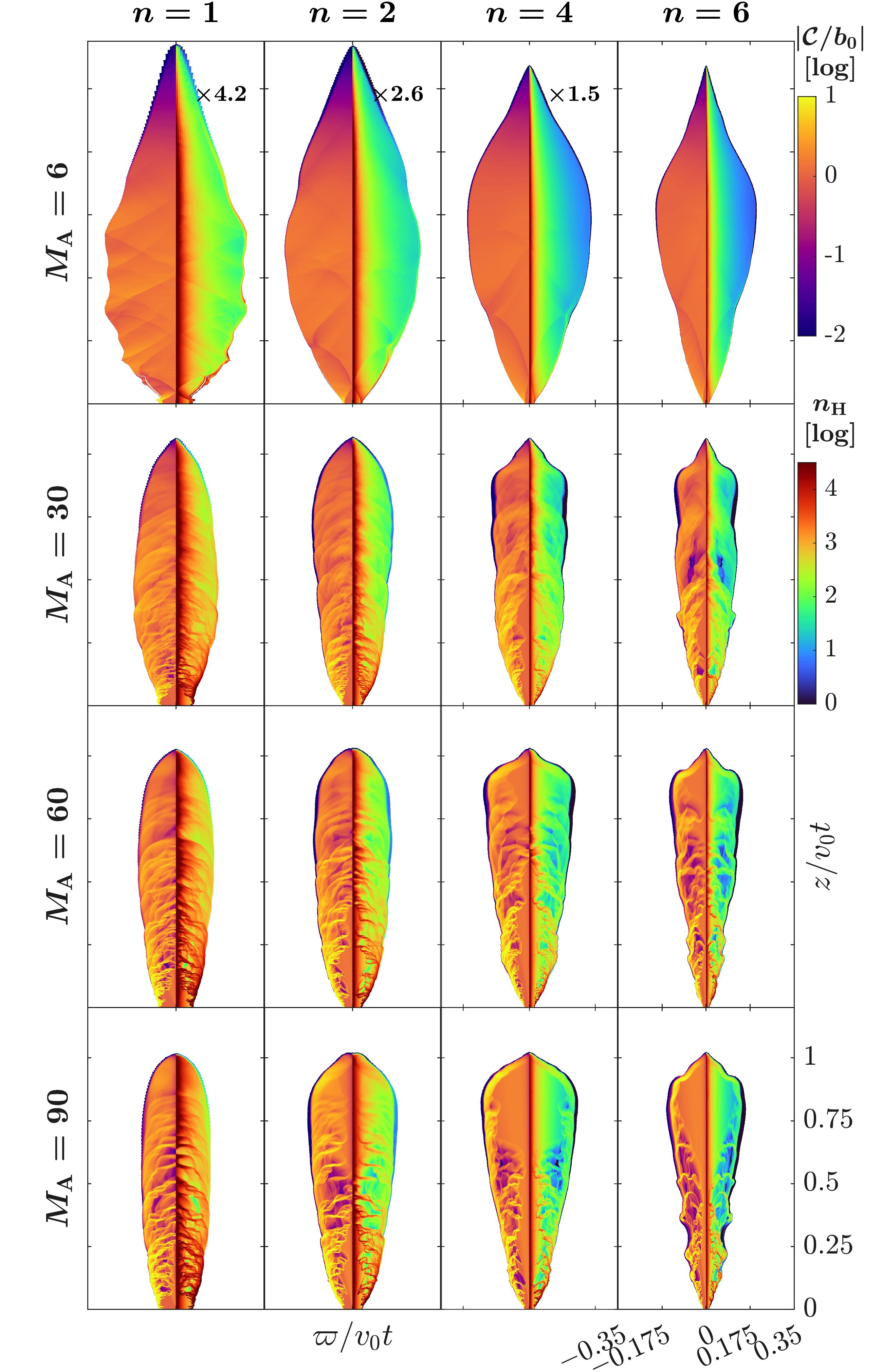}
\epsscale{1.0}
\caption{
(a) For $\alpha_b=1$ cases, the absolute values of $\Cfunc$ function, normalized to $b_0$, are shown on the left halves and number density $n_\mathrm{H}$, rescaled and spanning 4.5 dex, on the right halves in individual panels. Both quantities are shown in logarithmic scales and filtered with a threshold of wind mass fraction $f>10^{-4}$. The $\varpi$-axis has been exaggerated to show detailed structures. The exaggeration factors are $4.7$, $2.7$, $1.5$, and $1.0$ for $n=1$, $2$, $4$, and $6$, respectively.}
\label{fig:2}
\end{figure*}
\begin{figure*}
\figurenum{\arabic{figure}}
\centering
\epsscale{0.95}
\plotone{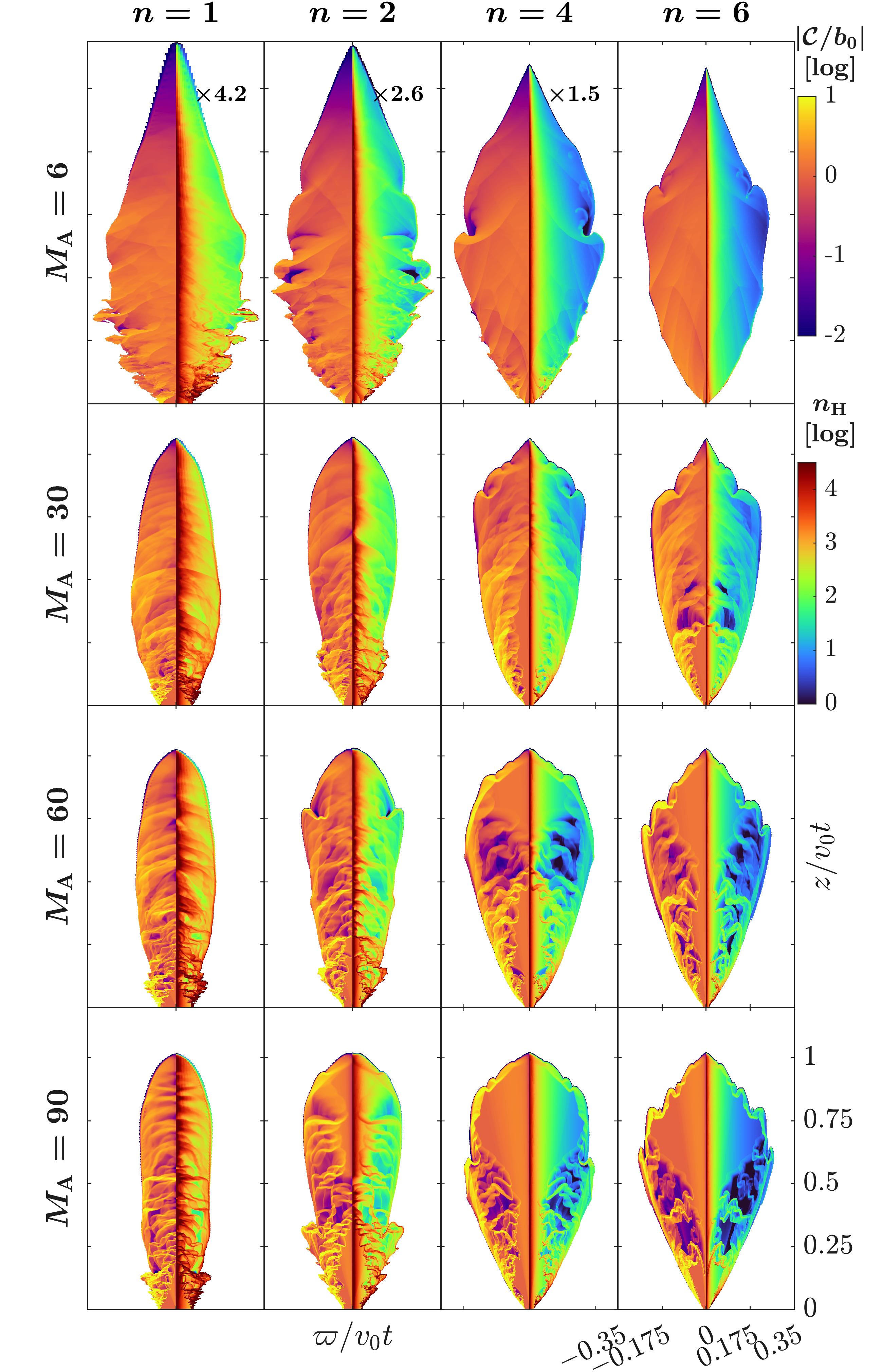}
\epsscale{1.0}
\caption{
(b) Same quantities ($|\Cfunc|$ and $n_\mathrm{H}$) as in Figure \ref{fig:2} (a), but for $\alpha_b=0.1$ cases. The $\varpi$-axis has been exaggerated to show detailed structures. The exaggeration factors are $4.2$, $2.6$, $1.5$, and $1.0$ for $n=1$, $2$, $4$, and $6$, respectively.}
\end{figure*}
\begin{figure*}
\figurenum{\arabic{figure}}
\centering
\epsscale{0.95}
\plotone{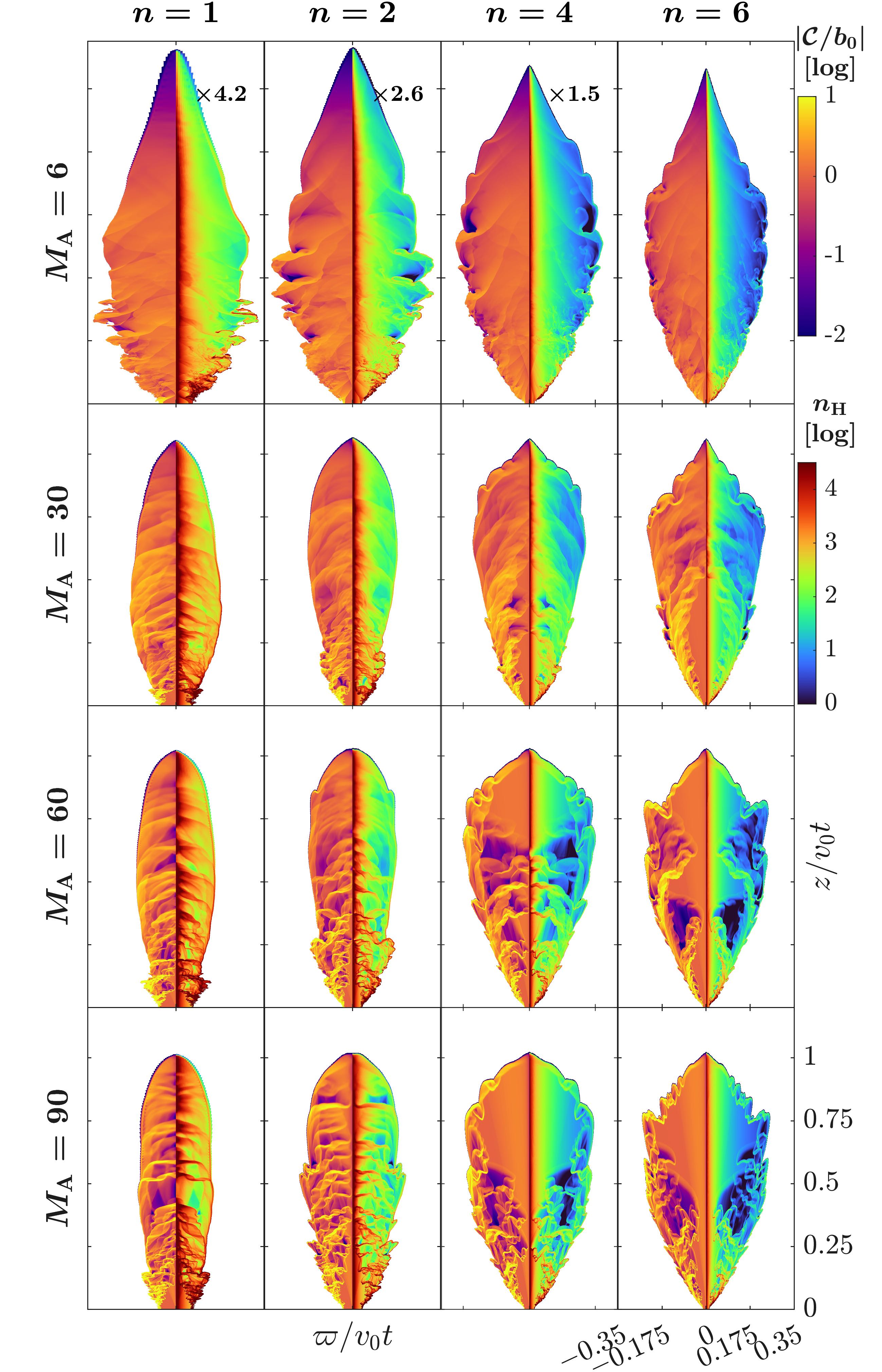}
\epsscale{1.0}
\caption{
(c) Same quantities ($|\Cfunc|$ and $n_\mathrm{H}$) as in Figure \ref{fig:2} (a), but for $\alpha_b=0$ cases. The $\varpi$-axis has been exaggerated to show detailed structures. The exaggeration factors are $4.2$, $2.6$, $1.5$, and $1.0$ for $n=1$, $2$, $4$, and $6$, respectively.}
\end{figure*}

\addtocounter{figure}{1}
\begin{figure*}
\figurenum{\arabic{figure}}
\centering
\epsscale{0.95}
\plotone{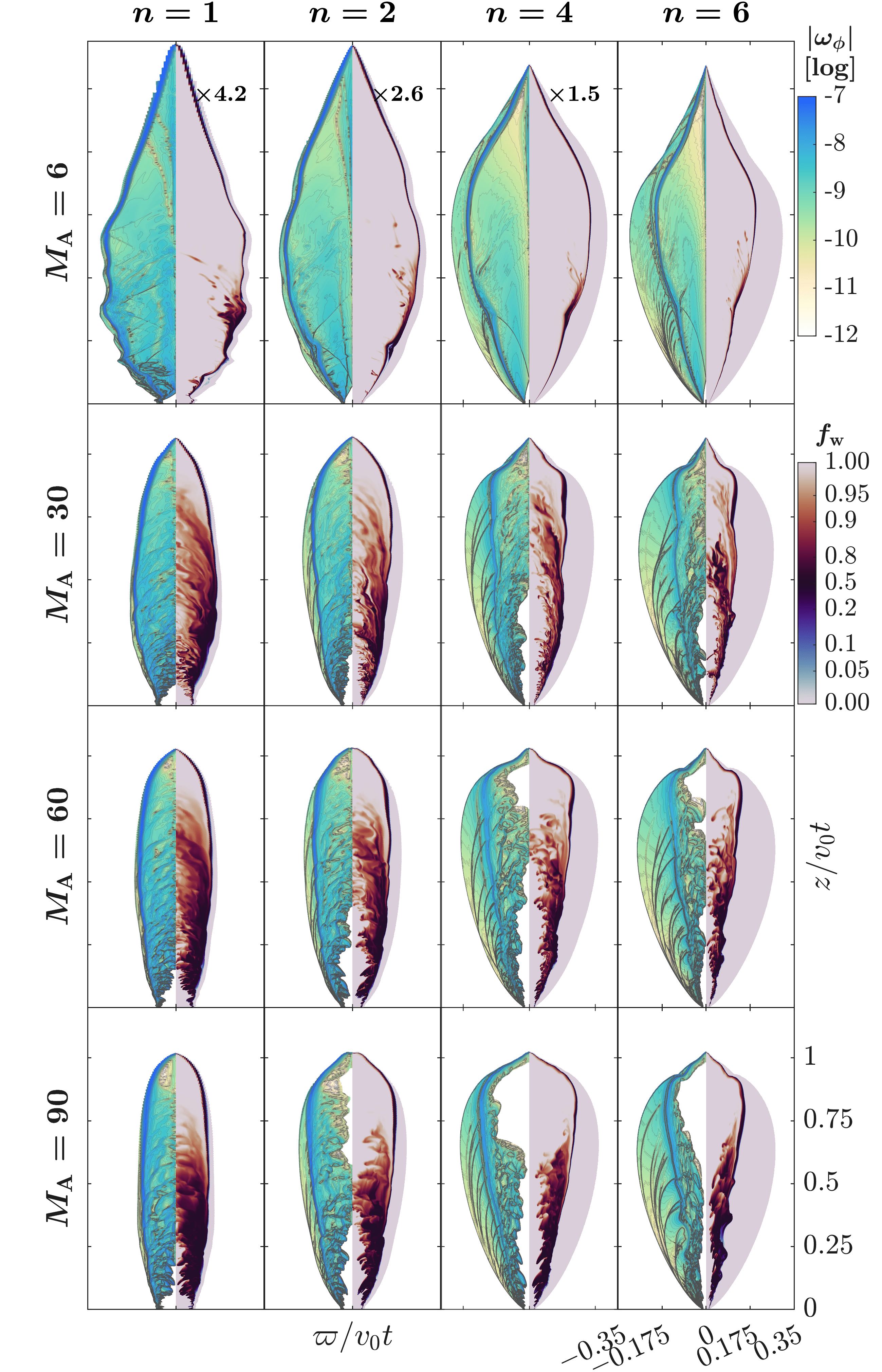}
\epsscale{1.0}
\caption{
(a) For $\alpha_b=1$ cases, the azimuthal vorticity $|\omega_\phi|$ in logarithmic scales is shown on the left halves and wind mass fraction $f$ on the right halves of individual panels. The $\varpi$-axis has been exaggerated to show detailed structures. The exaggeration factors are $4.2$, $2.6$, $1.5$, and $1.0$ for $n=1$, $2$, $4$, and $6$, respectively.}
\label{fig:3}
\end{figure*}
\begin{figure*}
\figurenum{\arabic{figure}}
\centering
\epsscale{0.95}
\plotone{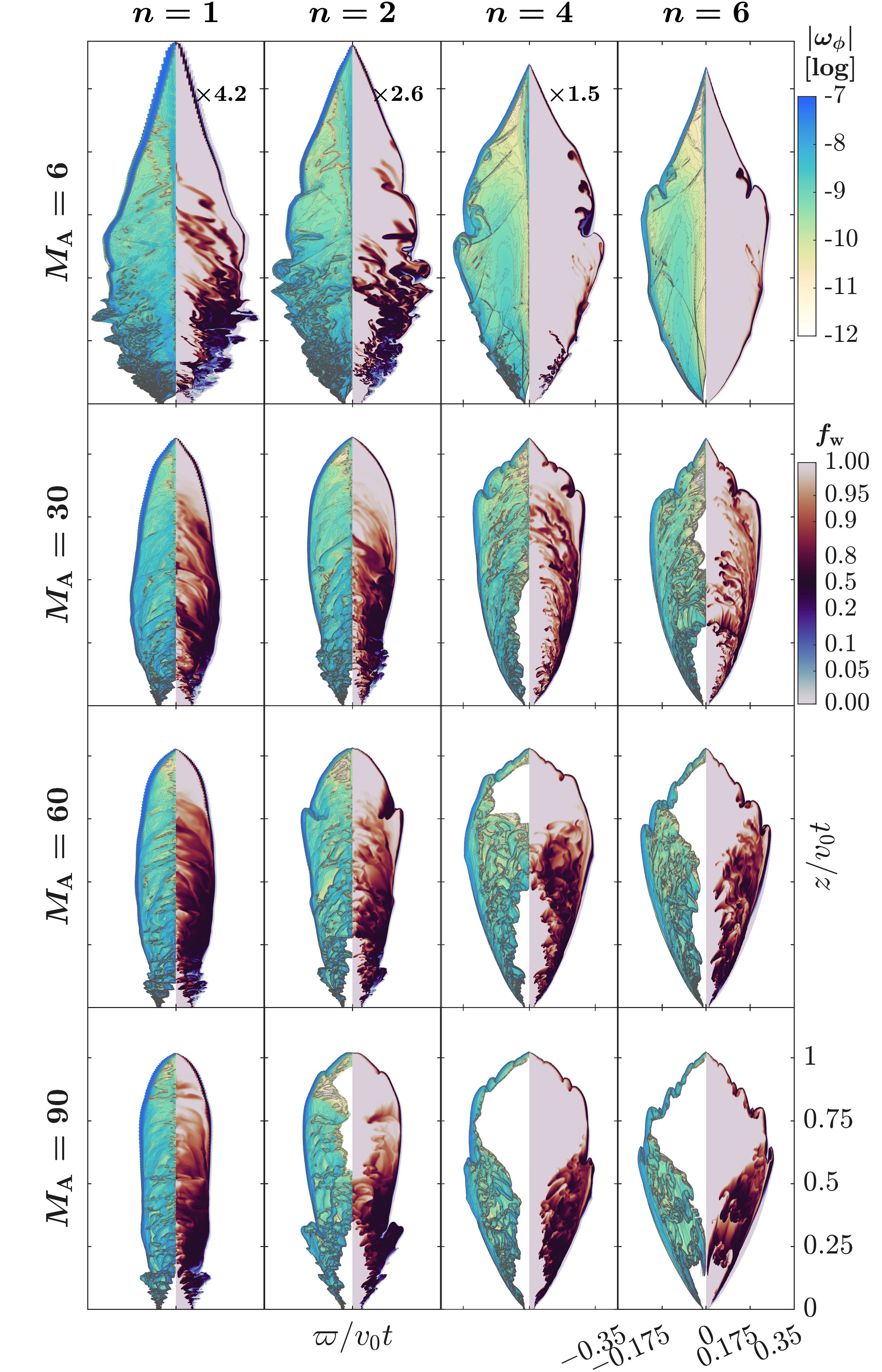}
\epsscale{1.0}
\caption{
(b) Same quantities ($|\omega_\phi|$ and $f$) as in Figure \ref{fig:3} (a), but for $\alpha_b=0.1$ cases. The $\varpi$-axis has been exaggerated to show detailed structures. The exaggeration factors are $4.2$, $2.6$, $1.5$, and $1.0$ for $n=1$, $2$, $4$, and $6$, respectively.}
\end{figure*}
\begin{figure*}
\figurenum{\arabic{figure}}
\centering
\epsscale{0.95}
\plotone{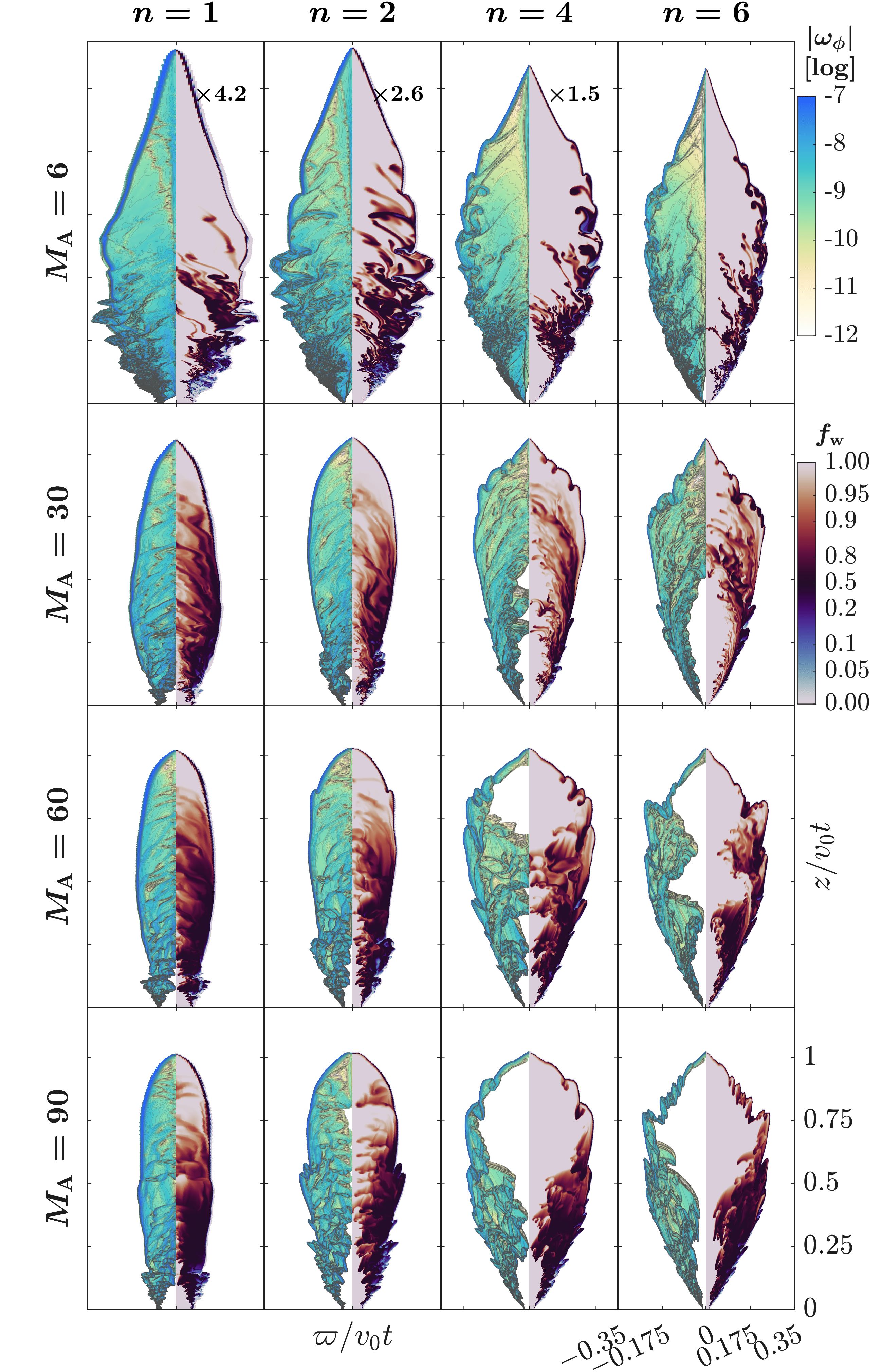}
\epsscale{1.0}
\caption{
(c) Same quantities ($|\omega_\phi|$ and $f$) as in Figure \ref{fig:3} (a), but for $\alpha_b=0$ cases. The $\varpi$-axis has been exaggerated to show detailed structures. The exaggeration factors are $4.2$, $2.6$, $1.5$, and $1.0$ for $n=1$, $2$, $4$, and $6$, respectively.}
\end{figure*}

\section{Numerical Setups}
\label{sec:setup}

\subsection{Simulations}
\label{subsec:simulations}

The setups of numerical simulations are described in detail in \citetalias{shang2020}.  We summarize key ingredients here for the numerical data utilized in this work, and their associated numerical values can be found in Table \ref{tab:quantities}. 

The parameter space originally explored in \citetalias{shang2020} is huge.
The wind is parameterized with seven $\MA=6$, $30$, $60$, $90$, $180$, $600$, and $\infty$ values, and two velocities of $v_0=50$ and $100\kms$ (a total of 14 wind combinations explored).
Initial ambient medium is parameterized with $\alpha_\rho=1$, four toroid configurations $n=1$, $2$, $4$, and $6$, and three levels of ambient toroid magnetization $\alpha_b$. Adding the four no-toroid configurations with $\alpha_\rho=0$ but without $n$ and $\alphab$, a total of $4\times3+1=13$ ambient configurations were explored.
For each of these $14\times13$ parameter combinations, two numerical resolutions of $3200\times1008$ and $6400\times2016$ are available in the archive.

We focus on the $v_0=100\kms$ velocity and the higher-resolution $6400\times2016$ 
cases in this work. An additional set of hydrodynamic ($\MA=\infty$, $\alphab=0$) cases is performed for $v_0=100\kms$ at $6400\times2016$ resolution, with $v_\theta=0$ to imitate an equivalent 1D bubble in $r$-coordinate extended to 2D at every $\theta$ by independent variation of $\theta$. This set serves as our reference for smooth $v_\theta=0$ hydrodynamic models without the onset of the KHI\@.

\subsection{\textbf{\textsl{Synline}} Data Post-processing}
\label{subsec:method_synthetic_images}

The numerical data are post-processed through the tool,
\textbf{\textsl{Synline}}, developed in our data pipeline locally by the team. \textbf{\textsl{Synline}}
is a package generating synthetic observables from two-dimensional (2D) axisymmetric simulation data. The 2D data blocks in desired properties such as the number density and velocity are first rotated about the symmetry axis, interpolated and mapped onto a new three-dimensional (3D) Cartesian uniform grid. Inclination of the system to the plane of the sky is incorporated during the construction of the new 3D data cubes. The synthetic observables are produced by performing line-of-sight integration of the emissivities of local cells at the optically thin limit. 

In the \Synline{} package, one has the ability to calculate the emissivity by a selection of line transition and its associated level populations with constants out of a database included in the package, under the non-LTE assumption with temperature as an input field.
This final module of \Synline{} is not used in our work, as we generate only column density maps without calling the emissivity routines. The reason for this is that our simulations do not provide a point-to-point map of the local temperature, which would be necessary to estimate the emissivity.

Column density maps are generated without calling the emissivity routines by directly summing up the number densities along the line of sight with the proper inclination angles. A position--position--velocity (PPV) data cube in column density is generated by binning the local line-of-sight velocities at an (adjustable) interval of $0.06\kms$. A thermal line profile is applied according to the local sound speed $a^2 = f \aw^2 + (1-f) \aamb^2$ for the $\sim 100\K$ wind and the $\sim10\K$ toroid in our setup \citepalias[see][]{shang2020}. Subsequent production of the PV diagrams of column density (PVDCD) follows by manipulating the data cubes parallel and perpendicular to the outflow axis in the projected plane of sky.

\textsl{We stress once again that the resulting diagrams that illustrate the column densities mapped in the PV space are not directly analogous to the PV diagrams as resulting from observations, which are based on surface brightness of emission lines. The line excitation depends strongly on the temperature, and the resulting surface brightness depends on the emission line selected, and on its appropriate radiative transport: for forbidden lines, it results just as a sum of the emissivities of the volume elements along the line of sight, while for permitted lines, the full radiative transport has to be taken into account, as the emission is mediated by absorption and re-emission at Doppler shifts corresponding to the local radial velocity.}

The self-similar nature of the evolution as demonstrated in \citetalias{shang2020} allows a self-similar presentation of the kinematic features. The PVDCDs are shown in the self-similar units normalized to $v_0t$ and $v_0$, respectively, at a representative epoch of $100\yr$. The spatial position is convolved with a Gaussian profile of $0.01v_0t$ for a smoother appearance. Different criteria can be applied prior to the density integration to extract different physical conditions from the data cubes. The range of the wind mass fraction $0<f<1$ is used to identify the contribution from the compressed wind and compressed ambient material confined by the RS and FS\@. The criterion, $v_p/\aamb > 1$, is implemented as the outer boundary across the FS\@.

\section{Structures of 2D Elongated Hydrodynamic Bubbles with Locally Oblique Shocks}
\label{sec:2Dinterior}

We establish in this section the basic structures of the 2D momentum-conserving hydrodynamic bubbles using semianalytic procedures. The procedures are novel in the construction of analytical expressions for the shell thickness across the RS, CD, and FS, the extension to 2D with locally oblique shocks, and development of a shear profile within the post-shocked regions. This whole set of procedures gives the reference thin-shell models of 2D hydrodynamic nonspherical bubbles with desired shear profiles without the turbulence induced by the KHI nor the thick post-shocked regions. The resulting kinematic signatures will serve as a baseline for detecting those generated by the KHI and the thick extended magnetized shocked-wind region. 

\paragraph{The obliqueness angle}
We define the obliqueness $\chi$ locally in terms of the angle that the radial wind makes to the normal to the  shape curve $\rsh(\theta)$:
\begin{equation}
\label{eqn:chi:arcsin}
\chi=\arcsin(\vhat{n}\cdot\vhat{\theta})=\arcsin(-\rsh^\prime / {(\rsh^{\prime 2}+\rsh^2)^{1/2}}),
\end{equation}
defined so that minimum obliqueness (such as in a spherical shape curve) corresponds to $\chi=0\arcdeg$, and maximum obliqueness $\chi=90\arcdeg$ corresponds to a locally radial tangent to the shape curve. 
In our most usual case, both in simulations and in analytical models, the obliqueness $\chi$ is in the first quadrant, allowing us to write it as $\arccos(\vhat{n}\cdot\vhat{r})$. Occasionally in simulations there are a few cases where the normal is closer to the axis than the radial line, moving that $\chi<0$ into the fourth quadrant at that place. The fact that typically $\chi>0$ implies that the equatorial radius of these outflows is narrower than the axial radius, giving them an overall elongated shape.

\subsection{Construction of Smooth Hydrodynamic Models}
\label{subsec:HDsmooth}

We lay out the conceptual development in four progressive steps to demonstrate how the structures are constructed. For completeness and for reading as a unit, the entire package of the procedure is described in more details in Appendix \ref{sec:appendix:steps}. Here we outline the four steps in sequence and their respective methodology.

We begin with the time-evolving position of the MC shape, which traces a curve in space for each of the $n=1$, $2$, $4$, and $6$ nonmagnetized toroids, with an input wind of $\MA=\infty$. Procedures to generate these curves are given in Section \ref{sec:MCcurves} and Appendix \ref{sec:appendix:steps}, and curves are shown in the upper-left panel in Figure 22 of \citetalias{shang2020}. This is referred to as Step 1 in Appendix \ref{sec:appendix:steps}, and in Equation \ref{eqn:mc_eqn} of this work.

In Step 2, the locations of the reverse and forward shocks are constructed into the 1D flow solutions with the proper obliqueness in the 2D flow without the complexity encountered in the real numerical simulations. We use these solutions as the clean and smooth background references.
The local obliqueness $\chi(\theta)$ at the position $\rsh(\theta)$ of the momentum-conserving curves found in Step 1 can be incorporated into the construction process.

\begin{figure*}
\centering
\plotone{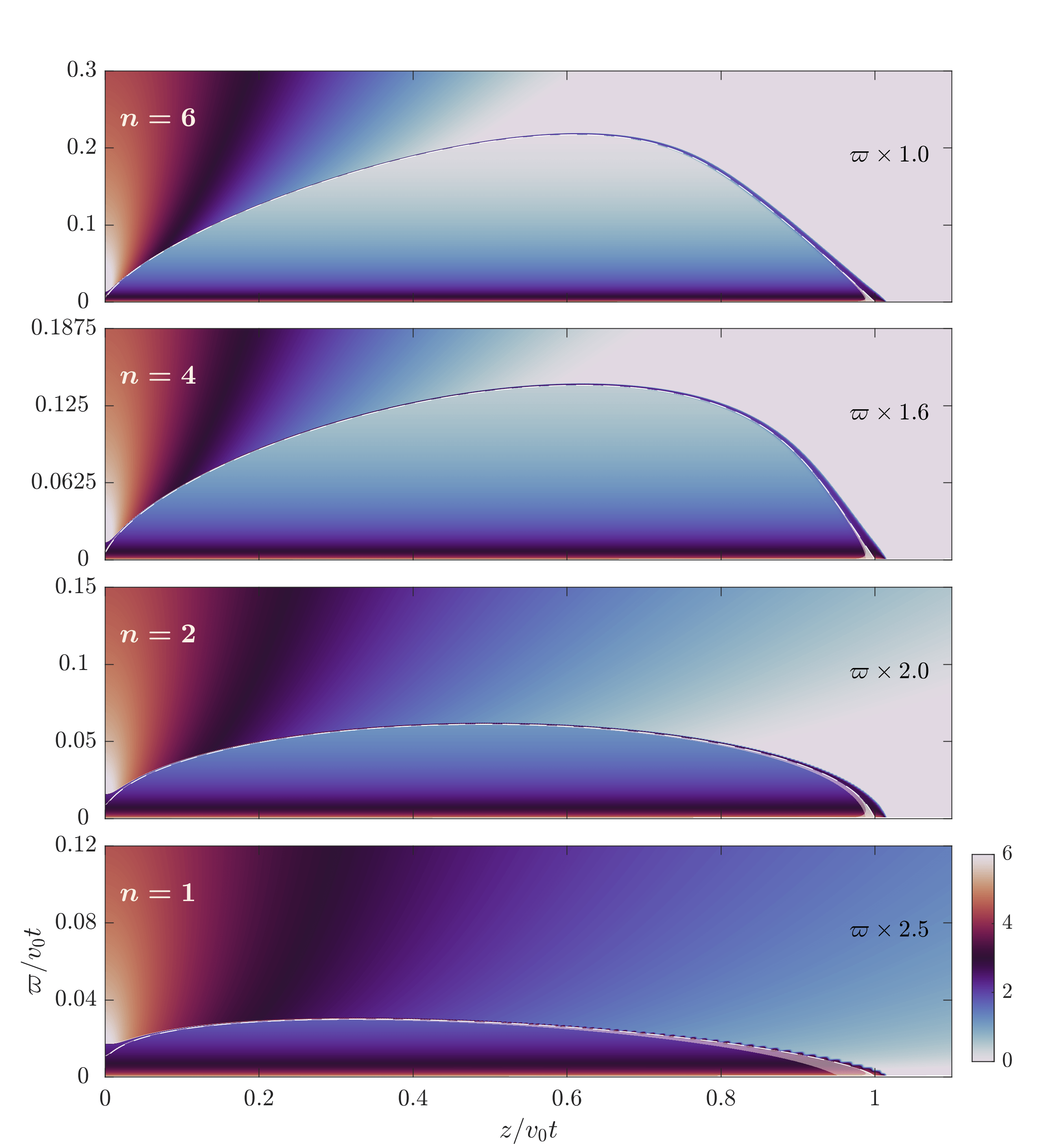}
\caption{MC curves and their respective shock positions overlaid on top of number density $n_\mathrm{H}$ distributions of the no-$v_\theta$ ($v_\theta=0$) numerical smooth reference calculations with $n=1$, $2$, $4$, and $6$ toroids (bottom to top). The rescaled logarithmic number density spans 6 orders of magnitude. The white dashed lines are the momentum-conserving curves obtained with Equation (\ref{eqn:mc_eqn}). The inter-shock regions of the model of Appendix \ref{sec:appendix:steps} are shaded. For better viewing, the $\varpi$-direction has been magnified for different $n$ values.}
    \label{fig:4}
\end{figure*}

A hydrodynamic 1D Riemann problem can be defined, having its left-side state with the density, projected velocity and sound speed of the wind, and its right-side state the with ambient density and sound speeds, with a velocity of zero.
When solving this Riemann problem, an EOS with $\gamma=1.0001$ is adopted for simplicity and for its close resemblance to our two-temperature equation of state, as shown in Appendix A of \citetalias{shang2020}.
The solution to the 1D Riemann problem gives the three discontinuities (or signal speeds) and is adopted to construct the locations of the RS, CD, and FS\@. The respective shock surfaces are approximated locally as two plane-parallel waves to allow for such choice of methodology. Depending on how the propagation of the shocks is calculated, three models of thickness can follow with various levels of fitting and projection of vectors (the normal, radial, and hybrid models; see Step 2 of Appendix \ref{sec:appendix:steps}).
The shaded areas in Figure \ref{fig:4} show the compressed (shocked) regions between the positions of RS and FS obtained by the Riemann problem, on top of the ``no-$v_\theta$'' ($v_\theta=0$) numerical smooth reference calculation.

In Steps 3 and 4, we utilize the analytic formulation of the jump conditions and behaviors of the parallel oblique shocks, obtained previously from \citetalias{shang2020}, and construct analytically a new family of shear profiles in Step 4. Step 3 requires detailed construction of shock velocities in different frames of reference, and in their respective normal and tangential directions. Therefore, we build upon the understanding of parallel oblique shocks and their respective jump conditions at the RS and FS, as shown in Appendix C of \citetalias{shang2020}. The detailed transformations among the frames of reference in RS and FS, pre-shock, post-shock and fixed frames, velocities normal or tangential to the shock surfaces, are constructed and implemented in Step 3 of Appendix \ref{sec:appendix:steps}.

In Step 4, an analytical expression further completes
the local velocity due to shear for an arbitrary point residing within the post-shock region bounded by the RS and FS\@. This step extends the knowledge on the total shear across the plane-parallel oblique shocks derived in Appendix C of \citetalias{shang2020}.
We construct the shear profiles for any point between a pair of locally parallel oblique shocks, once the end points on the RS and the FS can be identified. Figure \ref{fig:shear} shows the shear profiles for the shocked region in self-similar coordinates.
The patterns across the RS of the deflected flow directions from the original incoming radial wind are evidently demonstrated.

\subsection{Obliqueness of Thin Hydrodynamic Shells}
\label{subsec:chi_limit}

The obliqueness angle $\chi$ has positive values (in the first quadrant) in our usual case in which $\vhat{n}$ is further away from the axis than $\vhat{r}$, and negative values in the opposite case, which can happen only if $\rsh^\prime>0$ (not in our set of momentum-conserving examples). Hence, it is safe to define the shape curve S as that of the momentum-conserving model, with positions $\rsh=r\MC$ and normal vectors $\vhat{n}=\vhat{n}\MC$ as in Steps 1 and 2 of Appendix \ref{sec:appendix:steps}. The limiting behavior of $\chi$ near the axis (small values of the coordinate angle $\theta$), near the tip of the outflow, is a peculiar property of the detailed hydrodynamic momentum-conserving models.

For the obliqueness at the tip, we show in Appendix \ref{sec:appendix:chi_limit} the derivation of the small $\theta$ behavior with the density ratio $\delta=\rhow/\rhoa$. Using Equation\ (\ref{eqn:chi:arcsin}), one can obtain the obliqueness for small angles in Equations (\ref{eqn:chi_p:orig})--(\ref{eqn:chi:small_theta_zerok:orig}), and apply them to both the untapered and tapered toroids with the background scaled density $D^S$ (see Equation (\ref{eqn:rhoa})).

For the tapered toroids, as we have adopted primarily throughout \citetalias{shang2020} and this work, in which $D^{S}>0$, one reaches a $\sin\chi$ that is independent of $n$. For our usual parameter values as shown in Table 1 of \citetalias{shang2020}, this leads to a modest value $\sim 47\arcdeg$ with our usual winds and setup. Such behaviors can be observed in Figure \ref{fig:9}, and in the limiting values of $\chi$ in Figure 22 of \citetalias{shang2020}.
The numerical values of $\chi$ indeed approach the same $\approx 47\arcdeg$ for the constructed MC curves, and numerical calculations without $v_\theta$ ($v_\theta=0$) and with regular, unrestricted $v_\theta$ simulations. A scan of the $n$-independent $\chi$ values for a broad range of $D^S$ can be seen in Figure \ref{fig:chi_DScomparisons}.
Figure \ref{fig:MC_DScomparisons} shows a collection of detailed resultant MC curves of hydrodynamic bubble shapes associated with the varying $D^S$ and their obliqueness in the zoomed-in tips.
As $D^S$ decreases, the medium is less dense and resists the wind less. For $n>0$ and $D^S=0$, the opening region of the toroid is so light that it allows the shell velocity $v\MC$ to approach its maximum value $v_0t=r\MC/t$. The zoomed-in tips show this locally flat ($\chi=0$) limiting shape for small $\theta$ values, within the toroid opening for each $n$ case. As $D^S$ increases, the shell velocity decreases, the shapes become narrower, and acquire a pointy tip ($\chi>0$) at the axis $\theta\rightarrow0$ as shown in Figure \ref{fig:chi_DScomparisons}. The $n=0$ toroid is different because it has no opening, even at $\theta=0$, leading to a shape that is always narrow with a pointy tip.
We note that the match of behaviors of $\chi$ at small (including $\theta\rightarrow0$) angles with the opening angles at the bottom indeed decides the overall shapes of elongated bubbles.

\section{Internal Structures of the Elongated Nonspherical Hydromagnetic Bubbles}
\label{sec:struc_hydromagnetic_bubbles}

\subsection{Multicavities Revealed}
\label{subsec:multicavities}

We reviewed the conceptual development behind the ``multicavities'' in Section \ref{subsubsec:hydromagnetic_bubbles} of the physical framework.
The primary wind and the ambient nonperturbed toroids are vorticity-free. The vorticity $\omv$ is generated behind the shocks and further enhanced by magnetic forces. The $\Cfunc$ function varies abruptly across the shocks and boundaries into the nonmagnetized regions, and $v_\theta$ becomes nonzero and varies in the post-shocked regions.
Figures \ref{fig:2} and \ref{fig:3} illustrate the structures and the sense of ``multiple'' cavities out of the thick and extended mixing region and the magnetized ambient medium. An innermost cavity forms inside the loci of the reverse shock, in which the primary wind is unshocked and unperturbed. The outermost cavity forms with the magnetic cocoon by the surrounding ambient poloidal field and the compressed ambient medium exterior to the compressed field. Such layered structures could be interpreted as the layered shells from episodic ejections. The nonlinear growth and coalescence of the KHI modes further complicate the apparent presentation of the structures as a mixture of large and small density concentrations resembling multiple large and small shells. These structures are not expected of the thin-shell models. A magnetized outflow is an elongated bubble filled with internal structures, not an empty cavity.

\begin{figure*}
\epsscale{0.95}
\plotone{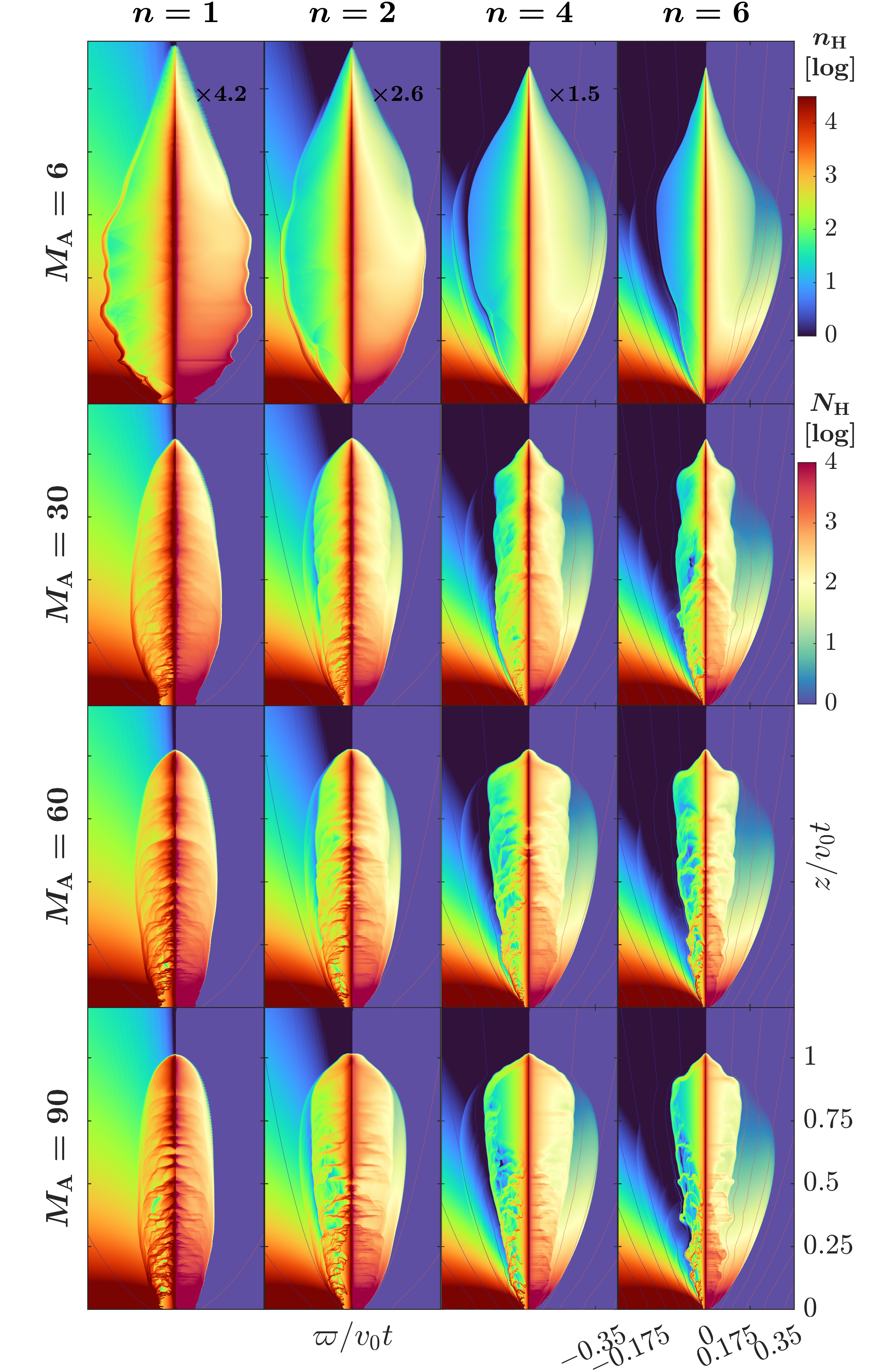}
\caption{
Distribution of the number density (left) and the column density integrated over regions with $v_p>\aamb$ (right) for magnetized ambient media with $\alpha_b=1$. The rescaled logarithmic number density has a range of 4.5 orders of magnitude, while the also rescaled logarithmic column density has a range of 4 orders of magnitude. The ambient $b_p$ field lines have been drawn on top, as in Figure 4 of \citetalias{shang2020}. The $\varpi$-direction of these panels is stretched to reveal more enlarged views of the detailed structures. The exaggeration factors are $4.2$, $2.6$, $1.5$, and $1.0$ for $n=1$, $2$, $4$, and $6$, respectively. 
}
\epsscale{1.0}
\label{fig:5}
\end{figure*}

\begin{figure*}
\epsscale{0.95}
\plotone{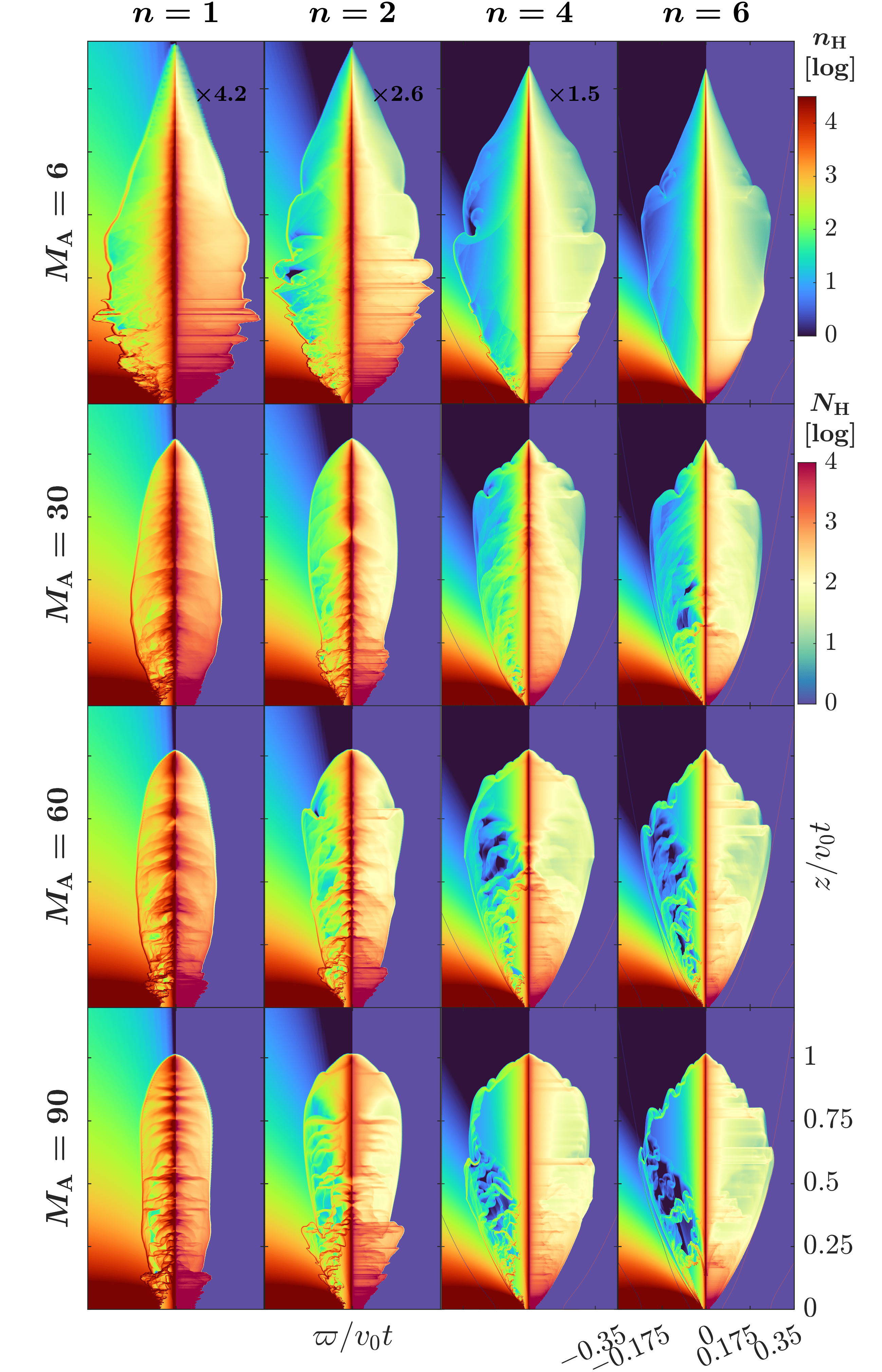}
\epsscale{1.0}
\caption{
Distribution of the number density (left) and the column density integrated over regions with $v_p>\aamb$ (right) for magnetized ambient media with $\alpha_b=0.1$, both rescaled and shown as in Figure \ref{fig:5}. The $\varpi$-direction of these panels is stretched to reveal more detailed structures. The exaggeration factors are $4.2$, $2.6$, $1.5$, and $1.0$ for $n=1$, $2$, $4$, and $6$, respectively. 
}
\label{fig:6}
\end{figure*}

We illustrate in Figures \ref{fig:5} and \ref{fig:6} density distributions of internal structures pertaining to the hydromagnetic bubbles within the outflow lobes surrounded by a moderate to strong ambient magnetic field. These figures reveal more details at a resolution $\sim 2\times2$ higher than that of Figure 18 of \citetalias{shang2020}, and scaled to the self-similar coordinates. 
The column and number densities juxtaposed naturally cast the visual impressions of nested ``multicavities'' from the inner to outer cavities and the apparent extended shells mimicking the episodicity. The visual impression of the outermost cavities, however, only extends to partial lengths of the outflow lobes. The contributions to column density pass through the density-depleted gap region filled with the ambient poloidal field lines, which is wider in horizontal scales for the larger $n$ values. 
This feature is enhanced at higher resolution, and forms the basis of the interpretation of an extended low-velocity outflowing gas from the base of the outer outflow lobes.
The newly detected structures around HH 212 (reported in \citealt{lee2021_hh212}; see Section \ref{subsec:HH212} below) may be an example of the manifestation.

\subsection{Velocity Structures of the Multicavities within the Magnetized Bubbles}
\label{subsec:velocity_multicavities}

Here we reveal the velocity profiles and patterns appearing in the internal structures of the compressed layers in the elongated hydromagnetic bubbles. As demonstrated using plane-parallel hydrodynamic oblique shocks in Section \ref{subsec:HDsmooth} and Appendix \ref{sec:appendix:steps}, substantial shear is present across the shocked region bounded by the RS and FS\@. The hydrodynamic shear profiles for smooth analytical models are illustrated in Figure \ref{fig:shear}. The directions and magnitudes of velocity vectors vary as the flows cross the reverse shock surfaces by following the jump condition at the RS\@. This property has significant observational implications for wind diagnostics.

\begin{figure*}
\epsscale{0.95}
\plotone{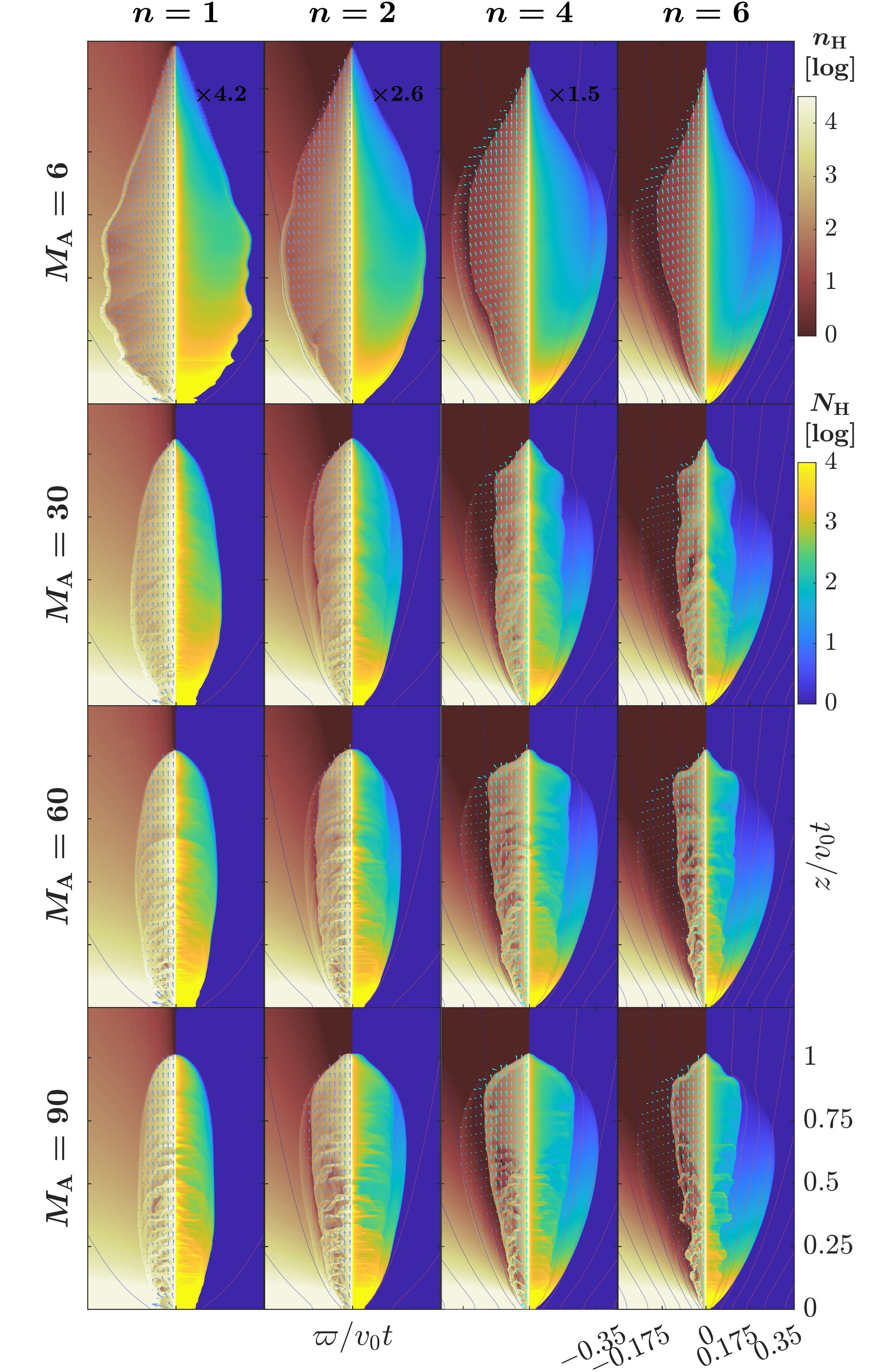}
\epsscale{1.0}
\caption{
Velocity vectors and flow patterns are shown for $\MA=6$, $30$, $60$, and $90$ with a magnetized medium of $\alpha_b=1$. These vectors are normalized to the nominal wind speed $v_0$. They are plotted on top of the number density on left, along with the column density integrated over regions with $v_p>\aamb$ shown on the right, both rescaled and shown as in Figure \ref{fig:5}. Both the $\varpi$-direction of these panels and $\varpi$-component of the vectors are stretched, with the same $n$-dependent exaggeration factors as in Figure \ref{fig:5}, to reveal more detailed structures. The ambient $b_p$ field lines are shown as in Figure \ref{fig:5}.
}
\label{fig:7}
\end{figure*}

\begin{figure*}
\epsscale{0.95}
\plotone{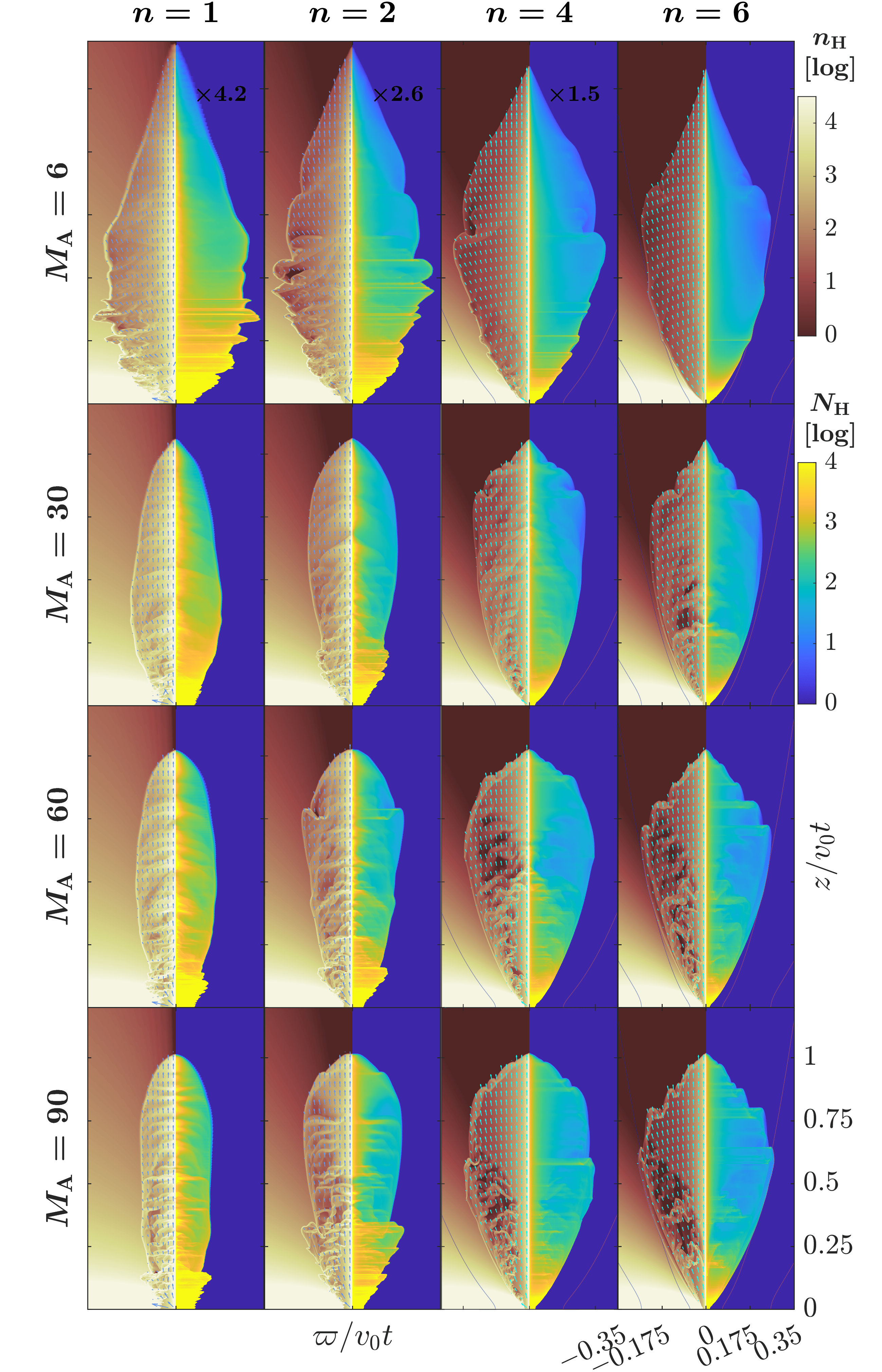}
\epsscale{1.0}
\caption{
Velocity vectors and flow patterns are shown for $\MA=6$, $30$, $60$, and $90$ with a magnetized medium of $\alpha_b=0.1$. These vectors are normalized to the nominal wind speed $v_0$. They are plotted on top of the number density on left, along with the column density integrated over regions with $v_p>\aamb$ shown on the right, both rescaled and shown as in Figure \ref{fig:6}. Both the $\varpi$-direction of these panels and $\varpi$-component of the vectors are stretched, with the same $n$-dependent exaggeration factors as in Figure \ref{fig:6}, to reveal more detailed structures.
}
\label{fig:8}
\end{figure*}

In Figures \ref{fig:7} and \ref{fig:8}, we track flow properties in the extended compressed regions between the RS and the FS\@. Our setup of a constant-velocity primary wind allows an easier tracking of the change of directions across the oblique shocks. The generation of vorticity through magnetic forces, oblique shocks, and nonlinearly grown modes of the KHI in the toroidally magnetized winds introduces local perturbations to the bulk flow. However, the overall patterns of bent and deflected flow vectors relative to the original radial free-wind are self-evident. The resultant lobe-scale flow patterns appear as emerging from near the base of the outflows, moving around the widest waist, then converging to the top. This apparent \textbf{bulk} flow motion can give an impression of a secondary wind rising on top of the ``disk'' or ``disk atmosphere'' as if it were an extended disk wind (EDW), which is launched from a few to tens of astronomical units from the inner launch loci of the primary wind. The inference of a separate launch region of tens of astronomical units by the reduced post-RS speed and deflected direction is a risky application of the connection between the terminal velocities and the launch radii, and a misuse of the formula derived in \citet{anderson2003}. Further discussion of the conceptual misunderstanding continues in Sections \ref{subsec:velocity_components} and \ref{subsec:notsosmallradii}.

Theoretically, these complex nested velocity shells simply occur naturally in elongated toroidally magnetized wind-blown bubbles as part of the physical mechanisms. A separate EDW launched slightly outside of and surrounding the primary wind is not required to generate the range of the extended intermediate velocities. The spatial correlation of observed PV diagrams with the illustrated column density maps provides a hint for the nature of the occurrence of the different velocity components between the jet and the very low shell velocities near the outflow base (see Section \ref{subsec:velocity_components}).

\subsection{Magnetic Pseudopulses}
\label{subsec:pseudopulses}

Here we revisit the generation of ``pseudopulses'' in the very extended compressed wind region. These pseudopulses were called magnetic pulses in \citetalias{shang2020}. 

In a toroidally magnetized compressed wind medium, vorticity can be generated by magnetic forces and fed back into the already grown KHI modes. Likewise, the enhanced vorticity drives the already pulsed $b_\phi$ with stronger amplitude, allowing the magnetic field to compress and relax further in the $z$-direction. This feedback loop produces self-generating pseudopulses cascading the magnetic energy in this region. Compounded by density stratification near the jet axis, it creates the impression of pulsed jets in the compressed wind region.

In a magnetized medium, the vorticity, $\omv=\nabla\times \vv$, can be generated by magnetic forces:
\begin{equation}
    \begin{split}
        \partial_t\omv &+(\vv\cdot\nabla)\omv
        -(\omv\cdot\nabla)\vv
        +\omv(\nabla\cdot\vv)=\\
        &=
        \frac{\nabla\rho\times\nabla p}{\rho^2}
        +\nabla\times\left(\frac{\jv\times\bb}{\rho}\right)
        \ ,
    \end{split}
    \label{eqn:vorticity_generation}
\end{equation}
where the terms on the right are the baroclinic term and the effects of magnetic forces. Inside the unmixed wind, where $\bb_p$ is zero or tiny, Equation (\ref{eqn:torque_bphi}) shows the dominant term (due to $b_\phi$) of this magnetic torque, $-\nabla\left(\lfrac{1}{\rho\varpi^2}\right)\times\nabla\Lfunc$.

The magnetic accelerations and the induced variations of $v_\theta\neq 0$ in turn feed into the induction equation allowing acceleration and deceleration in the $r$-direction, which then causes the variation of $v_r$, leading to the presence of pseudopulses in the radial direction. They leave even stronger marks at the higher-resolution ($6400\times2016$) set.  The connections of $f_r$ to $v_r$ and $f_\theta$ to $v_\theta$ are shown in Figure \ref{fig:mag_forces} in Appendix \ref{sec:appendix:mag_forces}.

\section{Lobe Shape Curves in PV Space}
\label{sec:MCcurves}

We start with producing curves, one-dimensional loci of the MC shapes, in axisymmetric position space. Then we project those MC curves onto PV space, without line-of-sight integration. Subsection \ref{subsec:MHD_Wind_curves} continues this for magnetized \MOS{} models.

\subsection{Momentum-Conserving Curves in PV Space}
\label{subsec:MCcurves_PV}

The self-similar MC curves in $r$--$\theta$ axisymmetric position space are computed in Section \ref{subsubsec:thinshells}, Equation (\ref{eqn:mc_eqn}). Projecting them onto axisymmetric $\varpi$--$z$ position coordinates, we build MC curves in $\varpi$--$z$ space or plane:
\begin{equation}\label{eqn:mc_varpiz}
    \begin{split}
    \varpi\MC(\theta) &=
    r\MC(\theta)\sin(\theta)=
    v\MC(\theta)\sin(\theta) t
    \\
    z\MC(\theta) &=
    r\MC(\theta)\cos(\theta)=
    v\MC(\theta)\cos(\theta) t
    \ .
    \end{split}
\end{equation}
Usually the coordinates $\theta$ and $\varpi$ are restricted to positive values. However, mirror axisymmetry of the MC curve in the 2D position space ($\varpi$--$z$ or $r$--$\theta$) allows an extension of $\varpi$ and $\theta$ into negative values.

Following the curve defined by Equation\ (\ref{eqn:mc_varpiz}) for a range such as $-180\arcdeg\leq\theta\leq+180\arcdeg$ parameterizes a continuous curve representing the two lobes, with lobe tips located at $\theta=0\arcdeg$ and $180\arcdeg$.
Inside our $n$-range, the equatorial wind/ambient density ratio is very small, giving the curves a narrow waist or ``neck'' at the equator, located at $\theta=\pm90\arcdeg$.
The $-90\arcdeg\leq\theta\leq+90\arcdeg$ range suffices to parameterize a single lobe.

\begin{figure*}
\plotone{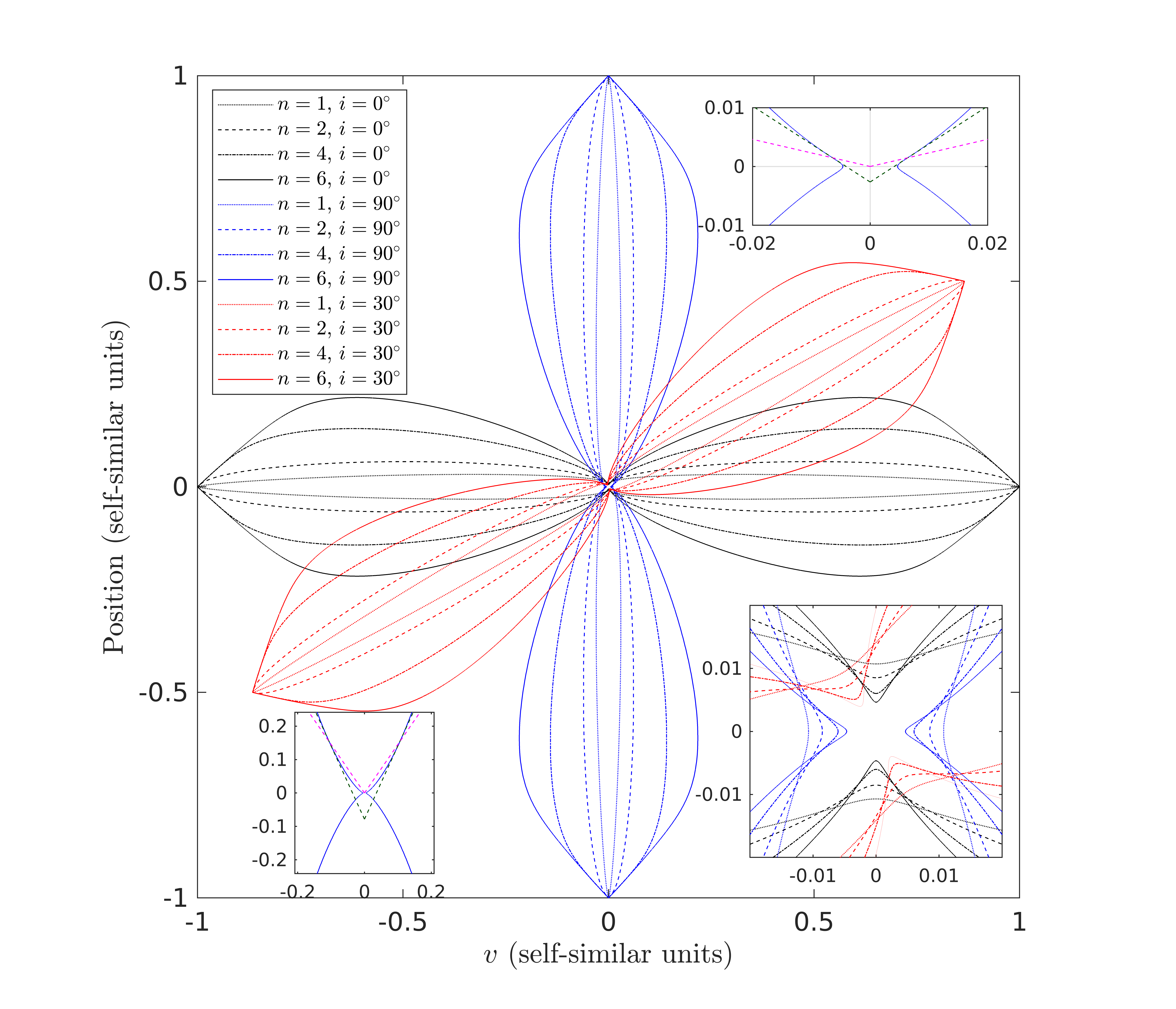}
\caption{Main figure: PV-space curves based on the MC velocity $\vMC$. Values of inclination angle are $i=0\arcdeg$ (black), $30\arcdeg$ (red), and $90\arcdeg$ (blue, this last curve equivalent to $\varpi$--$z$). Values of toroid parameter $n=1$, $2$, $4$, and $6$ are shown in dotted, dashed, dash-dotted, and solid lines. Bottom-right inset: zoom-in of the main figure's mathematical curves near the base. Top-right inset: zoom-in of the $n=6$, $i=90\arcdeg$ curve, with tangent lines (dark green) at the inflection point of the curve, and also secant lines (magenta) connecting that point to the origin. These pairs of lines define opening angles. Bottom-left inset: similar to the top-right inset, but with a less extreme zoom-in. The tangent (dark green) and secant (magenta) lines are now drawn at a point of the curve near $v=0.1$, $p=0.15$ (in self-similar units), instead of the inflection point.
\label{fig:9}}
\end{figure*}

We now construct the MC curves in PV space. By defining a line of sight with inclination angle $i$, the MC curves of Equation (\ref{eqn:mc_varpiz}) can be projected onto the PV plane (velocity $v$, position $p$) as
\begin{equation}\label{eqn:mc_pv}
    \begin{split}
    v_\mathrm{MC,PV}(\theta, i) &=
    v\MC(\theta) \cos(i+\theta+\pi )\\
    p_\mathrm{MC,PV}(\theta, i) &=
    v\MC(\theta) \sin(i+\theta+\pi ) t_\mathrm{sim}\ ,
    \end{split}
\end{equation}
showing perfect self-similarity.
The inclination angle $i=0\arcdeg$ corresponds to a bipolar outflow axis aligned along the line of sight, with $|\theta|<90\arcdeg$ corresponding to the blueshifted side (negative $v$ values), and $90\arcdeg<|\theta|<180\arcdeg$ corresponding to the redshifted side. The redshifted side can be alternatively parameterized with $i=180\arcdeg$ and $|\theta|<90\arcdeg$. A bipolar outflow axis lying on the plane of the sky corresponds to $i=90\arcdeg$.
Figure \ref{fig:9} illustrates the relationship of the inclination angles, base opening angles, and the toroid $n$ values,
showing a consistent trend of shapes and orientations with respect to the evolution in the $n$ values.
Figure \ref{fig:9} and Equation (\ref{eqn:mc_pv}) show that when using self-similar units ($v_0$ for velocity and $v_0 t$ for position) a change of inclination angle $i$ corresponds to a simple rotation of the MC curves in PV space. We also note that the mathematical shapes of the curves traced by Equations\ (\ref{eqn:mc_varpiz}) and (\ref{eqn:mc_pv}) are identical. Both of these remarkable properties are related to self-similarity. \textbf{The mathematical identity and rotation symmetry allow to use Figure \ref{fig:9} not only to represent curves in PV space as labeled, but also the curves in \text{\boldmath$\varpi$\unboldmath}--\text{\boldmath$z$\unboldmath} space.}

These MC space curves (both in PV and $\varpi$--$z$ spaces) are elongated oval-shaped contours, symmetric across an equatorial line, and also around an axis connecting the two apices (each apex touching the tips of the jets).
The series of oval curves follows the shapes of the individual lobes labeled by the toroid index $n$, and is further connected to the curve family with varying $\MA$ discussed in Section \ref{subsec:MHD_Wind_curves}.
For typical (not too small) values of $n$, the base of the curves near $|\theta|\sim 90\arcdeg$ has a ``neck'' that gets very close to the origin of coordinates, but the curves are smooth and do not cross each other before moving into other quadrants of the plane. This situation can be seen in the insets of Figure \ref{fig:9}.

The base region of the outflow can be described in terms of the MC curves near $|\theta|=90\arcdeg$ as shown in the insets. A useful parameter to describe that base region is an opening angle $o$ between the two tangent (or secant) lines to opposite sides of the lobes. The inflection point of the MC curves provides a maximum opening value, which can be used to parameterize the base of the outflow at very low heights (top-right inset of Figure \ref{fig:9}). Alternatively, a tangent or secant connected to the larger scale of the curves (bottom-left inset) may tell more about the overall outflow structure and observed images.
The opening angle based on the idealized MC curves in self-similar units has the same value in either $\varpi$--$z$ coordinates, or on PV space for any inclination angle $i$. 
This is because of self-similarity making the curves mathematically identical between these two spaces and responding to a change in $i$ as a rotation in PV space.

\subsection{Magnetized Outflow Shape Curves in PV space}
\label{subsec:MHD_Wind_curves}

We consider now the regimes of hydromagnetic winds interacting with magnetized or nonmagnetized ambient toroids. When the winds are toroidally magnetized, new phenomena arise with the thick and extended compressed regions, the KHI and pseudopulses feedback loop with generation of vorticity. This motivates us to compute magnetized outflow shape curves for different $\MA$ values using Equation (\ref{eqn:shape:vs_windbphibp}).

Equation\ (\ref{eqn:shape:vs_windbphibp}) gives the \MOS{} (magnetized outflow shape) of a hydromagnetic wind carrying a $b_\phi$ interacting with a possibly magnetized toroid carrying a $b_p$.
This is a magnetized analog of Equation\ (\ref{eqn:shape:ram}). Equation\ (\ref{eqn:shape:vs_windbphibp}) is exactly as self-similar and mirror-symmetric as the hydrodynamic model of Section\ \ref{subsec:MCcurves_PV}, and therefore equations similar to Equations (\ref{eqn:mc_varpiz}) and (\ref{eqn:mc_pv}) can be written, sharing thus in symmetry properties such as the identity of mathematical shapes traced in PV and $\varpi$--$z$ spaces, and changes of inclination angle $i$ being equivalent to rotations of the curves in PV space.

\begin{figure*}
\plotone{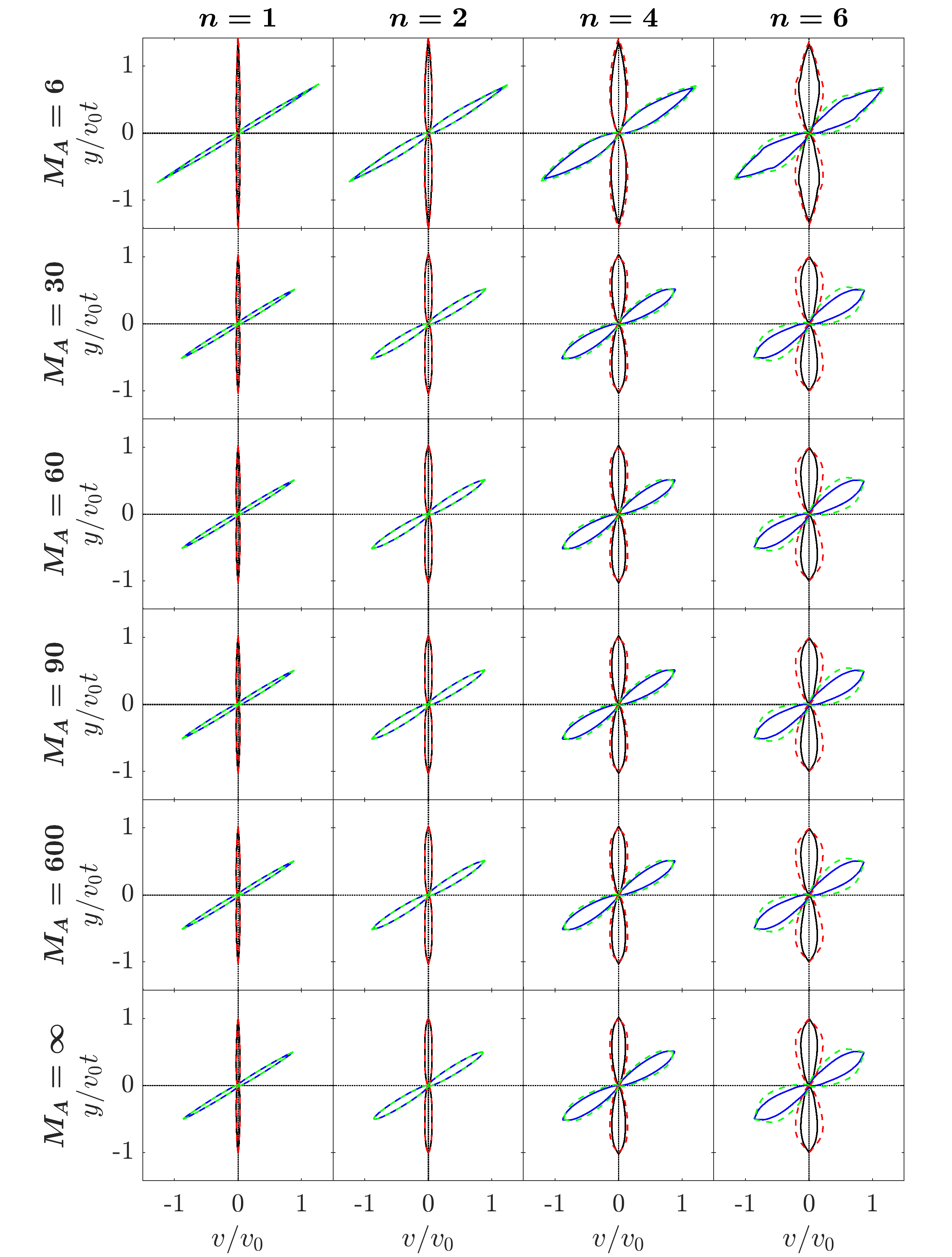}
\caption{
PV space curves based on the \MOS{} Equation (\ref{eqn:shape:vs_windbphibp}) with the use of factors $\beta(\theta)$ and $e_p$ as discussed in Appendix \ref{sec:appendix:ep_beta}\@. Each panel shows two values of $\alpha_b$ (solid line $\alpha_b=1$, and dashed line $\alpha_b=0$). Panels are presented for $n=1$, $2$, $4$, and $6$ (left to right) and $\MA=6$, $30$, $60$, $90$, $600$, and $\infty$ (top to bottom). The panels are shown in PV space, projected for inclination angles of $i=90\arcdeg$ (edge-on), and $30\arcdeg$.}
\label{fig:MC_MA_i}
\end{figure*}

Figure \ref{fig:MC_MA_i} summarizes such \MOS{} shape curves.
The variation of the contours of \MOS{} curves with different $n$ values is evident for the entire $\MA$ range. Prominent dependence of shape curves on the magnetic parameters is salient for the more magnetized cases, the $\MA \lesssim 30$ range for the wind, and $\alpha_b=1$ for the medium.
Detailed properties derived from Equation\ (\ref{eqn:shape:vs_windbphibp}) are given in Appendix \ref{sec:appendix:ep_beta}\@.

\section{The Position--Velocity Diagrams}
\label{sec:PVD}

In this section, we build signatures of each featured component of the integrated hydromagnetic outflow bubble in position--velocity diagrams of column density (PVDCD)\@. We present the parallel PVDCDs, in which the position $y$ represents the distance along the projected outflow axis.

We note that by producing the column density maps in PV space,
it is difficult to recover identical information from real observations without including specific tracers and their line excitation.
Different tracers probe different physical conditions, and tracers are subject to thermochemistry and chemical evolution, whose detailed conditions are beyond the scope of current work.

We have assumed an inclination angle of $45\arcdeg$ for PVDCD maps in this work. At this angle, the signatures of each of the components are separated the farthest. However, in real systems, the signatures can be embedded in different inclination angles from $0\arcdeg$ to $90\arcdeg$, and they will create different impressions. Modeling of 
real systems requires the knowledge of their inclination angles and of their physical conditions, and comparison should include a calculation of the emissivity maps. We leave such an effort to a future work.

\subsection{The Hydrodynamic No-\texorpdfstring{$v_\theta$}{v\_theta} Models}
\label{subsec:HD_noslip_PV}

\begin{figure*}
\plotone{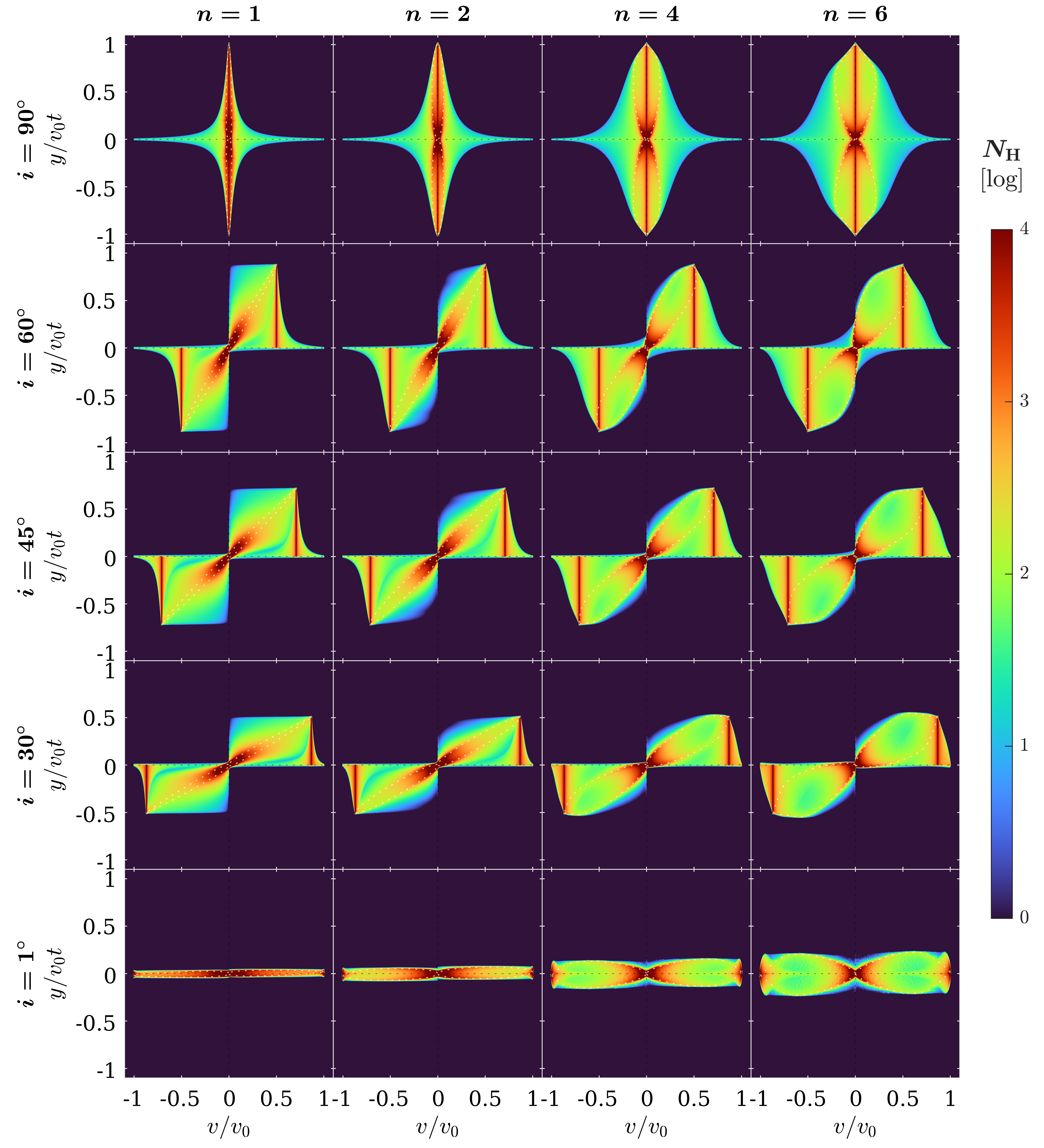}
\caption{Position--velocity diagrams of column density (PVDCD) of the no-$v_\theta$ ($v_\theta=0$) models for $n=1$, $2$, $4$, and $6$ toroids (left to right) at inclinations of $90\arcdeg$, $60\arcdeg$, $45\arcdeg$, $30\arcdeg$, and $1\arcdeg$ (top to bottom). Column density is rescaled logarithmically for the 4 orders of magnitude that are covered. The unshocked ambient material with $v_p < \aamb$ has been filtered out while producing the PVDCDs. Momentum-conserving curves are overlaid as white dashed lines.}
\label{fig:HDshell_pv_no-slip}
\end{figure*}

In this subsection, we build four smooth reference PVDCDs of the ``no-$v_\theta$'' ($v_\theta=0$) calculation runs for the hydrodynamic $\MA=\infty$ and $\alpha_b=0$ cases, one for each of our four $n$ values. The four runs represent at every $\theta$ angle a 1D radial bubble. This series with a restricted $v_\theta$ value will be contrasted against the cases with regular $v_\theta$ simulation treatment in subsections that follow.

The no-$v_\theta$ models are expected to preserve the basic features of the dual jet plus thin-shell structures without the complication from the presence of the KHI, despite the presence of shear. The central jet is the densest part of the underlying wind with a $\varpi^{-2}$ profile, which straddles a velocity centroid with broad wings extended to the positive and the negative ranges. The divergence of the radially pointed flow naturally produces a velocity spread projected onto the line of sight, as demonstrated in Figures 1 and 2 of \citetalias{shang2020}.
The flow near the base has close to equal projections toward and away from the observer at a full self-similar velocity $|v/v_0|=1$, giving rise to a total velocity width of $\sim 2$ near the base. Moving up the length of the jet, the projected velocity along the line of sight decreases, and the radial velocity vectors narrow down to the vicinity of the projected velocity centroid, causing the widths to decrease.

The PVDCDs of the smooth reference no-$v_\theta$ models at different inclination angles $i$ with the full coverage of both lobes in all four quadrants are depicted in
Figure \ref{fig:HDshell_pv_no-slip}.
These panels illustrate the variation of the PVDCD patterns with $n$ and the inclination angle $i$, along with the thin-shell MC curves in PV space. The four quadrants are needed to show the positive and negative projected velocities across the hemispheres due to inclinations. Clearly the two jet--shell components are distinctly visible. The densest part of the wide-angle wind maintains well the jet appearance at the projected velocity $v_0\cos i$ while the radial part of the flow is projected around the jet velocity centroid, giving a width of $2v_0\sin i$ at the base. The width varies with the thin-shell opening confined by the ambient toroid, leaving fast narrowing of the width for the small $n$ but gradual narrowing for the larger $n$.

The divergent radial flow with the density profile of $\varpi^{-2}$ allows the integrated column density to drop off approximately as $\varpi^{-1}$.
The MC curves in PV space, as in Section \ref{subsec:MCcurves_PV}, tracing the thin-shell cavities naturally lay on top of the projected jet--wind features in the PVDCD of these no-$v_\theta$ models. The bottom portion of the MC curves coincides with the base of the thin-shell cavities very well. These cases remain well-traced by the MC curves.
It is expected that MC thin shells would become more conical for larger $n$ and more collimated for smaller $n$, because the toroid density functions for larger $n$ are also further away from being spherical. 
The face-on situation at $1\arcdeg$, equivalent to the situation of a cone with small to large opening angles moving at $v_0$, is shown at the bottom row of Figure \ref{fig:HDshell_pv_no-slip}.
The small to large opening angles of the axisymmetric cones give the vertical width in $y/v_0t$ located at $v/v_0=\pm 1$. 

\begin{figure*}
\centering
\plotone{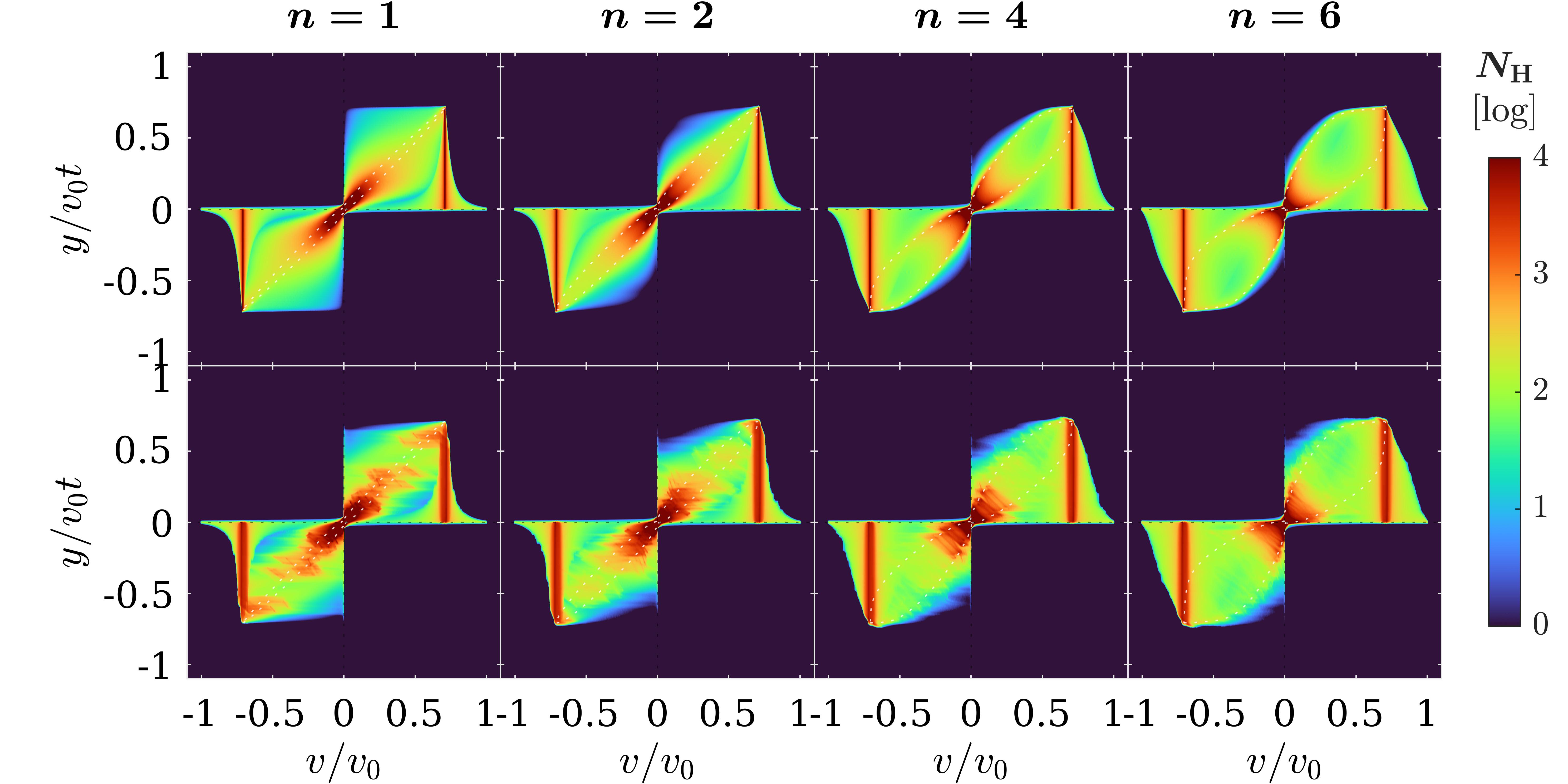}
\caption{Position--velocity diagrams of column density (PVDCD) of the no-$v_\theta$ ($v_\theta=0$) HD calculations (top) and regular HD simulations (bottom, any $v_\theta$), both with $\alpha_b=0$ and $\MA=\infty$, for $n=1$, $2$, $4$, and $6$ toroids (left to right) at an inclination angle of $45\arcdeg$. Column density is rescaled logarithmically, and 4 orders of magnitude are covered. The unshocked ambient material with $v_p < \aamb$ has been filtered out while producing the PVDCDs. The MC curves are overlaid as white dashed lines.}
\label{fig:HDshell_pv_slip}
\end{figure*}

\subsection{Hydrodynamic Simulations with Regular \texorpdfstring{$v_\theta$}{v\_theta}}
\label{subsec:HD_slip_PV}

To enter the more realistic regime in which shear is allowed to induce KHI and other complexities, we first make a comparison between hydrodynamical simulations with regular $v_\theta$ and those calculations with the no-$v_\theta$ ($v_\theta=0$) constraint.
The comparison of the two sets of models is made for the hydrodynamic cases, having control parameter values $\MA=\infty$ and $\alpha_b=0$. Figure \ref{fig:HDshell_pv_slip} illustrates the differences the presence of KHI makes for one inclination angle ($45\arcdeg$). Significant changes can be seen clearly in the bottom panels. The changes in the distribution of column density with the position and velocity are considered to come from the physical effects of active shear within the thin layer between the reverse and forward shocks. 

Within models with active shear (a contrast with the no-$v_\theta$ models in Section \ref{subsec:HD_noslip_PV}, which do have intense shear, but ``inactivated'' to induce a KHI), patches of feather-like structures extend across the ellipsoidal region of the cavity walls between the two wide-angle jet features. This action smears the thin-shell features traced by the MC curves. These patches arise from the large spread of velocity across the post-shock regions, produced by the large ``spikes'' or the large vortices aligned along the spikes present in the HD models illustrated in Appendix D of \citetalias{shang2020}.
The fuzzy threads protruding the MC curves trace the locally parallel shock fronts making different obliqueness angles across the shocked shells from the interiors of the physical cavities, through the outer surfaces, then into the ambient medium.

\begin{figure*}
\plottwo{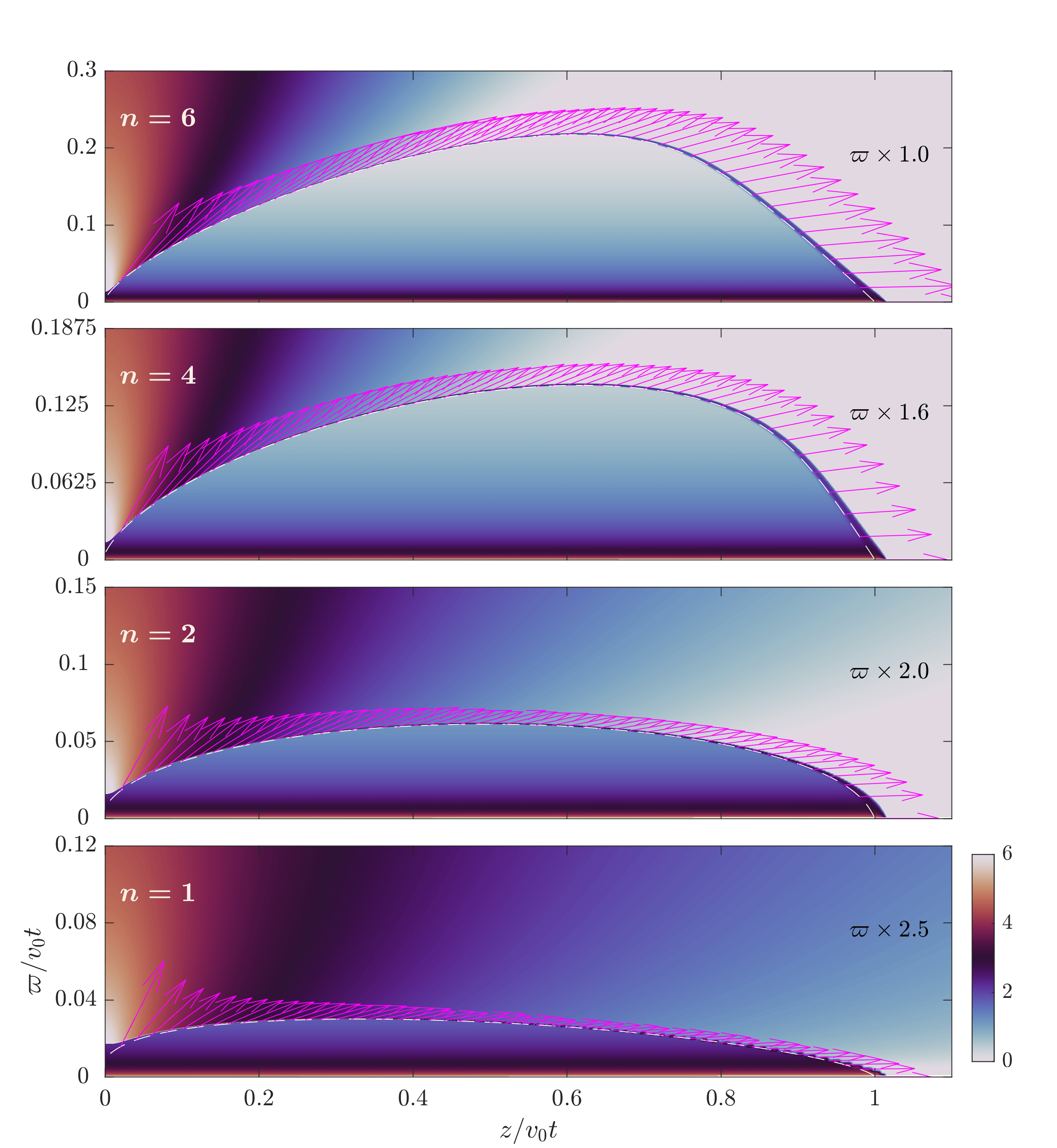}{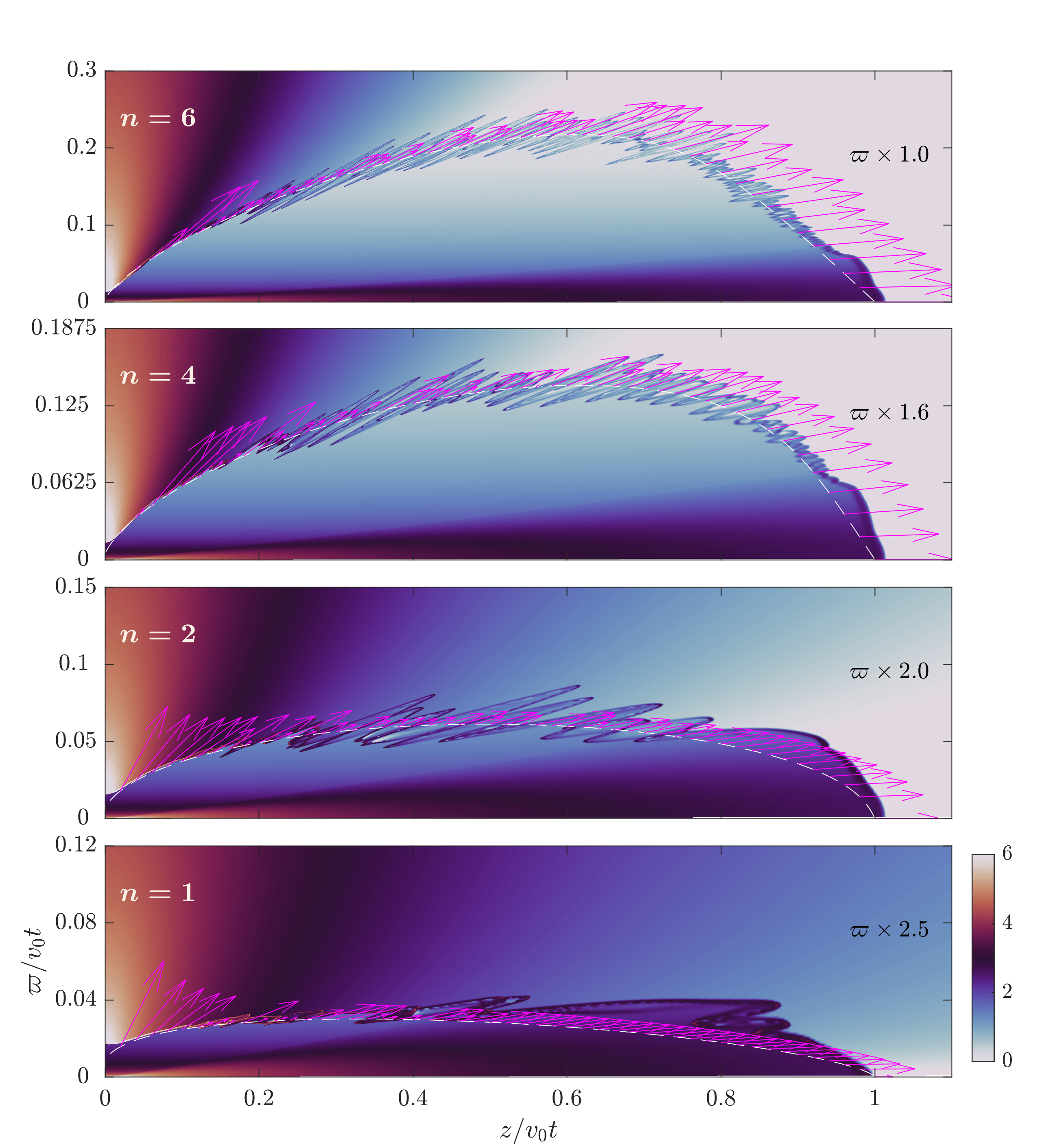}
\caption{Velocity structures close to the hydrodynamic outflow ($\MA = \infty$, $\alpha_b = 0$) no-$v_\theta$ calculations ($v_\theta=0$, left column) and regular HD simulations (any $v_\theta$ value, right column), on top of the rescaled logarithmic number density as shown in Figure \ref{fig:4}, and with the same $n$-dependent $\varpi$ magnification for better viewing. The velocity vectors are shown along the loci of the momentum-conserving thin-shell model curves traced by the white dashed lines.}
\label{fig:vel_HDshell}
\end{figure*}

The velocity vectors moving along the shell surfaces are compared in Figure \ref{fig:vel_HDshell} between the no-$v_\theta$ (left) and regular $v_\theta$ (right) cases. The velocity vectors from the simulations are shown along the positions on the theoretical MC curves. Variations of the velocities can be seen as a result of the development of KHI\@. In the no-$v_\theta$ calculations, the resulting shells follow more or less the MC curves, and the velocity variations appear to be smooth. In the simulations with regular $v_\theta$, the MC curves pass through the instability patterns. The vectors follow zigzag paths, resulting in fluctuating directions and magnitudes.

\subsection{Kinematic Signatures of Outflows Formed by Magnetized Winds}
\label{subsec:MHD_Wind_PV}

\begin{figure*}
\centering
\plotone{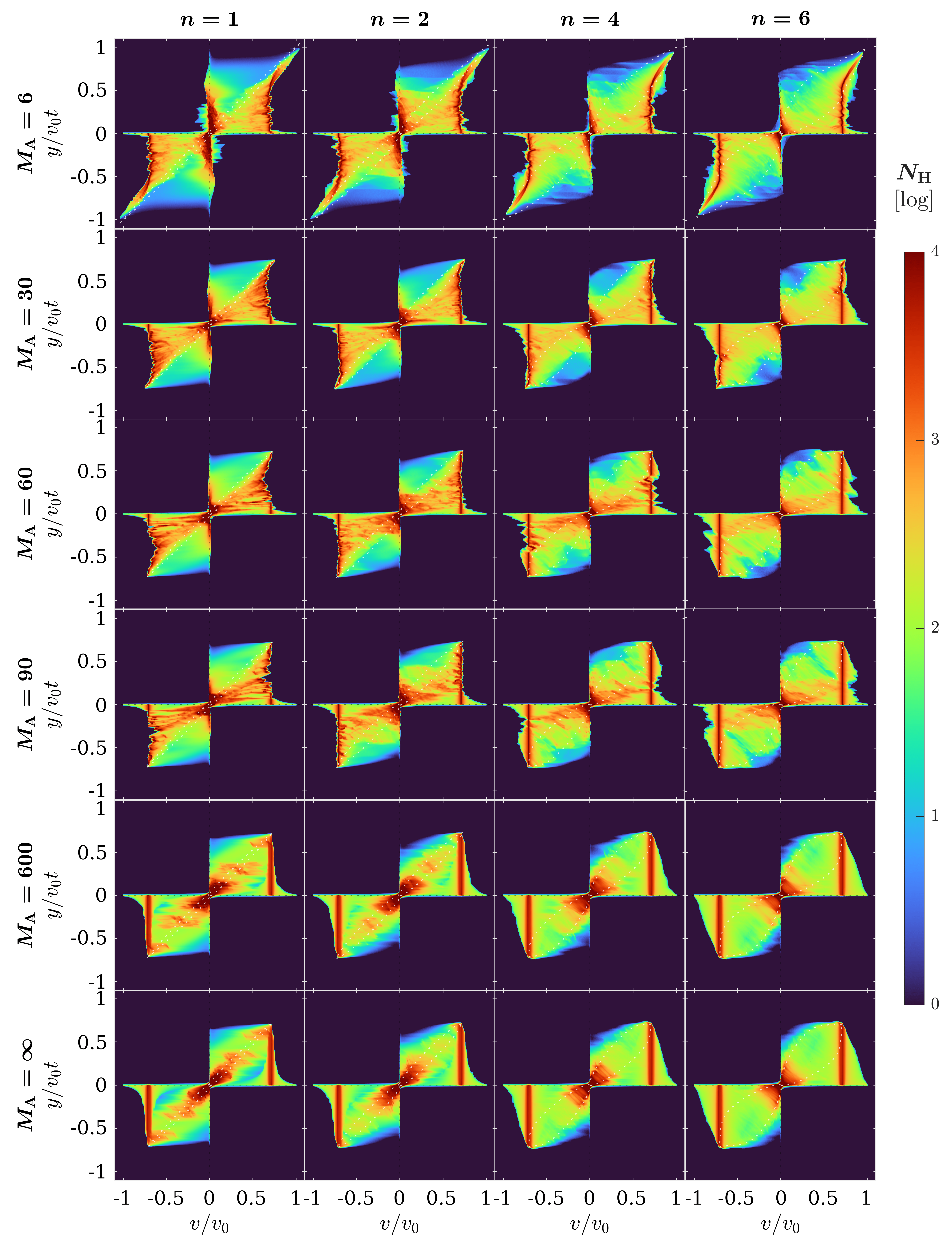}
\caption{Position--velocity diagrams of column density (PVDCD) of the magnetized winds of $\MA=6$, $30$, $60$, $90$, $600$, and $\infty$ (top to bottom) for the nonmagnetized ($\alpha_b = 0$) $n=1$, $2$, $4$, and $6$ toroids (left to right), at an inclination angle of $45\arcdeg$. The rescaled logarithmic column density covers 4 orders of magnitude. The unshocked ambient material with $v_p < \aamb$ has been filtered out while producing the PVDCDs. The \MOS{} curves are overlaid as white dotted lines. PV coordinate axes are shown in black dotted lines.}
\label{fig:PV_vw100_Bp0p0_45deg}
\end{figure*}

\begin{figure*}
\centering
\plotone{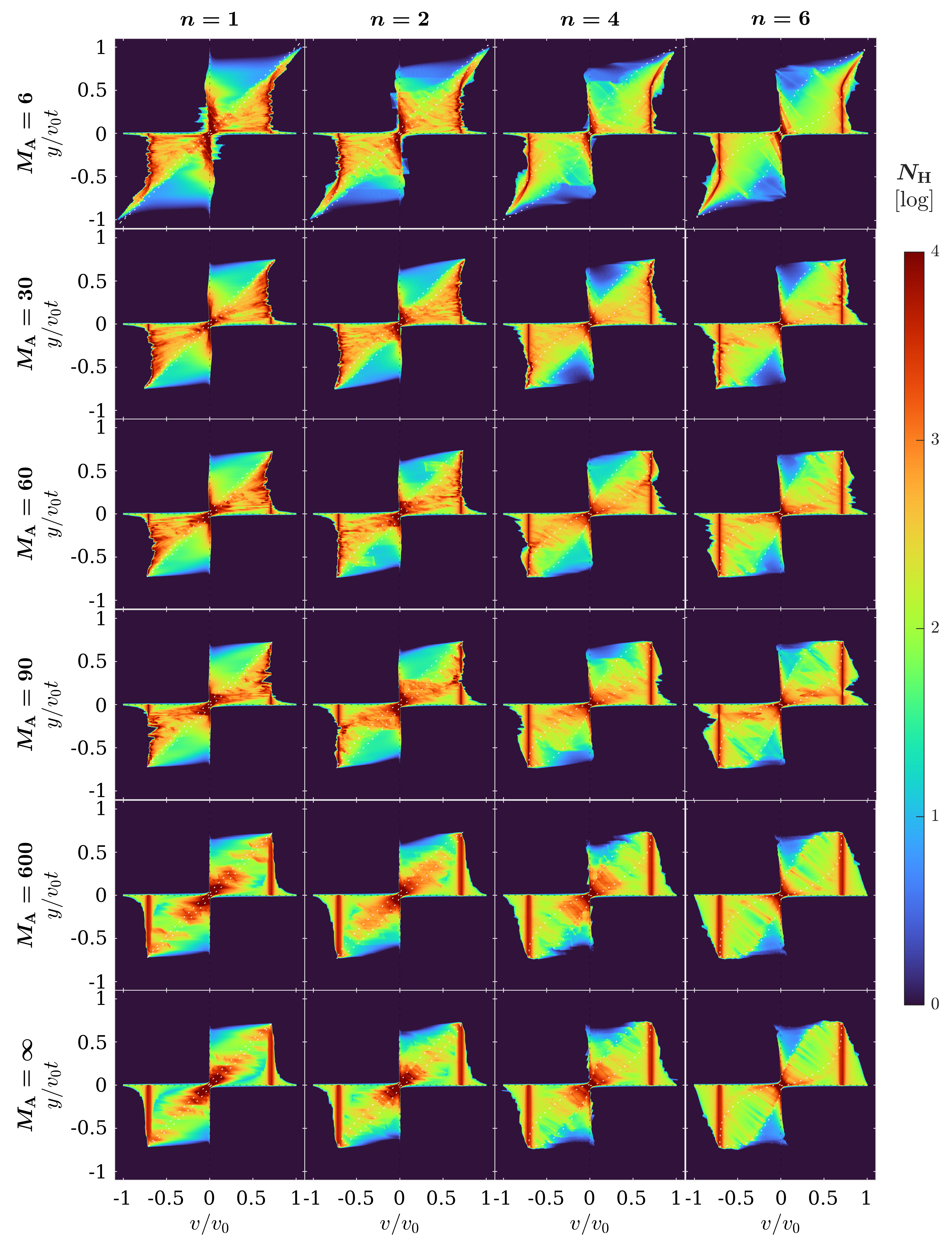}
\caption{Position--velocity diagrams of column density (PVDCD) of magnetized winds $\MA=6$, $30$, $60$, $90$, $600$, and $\infty$ (top to bottom), and weakly magnetized ($\alpha_b = 0.1$) $n=1$, $2$, $4$, and $6$ toroids (left to right), at an inclination angle of $45\arcdeg$. The rescaled logarithmic column density covers 4 orders of magnitude. The unshocked ambient material with $v_p < \aamb$ has been filtered out while producing the PVDCDs. The \MOS{} curves are overlaid as white dotted lines. PV coordinate axes are shown in black dotted lines.}
\label{fig:PV_vw100_Bp0p1_45deg}
\end{figure*}

\begin{figure*}
\centering
\plotone{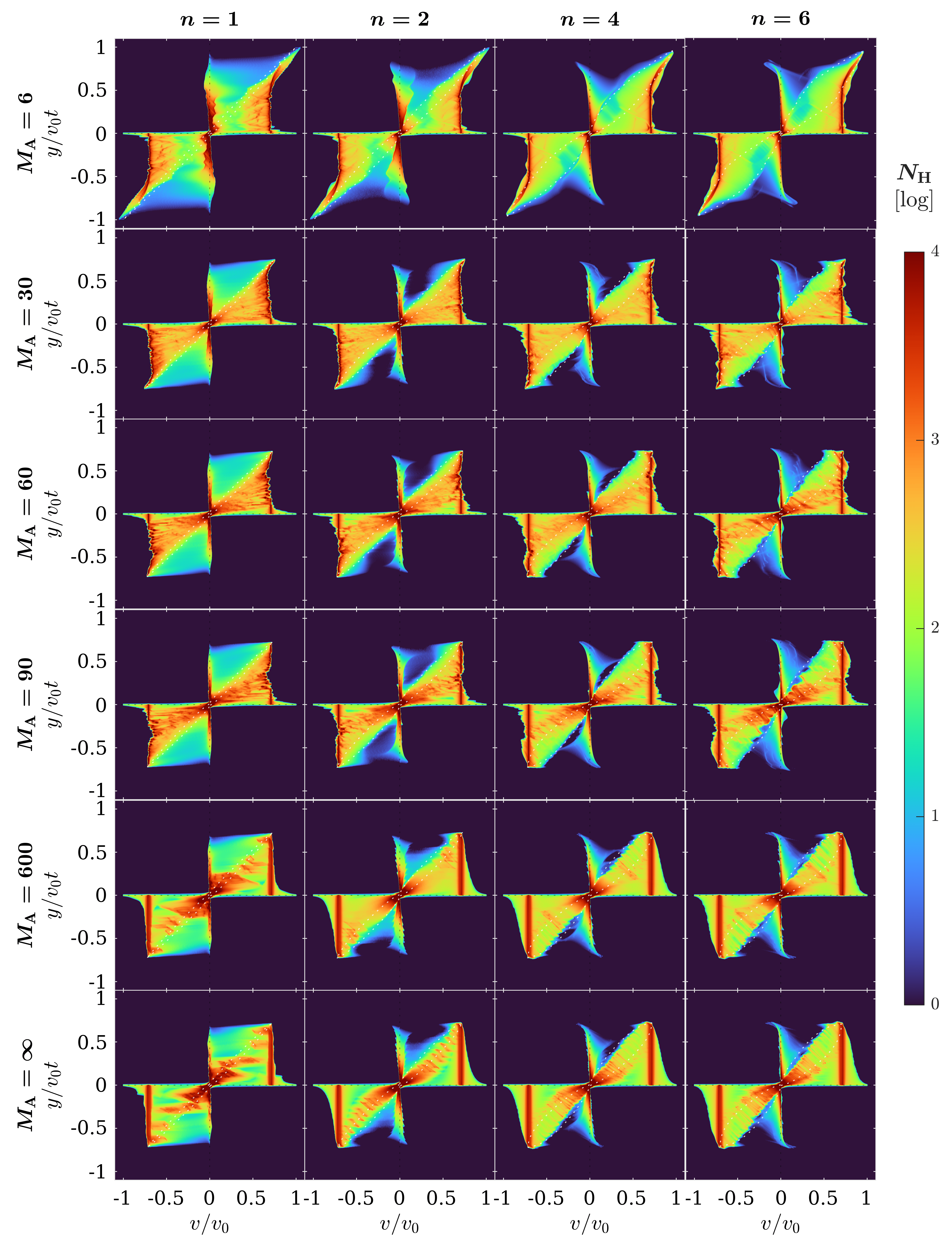}
\caption{Position--velocity diagrams of column density (PVDCD) of winds of various $\MA=6$, $30$, $60$, $90$, $600$, and $\infty$ (top to bottom) for the strongly magnetized ($\alpha_b = 1$) $n=1$, $2$, $4$, and $6$ toroids (left to right),  at an inclination angle of $45\arcdeg$. The rescaled logarithmic column density covers 4 orders of magnitude. The unshocked ambient material with $v_p < \aamb$ has been filtered out while producing the PVDCDs. The \MOS{} curves are overlaid as white dotted lines. PV coordinate axes are shown in black dotted lines.}
\label{fig:PV_vw100_Bp1p0_45deg}
\end{figure*}

For the synthetic PVDCDs of outflows made with column densities, we integrate densities along different individual lines of sight and collect the binned velocities. These PVDCD panels are shown in Figure \ref{fig:PV_vw100_Bp0p0_45deg} for the nonmagnetized ambient media and in Figures
\ref{fig:PV_vw100_Bp0p1_45deg}
and \ref{fig:PV_vw100_Bp1p0_45deg} for the respective weakly and strongly magnetized media. In this subsection, we demonstrate signatures produced by magnetized winds, free or compressed, from their interactions with the ambient media.

The PVDCDs derived from the hydromagnetic winds constructed by our models contain two intrinsic components: the jet and the wide-angle radial velocity vectors, and the structures induced by their interactions with the ambient medium.
For the free-wind models, as shown in Figures 1 and 2 of \citetalias{shang2020}, the jet collimated with the density profile $\rho\propto \varpi^{-2}$ appears at the velocity peaked as $v_0\cos i$, 
as the peak of emission near the base,
surrounded by the broad wings produced by the radial velocity vectors of the wide-angle wind. The outlining contours of the column density imitated emission (CDIE) originating from the highest and lowest projected velocities move up with the length of the jet and converge gradually near the tip of the jet.
The basic feature of the outflow cavity produced by the wide-angle portion appears near the origin of the PV coordinates. They vary in physical extent from $n=1$ to $n=6$, in their respective spread from the tip of their respective \MOS{} cone that forms from the revolution of the \MOS{} curves.

These jet plus wide-angle signatures are clearly distinguishable from the \MOS{} cone on the PV plane, similar to the MC analogs in the PVDCD of Figure \ref{fig:HDshell_pv_slip}, and
in transition to the \MOS{} in the bottom two rows of Figures \ref{fig:PV_vw100_Bp0p0_45deg} and \ref{fig:PV_vw100_Bp0p1_45deg}.
The \MOS{} curves touch the origin of the PV diagram coordinates on one side, with the other side touching the jet tips at the converged tops of the broad velocity dispersion from the bases. In addition to these general sketches, the picture is enriched by the PVDCD representations of the MS ellipsoids, which morph into different complex structures as the $\MA$ values decrease with increasing strengths of wind magnetization. This is the most evident change for the upper four rows in the Figures 
\ref{fig:PV_vw100_Bp0p0_45deg}, \ref{fig:PV_vw100_Bp0p1_45deg}
and \ref{fig:PV_vw100_Bp1p0_45deg} as the strength of $b_\phi$ increases.

The interactions of the magnetized wind with ambient media add the thick and extended compressed wind regions to the characteristics of the PVDCDs,
as threads connecting the PV origin and the jet features. The jets with wide-angle wind features are confined under smaller triangular-shaped areas following the boundaries of the RS\@. (We refer to this triangular-shaped area as the ``RS cavity''.)
On top of the jet tips, the jet portion of the compressed winds near the jet axes is
associated with clear but wiggled velocity centroids around the original values. The wiggled jet features trace the influence of the magnetic pseudopulses in the jet portion.
These kinematic features are evident in Figures \ref{fig:PV_vw100_Bp0p0_45deg}, \ref{fig:PV_vw100_Bp0p1_45deg}, and \ref{fig:PV_vw100_Bp1p0_45deg}.
The CDIE distribution appears extended in rectangular or butterfly-like areas between the systemic and the projected jet velocities as extended intermediate to low velocities, which are linked to the nested velocity features discussed in Section \ref{subsec:velocity_components}.

Small to large patches and threads of velocity patterns across the jet--shell trace the underlying shear and instabilities within the shocked regions. This feature is known to follow the occurrence of the KHI from Sections \ref{subsec:HD_noslip_PV} and \ref{subsec:HD_slip_PV}, and Figure \ref{fig:HDshell_pv_slip}.
It increases in PV coverage as the wind magnetization grows stronger until $\MA\approx30$. The patches in $\MA\lesssim 600$ spread wider toward the jet features. The patches in the smaller $n$ cases are more horizontally intertwined than the larger $n$ cases, while the $n=6$ case has less of this effect due to their wide-opened base cavities.

The presence of these features is consistent with the origins of KHI in the extended compressed wind regions. The boundaries between the \MOS{} cone and those of the jet plus wide-angle wind are blurred; however, the cavity formed by the RS is distinctly traceable and confined by these intertwined patches on the PVDCD\@. The patchy appearance of these features suggests the likely association with mild KHI shocks. In real PVs, they may appear as fainter, scattered emissions with different local excitation conditions.

Another noticeable feature associated with the jet component arises on the PVDCD for a strongly magnetized wind at large positions.
This \textbf{apparent acceleration} exists beyond a certain $y/v_0t$ from the top portion of the compressed wind region, for the strongly magnetized $\MA=6$ and the moderately magnetized $\MA=30$ cases. Some tiny tips remain all the way to $\MA=60$. We note that this increase of velocity only occurs at the outer boundary of the compressed wind region, which is far beyond the RS enclosing the free-wind propagation region. The nature of this feature is to be explored in depth in \citetalias{shang_PIII}\@. 

\subsection{Signatures of the Magnetized Mixing Regions}
\label{subsec:PV_magnetized_interplay}

Here we highlight the kinematic features that arise from the magnetic interplay. These include structures originating in the interacting compressed wind and ambient regions, which cover a substantial volume of mixing and entrainment, the jet fluctuations caused by the pseudopulses, the patterns of the nonlinear KHI, the multicavities, and the RS cavity enclosing the primary wind with broad velocity widths near the base of the jet. Naturally, the resultant PVDCDs would contain the imprints of these ingredients in their respective regions that alter the simple smooth thin-shell systems.

\begin{figure*}
\centering
\plotone{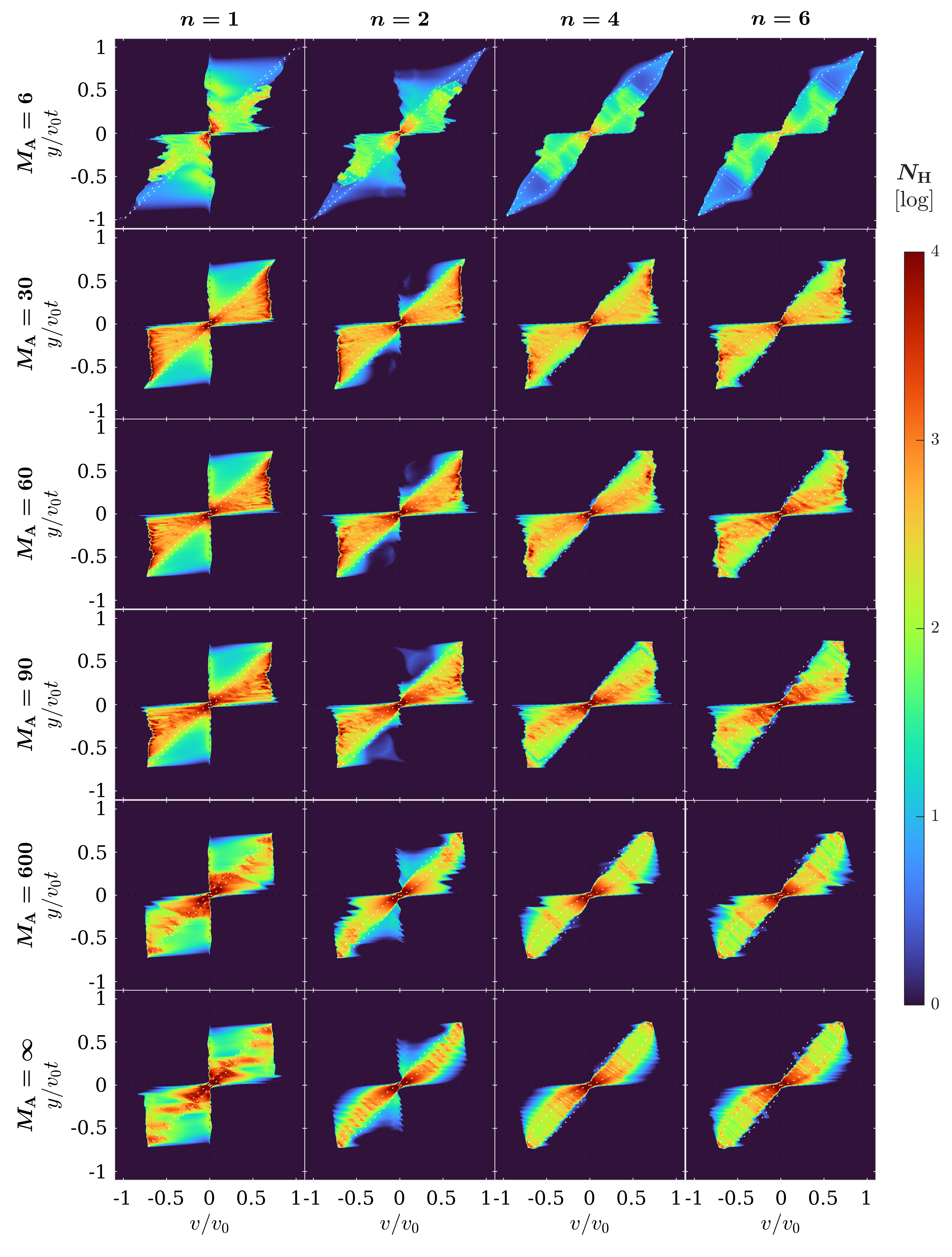}
\caption{Position--velocity diagrams of column density (PVDCD) at an inclination $i=45\arcdeg$ showing the post-shock regions with mixing ($0<f<1$, i.e., wind mass fraction $f$ not too close to pure wind $f=1$ or pure medium $f=0$),
for $\alpha_b=1$. The rescaled logarithmic column density covers 4 orders of magnitude. The \MOS{} curves are overlaid as white dotted lines. PV coordinate axes are shown in black dotted lines.}
\label{fig:PV_vw100_1Bp_45deg_CompressedWA}
\end{figure*}

The signatures of the compressed magnetized winds appear on the PVDCD maps as extended threads of CDIEs connecting the outskirts of the ``jet'' with wiggling velocity centroids and the ``sporadic'' patches around the origin and the $y$-axis, confining the broad velocity feature at the jet base across the RS cavity. These are nonexistent in Figure \ref{fig:HDshell_pv_no-slip}.
These features arising from the compressed regions with mixing can be viewed on the PVDCD maps filtered with the $f$ values between $0$ (pure ambient material) and $1$ (pure wind material). With this procedure, the gas of the primary wind and the ambient toroid is subtracted from the contributions to the PVDCDs maps. One can probe directly the PV patterns contributed by mixing. Figure \ref{fig:PV_vw100_1Bp_45deg_CompressedWA} is one such map for the strongly magnetized ambient medium $\alpha_b=1$, with both the pure wind $f$ very close to $1$ and pure ambient material $f$ very close to $0$ subtracted. The panels capture the spatial and spectral distributions of mixing resulting from the interplay.

These patterns suggest that identifying the patchy or feather-like structures by their additional occurrence in the PVDCDs compared with the reference ones is appropriate and consistent with their proposed origins in the internal structures of magnetized elongated bubbles. The intermediate velocity components (IVCs) on the PVDCDs may appear as part of these extended threads of CDIEs when local excitation conditions meet the requirements for the tracing molecular or atomic lines for the jets and winds in Section \ref{subsec:velocity_components}. These can further connect to the even lower-velocity components (below).

An extended patch of low-velocity features is naturally present in Figures \ref{fig:PV_vw100_Bp0p0_45deg}, \ref{fig:PV_vw100_Bp0p1_45deg},  \ref{fig:PV_vw100_Bp1p0_45deg} and \ref{fig:PV_vw100_1Bp_45deg_CompressedWA}
throughout the parameter space. Because of the conventional knowledge of the MC thin-shell models in the HD regime, the occurrence of low-velocity components (LVCs) near the origin of the PV coordinates is always expected. The LVCs near the origin of PV coordinates are thus considered to be part of the traces for the MC or MS contours. Distributions of the low-velocity profiles outside of the elliptical contours of the MC/MS curves along the $y$-axis increase with the magnetization of the winds, which are more evident for the $n=1$ and $2$ systems and more in the $\alpha_b>0$ than the $\alpha_b=0$ ones. The magnetized gap regions appear spur-like along the $y$-axis for the strongly magnetized ambient $\alpha_b=1$, indicative of the vortices aligned to the lobe surface confined by the ambient poloidal cocoons.

We note that the disappearance of the observed increase of velocity in the jet component in Figure \ref{fig:PV_vw100_1Bp_45deg_CompressedWA} suggests the phenomenon is connected to the intrinsic wind, free of ambient materials. That is, this acceleration phenomenon is not caused by mixed or entrained material.
We note this apparent acceleration is connected to the needle-like appearance at the tip in the 2D density distribution in our Figures \ref{fig:2}, \ref{fig:3}, and \ref{fig:5}--\ref{fig:8}, at the far end of the post-RS regions before hitting the FS\@. The needles were attributed to magnetic forces that accelerate $v_r$ while $b_\phi$ smoothly decreases inside the tip region. The last paragraph of Section 5.2 in \citetalias{shang2020} gives a first account of such possibility, which we will explore further in Paper III\@.

In the magnetic interplay, pseudopulses are generated in the post-RS compressed wind regions. 
This is evident in PVDCDs as oscillations in jet velocities downstream behind the RS cavities but upstream of the FS\@. They appear and operate cooperatively with the generation of vorticity through magnetic forces as described in Section \ref{subsec:pseudopulses} and in Figure \ref{fig:mag_forces} shown in Appendix \ref{sec:appendix:mag_forces}. 
When the fingers grow sufficiently large around the RS cavity as they advect, they generate the impression of multiple converging ``zigzag'' shells mimicking \textbf{episodicity} along the cavity as shown in Figures \ref{fig:5} and \ref{fig:6}. In the extended compressed wind region, the KHI fingers advect and coalesce, and larger interconnected fingers form, as part of the feedback loop in Equation (\ref{eqn:vorticity_generation}). When the large fingers merge near the jet proper, the feedback loop operates across the whole width of the lobe, leading to perturbations of density and velocity on the jet. Variations in density and the radial velocity component $v_r$ advected, projected onto the PVDCD, will appear as small quasiperiodic wiggles around the jet velocity centroids propagating toward the tip. Such variation in $v_r$ along the axis is mild, and occurs with shorter periodicity, extended along the upper portion of the whole outflow lobe. As the actual patterns depend on local properties as discussed in Section 8 of \citetalias{shang2020}, asymmetry could occur across the lobes.

Real pulses are varying incoming wind velocities or mass-loss rates from the base wind launching region, likely caused by episodic accretion and ejection in the disk or magnetic star--disk interaction, which will form their respective projected velocity centroids on PV diagrams. The expected amplitudes and frequencies of the episodic ejections may be different from those of the pseudopulses. The new ejections will make multiple larger direct shifts in the projected jet intensity peaks near the base as distinct knots or blobs, rather than occurring post-RS at some large distances. The shapes and velocity patterns are best manifested in Figures 7 and 8 of \citet{wang2019}, where the ejected knots can jump around distinct velocity centroids, interact and coalesce to form large-scale patterns, as in the case of IRAS 04166+2706 \citep{tafalla2010}.
We note that in a real series of pulses, a new reverse shock cavity forms near the base with each single ejection, and due to the density profile, the perturbations appear strongest on the axis rather than on the wide-angle portion. Details of the physics will be explored in depth for the coverage of parameter space at high numerical resolutions in follow-up publications.

\section{The Transverse Position--Velocity Diagrams of Column Densities}
\label{sec:perp_PV}

In this subsection, we explore the PVDCDs from cuts made perpendicular to the jet axes. Such PVDCDs are particularly interesting in probing the kinematic features caused by asymmetry in flow kinematics, especially those due to rotation. Despite the essential absence of rotation, the internal structures of the axisymmetric elongated magnetized bubbles can carry peculiar kinematic signatures revealing the nested layers of multicavities in the transverse PVDCDs generated.

The bubble signatures in these perpendicular PVDCDs serve as nonrotating baselines to avoid misinterpretation of the observational patterns. We note that the magnitude of poloidal velocity dominates the major features explored here, and the presence of rotation will not alter these results. On these scales, effects of rotation from launch and those of the collapsing rotating toroids will be small. These may likely complicate the signatures with small systematic skews or twists in the final patterns in real maps. 
The patterns displayed by the PVDCDs vary according to the internal structures traversed and sampled along the line of sight with each of the cuts. 
The maps are sensitive to the loci of the cuts on the jet axes in spite of the systematic self-similarity.

\subsection{Perpendicular PV Maps Made with Multicavities}

\addtocounter{figure}{1}
\begin{figure*}
\figurenum{\arabic{figure}}
\epsscale{0.8}
\centering
\plotone{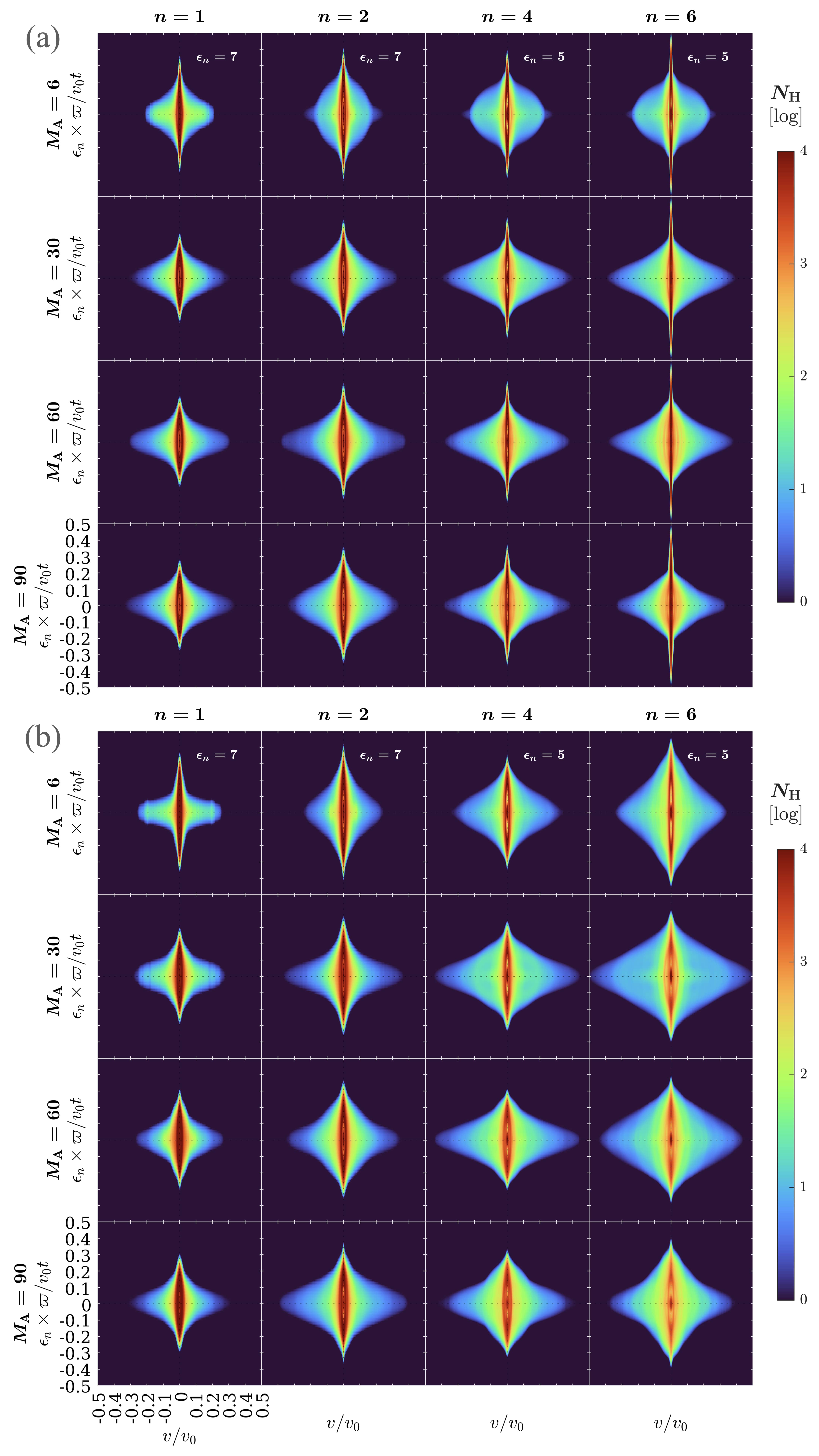}
\epsscale{1.0}
\caption{
Position--velocity diagrams of column density (PVDCD) perpendicular to the jet axis at a location of $z/v_0t=0.05$. The PVDCDs are shown at an inclination angle of $i = 90\arcdeg$ for (a) $\alpha_b=1$ (top panels) and (b) $\alpha_b=0$ (bottom panels) cases with $n=1$, $2$, $4$, and $6$ toroids (from left to right) and $\MA=6$, $30$, $60$, and $90$ winds (from top to bottom). The unshocked ambient material with $v_p < \aamb$ has been filtered out while producing the PVDCDs. The rescaled logarithmic column density covers 4 orders of magnitude. For each $n$-column, an associated exaggeration factor $\epsilon_n$ magnifies $\varpi$ to show detail structures ($\epsilon_n=7$ for $n=1$, $2$, and $\epsilon_n=5$ for $n=4$, $6$). PV coordinate axes are shown in black dotted lines.
}
\label{fig:perpPV_vw100_1Bp-0Bp}
\end{figure*}
\begin{figure*}
\figurenum{\arabic{figure}}
\epsscale{0.8}
\centering
\plotone{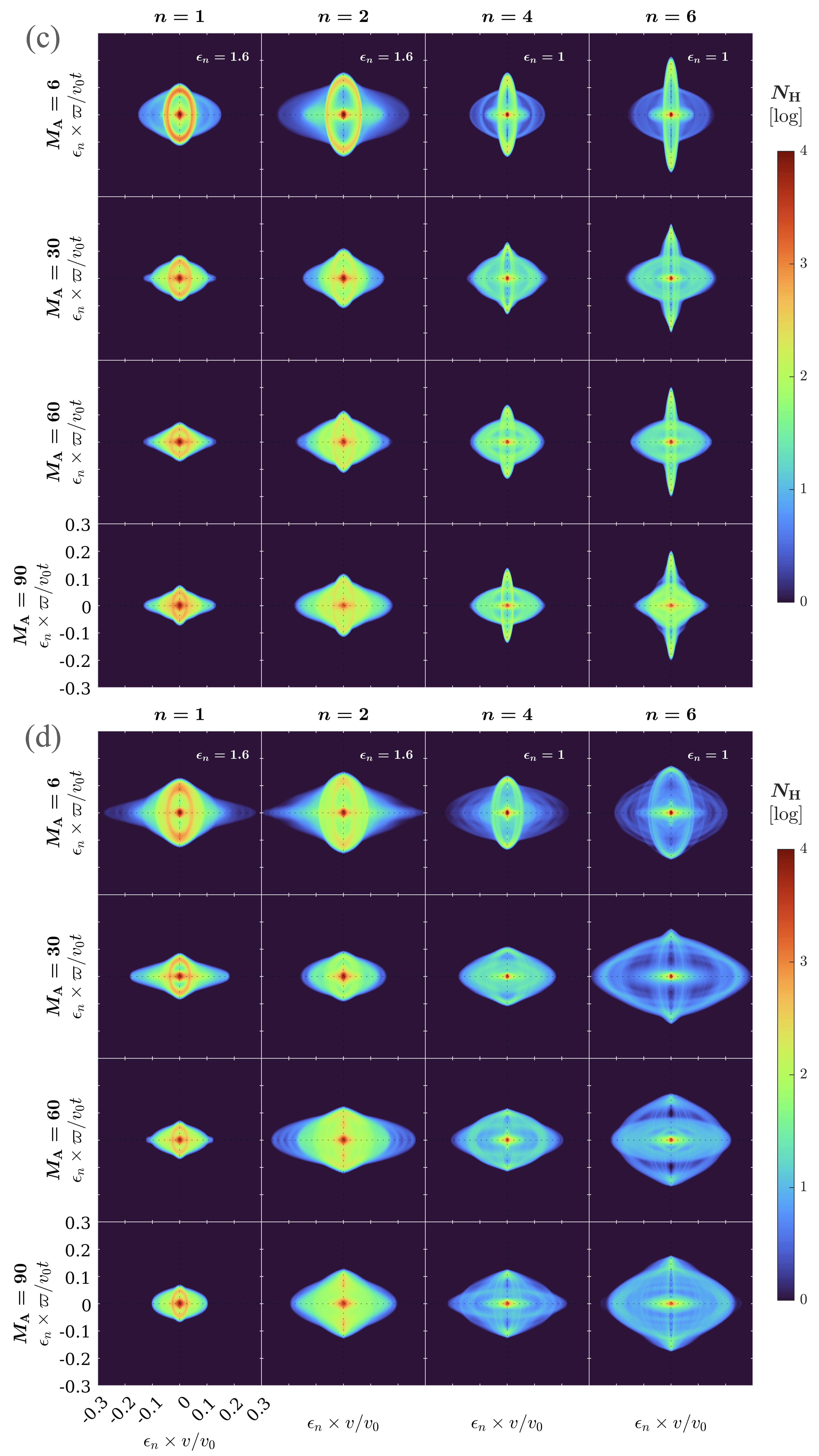}
\epsscale{1.0}
\caption{
PVDCDs perpendicular to the jet axis at a location of $z/v_0t=0.25$, shown at an inclination angle of $i = 90\arcdeg$ for (c) $\alpha_b=1$ (top panels) and (d) $\alpha_b=0$ (bottom panels) cases with $n=1$, $2$, $4$, and $6$ toroids (from left to right) and $\MA=6$, $30$, $60$, and $90$ winds (from top to bottom). The unshocked ambient material with $v_p < \aamb$ has been filtered out while producing the PVDCDs. The rescaled logarithmic column density covers 4 orders of magnitude. For each $n$-column, an associated exaggeration factor $\epsilon_n$ magnifies both $\varpi$ and $v$ to show detail structures ($\epsilon_n=1.6$ for $n=1$, $2$, and $\epsilon_n=1$ for $n=4$, $6$). PV coordinate axes are shown in black dotted lines.
}
\end{figure*}
\begin{figure*}
\figurenum{\arabic{figure}}
\epsscale{0.8}
\centering
\plotone{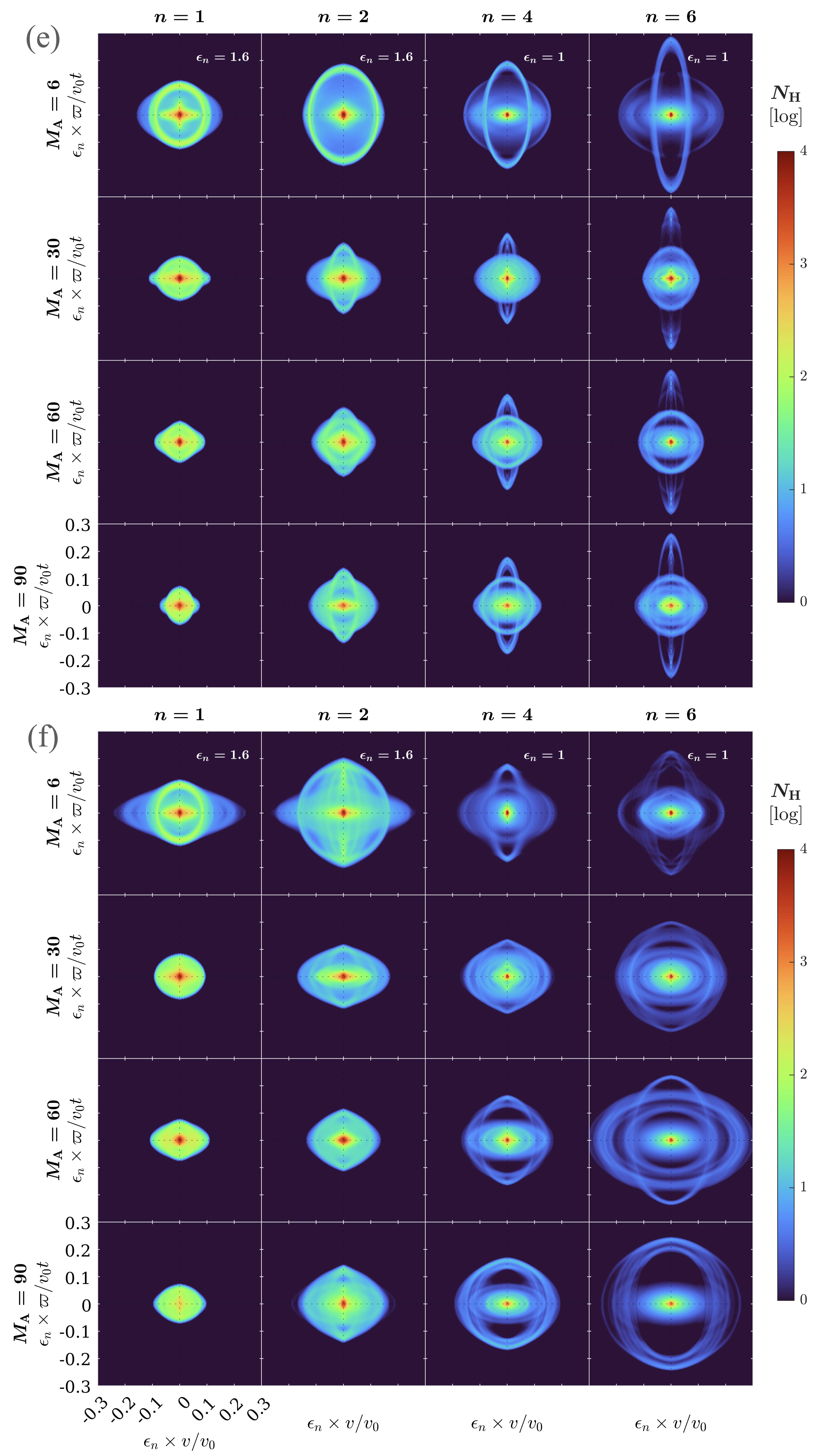}
\epsscale{1.0}
\caption{
PVDCDs perpendicular to the jet axis at a location of $z/v_0t=0.5$, shown at an inclination angle of $i = 90\arcdeg$ for (e) $\alpha_b=1$ (top panels) and (f) $\alpha_b=0$ (bottom panels) cases with $n=1$, $2$, $4$, and $6$ toroids (from left to right) and $\MA=6$, $30$, $60$, and $90$ winds (from top to bottom). The unshocked ambient material with $v_p < \aamb$ has been filtered out while producing the PVDCDs. The rescaled logarithmic column density covers 4 orders of magnitude. For each $n$-column, an associated exaggeration factor $\epsilon_n$ magnifies both $\varpi$ and $v$ to show detail structures ($\epsilon_n=1.6$ for $n=1$, $2$, and $\epsilon_n=1$ for $n=4$, $6$). PV coordinate axes are shown in black dotted lines.
}
\end{figure*}

We illustrate the situation using the cases of strongly magnetized $\alpha_b=1$ and nonmagnetized $\alpha_b=0$ ambient media with cuts made at three different positions in the self-similar coordinate $z/v_0t$ with widths of $0.05v_0t$. The line-of-sight inclination is chosen perpendicular to the jet axis.
Each cut passes through different parts of the outflow structure. The jet feature is concentrated to $\varpi=0$ and $v=0$ as a red dot surrounded by concentric shadows around it.

A cut passing through the lower portion of the outflow structure at $z/v_0t=0.05$ encounters the densest part of the multicavities, which include the compressed ambient material that ends with the FS of largest $\varpi$-extent as the elongated narrow rings along the position $\varpi$, and the compressed wind regions that form ovals gradually extending horizontally along the $v/v_0$ direction. 
The cut also passes through
the lines of sight that will penetrate the lowest part of the primary RS cavity that contains the fast-moving wide-angle free wind with largest velocity dispersion. This is reflected in the largest velocity width (in $v/v_0$), as shown in Figure \ref{fig:perpPV_vw100_1Bp-0Bp} (a) (for $\alpha_b=1$) and (b) (for $\alpha_b=0$). The very faint CDIE at this largest width at $\pm v/v_0$ gives the sense of the free wind.

The cut made at $z/v_0t=0.25$ (Figure \ref{fig:perpPV_vw100_1Bp-0Bp} (c) and (d)) has less dispersion in $v/v_0$ as the lines of sight usually traverse the upper portion of the free wind, which has mostly upward-pointing velocity vectors, and always traverse the compressed wind regions. The velocity vectors are deflected after they pass through the reverse shocks. The traversed regions cover a significant fraction of the multicavities, which are the sites of the instabilities, the compressed wind, the compressed ambient medium, and outer FS\@. The ring contributed by the compressed ambient medium is elliptical in $\varpi$, and apparently wider in $v$ than at the lower $z$-cut. Another ellipse, corresponding to the compressed wind, is even more extended in $v$, and may be visible for larger $n$ cases.

Moving further up in $z/v_0t$, the lines of sight traverse the middle of the outflow lobe at $z/v_0t=0.5$ (Figure \ref{fig:perpPV_vw100_1Bp-0Bp} (e) and (f)). At this height, the lines of sight pass through the pseudopulses, wind-bubble edge, multiarcs, and instabilities, integrating them into the generated PVDCDs. Ovals corresponding, respectively, to the compressed ambient medium and compressed wind are present in our images. For larger $n$, both ovals drop in CDIE with respect to lower heights, even more so for the ambient oval, which is also stretched in the $\varpi$-direction. For smaller $n$, the two ovals become virtually indistinguishable.

The outer $\varpi$ edge of the compressed wind is visible in the PVDCDs as an oval region. At lower heights, some of the free wind is present in our cuts, and its feature in our PVDCDs is often narrow in $\varpi$ and very extended in the $v$-direction, and typically corresponds to low (sometimes very low) CDIE values.
 In many practical cases, only the higher CDIE portions or features will be observationally detectable.

One signature that emerges when one compares the $\alpha_b=1$ and $\alpha_b=0$ cases is the feature of the ``gap'' region threaded by the compressed strong ambient poloidal field. The prominent CDIE peaks protrude out on the $\varpi$-axis for $\alpha_b=1$, but are absent for the $\alpha_b=0$ cases. This feature appears around the zero velocity. It reflects the compressed ambient regions that are cushioned by the ambient poloidal field near the base, projected toward and away along the line of sight. This is most evident in the configurations $n=4$ and $n=6$, for all $\MA$ values; although for smaller $n$, it may be difficult to distinguish it from the wind ovals.

The restriction to brighter CDIE parts still makes the compressed wind and ambient features valuable for PV observations, applicable both to systems with and without rotation. For example, Figure A.2 of \citet{louvet2018} shows the expected characteristics in a series of transverse PV diagrams of the edge-on system HH 30.
These panels,
if rotated $90\arcdeg$ to show velocity axis horizontally and position axis vertically,
shown from $y=0\farcs35$, then to $y>1\arcsec$, and up to $y\approx2\farcs5$, 
appear to follow the general features of the height variations of compressed materials shown in the different cuts of transverse PVDCDs for the relevant $n\gtrsim6$.
They vary from filled oval shapes to relatively unfilled ring-like shapes as one examines from the bottom to about the mid-height of the outflow lobe.

\subsection{A Schematic Two-Shell System}
\label{subsec:perpPVDCDsketch}

\begin{figure*}
\plottwo{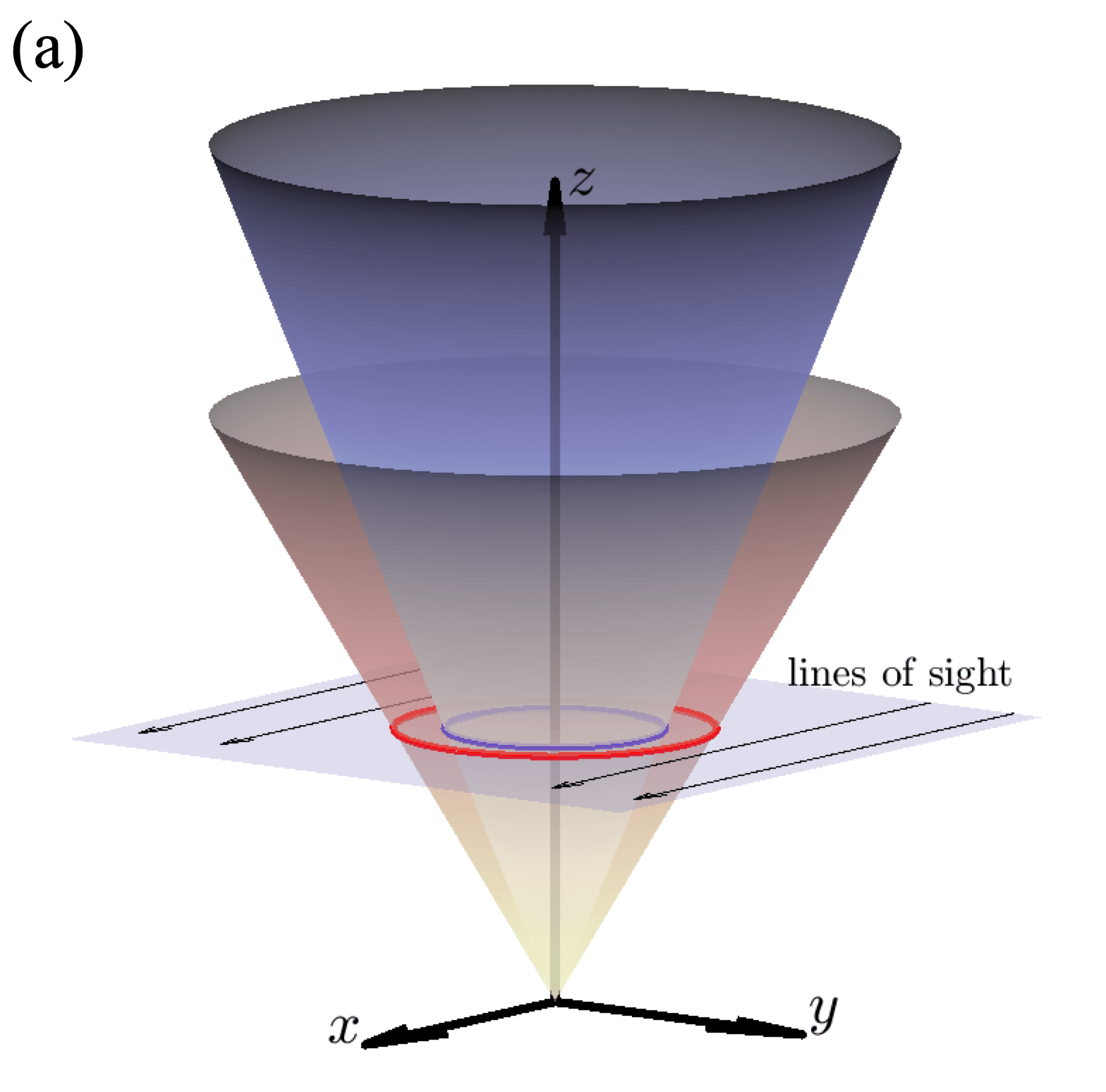}{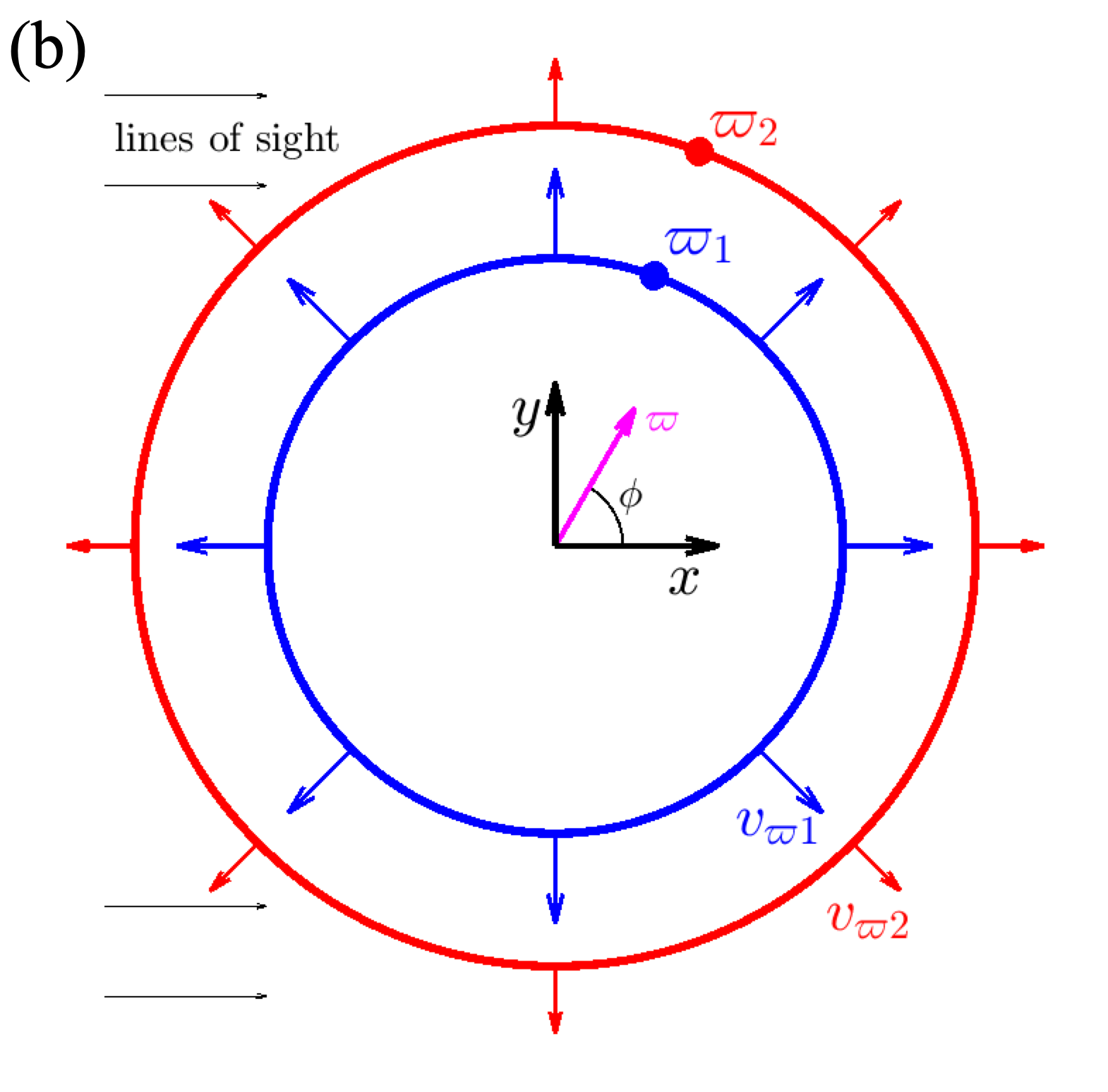}\\
\plottwo{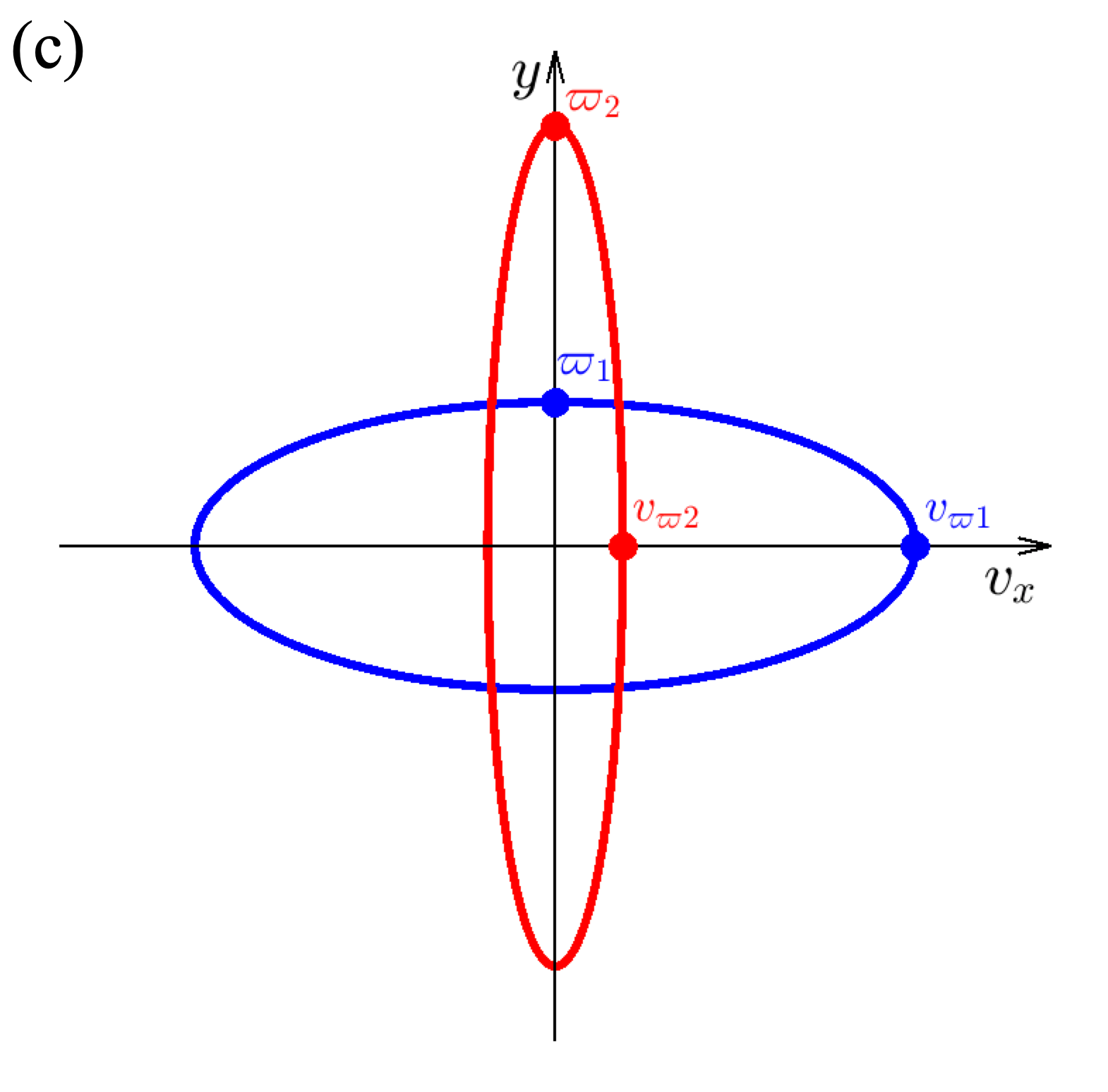}{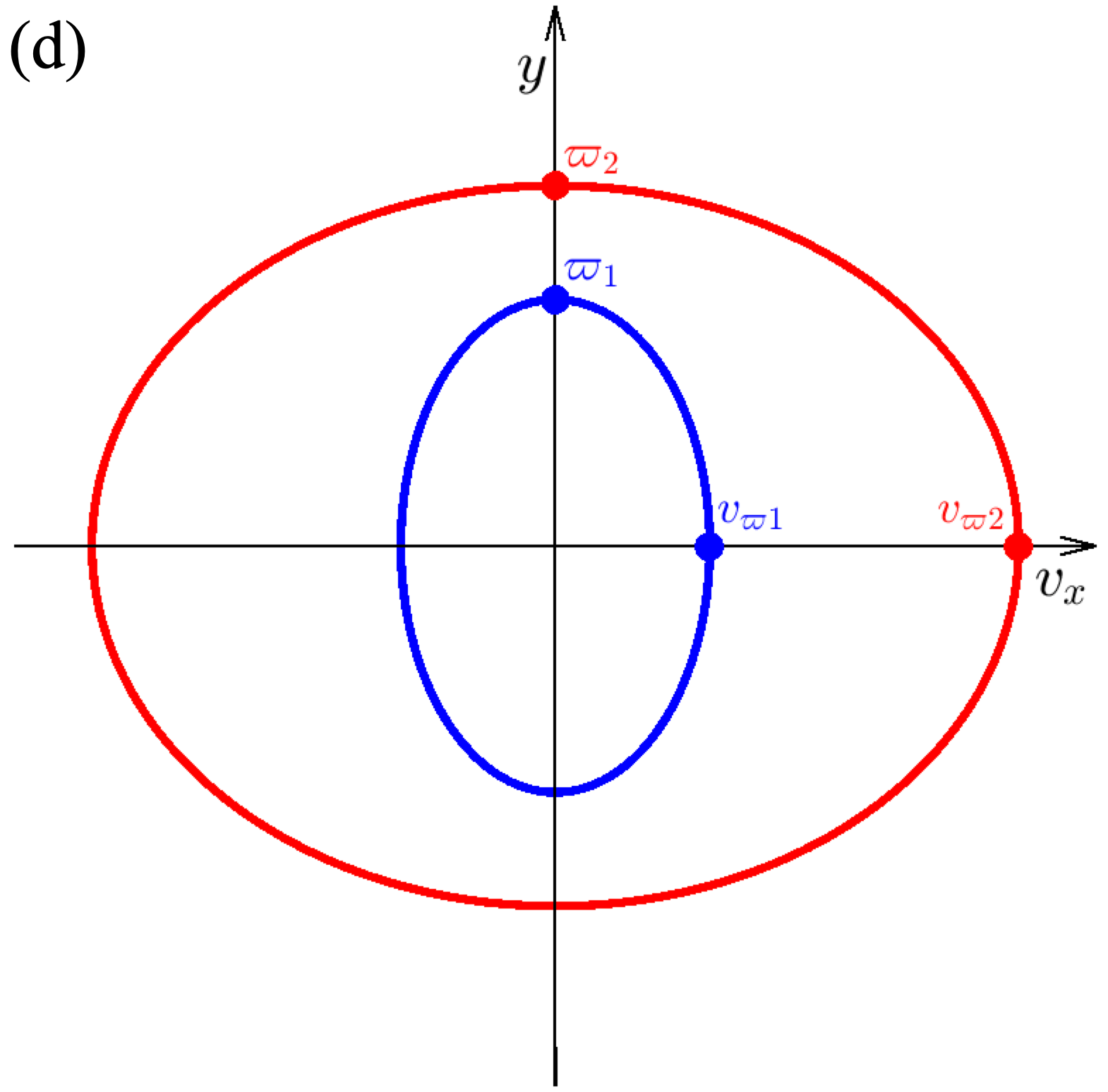}\\
\includegraphics[width=0.22\textwidth]{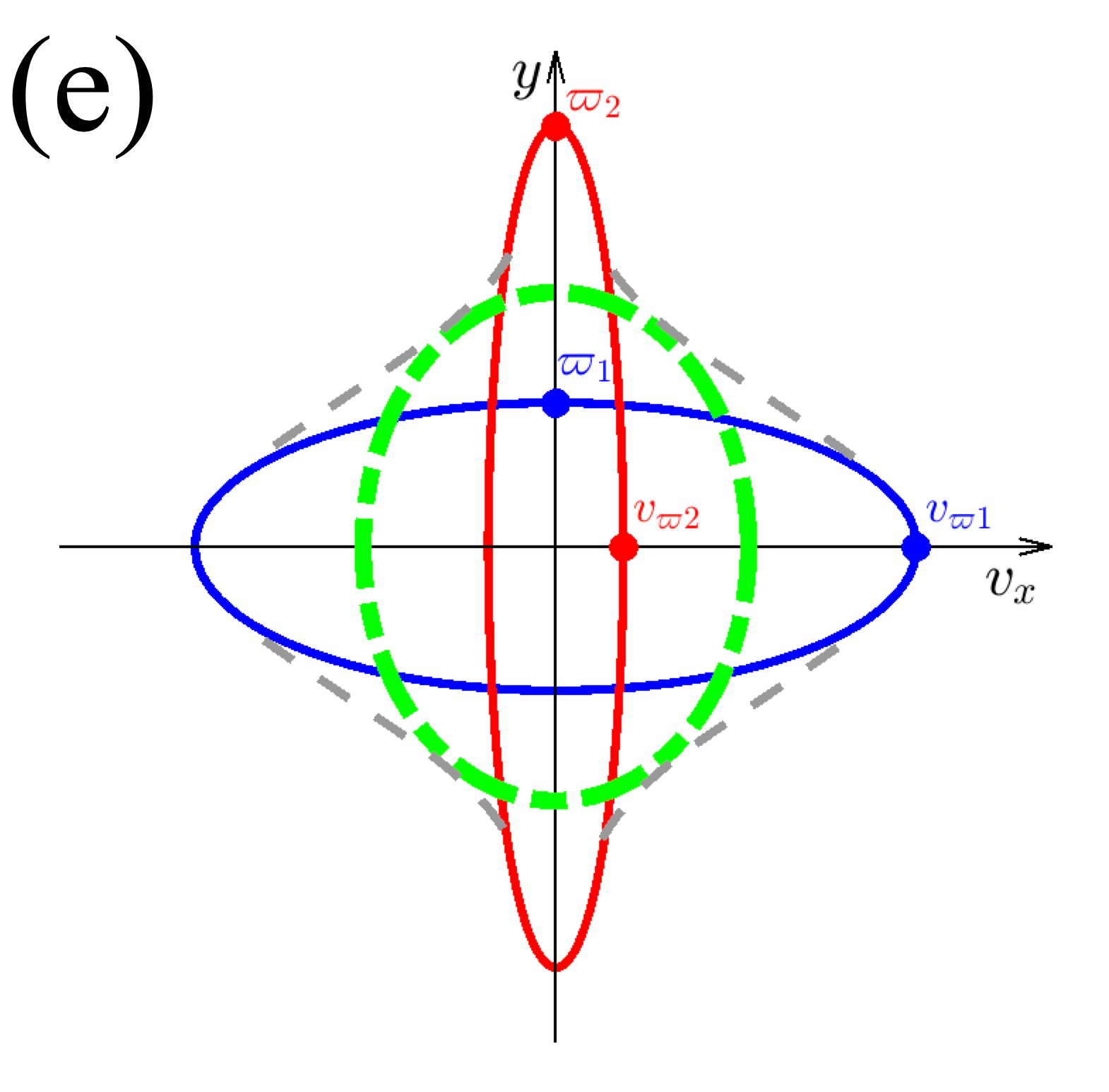}\
\includegraphics[width=0.22\textwidth]{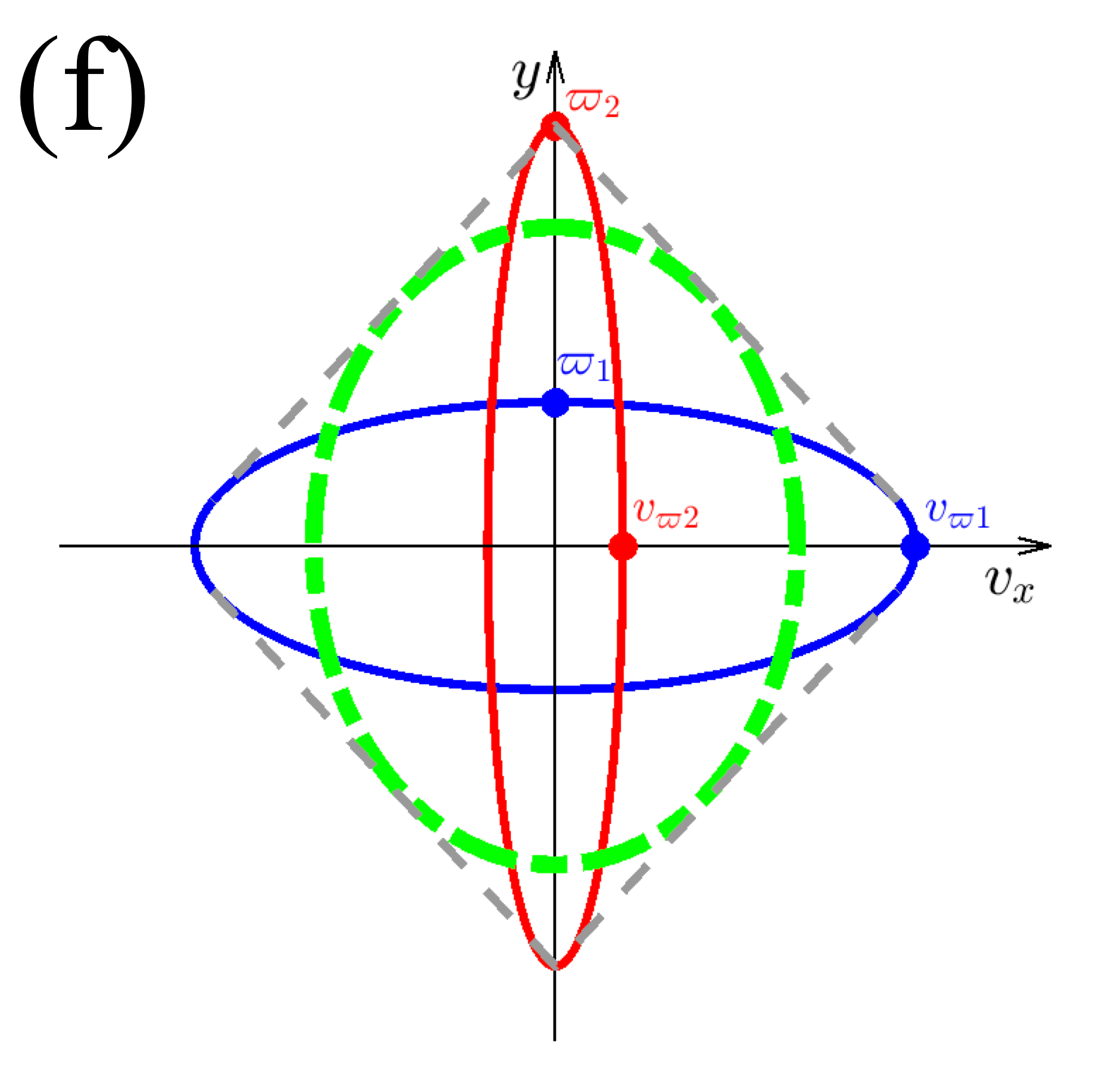}\
\includegraphics[width=0.22\textwidth]{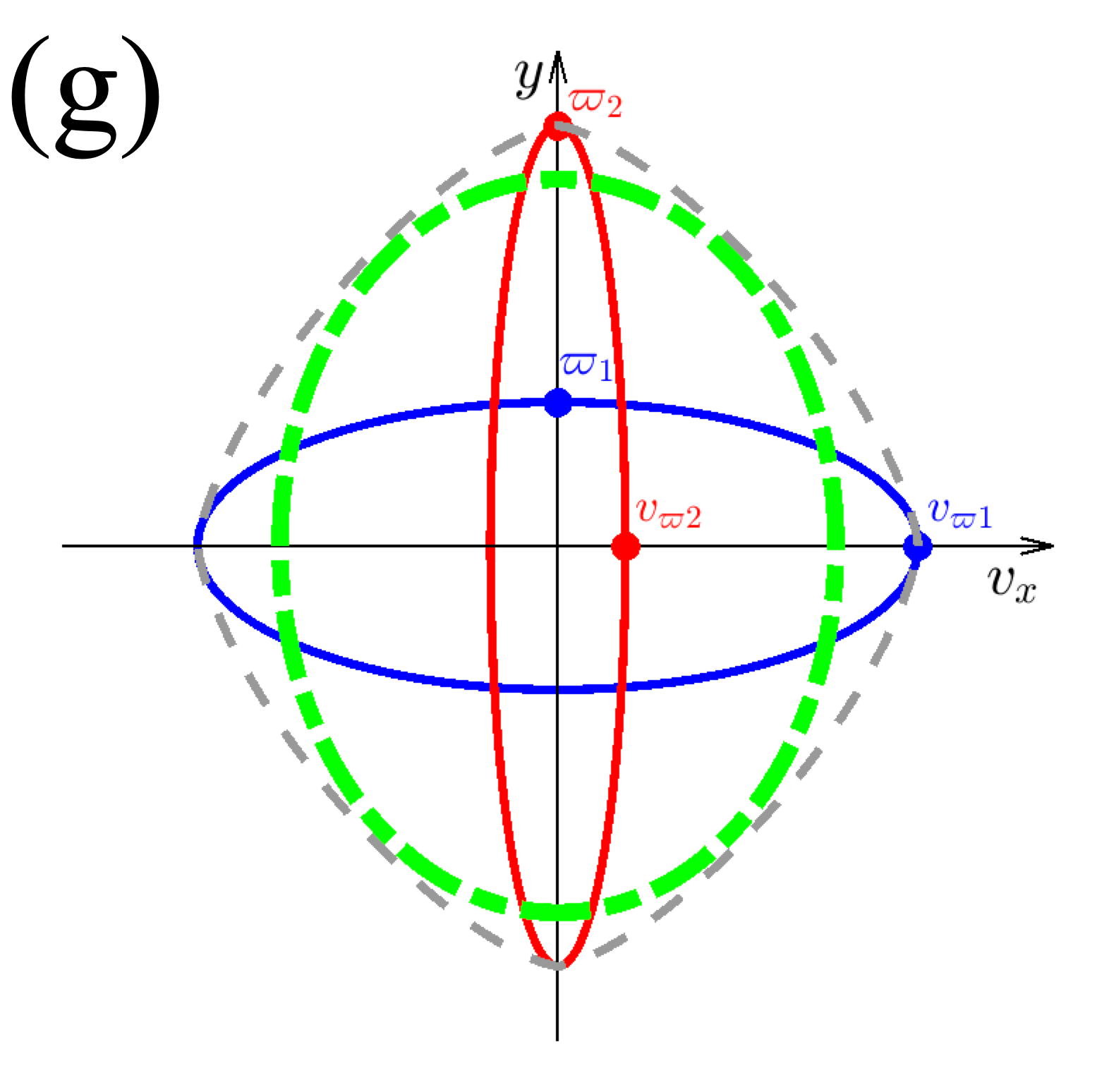}\quad\quad\quad\quad
\includegraphics[width=0.22\textwidth]{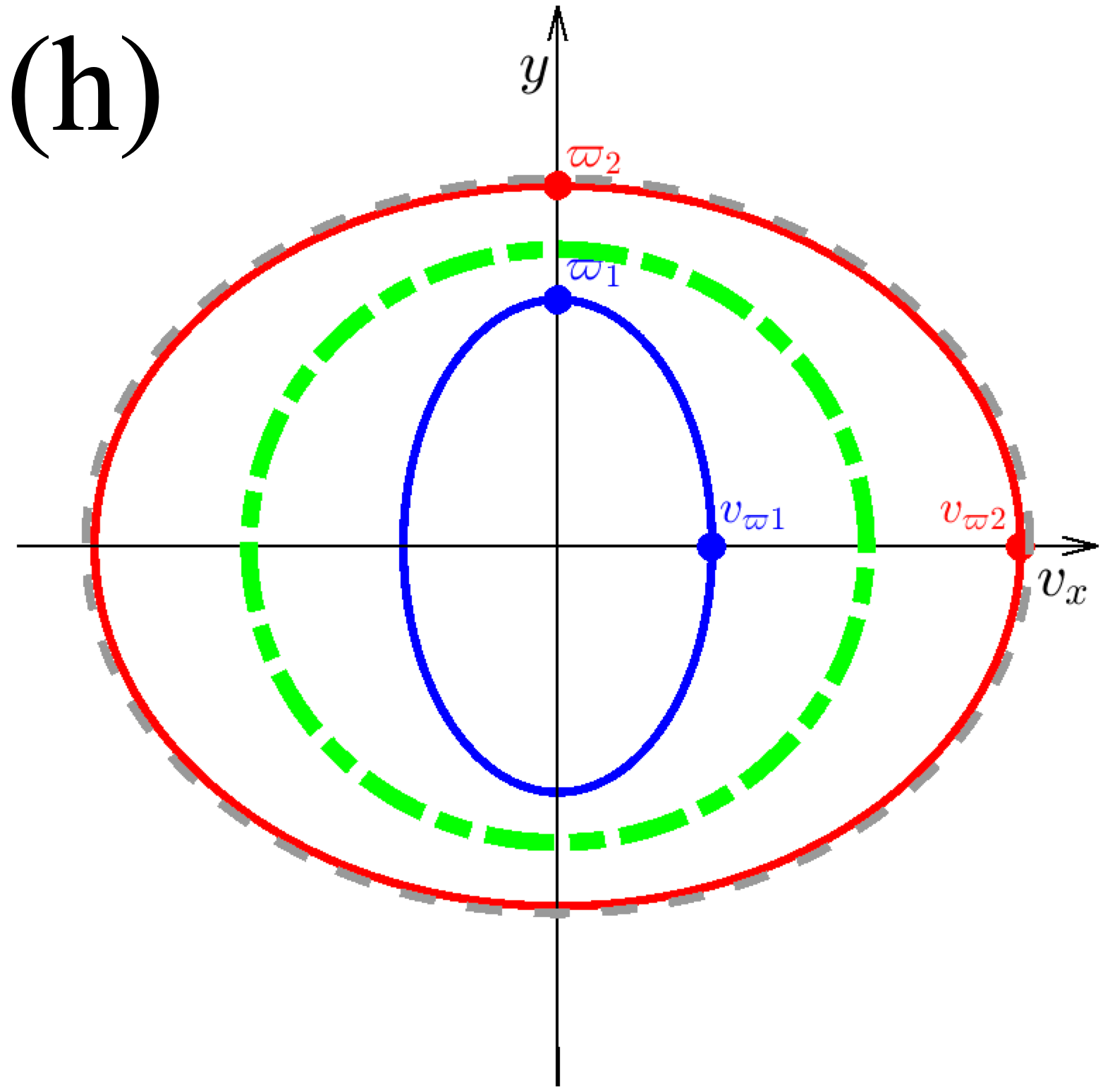}
\caption{Schematic illustrations of transverse PVDCDs. Panel (a) is a 3D sketch of a transverse slit cut through an outflow lobe having two nested cones (representing outflow features). Blue and red circles mark the intersections of 3D cones with the plane cut. In this figure, the outflow axis is along $z$, the line of sight is along $x$, and $y$ is the position axis of the transverse PVDCDs. In panel (b) the two cones are seen on the plane of the cut. The inner and outer cones are, respectively, cut at the cylindrical radii $\varpi_1$ and $\varpi_2$, with $\varpi$ velocity components $v_{\varpi1}$ and $v_{\varpi2}$. The transverse PVDCDs of (b) are shown in two cases: (c) $v_{\varpi1}\!>\!v_{\varpi2}$ (two intersecting ellipses) and (d) $v_{\varpi1}\!<\!v_{\varpi2}$ (two nested nonintersecting ellipses). Panels (e)--(h) show the results of adding a hypothetical cone (green dashes) representing a wide feature spreading from blue to red (details in Section \ref{sec:perp_PV}). The addition of the green series leads from (c) either to (e) crossed ovals, (f) a rhombus, or (g) a single oval envelope of intercrossing ellipses, and then from (d) to (h) concentric ellipses. The total sum of the blue-green-red series is indicated in light-gray dashes.}
\label{fig:transversePV_schematics}
\end{figure*}

We further clarify the situations conveyed in the perpendicular cuts made at different heights by a schematic illustration in Figure \ref{fig:transversePV_schematics} for the expected appearance of PV density diagrams for a multiple shell. We start from two shells by drawing two nested cones with poloidal velocity higher for the inner one, and sketch the expected PVDCDs for an imaginary slit perpendicular to the flow axis.
Panels (a) and (b) in Figure 
\ref{fig:transversePV_schematics} 
illustrate in 3D the two shells simplified as cones, cut by a slit at constant height $z$ and viewed along the $x$-axis, where the resulting cut is shown in 2D position space in panel (b).
The two curves produced by the cut are represented by two circles, inner and outer, labeled as 1 (blue) and 2 (red). Here, the position coordinates on these circles are simply called $x=\varpi\cos\phi$ and $y=\varpi\sin\phi$. The velocity component of this nonrotating flow in the plane of the panel is $v_\varpi=v_r\sin\theta+v_\theta\cos\theta$, often $v_\varpi\approx v_r\sin\theta$ for the flows considered here, $v_r\gg v_\theta$. 

Projecting the circles in panel (b) along the line of sight into the position--velocity space leads to the situations in panels (c) and (d) of Figure \ref{fig:transversePV_schematics}. Here the position coordinate (labeled as $y$) is $y=\varpi\sin\phi$, and the velocity coordinate is $v_x=v_\varpi\cos\phi$, which give the mathematical expression of an exact ellipse with the corresponding semiaxes $\varpi$ and $v_\varpi$. The ellipses would cross each other as in panel (c) or not, as in panel (d), depending on whether the velocity component $v_{\varpi2}$ of the outer red cone is smaller or larger than that of the inner blue cone, $v_{\varpi1}$. The function $v_\varpi\approx v_r\sin\theta$ can grow with $\theta$ because of the sine factor (leading to case (d)), but it can also decrease with $\theta$ when the poloidal velocity decreases fast enough from the inner to outer parts inside the compressed flow region (leading to case (c)), a typical effect in a compressed wind bubble. Both cases may be observed in perpendicular PVDCDs, for different slits, cases, or features. The inner (blue) and outer (red) cones may be separate features, in which case the final PVDCD is either as two crossed ellipses $\,\ellipsesc\,$ in panel (c), or as two concentric ellipses $\!\ellipsesd\!$ or $\ellipsesdv$ in panel (d).

We also consider the alternative case in which the two cones are the inner and outer end of a continuous structure of cones, where $v_\varpi$ is either monotonically decreasing with $\varpi$, or monotonically increasing. Panels (e)--(h) in Figure 
\ref{fig:transversePV_schematics} explore different cases, with a green dashed ellipse exemplifying the cones intermediate to the blue and red boundaries. The overall shape summing the set of intermediate ellipses is indicated with light-gray dashed lines of various shapes. If $v_\varpi$ decreases with $\varpi$, the intersecting blue and red ellipses will be located as in panel (c). The intermediate ellipses may be rather small in PV space, leading to an
overall shape enveloping the crossed blue, green, and red ovals in panel (e). The intermediate ellipses may be rather large, leading, in panel (g), to the overall
shape of a bulging oval enveloping the set of large intercrossing ellipses.
Panel (f) shows the intermediate case for which the envelope of the ellipses takes the shape $\blacklozenge$ of a rhombus.
For the situation in which $v_\varpi$ increases with $\varpi$, the blue and red ellipses will be located as in panel (d) in Figure 
\ref{fig:transversePV_schematics}.
The intermediate ellipses would also be noncrossing and concentric, leading to the overall shape of a set of concentric ellipses shown in panel (h). Within the overall shapes described in (e)--(h), a hollow region may exist, large or small according to the behavior of the set of ellipses at small values of the PV coordinates.

Examples of these overall shapes are present in different features and panels of Figure \ref{fig:perpPV_vw100_1Bp-0Bp}.
For example the rhombus shapes are present in Figure \ref{fig:perpPV_vw100_1Bp-0Bp}(c) and (d) for most panels with $n=2$ and nearly all panels for (d). The intersecting ellipses or ovals are present in Figure \ref{fig:perpPV_vw100_1Bp-0Bp}(a), (c), and (e) for most of the panels with $n=4$ and $6$. More examples are shown in Appendix \ref{sec:appendix:perpPVDCDshapes}.

In real observations, some of these ellipses, crosses, and rhombuses may be seen only in parts due to natural limitations of the slit length or width, and of the excitation conditions of different parts of the source. For instance, the shapes like parentheses `( )' in \citet{louvet2018} may be explained as ellipses clipped by limits of position or velocity coverage. Inclinations different from $90\arcdeg$ may also introduce complexity of shapes due to projection effects. The presence of $v_\phi$ components in either wind or ambient material also leads to a more complex picture, such as in the ellipses tilted with respect to the PV axes shown in the diagrams of Appendix B in \citet{tabone2020}, or other factors such as asymmetry in the flow structures.

\section{Observational Implications and Discussions}
\label{sec:discussions}

In what follows, we discuss the theoretical interpretations and the implications of our results, in the context of existing models and of recent observational results.

\subsection{The Thin-Shell Models}
\label{subsec:thin_shell_models}

We start with the thin-shell models. In our \citetalias{shang2020} and this work, we have demonstrated important internal structures as bubbles exist inside the outflow lobes. This holds true for the thin shells, which were the state-of-the-art for the hydrodynamic regimes of the smooth model assumptions. 

In generating the hydrodynamic MC curves, we utilize a hydrodynamic deduction based on mass and momentum conservation inside a thin shell, but for a general ratio of wind--ambient mass densities. Despite the different deduction, the first appearance of Equation\ (\ref{eqn:mc_eqn}) is notably shown with Equation\ (12) in \citet{schwartz1978}, starting from the discussion below Equation\ (7) in \citet{mckee_cowie1975}, and was nonetheless attributed to \citet{mckee_cowie1975}. Likewise, Equations (13)--(17) in \citet{lee2001}, in which wind mass is included into \citetalias{shu1991}, lead to the same expression Equation (\ref{eqn:mc_eqn}) as discussed above.
These thin-shell MC curves are often good predictors of the simulated lobe shape curves (including some of the cases in which the wind-bubble shell is not so thin). In this work, we utilize them to construct and to compare with HD models in Sections \ref{subsubsec:thinshells}, \ref{sec:2Dinterior}, \ref{subsec:MCcurves_PV}, and Appendix \ref{sec:appendix:steps}. We also use them as a baseline to compare with simulations of PVDCDs for the HD cases in Sections \ref{subsec:HD_noslip_PV} and \ref{subsec:HD_slip_PV}. 

These thin shells with the locally oblique structures we implemented in Section \ref{sec:2Dinterior} serve as the HD limit for bubbles with smooth and laminar flow across the shocked regions bounded by the RS and FS without the onset of KHI\@.
These treatments are fundamentally different from driving an outflow by entrainment, and the same applies to all of our MHD models in this work.

\subsection{The High-, Intermediate-, Low-Velocity Components}
\label{subsec:velocity_components}

In Section \ref{sec:intro} and the Introduction of \citetalias{shang2020}, we review the observational evidences of the jets and outflows from the deeply embedded protostars to the optical jets of the revealed T-Tauri systems, based on which our theoretical framework has been built and developed (references therein). Notably there are those very collimated molecular jets and outflows with prominent SiO and CO jet emissions from Class 0 protostars (e.g., HH 211, HH 212, L1448C(N), and IRAS 04166+2706), while losing the SiO emissions before entering the Class I phase. The wide cavities, the highly collimated shapes, and large terminal and internal bow shocks establish the foundations of the outflow paradigms. 

Kinematically, the features of the wind-driven and jet-driven paradigms of the protostellar outflows are best summarized in the velocity components. The jet--shell signatures are identified as the extremely--high-velocity (EHV) and low-velocity (LVC) components with the jet signatures associated to the EHV, and the thin shells to the LVC, respectively. Between the EHV ($\sim100\kms$, used interchangeably with the HVC) and the LVC, there exists the standard--high-velocity (SHV) component (a few to $\sim20$--$50\kms$, also as the IVC here), and extended LVC (a few kilometers per second down to systemic velocity) that are often present in Class 0/I sources \citep[e.g.,][]{bally2016,bachiller1996}.
These are what may link to the signatures of the magnetic interplay identified in this work (Section \ref{sec:PVD}). All of these components appear in our predicted parallel and transverse PVDCDs as signatures of jet-bearing elongated magnetized bubbles.

Recently, evidence has been reported of another phenomenon, the so-called ``slow molecular winds'' arising in Class 0/I sources at scales of $\lesssim 2000\au$, and apparently ``rotating'' in the same sense as their gaseous disks. They are often inferred to root at radii $\sim 10$--$100\au$ from the protostars, e.g.\ the review of \citet{pascucci2022_PPVII}. They are often associated with cavities or shells nested like onions. These systematic characters match well with the morphological and kinematic features of the elongated magnetized ``bubbles'', especially the multicavities in \citetalias{shang2020} and the extended IVC and LVC connecting the HVC (EHV) and the systemic velocity demonstrated in this work. The CO $J=2-1$ emission from TMC-1A appears to open up with increasing angles as the velocities approach the systemic velocity \citep{bjerkeli2016}, with ``apparent footpoints'' ranging between $5$ and $25\au$ based on \citeauthor{anderson2003} (\citeyear{anderson2003}; see Section \ref{subsec:notsosmallradii}).
\citet{harsono2021} resolved the molecular line emission SO, HCN, and HCO$^+$ in TMC-1A, and their Figures 5, 11, and 12 outline structures similar to multicavities.    
The multicavity of HH 212 is suggested in \citetalias{shang2020}. We compare additional kinematic features traced by SO/CO/SiO in Section \ref{subsec:HH212}.

Some T-Tauri sources also have visible multicavity structures surrounding the central optical/IR jets, by CO/H$_2$ nested cones, with ``nested'' velocity structures. We note the presence of the nested multicavities in the source DG Tau B \citep{devalon2020} in \citetalias{shang2020}, and connect them with the image contrast in the column density maps \citepalias[Figure 18,][]{shang2020}\@. It displays onion-like cavities inside-out, with a gap potentially filled with magnetic field lines, as a more evolved version from that shown in HH 212 (see Section \ref{subsec:HH212}). Similar structures exist in DG Tau A \citep{agra-amboage2014,guedel2018} and HH 30 \citep{louvet2018}. The slower molecular cones surrounding the jet may reveal the compressed wind as their flow directions deflected (Section \ref{subsec:velocity_multicavities}) and velocity magnitude reduced after passing the RS\@. Combined with the multicavity morphology, the nested jet--slower cone--cavity layers do support the structures of magnetized bubbles produced by a jet-bearing (unseen) wide-angle (X-)wind from the innermost regions.

It may not be surprising that the magnetized outflow bubble phenomena can last into the Class II phase. Remnant cavities in H$_2$ are known to exist, around T Tauri stars, such as FS Tau B \citep{liu2012} and T Tau \citep{saucedo2003,beck2008}. Multiple velocity components do exist in these more evolved systems, also identified as the high-velocity component (HVC, $\gtrsim 100\kms$), and the low-velocity component (LVC, $\lesssim 30\kms$) in optical/IR forbidden lines \citep{bachiller1996,bally2016}. The LVC as a whole is usually interpreted as the presence of a lower-velocity ``(extended) disk wind'' launched farther away from the central source (see further discussions in Sections \ref{subsec:DW_Scenarios} and \ref{subsec:notsosmallradii}; \citealt{HEG1995,KT1995,shang2007_PPV}). The HVCs are connected to the microjets of the T Tauri stars. However, the origins of the LVCs and their connections to the HVCs have remained evasive and controversial. Our results here can provide insights.

In recent years, some LVCs, seen in atomic [\ion{O}{1}] $\lambda6300$, can be further decomposed into two distinct components \citep[e.g.,][]{rigliaco2013,simon2016,fang2018,banzatti2019}, one broad (BC), and one narrow (NC)\@. The two-peak profiles of the LVCs are more frequently associated with higher accretion rates \citep{banzatti2019}. The BCs and NCs are seen as two distinct kinematic features in the systems, often referred to as ``slow winds'' \citep{pascucci2022_PPVII} in a similar fashion as those from the Class 0/I sources mentioned earlier. The BCs can have highest velocities of $30$--$40\kms$, and have broader widths than the NCs. Conventional application of 
\cite{anderson2003} has led to interpret the BCs as arising from inner ($<0.5\au$), and outer as $0.5$--$5\au$, in disk radii, with the former being faster (see discussion in Section \ref{subsec:notsosmallradii}).

Like the Class 0/I counterparts, these broad LVCs may be the kinematic manifestations of the more evolved (larger $n$) magnetized bubbles, but in different atomic tracers. The appearance of LVC-BC correlates strongly with the HVC\@. As disks disperse, in [\ion{O}{1}], the HVC fades away eventually as LVC-BC becomes nondetectable \citep{simon2016,EP2017,mcginnis2018,banzatti2019}. The LVC-NC persists. The connection to the nested velocity structure appears to persist in this phase. If the HVC is identified with the jet, the identified broad LVC(-BC) may arise from the compressed wind across the RS from the bubble interior. The LVC-NC persists as the compressed and entrained ambient material all the way out to the FS\@.

The DG Tau A system is extensively studied in the optical and near-infrared wavelengths for its multiple velocity components and line profiles. Onion-like velocity components are best seen in [\ion{O}{1}], [\ion{S}{2}], [\ion{Fe}{2}], and [\ion{Ne}{3}] showing HVC and LVC peaks \citep{HEG1995,bacciotti2000,agra-amboage2011,liu2016}, and the [\ion{O}{1}] LVC can be further decomposed into a BC and an NC \citep{simon2016,banzatti2019}. 
Morphologically, these velocity components are encompassed by an even lower-velocity wind (of $\lesssim10\kms$) traced by H$_2$ $2.12\Micron$ \citep{takami2004,beck2008,agra-amboage2014}, with an apparent trend of decreasing vertical extent, followed by the CO $J=2-1$ emission with a speed of $\sim3.1\kms$ \citep{guedel2018}. These evidences suggest they manifest the multicavities.

In our framework, we can reproduce the kinematic features seen in systems from Class 0, I, and II phases as different velocity components in tracers ranging from molecular lines to forbidden lines. The known observational features are consistent with the natural signatures produced by jet-bearing wind-driven magnetized bubbles demonstrated in this work.

Our framework offers a different but self-consistent interpretation of the inferred presence of outflowing gas with slower velocities confined within multicavities, the so-called ``slower (molecular) disk winds,''
actually as evidence of the compressed bubble structures revealed.

In Sections \ref{subsec:DW_Scenarios} and \ref{subsec:notsosmallradii}, we discuss conventional assumptions and approaches 
adopted to explain the observational data,
and their limitations and uncertainties. 
The magnetic interplay of the primary jet-bearing X-wind or an X-wind resembling inner disk wind (IDW) from the innermost regions, with the ambient media and, or with an outer Extended Disk Wind (EDW) would lead to extensive shocked and compressed regions, 
making the hypothesis of existence of an outer EDW less solid.

\paragraph{Note} 
The new JWST image of L1527 is one that captures the large KHI fingers generated by magnetic interplay when a magnetized wide-angle wind interacts with its surrounding ambient toroid. The structures resemble those produced in the evolved cavity formed with a moderately magnetized wind ($\MA \gtrsim 60$) interacting with an $n\gtrsim 6$ toroid in a mildly magnetized ($\alphab\sim 0.1$) ambient medium. These structures originate deeply from the outflow base as part of the magnetized bubble, grow, and advect to large scales with the underlying flow.

\subsection{The Disk Wind Scenarios}
\label{subsec:DW_Scenarios}

Historically, the general physics of \textbf{magnetocentrifugal launch} of winds from disks is well recognized as proposed in \citet{blandford1982}, the \textbf{BP mechanism}. This is the original definition of the term ``disk winds'' adopted for the context in star formation of the earliest phases \citep{konigl1993,konigl2000}, and in reference to the X-wind models \citep{shu1994_XWI,shu1994_XWII,shu2000}.

Nowadays, the term ``disk winds'' as presented has become analogous to phenomena of outflowing gas from the general protoplanetary disks or disk-like structures of all scales and of all evolutionary stages. They facilitate transport of angular momentum from the disk to the wind that allows or helps disk accretion. This term has become too loosely defined to convey specifics of basic physical mechanisms.

The BP mechanism applies in principle to all radii on the disk from the inner to the outer parts, because the effective gravitational potential (gravitational plus centrifugal) is able to accelerate gas along inclined poloidal field lines at any given disk radius (Figure 1 in \citealt{blandford1982}). The effective potential for gas attached to a poloidal field line has a gradient that accelerates gas in two regions between the equator and an angle of $60\arcdeg$ from it, allowing the gas to launch, while for angles closer to the vertical than $30\arcdeg$, the effective potential has the opposite gradient sign and does not allow magnetocentrifugal launching. Once launched, the mechanism allows gas to accelerate due to the projection of the centrifugal force onto the poloidal field line attached to the gas (typically with some additional push by projected gradients of gas pressure and of $b_\phi$ magnetic pressure). These processes can happen in principle at any radius, either inner or outer, including in particular the inner disk radius.
Gas in open field lines of the acceleration region leaves the disk into larger $\varpi$ values and eventually forms part of disk winds of different categories discussed below.

Field lines that are loaded with matter (as allowed by gas--magnetic attachment and field line angle) can happen at any location. Disk-wind scenarios are classified according to the launching location along the disk.
The range of launching locations can be extended and the mass loading be distributed along that range, leading in this case to an extended disk wind (EDW) scenario.
Alternatively, mass loading can be either limited to a narrow inner range of locations or have a mass loading concentrated toward the inner edge of the disk, leading in both cases to an inner disk wind (IDW) scenario (including the wind portion of the X-wind model as a particular case).
Both scenarios are equally able to participate in the BP mechanism. The general EDWs possess a radius-dependent profile of their terminal velocities, $v_{p,\infty}=v_\mathrm{K,0}(\varpi_0)[2(\varpi_\mathrm{A}/\varpi_0)^2-3]^{1/2}$ \citep{blandford1982,shu1995} where the ratio $(\varpi_\mathrm{A}/\varpi_0)$ depends on the specific model.
This ratio, the terminal velocity and their respective density profiles are related to the local mass loading onto each of the individual streamlines which characterize a given disk-wind model. Different from the EDWs, the IDWs have finite inner and outer disk radii, launched from small inner disk regions that can reach high escape velocities to account for those observed in jets, and usually identifiable with the HVC\@.

The X-wind models launch a disk wind from a region located close to the inner edge of the disk, connected also to the stellar magnetosphere \citep{shu1994_XWI,ostriker1995_XWIV}.
Therefore, X-wind models launch the disk wind from the deepest part of the gravitational potential in the system and can drive jets reaching the highest terminal velocities. 
Note that both the X-winds and the IDWs in their original forms have their respective velocity profiles as shown in \citet{shang1998},  \citet{shu2000}, \citet{shang2000,shang2010}, \citet{LS2012}, and, e.g.,\ \citet{krasnopolsky2003} and \citet{anderson2005}. The details of the velocity profiles pertaining to the launch will be reflected in the associated HVC\@.
The formulation shown in \citetalias{shang2020} is a simplified solution for lightweight computation for otherwise complex physics and expensive cost for the focus of the larger-scale bubble structures they form.

In IDWs, asymptotic behavior for large $z$ with vertical isodensity contours is regularly observed, as seen in \citet{krasnopolsky2003}, Figure 6, and in \citet{anderson2005}, Figure 12 (dash-dotted line). Simulations having mass loading concentrated toward the inner radius approach nearly vertical isodensity contours for all $z$, and in that regard, they resemble X-wind disk-launching models.

In the context of magnetized gravitational collapse, magnetic braking can slow down rotation of material in an infalling core through the emission of torsional Alfv\'en waves to outer parts of the environment \citep{basu1994,krasnopolsky2002}, in some circumstances leading to values of $v_\phi$ problematically small for large disk formation (the ``magnetic braking catastrophe''; e.g.\ \citealt{mellon2008,mellon2009,krasnopolsky2010}). The mechanism operates through poloidal field lines connecting an outer environment having slower angular rotation to a magnetically braked inner part. In principle, this can include inner parts of collapsing cores, infalling pseudodisks, and also parts of the disk proper.

At a later evolutionary stage, when the disk can evolve self-consistently, a new generation of application of the EDW in the context of ``wind-driven accretion'' is gaining popularity for the consideration of angular momentum transport in the disk.
The magnetorotational instability (MRI) can generate the anomalous viscosity required to transport angular momentum within the disk by the $\alpha$-disk model for accretion.
The alternative mechanism utilizes the capabilities of the magnetic torque in EDW to carry angular momentum out of this disk in place of the MRI\@.
These models are considered with different ingredients of the nonideal MHD terms for their respective physical mechanisms \citep[e.g.][]{bai2016}. Alternatively, these EDWs are connected to the spontaneous formation of structures \citep{suriano2017,suriano2018,suriano2019}, leading to varieties of rings and reconnection islands in the planet-making disks. These lighter EDWs share some common features that differentiate them from the original cold EDWs, as they require help from thermal effects and large magnetic pressure for the successful launch at larger radii.

Other flows within the new ``disk-wind'' category can also be launched thermally by photoevaporation, ``photoevaporative winds'' \citep{hollenbach1993,hollenbach1994,gorti2009,gorti2015}, from the extended radial regions farther away from the central source out of a transitional or debris disk at a much later evolutionary stage. These photoevaporative flows (mostly features of flow close to the upper disk layers) are not large-scale winds in the same sense as IDW or EDW\@. Additionally, because they are not magnetized winds, they cannot extract angular momentum from the disk.

\textbf{Regardless of the underlying wind-driving mechanism, the basic structure of free wind, wind shock, shocked wind, contact, shocked ambient medium, and free ambient medium, should apply.}

\subsection{Inferences of Slower Disk Winds Launched from an Extended Disk Region}
\label{subsec:notsosmallradii}

In this subsection, we discuss the theoretical basis, validity, and limitation of the conventional approach to infer the presence of extended disk winds (EDWs). We will make cautionary remarks for the inferred launch radii of an EDW at large, and the presence of the so-called ``slow molecular winds''.  

The standard formulation of the magnetocentrifugal winds follows a number of conservation laws, which, in axisymmetric steady state, lead to a number of conserved quantities, in the theory of the BP mechanism \citep{blandford1982}. Applying the conserved quantities at relatively large distances within the wind would allow, in principle, to infer knowledge about conditions at the launching point, as shown in \citet{anderson2003}, provided strict requirements are fulfilled. These requirements, especially regarding accurate data on rotation velocity, have often been ignored when the methodology of \citet{anderson2003} has been utilized. The methodology is based on the  ``fieldline angular velocity'' $\Omega\equiv(v_\phi - b_\phi v_p/b_p)/\varpi$, and the Bernoulli constant
$H\equiv\lfrac{v^2}{2} + h + \Phi_g - \Omega\varpi v_\phi$
(e.g.,\ \citealt{shu1994_XWII} or \citealt{lovelace1986}), both conserved along field lines of an ideal axisymmetric steady-state pristine wind. Under those conditions, the values of the Bernoulli constant $H$ are equal at the observing and launching points $P_\infty$ and $P_0$. A number of simplifying assumptions is then applied (cold wind allowing setting enthalpy $h=0$, Keplerian value $v_K=(G\stellarmass/\varpi)^{1/2}=(G\stellarmass\Omega_{0})^{1/3}$ for both $\varpi\Omega$ and $v_\phi\ (\gg v_p)$ at the launching point, and negligible $\Phi_g$ at the observation point), together writing an easily solvable cubic equation
\begin{equation}
\varpi_{\infty}v_{\phi,\infty}\Omega_{0} - \frac{3}{2}(G\stellarmass)^{2/3}
\Omega_{0}^{2/3} - \frac{v_{p,\infty}^{2}+v_{\phi,\infty}^{2}}{2}
\approx 0
\label{eqn:anderson_2003_final2}
\end{equation}
giving $\Omega_0^{1/3}$ and the location as $\varpi_0\approx(G\stellarmass/\Omega_0^2)^{1/3}$. 

Another assumption of validity is that the input data must be purely from the \textbf{free wind}, without involving the ambient medium directly. This is because the ambient data are not connected to the conserved quantities. That is, the region where the data are used and extrapolated, must be far enough from the region, which has strong interaction with the surrounding medium. Avoidance of the known shocked region should be exercised. The interaction may break axisymmetry of rotation confusing the measurements of rotation needed to apply \citet{anderson2003} if wrong assumptions are taken.

Application of \citet{anderson2003} requires resolved pristine-wind data so as to trust values of $\varpi_\infty$ and $v_{\phi,\infty}$. Therefore, the use of data from under-resolved sources \citep{pascucci2022_PPVII} can lead to uncertain values of $\varpi_0$, depending on the details and purpose of data management. Since typically $v_{\phi,\infty}\ll v_{p,\infty}$, distinguishing these two components in order to measure $v_{\phi,\infty}$ can be sensitive to the estimate of source inclination $i$, and to minor asymmetries in the source. 
Interaction and interplay between the magnetized wind and the ambient medium can reach much further inside, up to the RS\@. All models, without exception, are subject to the RS-limitations in comparing a pure-wind model against observational data unavoidably in interaction with the environment. This not only applies to the interactions with the ambient media but also to the interactions between the inner and outer winds if they coexist.

Observational data obtained between the RS and the FS can be too severely modified by the complex interplay to show the pristine conserved quantities along streamlines that are necessary to apply \citet[][]{anderson2003}. This can be observationally very demanding, as toroidally magnetized winds often have very thick shells moving the RS close to the origin, and presenting vortex structures caused by magnetization and by oblique shocks in parts of the wind region. Time variabilities can also break the assumptions for the valid application of \citet{anderson2003}. New shocked structures will alter the pristine environment required for the validity of the approach.

For the reasons presented above, identifying the observed shells with an EDW appears to be doubtful. On the contrary, the model presented in this work could justify the observations without the need of assuming a separate wind launched from the outer regions of the disk. We illustrate these points further with the ALMA observations obtained for the HH 212 in Section \ref{subsec:HH212}.

\subsection{The HH 212 System}
\label{subsec:HH212}

In \citetalias{shang2020}, the extensively studied HH 212 is inferred to be a candidate for a wide-angle wind interacting with a somewhat opened environment of $2\lesssim n \lesssim 4$, given the shape of the opening near the base, and the presence of an SiO jet and CO cavities. 
The molecules SO and SO$_2$ appear to also originate from the same location where the SiO jet appears. \citet{lee2018_hh212} resolved the SO emission into a collimated jet and a wide-angle rotating shell. They appear to be spatially surrounding the highly collimated, cylindrically narrow SiO jet. The slower SO emission appears to be tracing the wide-angle wind, with SO$_2$, connected to its inner densest core surrounding the highly excited SiO $J=8-7$ transition \citep{lee2018_hh212}. 

In \citet{lee2021_hh212}, additional shell structures are revealed in SiO and SO within $1400\au$ of the central protostar at 13 au resolution. In their Figure 1, the relative relationship between SiO and SO is shown, in which the SO is indeed aligned in jet with SiO but distributed in a wide-angle way more broadly around the core near the base, up to $\pm3\kms$ from the systemic velocity.
The spatial distribution confirms the suggestion in \citetalias{shang2020}.
Furthermore, the inner jet core is clearly enclosed by an extended shell, further surrounded by a wider more opened thick butterfly-like SO shell at $100\au$ scale (shown in their Figure 2). The cross-connected spatial morphologies in SiO and SO illustrate the nested-shell or multicavities identified in \citetalias{shang2020}, and as summarized in our Figures \ref{fig:2}(a), \ref{fig:3}(a), and \ref{fig:5}. The thick SO shell protruding from the base corresponds to the ``half-wall'' seen in the lower half of the column density passing through the ``gap'' region threaded by the compressed ambient magnetic field lines in Figure \ref{fig:5}. 

These features match well with the predicted structures from our framework, and serve as a manifestation of the physical mechanism elucidated. The nested SiO and SO emission structure of HH 212 is seen as an inner elongated shell surrounded by an X-shaped cavity with outward pointing velocity vectors. This situation is manifested in Figure \ref{fig:7} of this work. However, in \citet{lee2021_hh212}, such structure, as shown in their Figure 2, is interpreted as evidence of a jet interacting with a surrounding disk wind launched from $4\au$ all the way to $40\au$, following a previously reported slow wide-angle wind in SO and SO$_2$ by \cite{tabone2017}.  
\citet{lee2015} 
showed prominent ``\textbf{multicavities}'' of CO $J=3-2$ ``nested shells'' in channel maps (their Figure 1) and integrated intensity maps (their Figure 2). \citet{lee2017b} revealed a flattened envelope traced by HCO$^+$ $J=4-3$ in the absolute velocity range $\sim0.5$ to $\sim3.0\kms$ away from the systemic velocity (their Figure 1).

Our work can accommodate all of these observed features without the need for the presence of a separately launched disk wind from $4$--$40\au$, in one unified framework. The structure can be interpreted as part of the multicavities interior to the compressed magnetized ambient envelope (the cocoon). Their ``inferred'' SO/SO$_2$ disk wind shown in Figure 3 of \citet{lee2021_hh212} in fact can be interpreted as the compressed wind outside the RS cavity, whose flow velocities in the post-shock region have been reduced and deflected by the crossing of the RS at different local oblique angles, showing an optical illusion of direct launch from the disk surface. The outwardly pointing but decreasing velocities with values much reduced than those of the primary wind, extended to the compressed gap region threaded by strong ambient field lines, will result in a range of apparent launch by using the formulation of \citet{anderson2003}, as discussed in Section \ref{subsec:notsosmallradii}. The outer HCO$^+$ labels the infalling material outside the influence of the expansion, which represents the background infalling envelope.

We note that the overall SiO parallel PV diagram of HH 212 as reported in \cite{lee2017_NatAs} resembles our PVDCDs projected at the identical inclination angle of $\sim 86\arcdeg$. 
As shown in Supplementary Figure 2 of \cite{lee2017_NatAs} and further explored in Figure 5 of \citet{lee2022}, the velocity at the base of the SiO emission is broad, with an MS-type shell outlining the compressed wind across the RS-enclosed cavity, similar to our predictions. In Figure 1 of \cite{lee2021_hh212}, the relative positions of the SO HVCs and LVCs are shown with respect to SiO\@. The SO HVC traces the primary wind interior to the reverse shock, while the SO LVC traces the post-shock emission. Indeed, the radial wide-angle wind is seen through the detection of the wide velocity dispersion at the very base of the jet in Figure 1 of \cite{lee2022}, as predicted in this work.

\subsection{Extended Mixing Regions in CARMA-7}
\label{subsec:CARMA7}

The source CARMA-7 (C7) reported in \citet{plunkett2015_nat} is a Class 0 protostar that has 22 noted knot-like episodic features in $^{12}$CO\@. Some of these knots appear in the high-velocity portion of the PV diagram, and the rest appear in the low-velocity portion. According to Figure 2 of \citet{plunkett2015_nat}, the low-velocity points are located along the ``collimated'' axis near the systemic velocity, and the high-velocity ones are those located near the velocity maxima. The high-velocity points are near the base of the ``embedded'' unresolved blue-shifted and red-shifted jets. As we demonstrated in this work, the high-velocity component at a $\sim 90\arcdeg$ inclination angle traces the contribution from the wide-angle wind flowing along the line of sight, and the near systemic low velocity on the contrary arises from the density-collimated wind streamlines moving along the jet axis in C7 due to the effect of the inclination angle. This effect can be properly identified by comparison with its image in Figure 1 of \citet{plunkett2015_nat}. The overall outflow itself is very compact and jet-like near the base, indicating its youth in evolution \citepalias{shang2020}.

The reconstructed emission in Figure 2 of \citet{plunkett2015_nat} is consistent with the PVDCD features within this work in that a cavity feature resembling the image appears in both the PV and PP spaces, enclosing the triangular-shaped emission on the PV at the baseline of the offset. The high-velocity portion of the PV appears as zigzag patches tracing the contour of a cavity reminiscent of the \MOS{} curves (Section \ref{sec:MCcurves}) projected at the same inclination angle on the upper half \citep[beyond $\sim6\arcsec$ as shown in Figure 2 of][]{plunkett2015_nat}. The jet portion of the data was missing because of the combined effects of $^{12}$CO self-absorption and interferometric filtering \citep[][]{plunkett2015_apj}. Nonetheless, the velocity dispersion broadens toward the source (within $\sim6\arcsec$) clearly, especially on the blue side, suggesting that the emission originated in a wide-angle wind, which gives rise to a velocity with a relatively large magnitude and with a projected direction along the line of sight at $86\arcdeg$. The magnitude of $\gtrsim 15$--$20\kms$ is already significantly reduced from that of a fast free wide-angle wind that could have a velocity of $\sim 100 \kms$ as in a fast jet. Therefore, this emission mostly originates from the compressed wind to be become visible in $^{12}$CO on top of a faint potentially high-velocity background. A lobe-like contour encompassing the systemic velocity is also traceable, in both of the physical lobes and in PV, revealing the nature of multicavities.

The apparent episodicity of C7 supports the magnetic interplay, as pointed out in \citetalias{shang2020}. The presence of an extended region of entrainment was suggested in Figure 3 of the Extended Data in \citet{plunkett2015_nat}. The image shown in Figure 1 of \cite{plunkett2015_nat} can be compared with the moderately magnetized wind cases in Figures \ref{fig:5} and \ref{fig:6}. Within these complex structures, density enhancement and induced variation in velocity components in the multicavities can produce zigzag boundaries as part of the manifestation of the KHI within the extended mixing layer. The images separated in HV and LV panels appear nested with the HVC enclosed by the LVC\@. The PV patterns of C7 without the central jet can also be interpreted as originating from the multicavity of the mixing and gap structures surrounding it.

\section{SUMMARY}
\label{sec:summary}

In this work, we illustrate the kinematic features of the unified model of jet- and wide-angle--wind-driven outflow described in \citetalias{shang2020} and further developed here. The description makes use of synthetic Position--Velocity Diagrams of the column density (PVDCDs) formed in directions parallel and transverse to the elongated outflow, and explored in a wide parameter space.
We demonstrate different contributions and expected patterns in controlled steps. We start from basic analytic formulations with simple locally parallel oblique shock structures, followed by hydrodynamical numerical thin-shell calculations, producing reference smooth PVDCD signatures.
We proceed to the subsequent production of full PVDCD patterns from HD and MHD simulations that allow the development of KHI\@.
We demonstrate the corresponding changes in kinematic signatures generated in full simulations from HD to MHD, covering the parameter space adopted for different strengths of magnetization in winds and ambient toroids of various mass-to-flux configurations. The produced gallery can serve as a guide for observational comparisons for diagnosing the presence of outflows as magnetized elongated bubbles.

Our results suggest that bubble structures dominate the outflows driven by magnetized winds, and their interplay with the environment. These structures are thick and extended, contrary to the conventional expectations arising in thin-shell models.
One particular objective of the figure gallery is providing insights into the unavoidable bubble structures that result, with the jet-bearing inner disk winds possessing a strong wide-angle component. Real observations, if not very well-resolved, will typically show information connected to the large-scale nested-shell structure, and if better resolved, will reveal information in their respective multicavity layers and velocity components. Thus, observations require careful analyses to separate the contributions from individual signatures. These will have profound implications for outflow phenomena based on similar physical mechanisms of magnetized bubbles driven by magnetocentrifugal winds, as their basic principles hold in general.

Outflow bubbles are the ubiquitous, natural, and inevitable result of interaction between a wind and an ambient medium. They will be present, no matter the launching mechanisms, launching locations, or evolutionary stages. Interaction between the wind and the ambient medium provides the shape by solving the shock jump equations. The jet core is the densest part of the wind launched sufficiently close to the innermost region. Wide-angle wind is seen through the interacting shock layers, usually in the form of a compressed rather than pristine wind; however, it is often mistaken as a separate slower extended disk wind component. The more revealed and more resolved bubbles that formed in an environment with a much wider toroid of larger $n\gtrsim 6$ can give similar identification when signatures of the compressed interacting wide-angle wind are revealed in optical lines.

Here we summarize our highlights:

\begin{enumerate}

\item Schematic pictures of hydromagnetic elongated bubbles with extended compressed wind structures and KHI patterns are illustrated and explained.

\item The simple features of momentum-conserving shells are recovered in analytic and semianalytic formats. Hydrodynamical shear patterns can be constructed with semianalytic methods to illustrate simple HD elongated bubbles with oblique shocks.
Smooth hydrodynamic shells without KHI are constructed through zero-$v_\theta$ constrained simulations. The corresponding full simulations with regular $v_\theta$ have zigzag shells due to the presence of the KHI\@.
    
\item The basic jet--shell structures as seen in PVDCDs are shown in the hydrodynamic limit using different $n$ values of toroids.
The kinematic features are shown in PVDCDs with a $45\arcdeg$ inclination angle. This angle only serves as a demonstration where the contribution from the jet and the wide-angle wind can be separated the farthest to be identified in the PV plane.

\item The reverse shock has its own kinematic presence as the boundary across the change of behavior in the velocity widths of the wind projected on the PVDCD\@. 

\item The crossing of the reverse shock from free to compressed wind can produce deflection or rotation of flow velocity vectors that give an impression of the existence of a different layer of flow. This effect of deflection can imitate the behavior of a disk wind that may be interpreted and extrapolated to originate from somewhere inside the shell but outside of the innermost launch point of the primary fast wind. Because of the reduced poloidal flow velocity and a modified direction that is less radial compared to the original, the extrapolated location on the Keplerian disk may be farther out. The deflected flow direction inside the compressed wind could be misinterpreted as a separate slower ``disk wind'' originating from an outer region.
    
\item Signatures of KHI, compressed wind and compressed ambient media, magnetized or not, strongly or weakly, are directly revealed through the PVDCDs. They are juxtaposed for comparisons and identification in figure panels in this work for source-to-source comparisons. The notable presence in fact appears patchy and episodic as extended threads of emission features. These could be alternatively interpreted as \textbf{episodicity} in ejections, out of a smooth and steady flow.

\item
Magnetic pseudopulses are generated in the compressed wind regions. Pseudopulses appear and cooperate with the KHI fingers in an interlinked feedback loop of magnetic generation of vorticity. Jet pseudopulses are particularly evident in PVDCDs as fluctuating jet velocities on top of the RS cavities, that appear projected onto the PV diagrams as mild quasiperiodic wiggles around the velocity centroids propagating along the axis toward the tip. Pseudopulses can be distinguished from real wind pulsations, which create their own respective velocity centroids and have reverse shocks near the base, and by the association of the pseudopulses with magnetized KHI and magnetized vorticity generation within compressed wind.

\item The high-velocity component is easily identified as the EHV jet component, while the lower-velocity components can arise due to the interaction of wind with its surroundings, as the compressed ambient media, vortices aligned along the cavities. Multiple velocity peaks can arise that have broad and extended features in between the extremely high jet velocity, and the traditional low-velocity shell or the extremely low velocity near the systemic velocity reflecting the complexity of the flow patterns.

\item The velocity components arising from the multicavities follow a nested onion-like structure from the inside-out, displaying a profile of the various components as the extremely high (EHV), (standard) high or high velocity (SHV/HVC), intermediate velocity (IVC), the broad (BC) and narrow (NC) parts of the low velocity (LVC), and the very low velocity.

\item The presence of the broad low-velocity profiles found with atomic [\ion{O}{1}] forbidden line \citep{pascucci2022_PPVII} in T-Tauri stars can also be attributed to originate in the compressed wind resulting from the interplay with a less-dense ambient environment.

\item Transverse PVDCDs afford identification of compressed wind/ambient features and establish a 
nonrotating axisymmetric baseline in the comparisons between simulations and real data.

\item Our predicted model images and PVDCDs can provide an explanation to the observed nested bubbles and their associated velocity components in sources such as HH 212, HH 30, and DG Tau A and B. This demonstrates the power of the methods and models utilized.

\item Apparent acceleration can arise in the tip region when the toroidal magnetic field in wind decreases toward the tip. This can be observed in the PVDCDs for small enough $\MA$ values. The underlying physics will be followed in \citetalias{shang_PIII}\@.

\item We review assumptions and caveats needed for the application of \citet{anderson2003} to find the launch radii from the measurements of rotation in the flows. In this work we indeed demonstrate that the bubble interiors include compressed regions arising from the interaction with extended shocks. The applicability of the \citet{anderson2003} formulation appears thus to be questionable in such winds, as here the condition of laminar ideal flow is not verified.

\item We classify our results as the interaction between a toroid and a wide-angle X-wind or X-wind--like IDW, giving rise to
the specific form of density profiles that must be cylindrically stratified, at all vertical heights along the jet axis, and a wide-angle wind. Such winds belong to the X-wind Class. The X-wind naturally displays the strongest feature and has the widest angle in streamline distribution in the base when the launch region narrows to the neighborhood of the inner disk edge.

\item \textbf{Regardless of the underlying wind-driving mechanism, the basic bubble structure of free wind, wind shock, shocked wind, contact, and shocked ambient medium should apply.}

\end{enumerate}

The successful extended comparison with outflow sources from Class 0 to Class II phases, such as HH 212, HH 30, and DG Tau A and B demonstrated in our image and position--velocity maps has suggested a clear trace of bubble signatures existing in the known observational data.
Our framework provides a self-consistent and integrated understanding of the enigmatic presence of the lower-velocity components of the embedded and revealed outflows, which had not been properly explained in the literature since their discovery nearly $30\yr$ ago until our work presenting them as arising in the multicavities formed in the magnetic interplay.
Our gallery of results can serve as a guide for observational probes for the physical and dynamical situations inside the evolution of jet--outflow systems from Class 0, I, to II phases in molecular to atomic forbidden lines.

\begin{acknowledgments}

The authors would like to thank the anonymous referee profusely, for the critical comments and suggestions for improvements that significantly transformed the presentation and scope of this manuscript.

The authors acknowledge grant support for the CHARMS group under Theory from the Institute of Astronomy and Astrophysics, Academia Sinica (ASIAA), 
and the National Science and Technology Council (NSTC) in Taiwan through grants 109-2112-M-001-028-, 110-2112-M-001-019-, and 111-2112-M-001-074-. The authors acknowledge the access to high-performance facilities (TIARA cluster and storage) in ASIAA, and thank the National Center for High-performance Computing (NCHC) of National Applied Research Laboratories (NARLabs) in Taiwan for providing computational and storage resources. This work utilized tools (Zeus-TW, \Synline{}) developed and maintained by the CHARMS group. This research has made use of SAO/NASA Astrophysics Data System.

\end{acknowledgments}

\software{\Synline{} (this work) CHARMS (ASIAA), Zeus-TW \citep{krasnopolsky2010} CHARMS (ASIAA), MATLAB, Matplotlib \citep{hunter2007}.}

\newpage

\bibliographystyle{aasjournal}
\bibliography{Outflows}

\newpage
\appendix

\section{Background Formulation and Setups}
\label{sec:appendix:setups}

\begin{deluxetable}{cccc}
\tablecaption{Quantities Defined in Paper I and This Work}\label{tab:quantities}
\tabletypesize{\scriptsize}
\tablewidth{\textwidth}
\tablecolumns{4}
\tablehead{Symbol & Description & Definition & Adopted Value(s) }
\startdata
\hline
\multicolumn{4}{c}{coordinate systems} \\
\hline
$(r,\theta,\phi)$ & spherical polar coordinates & & \\
$(\varpi,z,\phi)$ & cylindrical polar coordinates & & \\
\hline
\hline
\multicolumn{4}{c}{vector quantities} \\
\hline
$\vv$  & flow velocity & & \\
$\bb$  & magnetic field in LH units & & \\
$\jv$  & current density in modified LH units & $\jv\equiv\nabla\times\bb$ & $\jv_p=(\lfrac{1}{\varpi})\vhat\phi\times\nabla\Cfunc$ \\
$\fv_c$  & specific magnetic force & $\jv\times\bb$ & \\
$\omv$ & vorticity & $\nabla\times\vv$ & \\
\hline
\hline
\multicolumn{4}{c}{scalar quantities} \\
\hline
$\rho$ & gas density & & \\
$f$    & wind mass fraction & $\rhow/\rho$ & \\
$a$    & local sound speed & $\left[f \aw^2 + (1-f) \aamb^2\right]^{1/2}$ & \\
$\vw$ & radial wind velocity & & $v_0$ at $r=\rin$ \\
$\vsh$  & radial shell velocity & & \\
$p$    & gas pressure & $a^2\rho$ & \\ 
$\mathcal{M}$ & mass per steradian & & \\
$\dot{\mathcal{M}}_\mathrm{w}$ & wind mass-loss rate & & $=3.6415\times10^{-6}\solarmassyr$ \\
$\Phi$ & magnetic flux function & & \\
$\Cfunc$ & current function & $-b_\phi\varpi$ & \\
$\Lfunc$ & force function & $\Cfunc^2/2$ & \\
$i$    & \begin{tabular}{c} inclination angle, between\\ the line of sight and outflow axis\end{tabular} & & $0\arcdeg/1\arcdeg$, $30\arcdeg$, $45\arcdeg$, $60\arcdeg$, $90\arcdeg$ \\
$\chi$ & obliqueness angle & Equation (\ref{eqn:chi:arcsin}) & \\
$\rho^S(r)$ & initial density tapering term & $D^Sr^{-2}$ & \\
\hline
\hline
\multicolumn{4}{c}{parameters} \\
\hline
\multicolumn{4}{l}{setup wind parameters (setting boundary conditions at $r=\rin$)}\\
\hline
$v_0$  & wind velocity at the inner boundary & Equation (\ref{eqn:initcon}) & $50$, $100\kms$ \\
$D_0$  & wind density constant & Equation (\ref{eqn:initcon}) & $2.0025\times10^{11}\gram\cm^{-1} \times(v_0/100\kms)$ \\
$\aw$    & wind sound speed & & $0.6\kms$ \\
$\rin$ & position of the inner radial boundary & & $1.5\au$ \\
$\MA$  & Alfv\'enic Mach number of the wind & $v_0\sqrt{D_0}/b_0$ & $6$, $30$, $60$, $90$, $180$, $600$, $\infty$ \\
\hline
\multicolumn{4}{l}{setup ambient parameters (setting initial conditions at $t=0$)}\\
\hline
$\aamb$ & ambient sound speed & & $0.2\kms$ \\
$n$    & opening of the initial toroid & $R(\theta)\propto\sin^n(\theta)$ for $\theta\rightarrow 0$ & $1$, $2$, $4$, $6$ \\
$\alphab$ & ambient poloidal field strength factor & Equation (\ref{eqn:alphab}) & $0$, $0.1$, $1$ \\
$\alpha_\rho$ & initial toroid density factor &  Equation (\ref{eqn:alpharho}) & $0$ (without toroid), $1$ (with toroid) \\
$D^S$  & tapering density factor for the toroid & Equations (\ref{eqn:alpharho},\ref{eqn:rhoa}) & $2.25\times10^{11}\gram\cm^{-1}$ (or zero) \\
\hline
\multicolumn{4}{l}{other setup parameters and chosen units}\\
\hline
$\rout$ & position of the outer radial boundary & & $10^5\au$ \\
$G$  & Newton's constant & & $6.67428\times10^{-8} \cm^3\gram^{-1}\second^{-2}$ \\
$1\au$ & approximate astronomical unit & & $1.5\times10^{13}\cm$\\
\hline
\multicolumn{4}{l}{dependent parameter of wind}\\
\hline
$b_0$  & wind toroidal magnetic field constant & Equation (\ref{eqn:initcon}) & $v_0\sqrt{D_0}/\MA$ \\
\hline
\hline
\multicolumn{4}{c}{angular functions for the initial conditions of toroids and boundary conditions of winds} \\
\hline
$R(\theta)$ or $Q(\theta)$ & ALS toroid density distribution function & $\rho^\mathrm{T}=(\aamb^2/2\pi G r^2) R_n(\theta)$ &\\
$\phi(\theta)$ & ALS toroid poloidal magnetic flux function
& $\Phi^\mathrm{T}=\lfrac{(a^2r)}{(4{\pi}G)^{1/2}}\phi(\theta)$ & (coefficient in LH units)\\
$a_0(n)$ & axial limit scale of $R_n(\theta)\theta^n$ &$\lim_{\theta\rightarrow0} [R(\theta)\theta^n]$ &\\
$b_0(n)$ & axial limit scale of $\phi_n(\theta)\theta^2$
&$\lim_{\theta\rightarrow0} [\phi(\theta)\theta^2]$ &\\
$P(\theta)$ & momentum input function of the wind & \multicolumn{2}{c}{$\rhow\vw^2$ cast into dimensionless units}\\
\hline
\hline
\multicolumn{4}{c}{other magnetized outflow shape model quantities} \\
\hline
$e_p(\theta)$ & magnetic efficiency factor & Equation (\ref{eqn:shape:vs_windbphibp}) & Appendix \ref{sec:appendix:ep_beta} \\
$\beta(\theta)$ & wind efficiency factor & Equation (\ref{eqn:shape:vs_windbphibp}) & Table \ref{tab:beta} \\
$\eta$ & magnetic/ram pressure ratio& $(b_p^2/2)\,/\,(\rho v_p^2)$ & quantity used in \citetalias{shang2020}\\
\hline
\hline
\multicolumn{4}{c}{subscripts and superscripts} \\
\hline
$q\wind$ & quantities of the wind & & \\ 
$q\amb$ & quantities of the ambient medium & & \\ 
$q^\mathrm{T}$ & quantities of the toroid & & \\
$q^S$ & quantities for tapering the toroid & & \\
\hline
\hline
\enddata
\end{deluxetable}

In this appendix, we summarize a few of the basic equations and simplifications leading to the major physical ingredients utilized throughout the work. More details are presented in Section 2 of \citetalias{shang2020}.
The formulations are based on the full set of MHD equations. In this work we write these equations in Lorentz-Heaviside (LH) cgs units (and lower case) for the magnetic field vector $\bb$ (and $\jv=\nabla\times\bb$ for the current density) as in the Zeus family of codes, such as the Zeus-TW code \citep{krasnopolsky2010}. The magnetic field in Gaussian cgs units (and upper case) can be found as $\Bv=\bb\sqrt{4\pi}$. 

In axisymmetry ($\partial_\phi=0$), the time-dependent MHD mass conservation, force, and induction equations give the time evolution of mass density $\rho$, velocity $\vv$ and momentum density $\rho\vv$, the magnetic field $\bb$:
\begin{equation}
    \label{eqn:mhd:vector:mass}
    \partial_t \rho + \nabla\cdot(\rho\vv) = 0
\end{equation}
\begin{equation}
   \label{eqn:mhd:vector:force}
    \rho
    \left(\partial_t \vv + (\vv\cdot\nabla)\vv \right) =
    -\nabla p
    +\jv \times \bb + \rho\gv
\end{equation}
\begin{equation}
    \label{eqn:mhd:vector:induction}
    \partial_t \bb = \nabla \times (\vv \times \bb ),
\end{equation}
where $\gv=-\nabla \mathcal{V}$ is the local acceleration of gravity, derivable from the potential function $\mathcal{V}$ which is a solution of the Poisson equation
$\nabla^2 \mathcal{V} = 4 \pi G \rho$ with appropriate boundary conditions.
The vorticity vector is defined as usual, $\omv \equiv \nabla\times\vv$.
The no-monopole constraint ($\nabla\cdot\bb=0$) is fulfilled at all times.
 Ideal MHD is assumed, and gravity and viscous forces are neglected. 
In \citetalias{shang2020} we show that the mass conservation (Equation\ (\ref{eqn:mhd:vector:mass})), and
the divergence-free (no-monopole) magnetic constraint
equations can be written in spherical components, under the assumption of axisymmetry as
\begin{equation}
\label{eqn:mhd:spherical:mass}
    \partial_t \rho + 
    r^{-2} \partial_r (r^2 \rho v_r) +
    \frac{1}{r\sin\theta}\partial_\theta(\rho v_\theta \sin\theta)
    = 0
\end{equation}
and
\begin{equation}
    \nabla\cdot\bb=
    r^{-2} \partial_r (r^2 b_r) +
    \frac{1}{r\sin\theta}\partial_\theta(b_\theta \sin\theta)
 =0\ ,
\end{equation}
allowing the definition of a magnetic flux function $\psi=\Phi/(2\pi)$ from which the poloidal components $b_r$ and $b_\theta$ can be derived as
\begin{equation}
\label{eqn:bp}
    \bb_p=\frac{1}{2\pi}\nabla\times(\frac{\Phi}{r\sin\theta}\ev_\phi)\ ,
\end{equation}
or
\begin{equation}
\label{eqn:bp_components}
b_r=\frac{1}{2\pi r^2\sin\theta}{\partial_\theta\Phi}
\ ,\ 
b_\theta=-\frac{1}{2\pi r \sin\theta} {\partial_r\Phi}\ .
\end{equation}

The force equation (\ref{eqn:mhd:vector:force})
and the induction equation (\ref{eqn:mhd:vector:induction}) can be simplified in the wind and in the (unperturbed ambient) toroid regions with their respective considerations.
The dimensional coefficient of the flux function $\Phi^\mathrm{T}$ has been divided by $(4\pi)^{1/2}$ in Equation (\ref{eqn:toroid_scaling}), consistent with the LH units as used in the code, while keeping the dimensionless flux function $\phi(\theta)$ unchanged as in Equation (2) in \citet{allen2003b}.
Both of these regions have $v_\phi=0$ and $\vv=\vv_p$, which, in combination with axisymmetry ($\partial_\phi=0$ and $\nabla=\nabla_p$), leads to a simplified expression for the vorticity vector, $\omv \equiv \nabla\times\vv=$
$\nabla_p\times\vv_p=\omega_\phi \ev_\phi$, leaving only one vorticity component in axisymmetric regions without $v_\phi$.
In this work, we refer to $\omega_\phi$ specifically
throughout.

In the limit of strongly radial ($v_r \gg v_\theta$) and cold wind, in the absence of rotation ($v_\phi=0$) and poloidal field ($b_p=0$), the steady-state model ($\partial_t=0$) equations can be further simplified to
\begin{equation}
  \label{eqn:wind:vtheta0:mass}
    \partial_r (r^2 \rho v_r)
    = 0\ ,
\end{equation}
\begin{equation}
  \label{eqn:wind:vtheta0:force_r}
   v_r\partial_r v_r
   = 
   -\frac{1}{\rho}\partial_r p
   -\frac{b_\phi}{r\rho}\partial_r(r b_\phi)\ ,
\end{equation}
\begin{equation}
  \label{eqn:wind:vtheta0:force_theta}
   0
   =
   - \frac{1}{r\rho} \partial_\theta p
   -\frac{b_\phi}{r\rho\sin\theta}\partial_\theta(b_\phi\sin\theta)\ ,
\end{equation}
\begin{equation}
  \label{eqn:wind:vtheta0:induction}
    0 =
    \partial_r(r v_r b_\phi )\ .
\end{equation}
This system of steady-state equations
(Equations (\ref{eqn:wind:vtheta0:mass})--(\ref{eqn:wind:vtheta0:induction}))
admits (for the $p=0$ case) solutions such that the three quantities
($v_r$, $\rho \varpi^2$, $|b_\phi| \varpi$) are constant.
These extremely simplified equations can be useful to describe the steady-state portions of wind, and have been used to design the boundary conditions for the wind at $r=\rin$, a region in which steady-state can be achieved, despite the fact that the model is not completely cold in that region, it does feature zero rotation and $b_p=0$ for our wind model.

\section{Semianalytic Shape Fitting of the Magnetized Bubbles}
\label{sec:appendix:ep_beta}
We adopt the following procedure to obtain the analytic expression for outflow shapes, leading to the final production of 
Figure \ref{fig:MC_MA_i}. The algebraic basis of the procedure is the pressure balance equation (Equation\ (\ref{eqn:shape:vs_windbphibp})), a quadratic equation in the velocity $\vMOS(\theta)$ that gives the shape $\rMOS(\theta)=t\vMOS$ of our self-similar outflow fitting.

The outflow is surrounded by a magnetic cocoon (Figure\ \ref{fig:1}) if $b_p$ is substantial, and can possess a tip if $b_\phi$ is very strong, for $\MA\lesssim 30$ (see PVDCDs for $\MA=6$ cases, and stronger ones in \citetalias{shang_PIII}).
Pressure balance equations can be written in all cases, and for small magnetization values, the resulting shapes do not deviate much from those obtained by the unmagnetized momentum-conserving Equation (\ref{eqn:mc_eqn}) \citep[see also][]{shu1991,KM1992b,lee2001}.

In \citetalias{shang2020}, the ratio $\eta=(b_p^2/2)\,/\,(\rho v_p^2)$ of local poloidal magnetic pressure to local total ram pressure is defined to measure the relative strengths of these quantities. Contours of constant $\eta$ are estimators of the location of the edge of the outflow (e.g.\ $\eta=0.5$ as experimentally adopted in Figure 14 of \citetalias{shang2020}).

The $\alpha_b=1$ cases contain substantial magnetic poloidal pressure, represented in Equation (\ref{eqn:shape:vs_windbphibp}) as a term $e_pb_p^2/2$. Here $e_p$ represents the amplification of the initial poloidal field due to compression.  Figure\ \ref{fig:1} shows how the wind regions drive out their initial $b_p$, compressing it into a fairly narrow magnetic cocoon surrounding the wind. Magnetic flux conservation allows us to estimate $e_p$ utilizing the cross-sectional area ratio between the cocoon+wind divided by the much narrower cocoon alone. This effect depends both on the parameter $n$ and $\theta$ along the outflow shape, but for the purposes of computing Figure\ \ref{fig:MC_MA_i} we consider it as only a function of $n$ ($e_p=40$, $20$, $9$, and $8$ for $n=1$, $2$, $4$, and $6$, respectively).

The factor $\beta(\theta)$ adjusts the wind velocity term in Equation (\ref{eqn:shape:vs_windbphibp}) for effects that either increase ($>1$) or decrease ($<1$) its strength.
The tips for $\MA<30$ are fit by locally increasing $\beta$ at small $\theta$. Deceleration inside the magnetized extended structures formed by compressed wind and ambient medium decreases $\beta$. Thickening of these same structures due to their magnetization and KHI is represented as a modest increase in $\rMOS$ and $\beta$. The adopted $\beta$ values (Table \ref{tab:beta}) combine these reductions and increases, often staying very close to unity.

Equation (\ref{eqn:shape:vs_windbphibp}) is a quadratic equation in $\vMOS$. For our experimental values of $\beta(\theta)$ and $e_p$, the equation discriminant can become negative if $\rhoa<\rhow$ (small $\theta$ values) and $b_\phi^2/2$ is large (small $\MA$ values). This occasionally happens in small tip portions of Figure \ref{fig:MC_MA_i}, and for such cases, the $\vMOS(\theta)$ values are carefully extrapolated from adjacent regions without complex roots, obtaining excellent fitting results.

The selected numerical values of $\beta(\theta)$ are summarized in Table \ref{tab:beta}. Monotone cubic Hermite interpolation of $\beta(\theta)$ is followed for all other intermediate $\theta$ values. 
\begin{deluxetable}{lcc|ccccccccccc|cccccccc}
\tablecaption{Look-up Table of $\beta(\theta)$ in Equation (\ref{eqn:shape:vs_windbphibp}) adopted for Figure \ref{fig:MC_MA_i}}\label{tab:beta}
\tabletypesize{\footnotesize}
\tablewidth{1.5\textwidth}
\tablecolumns{22}
\tablehead{&&&\multicolumn{11}{c}{$\MA=6$}&\multicolumn{7}{|c}{$\MA\geq30$}}
\startdata
$\alpha=1$ & $n=1$ & $\theta$ & $0\arcdeg$ & $1\arcdeg$ & $2\arcdeg$ & $3\arcdeg$ &&&&&&& $90\arcdeg$ & $0\arcdeg$ &&&&&& $90\arcdeg$ \\
           &       & $\beta(\theta)$ & $1.1$ & $1.1$ & $1.1$ & $1.1$ &&&&&&& $1.0$ & $1.0$ &&&&&& $1.0$ \\
           & $n=2$ & $\theta$ & $0\arcdeg$ &&&&&&&&&& $90\arcdeg$ & $0\arcdeg$ &&&&&& $90\arcdeg$ \\
           &       & $\beta(\theta)$ & $1.0$ &&&&&&&&&& $1.0$ & $1.0$ &&&&&& $1.0$ \\
           & $n=4$ & $\theta$ & $0\arcdeg$ & $5\arcdeg$ & $10\arcdeg$ && $18\arcdeg$ & $30\arcdeg$ & $40\arcdeg$ & $50\arcdeg$ & $60\arcdeg$ & $70\arcdeg$ & $90\arcdeg$ & 
           $0\arcdeg$ & $4\arcdeg$ & $10\arcdeg$ & $20\arcdeg$ & $30\arcdeg$ & $70\arcdeg$ & $90\arcdeg$ \\
           &       & $\beta(\theta)$ & $1.1$ & $1.0$ & $0.9$ && $0.9$ & $0.8$ & $0.75$ & $0.85$ & $1.0$ & $1.0$ & $1.0$ & 
           $1.02$ & $1.0$ & $0.85$ & $0.6$ & $0.5$ & $1.0$ & $1.0$ \\
           & $n=6$ & $\theta$ & $0\arcdeg$ & $5\arcdeg$ & $10\arcdeg$ & $12\arcdeg$ & $20\arcdeg$ & $30\arcdeg$ & $40\arcdeg$ & $50\arcdeg$ & $60\arcdeg$ & $70\arcdeg$ & $90\arcdeg$ & 
           $0\arcdeg$ & $5\arcdeg$ & $10\arcdeg$ & $20\arcdeg$ & $30\arcdeg$ & $70\arcdeg$ & $90\arcdeg$ \\
           &       & $\beta(\theta)$ & $1.1$ & $1.1$ & $1.0$ & $1.0$ & $0.6$ & $0.6$ & $0.7$ & $0.8$ & $1.0$ & $1.0$ & $1.0$ &
           $1.0$ & $1.0$ & $0.92$ & $0.6$ & $0.6$ & $1.0$ & $1.0$ \\
\hline
$\alpha=0$ & $n=1$ & $\theta$ & $0\arcdeg$ & $1\arcdeg$ & $2\arcdeg$ & $3\arcdeg$ &&&&&&& $90\arcdeg$ & $0\arcdeg$ &&&&&& $90\arcdeg$ \\
           &       & $\beta(\theta)$ & $1.2$ & $1.2$ & $1.1$ & $1.0$ &&&&&&& $1.0$ & $1.0$ &&&&&& $1.0$ \\
           & $n=2$ & $\theta$ & $0\arcdeg$ & $1\arcdeg$ & $2\arcdeg$ & $3\arcdeg$ & $4\arcdeg$ & $6\arcdeg$ & $10\arcdeg$ & $15\arcdeg$ &&& $90\arcdeg$ & $0\arcdeg$ &&&&&& $90\arcdeg$ \\
           &       & $\beta(\theta)$ & $1.05$ & $1.05$ & $1.05$ & $1.05$ & $1.05$ & $1.035$ & $1.01$ & $1.0$ &&& $1.0$ & $1.0$ &&&&&& $1.0$ \\
           & $n=4$ & $\theta$ & $0\arcdeg$ &&& $3\arcdeg$ && $6\arcdeg$ & $9\arcdeg$ &&&& $90\arcdeg$ & $0\arcdeg$ &&&&&& $90\arcdeg$ \\
           &       & $\beta(\theta)$ & $1.08$ &&& $1.08$ && $1.08$ & $1.0$ &&&& $1.0$ & $1.0$ &&&&&& $1.0$ \\
           & $n=6$ & $\theta$ & $0\arcdeg$ && $2\arcdeg$ &&& $6\arcdeg$ && $12\arcdeg$ & $30\arcdeg$ && $90\arcdeg$ & $0\arcdeg$ &&&&&& $90\arcdeg$ \\
           &       & $\beta(\theta)$ & $1.37$ && $1.24$ &&& $1.09$ && $1.002$ & $1.002$ && $1.002$ & $1.0$ &&&&&& $1.0$ \\
\enddata
\end{deluxetable}

\section{Construction of Hydrodynamic Shell Structure}
\label{sec:appendix:steps}

\begin{deluxetable}{ccc}
\tablecaption{Quantities Defined in Appendix \ref{sec:appendix:steps}}\label{tab:steps}
\tabletypesize{\scriptsize}
\tablewidth{\textwidth}
\tablecolumns{3}
\tablehead{Symbol & Description & Definition, Value(s) }
\startdata
\hline
\multicolumn{3}{c}{Momentum-conserving shell model:} \\
\hline
$v\MC(\theta)$ & radial velocity & Equation (\ref{eqn:app:mc_eqn})\\
$r\MC(\theta)$ & radial coordinate & Equation (\ref{eqn:app:mc_eqn})\\
$\vhat{T}\MC$ & unit tangent vector & Equation (\ref{eqn:tnvectors})\\
$\vhat{n}\MC$ & unit normal vector & Equation (\ref{eqn:tnvectors})\\
$\Pv\MC$ & position vector & $r\MC \vhat{r}$\\
\hline
\multicolumn{3}{c}{parameters} \\
\hline
$t_\mathrm{sim}$ & time & \\
$\gamma$ & \multicolumn{2}{c}{adiabatic index $\gamma=1.0001$. Used in the auxiliary Riemann problem.}\\
\hline
\multicolumn{3}{c}{Auxiliary Riemann problem symbols and values} \\
\hline
$\vhat{d}$ & unit vector used to set up the problem& either $\vhat{n}\MC$ or $\vhat{r}$\\
$v_d$ & wind velocity coefficient & $\vw(\vhat{r} \cdot \vhat{d} )$\\
$\vhat{p}$ & unit vector used to apply the results& either $\vhat{d}$ or $\vhat{r}$\\
$f_{pd}$ & hybrid factor & either $1$ or $(\vhat{d}\cdot\vhat{p})^{-1}$\\
\hline
\multicolumn{2}{c}{left side} & \\
$\rho_L$ & density & $\barrhow/\rsh^2$ \\
$a_L$ & sound speed & $\aw$ \\
$p_L$ & gas pressure & $a_\mathrm{L}\rho_L/\gamma$ \\
$v_L$ & velocity & $v_d$\\
\hline
\multicolumn{2}{c}{right side} & \\
$\rho_R$ & density & $\barrhoa/\rsh^2$\\
$a_R$ & sound speed & $\aamb$\\
$p_R$ & gas pressure & $a_\mathrm{R}\rho_R/\gamma$\\
$v_R$ & velocity & 0\\
\hline
\multicolumn{2}{c}{results} &\\
($s_L$, $s_M$, $s_R$) & \multicolumn{2}{c}{left, middle, and right velocity solutions of the Riemann problem}\\
\hline
\multicolumn{3}{c}{Reverse and forward shocks (as modeled here)} \\
\hline
$\Pv_\mathrm{RS}$ & reverse shock position & Equation \ref{eqn:step2_pos} \\
$\Pv_\mathrm{FS}$ & forward shock position & Equation \ref{eqn:step2_pos} \\
$\vv_\mathrm{RS}$ & reverse shock velocity & Equation \ref{eqn:step2_vel} \\
$\vv_\mathrm{FS}$ & forward shock velocity & Equation \ref{eqn:step2_vel} \\
$p_\mathrm{RS}^n(z_\mathrm{RS})=\varpi_\mathrm{RS}$ & \multicolumn{2}{c}{polynomial of order $n$ used to fit the reverse shock position}\\
$p_\mathrm{FS}^n(z_\mathrm{FS})=\varpi_\mathrm{FS}$ & \multicolumn{2}{c}{polynomial of order $n$ used to fit the forward shock position}\\
\hline
\multicolumn{3}{c}{Model choices} \\
\hline
normal model & \multicolumn{2}{c}{$\vhat{d}=\vhat{p}=\vhat{n}\MC$, $f_{pd}=1$}\\
radial model & \multicolumn{2}{c}{$\vhat{d}=\vhat{p}=\vhat{r}$, $f_{pd}=1$}\\
hybrid normal model & \multicolumn{2}{c}{$\vhat{d}=\vhat{n}\MC$, $\vhat{p}=\vhat{r}$,$f_{pd}=(\vhat{d}\cdot\vhat{p})^{-1}$}\\
\hline
\multicolumn{3}{c}{Construction of fields} \\
\hline
($\Pv_g=\Pv$, $z_g$, $\varpi_g$) & \multicolumn{2}{c}{generic point, with its cylindrical coordinates}\\
($\Pv_\mathrm{proj}$, $z_\mathrm{proj}$, $\varpi_\mathrm{proj}$)
& \multicolumn{2}{c}{projection of $\Pv_g$ onto the RS, with cylindrical coordinates}\\
($\Pv_\mathrm{fproj}$, $z_\mathrm{fproj}$, $\varpi_\mathrm{fproj}$)
& \multicolumn{2}{c}{``projection'' of $\Pv_g$ and $\Pv_\mathrm{proj}$ onto the FS, with cylindrical coordinates}\\
$m$, $d^2$ & \multicolumn{2}{c}{geometry elements used to compute the projections}\\
$M_\mathrm{eff}$ & \multicolumn{2}{c}{effective Mach number used to compute post-shock values}\\
$f_{RS}$ & \multicolumn{2}{c}{factor used to interpolate between FS and RS post-shock values}\\
$\vv(\Pv_g)$ & \multicolumn{2}{c}{model velocity field at $\Pv_g$} \\
\hline
\multicolumn{3}{c}{subscripts and superscripts} \\
\hline
$q\MC$ & momentum-conserving shell model & \\
$q_\mathrm{L}$ & left state of the Riemann problem & \\
$q_\mathrm{R}$ & right state of the Riemann problem & \\
$q_\mathrm{RS}$ & reverse shock model & \\
$q_\mathrm{FS}$ & forward shock model & \\
$\vv^\mathrm{ff}$ & velocity in fixed frame of reference & \\
$\vv^\mathrm{rsf}$ & velocity in RS frame of reference & \\
$\vv^\mathrm{fsf}$ & velocity in FS frame of reference & \\
$a_\mathrm{pre}$ & pre-shock quantities & \\
$a_\mathrm{post}$ & post-shock quantities & \\
\hline
\enddata
\end{deluxetable}
In this appendix, we devise four steps to demonstrate structures formed by an oblique bubble approximated by a pair of locally parallel oblique shocks. The detailed procedures are laid out for the conceptual development. Table \ref{tab:steps} compiles symbols and quantities used here.

\subsection{\textbf{Step 1: The Momentum-Conserving Shell}}

The position of the momentum-conserving shell found analytically has been previously demonstrated to well capture the behaviors of the thin shells. The first step of the development is obtaining the solution of the momentum-conserving shell. 

The (radial) velocity $\vsh=v\MC$ and position $\rsh=r\MC$ of the momentum-conserving (MC) shell are
\begin{equation}\label{eqn:app:mc_eqn}
    v\MC(\theta) = \frac{v_0}{1+\left(\lfrac{\barrhow}{\barrhoa}\right)^{-1/2}}\quad\text{and}\quad
    r\MC(\theta) = v\MC\times t_\mathrm{sim} + \rin\ ,
\end{equation}
which is equivalent to Equation\ (\ref{eqn:mc_eqn}).

\medskip
\subsection{\textbf{Step 2: Locations of the Reverse and Forward Shocks}}

In this step, we add the structure of an oblique bubble to the position $r_{\mathrm S}$ of the momentum-conserving shell. We construct the locations of the reverse and forward shocks to the thin-shell curve.

We start by constructing a system of unit vectors. The unit vectors tangent and normal to the MC shell are
\begin{equation}
    \label{eqn:tnvectors}
    \begin{split}
        \vhat{T}\MC &= \frac{r\MC}{\sqrt{r^2\MC+\dot{r}^2\MC}}\vhat{\theta} + \frac{\dot{r}\MC}{\sqrt{r^2\MC+\dot{r}^2\MC}}\vhat{r} \\
        &\equiv T_{\mathrm{MC}\_\theta}\vhat{\theta} + T_{\mathrm{MC}\_r}\vhat{r}, \\
        \vhat{n}\MC &= T_{\mathrm{MC}\_\theta}\vhat{r} - T_{\mathrm{MC}\_r}\vhat{\theta}.
    \end{split}
\end{equation}
The unit vectors in the $r$- and $\theta$-directions can be projected onto the unit vectors in the $\varpi$ and $z$-directions
$\vhat{z}$ and $\vhat{\varpi}$, resulting in
$\vhat{r}=\sin\theta\vhat{\varpi}+\cos\theta\vhat{z}$, and $\vhat{\theta}=\cos\theta\vhat{\varpi}-\sin\theta\vhat{z}$. The normal vector can also be similarly written as
$\vhat{n}\MC= n_{\mathrm{MC}\_\varpi}\vhat{\varpi} + n_{\mathrm{MC}\_z}\vhat{z}$. The derivative $\lfrac{dr\MC}{d\theta}\equiv \dot{r}\MC$ 
can be found numerically by finite differences at high $\theta$ resolution.

\begin{figure*}
    \centering
    \plotone{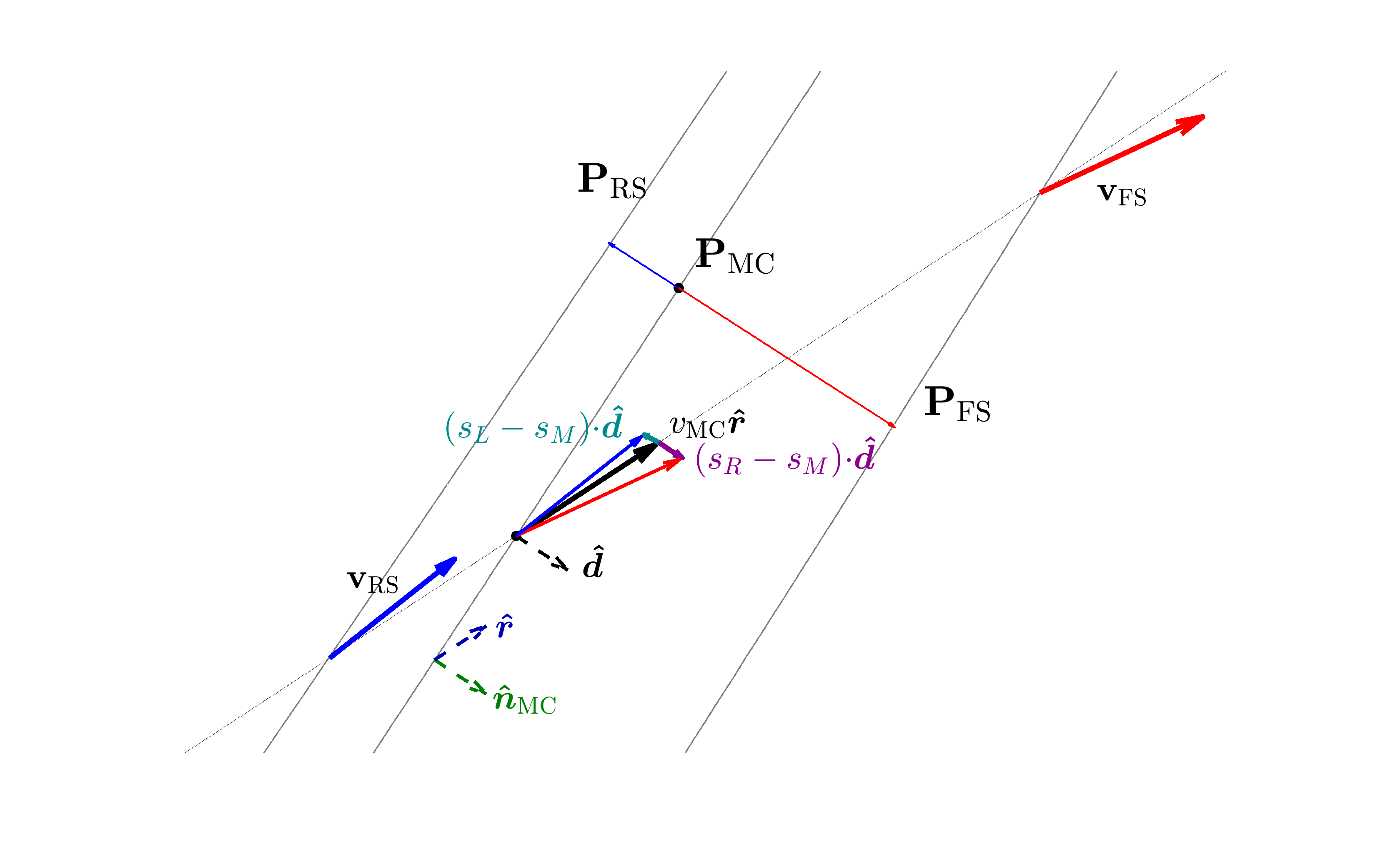}
\caption{Schematic illustration for Step 2, determining the shock (FS, RS) positions and velocities from the momentum-conserving (MC) shell solution, based on Equations (\ref{eqn:step2_pos}) and (\ref{eqn:step2_vel}).
The directional unit vector $\vhat{d}$ can be set to either $\vhat{n}\MC$ as in the normal model illustrated here, or $\vhat{r}$ as in the radial model.
}
\label{fig:step2}
\end{figure*}

We illustrate the schematic configuration in Figure~\ref{fig:step2}. 
The construction needs to choose a unit vector $\vhat{d}$ at each point $\Pv\MC=r\MC \vhat{r}$ located along the MC shell.
We will use either of the choices $\vhat{d}=\vhat{n}\MC$ or $\vhat{d}=\vhat{r}$.
The radial incoming wind velocity $\vw$ is projected along $\vhat{d}$ obtaining a set of values $v_d=\vw(\vhat{r}\cdot\vhat{d})$. A hydrodynamic 1D Riemann problem is then set, having, as its left-side state, this velocity $v_d$ with the density and sound speed of the wind, and, as its right-side state, the ambient density and sound speeds, with a velocity of zero, as follows:
\begin{equation}
    \label{eqn:leftright}
    \begin{split}
        \mathrm{left\;side}&:\;\; \rho_L = \barrhow/\rsh^2,\; a_L=\aw,\; p_L = \aw^2\rho_L/\gamma,\; v_L =\vw(\vhat{r}\cdot\vhat{d}); \\
        \mathrm{right\;side}&:\;\; \rho_R = \barrhoa/\rsh^2,\; a_R=\aamb,\; p_R = \aamb^2\rho_R/\gamma,\; v_R = 0\ .
    \end{split}
\end{equation}
A value of $\gamma=1.0001$ is set for the equation of state of the 1D Riemann problem, which is solved by the usual analytical methods, such as, e.g.,\ in Chapter 4 of \citet{torobook}. This value of $\gamma=1.0001$ has been shown to closely reproduce the results of two-temperature numerical simulations in \citetalias{shang2020}.

The solution of this 1D Riemann problem has three discontinuities, left, middle, and right, with speeds $s_L$, $s_M$ and $s_R$. A second unit vector $\vhat{p}$ is now chosen, equal to either of $\vhat{d}$ or $\vhat{r}$, and a factor $f_{pd}$ is defined to be either equal to $1$ or to $(\vhat{d}\cdot\vhat{p})^{-1}$.
The modeled positions of the reverse and forward shocks of the axisymmetric bubble (RS, FS) are then constructed as
\begin{equation}\label{eqn:step2_pos}
    \begin{split}
    \Pv_\mathrm{RS} &= \Pv\MC + f_{pd}\, t_\mathrm{sim}\, (s_L-s_M)\,\vhat{p}\\
    \Pv_\mathrm{FS} &= \Pv\MC + f_{pd}\, t_\mathrm{sim}\, (s_R-s_M)\,\vhat{p}\ .
    \end{split}
\end{equation}

The construction completed by Equation\ (\ref{eqn:step2_pos}) allows an oblique 2D axisymmetric flow to be represented by a set of 1D flows having a direction of motion determined by the choice of $\vhat{d}$ (either normal or radial). 

The left and right initial states (Equation\ (\ref{eqn:leftright})) of those 1D flows are built from the wind (on the left side) and the toroids (on the right side), with the wind velocity projected in the direction of $\vhat{d}$ and applied to the left side. Based on these initial states, the discontinuity structure of the 1D flows at an arbitrary time $t_\mathrm{sim}$ is found analytically, and it is expressed as the three speeds $s_L$, $s_M$, and $s_R$, corresponding to the reverse shock, CD, and forward shock of the 1D flows. These three speeds are now used to construct locations for the RS and FS of the 2D bubbles. The middle speed $s_M$ is made to correspond to the location $\Pv\MC$ of the MC shell, and the left and right discontinuity speeds are made to correspond to the RS and FS of the bubble. These correspondences are applied in the direction of the vector $\vhat{p}$, and with a scaling factor equal to the product of $f_{pd}$ and time: these notations are chosen for generality, but more frequently are applied simply as $\vhat{p}$ equal to $\vhat{d}$, and the factor $f_{pd}$ equal to 1.

The respective shock velocities for the RS and FS are estimated as 
\begin{equation}\label{eqn:step2_vel}
\begin{split}
    \mathbf{v}_\mathrm{RS} &= v\MC\,\vhat{r} + f_{pd}\, (s_L-s_M)\,\vhat{p}\\
    \mathbf{v}_\mathrm{FS} &= v\MC\,\vhat{r} + f_{pd}\, (s_R-s_M)\,\vhat{p}\ ,
    \end{split}
\end{equation}
which is based on a time derivative of Equation\ (\ref{eqn:step2_pos}). 
For relatively thin shells, the approximation $\vv_\mathrm{RS} = \vv_\mathrm{FS} = v\MC\,\vhat{r}$ can also be applied.

The positions $\Pv_\mathrm{RS}$ and $\Pv_\mathrm{FS}$ are often not used directly, but via a polynomial fit procedure, which gives as a result polynomials $p^n(z)$ ($n=8$ as of now) such that the point $\varpi_\mathrm{FS} = p^n_\mathrm{FS}(z_\mathrm{FS})$ is on the fitted forward shock and $\varpi_\mathrm{RS} = p^n_\mathrm{RS}(z_\mathrm{RS})$ is on the fitted reverse shock.

We have chosen the following configurations for the constructions of the positions of the reverse and forward shocks:
\renewcommand{\labelenumi}{(\alph{enumi})}
\begin{enumerate}
    \item 
      The normal model $\vhat{d}=\vhat{p}=\vhat{n}\MC$, $f_{pd}=1$: represents the 2D bubble flow as a set of 1D flows moving in the direction normal to the MC shell, and constructs the RS and FS positions based on that, motivating the particular choice of $\vhat{d}=\vhat{p}$, and $f_{pd}=1$ values;
    \item
      The radial model $\vhat{d}=\vhat{p}=\vhat{r}$, $f_{pd}=1$: represents the 2D bubble flow also as a set of 1D flows, but now moving in the radial direction. It is observed to lead to less separation between RS and FS than the normal model. The radial model is expected to be connected to the results of the $v_\theta=0$ numerical models, which are also a set of radial flow;
    \item
      The hybrid normal model $\vhat{d}=\vhat{n}\MC$, $\vhat{p}=\vhat{r}$, $f_{pd}=(\vhat{d}\cdot\vhat{p})^{-1}$: is intended as a mathematical simplification of the normal model, with 1D flows modeled to be in the normal direction, but with their results applied in the radial direction with the help of a projection factor $f_{pd}$ approximately restoring the intended normal direction of the 1D flows. Such a hybrid approximation is closer to representing the normal model in regions of the bubble having small curvature and small thickness.
\end{enumerate}

\begin{figure}
    \centering
    \includegraphics[width=0.98\textwidth]{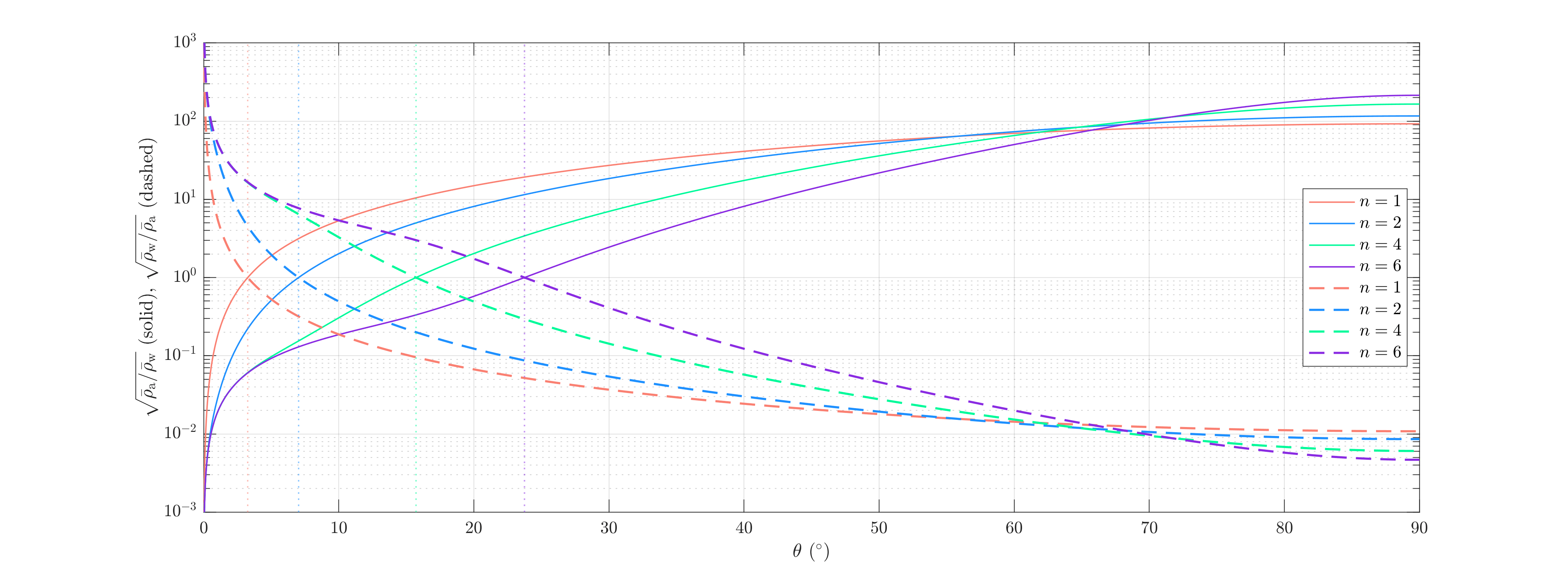} \\
    \includegraphics[width=0.98\textwidth]{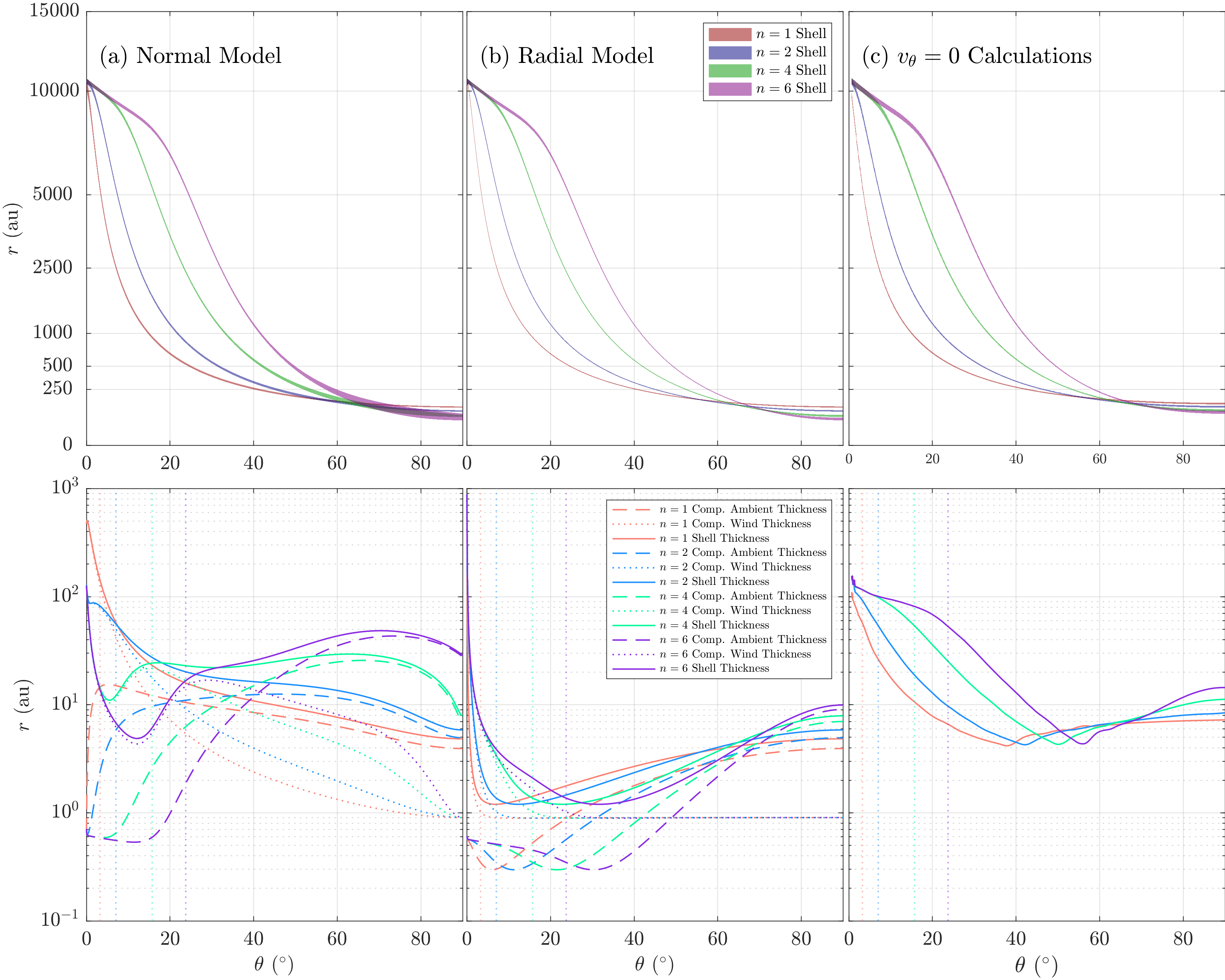}
\caption{{\it Top:} Density ratios of the wind and ambient material, used to construct the momentum-conserving shell solutions for $n=1$, $2$, $4$, and $6$ toroids. {\it Bottom:} Properties of the 2D hydrodynamic bubbles constructed from (a) normal model and (b) radial model by Equation\ (\ref{eqn:step2_pos}) in Step 2, and (c) the $v_\theta=0$ numerical model. The upper half of the panels shows the radial shell positions as a function of $\theta$, with the shades indicating the regions between the reverse shocks and forward shocks (whose positions have been stretched by $2\%$ for better visualization). The $r$-axis is nonlinearly stretched as a square root. The lower half shows the respective shell thickness measured radially between the RS and FS, the RS and the CD, and the CD and the FS\@, with a logarithmic $r$-axis.}
    \label{fig:shellprofstep2}
\end{figure}

The shock surfaces constructed by these models will be utilized in Step 3. Figure~\ref{fig:shellprofstep2} summarizes the constructed profiles for the positions and the thickness.

\medskip
\subsection{\textbf{Construction of the Field of Shear Velocities}}

In the following two steps (3 and 4), we will construct a field of velocity vectors for the whole flow, both inside and outside the shell of the outflow bubble. With this construction, the flow structures across the RS and FS can be analyzed analytically in the post-shock regions between the RS and FS\@.

Basically, starting from an arbitrary generic point $\Pv=\Pv_g$ given in cylindrical coordinates as $\Pv_g=z_g\vhat{z}+\varpi_g\vhat{\varpi}$, its gas velocity $\vv_\mathrm{gas}(\Pv_g) \equiv \vv(\Pv_g)$ is estimated in a sequence of calculations as follows, using the fitted curves of RS and FS found in Step 2. Two trivial cases are checked first: if $\varpi_g > p^n_\mathrm{FS}(z_g)$ then $\Pv_g$ is well inside the toroids with $\vv(\Pv_g) = 0$, and if $\varpi_g < p^n_\mathrm{RS}(z_g)$ then $\Pv_g$ is inside the pure-wind region with $\vv(\Pv_g) = v_0\vhat{r}$.

Otherwise, $\Pv_g$ is inside the shell of the bubble, located between RS and FS\@. Step 3 treats this case by finding two representative projection points for each $\Pv_g$, one at the RS and another one at the FS, as follows.

\subsection{\textbf{Step 3: Projections onto the shock surfaces}}

First, the point $\Pv_g$ is projected onto the RS at the place closest to it. If $d$ is the distance between $\Pv_g$ and a point on fitted RS curve
\begin{equation}
    d^2 = (z-z_g)^2 + (p^n_\mathrm{RS}(z)-\varpi_g)^2
\end{equation}
then the minimum of $d^2$ (as a function of $z$) would correspond to the projection point. This minimum is an internal extremum of a smooth function, and so it fulfills the usual zero-derivative criterion $\frac{d(d^2)}{dz}=0$
\begin{equation}\label{eqn:step3_proj}
    \begin{split}
        \frac{d(d^2/2)}{dz} &= \frac{1}{2}\frac{d}{dz} \left[(z-z_g)^2+(p^n_\mathrm{RS}(z)-\varpi_g)^2\right] \\
        &= (z-z_g) + (p^n_\mathrm{RS}(z)-\varpi_g) p^{n\prime}_\mathrm{RS}(z) = 0.
    \end{split}
\end{equation}
Therefore we need the real root of the polynomial
\begin{equation}
(p^n_\mathrm{RS}(z)-\varpi_g) p^{n\prime}_\mathrm{RS}(z) + (z-z_g)\ ,
\end{equation}
which finds the point $\Pv_\mathrm{proj} = z_\mathrm{proj}\vhat{z} + \varpi_\mathrm{proj}\vhat{\varpi}$.

We need also a representative point on the FS\@. We choose to find it as the intersection of the FS with a line connecting $\Pv_g$ to the projection point $\Pv_\mathrm{proj}$ (located on the RS)\@. The coordinates of the intersection point $\Pv_\mathrm{fproj}$ can be called $z_\mathrm{fproj}$ and $\varpi_\mathrm{fproj}\equiv p^n_\mathrm{FS}(z_\mathrm{fproj})$. We find this intersection point analytically by first computing the slope $m$ of the line
\begin{equation}
m = (\varpi_\mathrm{proj}-\varpi_g)/(z_\mathrm{proj}-z_g)\ ,
\end{equation}
and then finding the place where that line cuts the FS, using the polynomial fitting curve. That gives a polynomial equation
\begin{equation}\label{eqn:step3_fproj}
    p^n_\mathrm{FS}(z) - mz + (mz_g-\varpi_g) = 0\ ,
\end{equation}
whose relevant real root gives the value of $z_\mathrm{fproj}$.

Once the two representative points $\Pv_\mathrm{proj}$ and $\Pv_\mathrm{fproj}$ have been identified, we find their pre-shock and then their post-shock velocities.
The pre-shock (gas) velocities in the fixed frame (ff) are, at $\Pv_\mathrm{proj}$ and $\Pv_\mathrm{fproj}$,
\begin{equation}
 \vv^\mathrm{\,ff}_\mathrm{pre}(\mathrm{RS}) = v_0\vhat{r}\;\;\; \mathrm{and}\;\;\; \vv^\mathrm{\,ff}_\mathrm{pre}(\mathrm{FS}) = 0\ .
\end{equation}
The usual formulas to obtain post-shock velocities are given in frames comoving with the shocks and a simple transformation provides that, as follows.
The shock (pattern) velocities have been estimated above (Equation\ (\ref{eqn:step2_vel})) as
\begin{equation}\label{eqn:step2_vel_copy}
    \begin{split}
    \vv_\mathrm{RS} &= v\MC\vhat{r} + f_{pd}\, (s_L-s_M)\,\vhat{p}\\
    \vv_\mathrm{FS} &= v\MC\vhat{r} + f_{pd}\, (s_R-s_M)\,\vhat{p}\ .
    \end{split}
\end{equation}
The numerical values of these vectors are stored in arrays, which are then interpolated in $z(\theta)$.
Velocity subtractions such as
\begin{equation}
    \begin{split}
        \vv^\mathrm{\,ff}_\mathrm{pre}(\mathrm{RS}) - \vv_\mathrm{RS} &= \vv^\mathrm{\,rsf}_\mathrm{pre}(\mathrm{RS})\;\; \mathrm{and} \\
        \vv^\mathrm{\,ff}_\mathrm{pre}(\mathrm{FS}) - \vv_\mathrm{FS} &= \vv^\mathrm{\,fsf}_\mathrm{pre}(\mathrm{FS})
    \end{split}
\end{equation}
provide the transformations of the pre-shock gas velocities to the frames of the reverse and forward shocks (rsf and fsf).

In order to obtain the post-shock gas velocities, it is useful to separate the pre- and post-shock velocity components, respectively, tangential and normal to the shocks. That requires knowing the respective unit vectors tangential and normal to the shock. Simple differential geometry using the fitting polynomials gives the tangent and normal unit vectors at $\Pv_\mathrm{proj}$ and $\Pv_\mathrm{fproj}$, which we may call here $(\vhat{T}_\mathrm{RS},\vhat{n}_\mathrm{RS})$ and $(\vhat{T}_\mathrm{FS},\vhat{n}_\mathrm{FS})$.

These unit vectors are now used to obtain the tangential and normal pre-shock velocities in the shock frame.
The tangential components are unchanged before and after the shocks in the shock frames, whereas the normal component is altered based on shock jump conditions. 
The effective (normal) shock Mach numbers are now known,
\begin{equation}
    \begin{split}
        M_\mathrm{eff}(\mathrm{RS})&=\aw^{-1}(\vv^\mathrm{\,rsf}_\mathrm{pre}(\mathrm{RS})\cdot\vhat{n}_\mathrm{RS})\\
        M_\mathrm{eff}(\mathrm{FS})&=\aamb^{-1}(\vv^\mathrm{\,fsf}_\mathrm{pre}(\mathrm{FS})\cdot\vhat{n}_\mathrm{FS})
    \end{split}
\end{equation}
and applying formulas for oblique shocks such as Equations (94) and (133) in \citet{naca1953} yields the value of the post-shock normal components, which in the shock frames are equal to
\begin{equation}
    \begin{split}
    (\vv^\mathrm{\,rsf}_\mathrm{post}(\mathrm{RS})\cdot\vhat{n}_\mathrm{RS}) &=
    (\vv^\mathrm{\,rsf}_\mathrm{pre}(\mathrm{RS})\cdot\vhat{n}_\mathrm{RS})
    \frac{\gamma-1+2\,[M_\mathrm{eff}(\mathrm{RS})]^{-2}}{\gamma+1}\;\; \mathrm{and} \\
    (\vv^\mathrm{\,fsf}_\mathrm{post}(\mathrm{FS})\cdot\vhat{n}_\mathrm{FS}) &=
    (\vv^\mathrm{\,fsf}_\mathrm{pre}(\mathrm{FS})\cdot\vhat{n}_\mathrm{FS})
    \frac{\gamma-1+2\,[M_\mathrm{eff}(\mathrm{FS})]^{-2}}{\gamma+1}
    \end{split}
\end{equation}
which give the dependence of the ratios of normal components on the effective Mach numbers. Combining now the tangential and normal post-shock gas velocities in the shock frames gives the total velocity vectors in the shock frames, which can be simply transformed to the fixed frame of reference by adding back the respective shock velocities ($\vv_\mathrm{RS}$ and $\vv_\mathrm{FS}$), yielding as a result the post-shock gas velocities in the fixed frame at the two representative points.

\begin{figure*}
    \centering
    \plotone{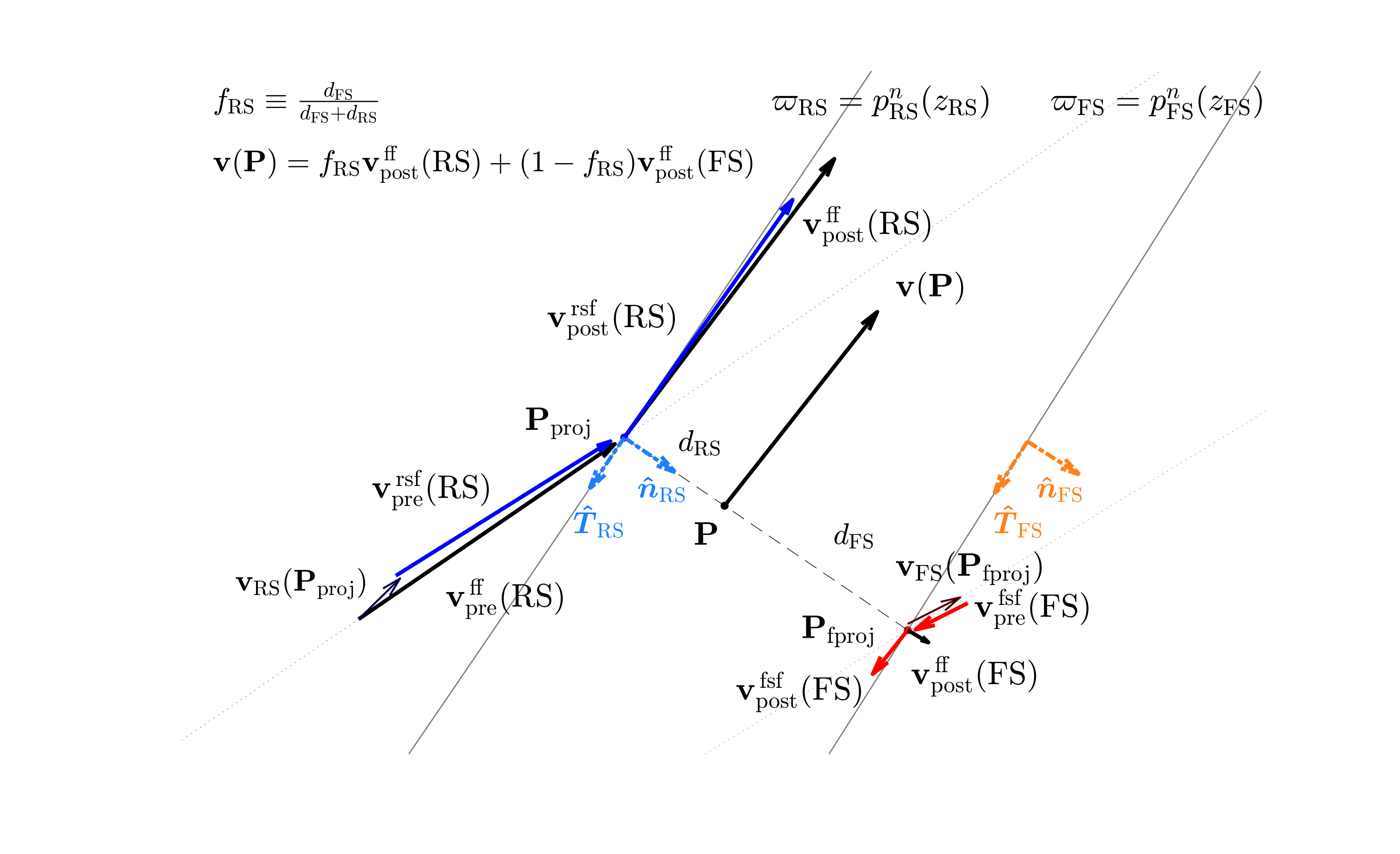}
    \caption{Schematic illustration for the procedures in Steps 3 and 4, for an arbitrary point $\Pv=\Pv_g$, determining the projection points on fitted RS and FS ($\Pv_\mathrm{proj}$ and $\Pv_\mathrm{fproj}$) using root finding, as in Equations (\ref{eqn:step3_proj}) and (\ref{eqn:step3_fproj}), solving post-shock velocities in the shock frames, and linearly interpolating in the fixed frame with Equation (\ref{eqn:step4_vP}).
    }
    \label{fig:step34}
\end{figure*}

\medskip
\subsection{\textbf{Step 4: Construction of the Shear Profile}}

The final step is to construct and identify the local velocity field for
any point $\Pv_g$ located between the RS and FS\@. The two points $\Pv_\mathrm{proj}$ and $\Pv_\mathrm{fproj}$ with fixed-frame post-shock velocities as illustrated in Figure \ref{fig:step34} are determined in Step 3. Those post-shock velocities are known as $\vv^\mathrm{\,ff}_\mathrm{post}(\mathrm{RS})$ and $\vv^\mathrm{\,ff}_\mathrm{post}(\mathrm{FS})$. The distances $d_\mathrm{RS}$ and $d_\mathrm{FS}$ from $\Pv_g$ to the projection points $\Pv_\mathrm{proj}$ and $\Pv_\mathrm{fproj}$ are also known.

\begin{figure*}
    \centering
\includegraphics[height=0.18\textheight]{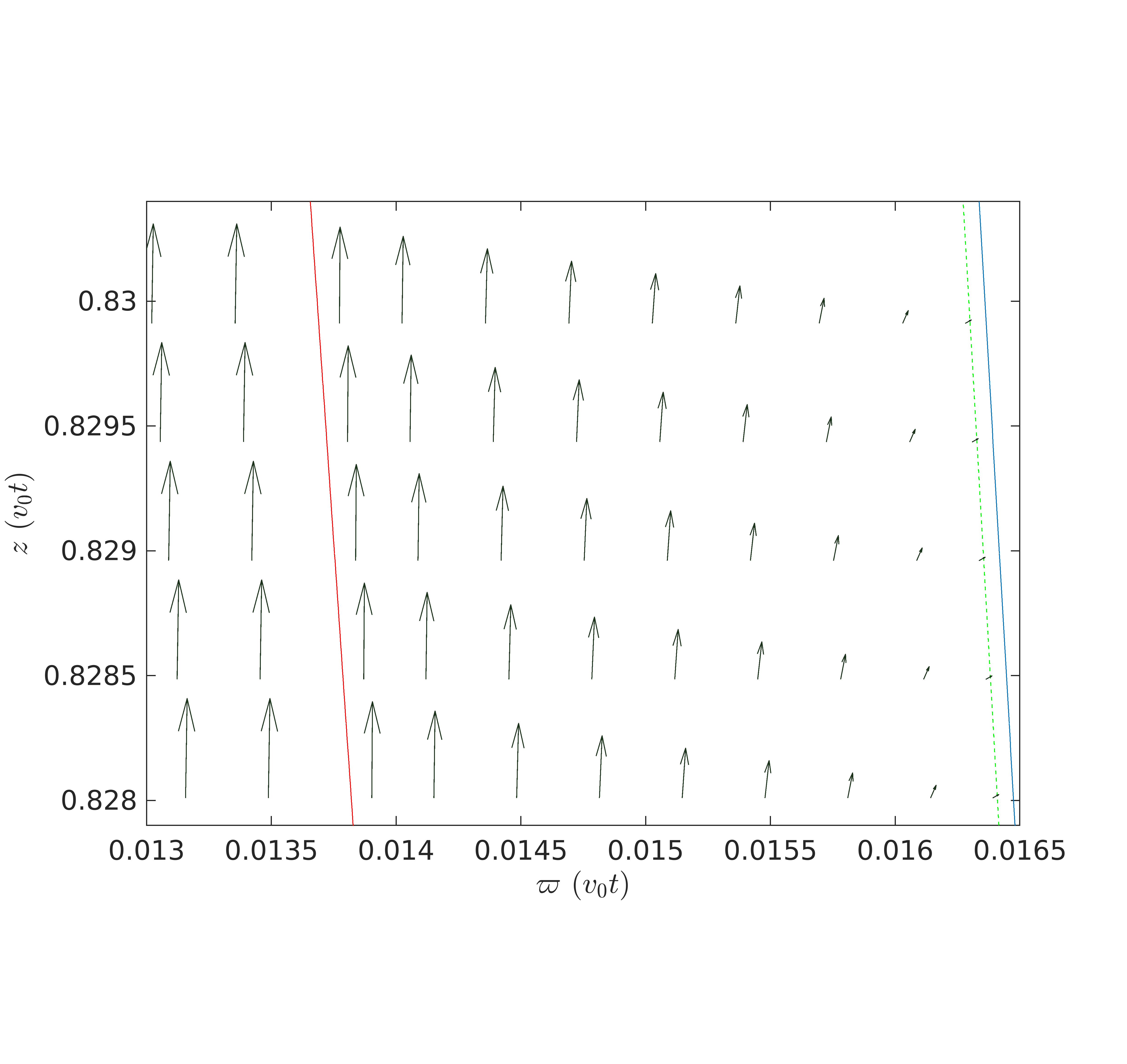}
\includegraphics[height=0.18\textheight]{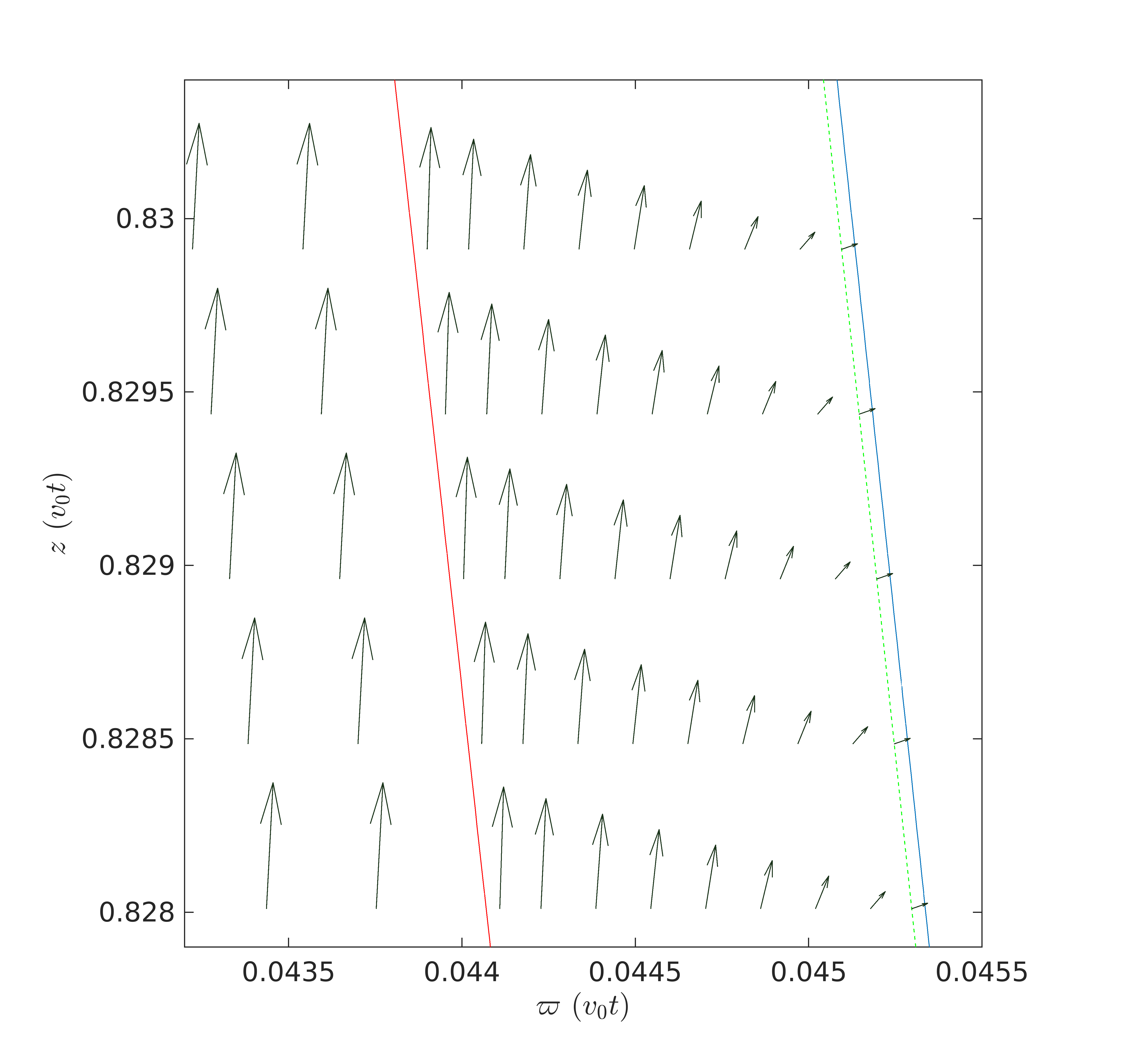}
\includegraphics[height=0.18\textheight]{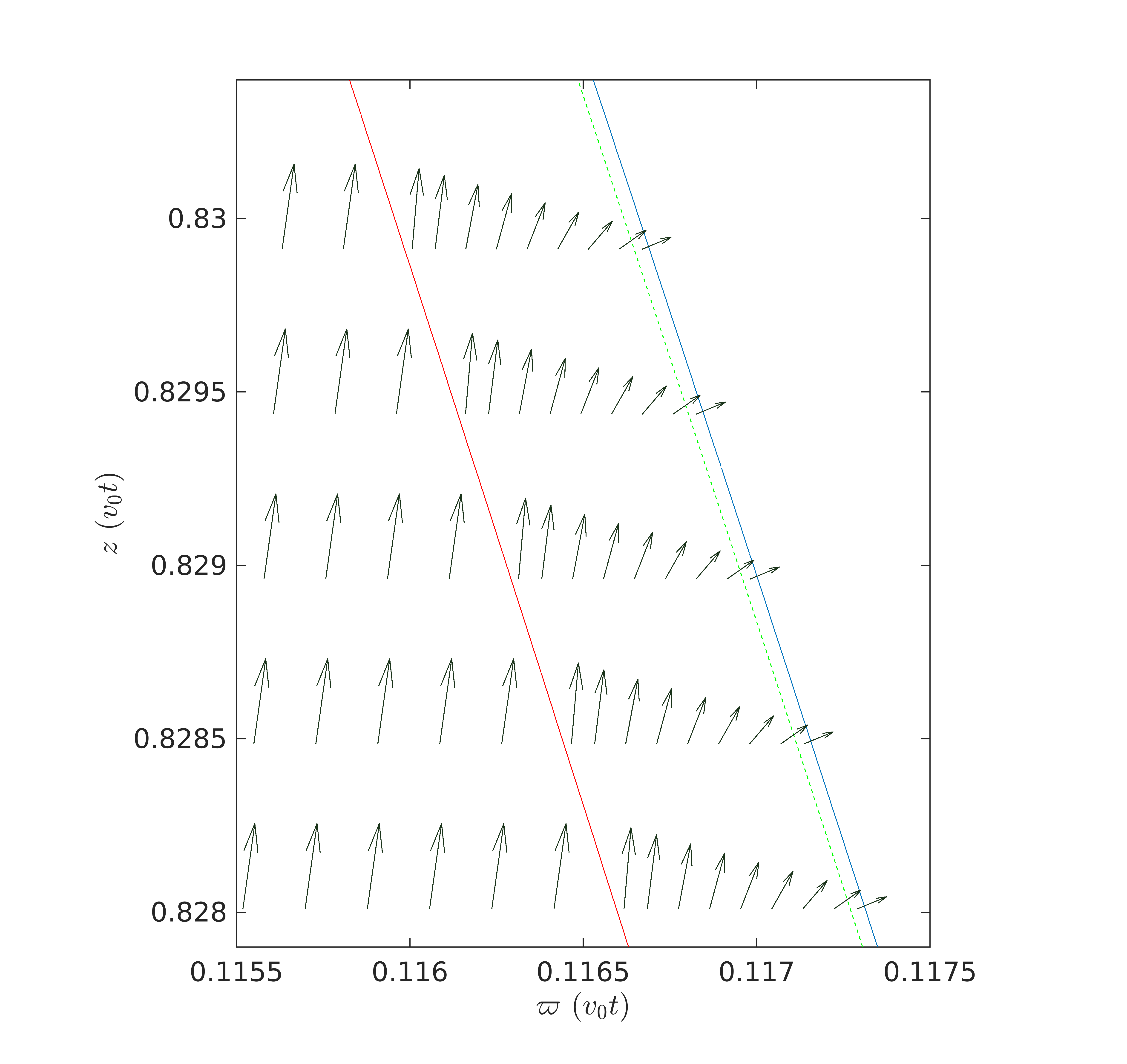}
\includegraphics[height=0.18\textheight]{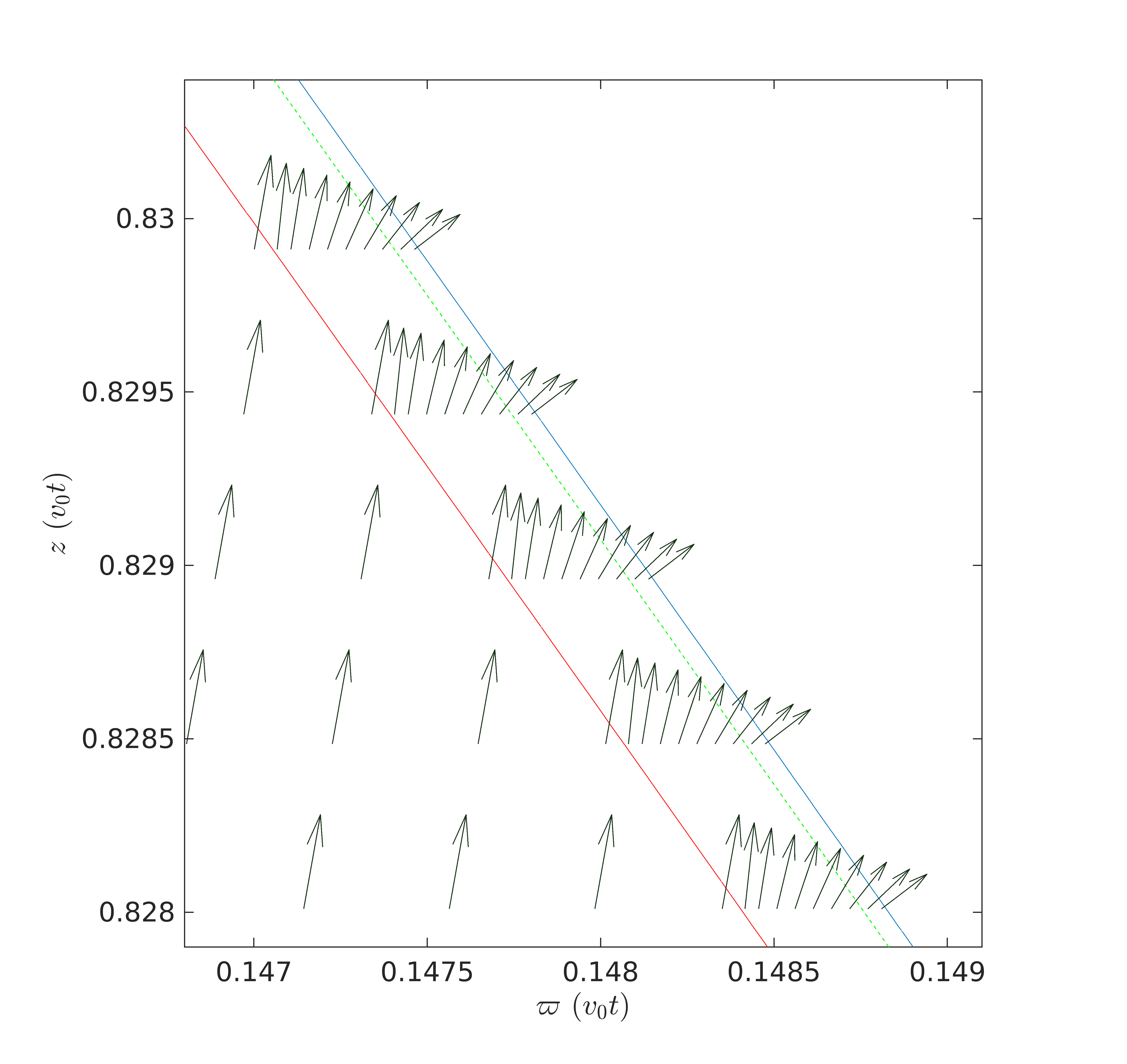}
\\
\includegraphics[height=0.18\textheight]{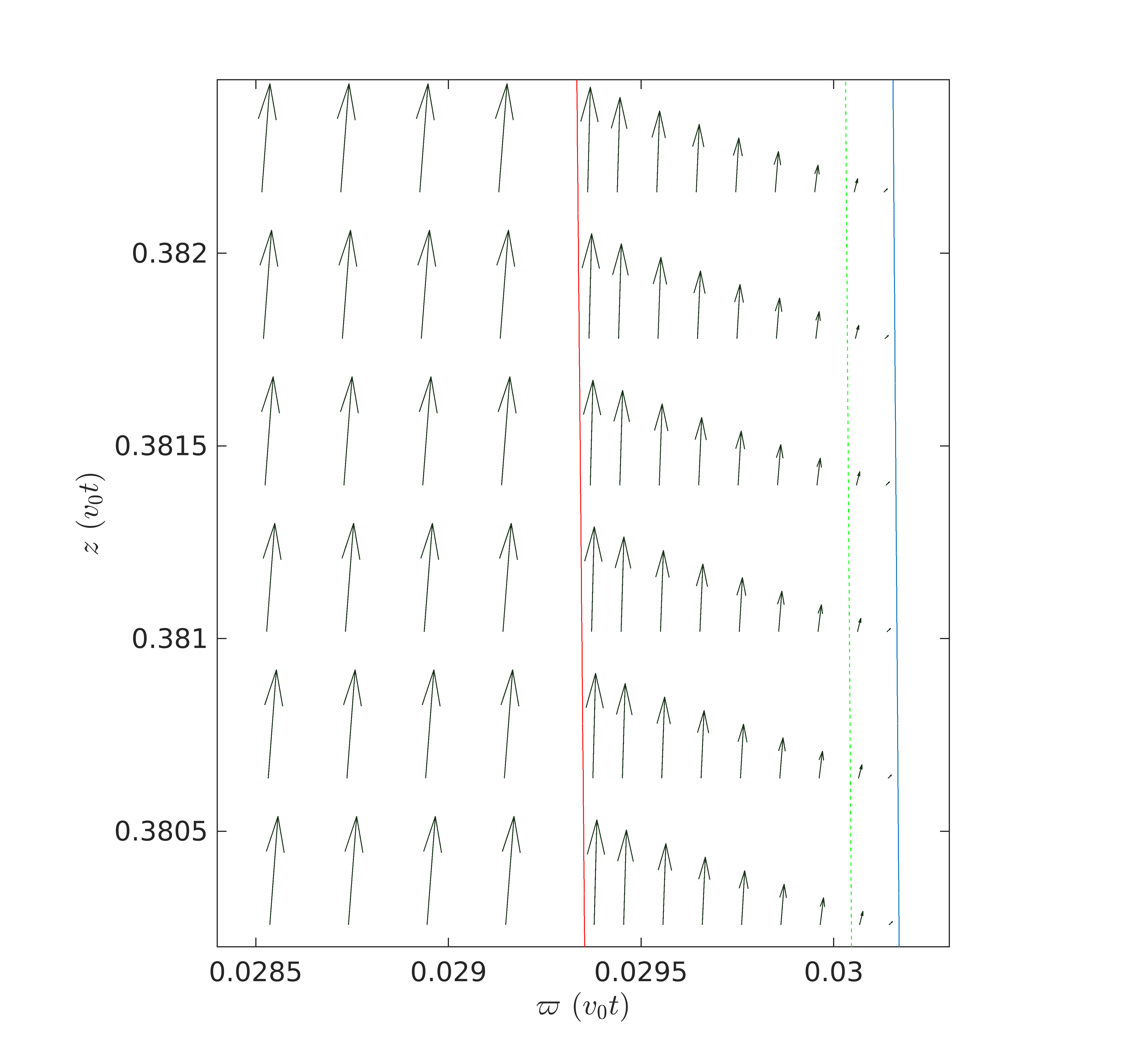}
\includegraphics[height=0.18\textheight]{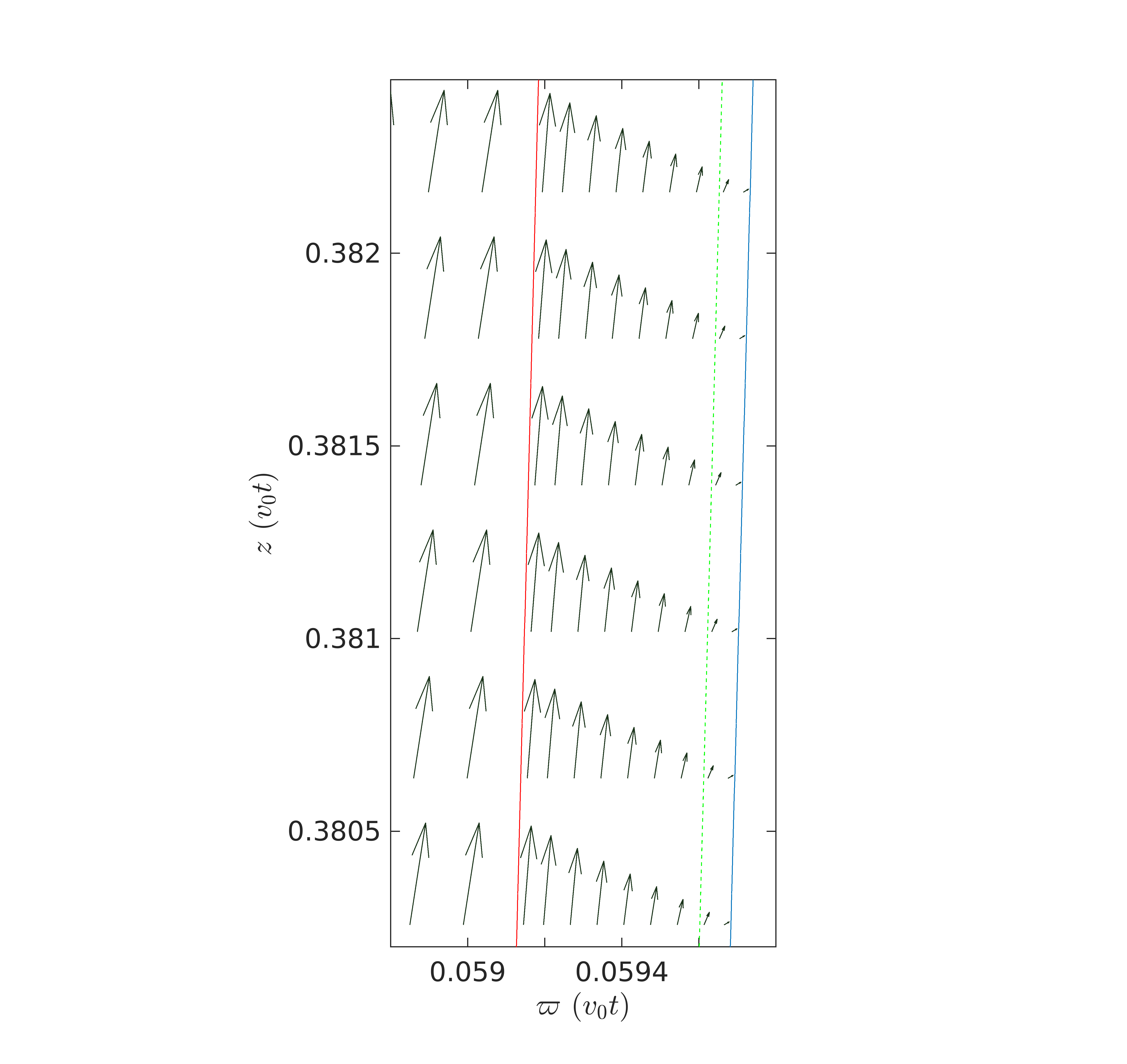}
\includegraphics[height=0.18\textheight]{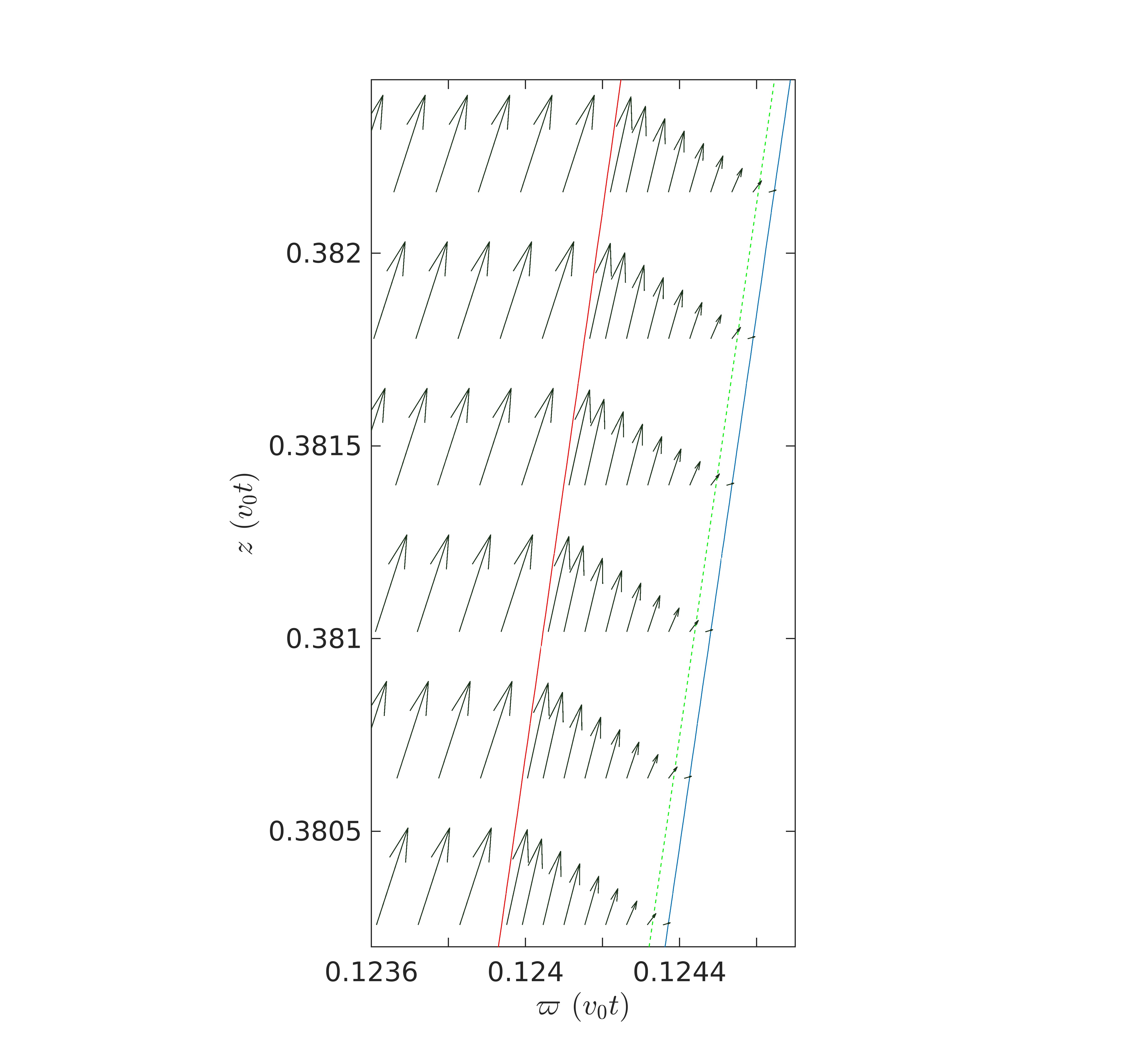}
\includegraphics[height=0.18\textheight]{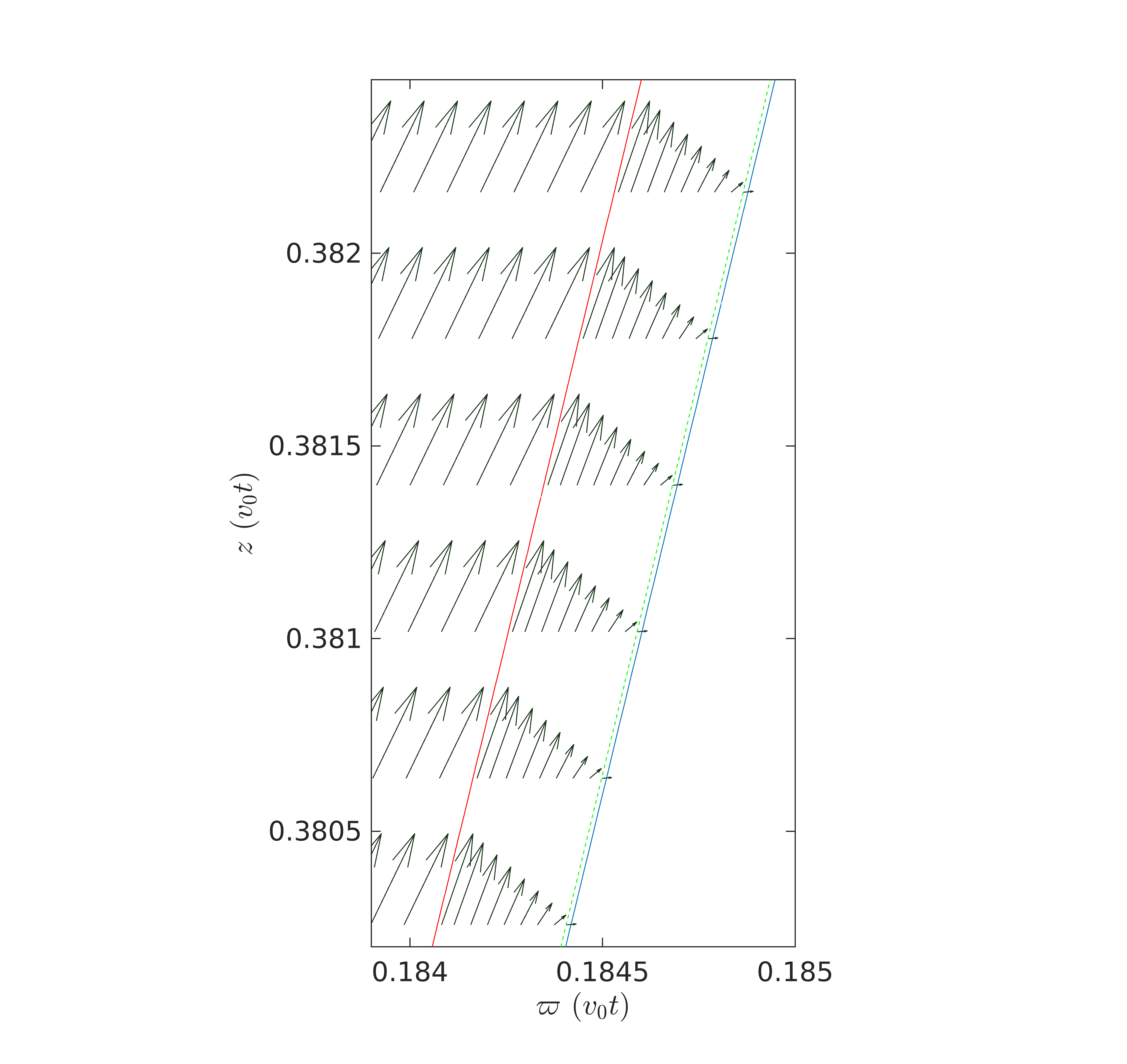}
\\
\includegraphics[height=0.18\textheight]{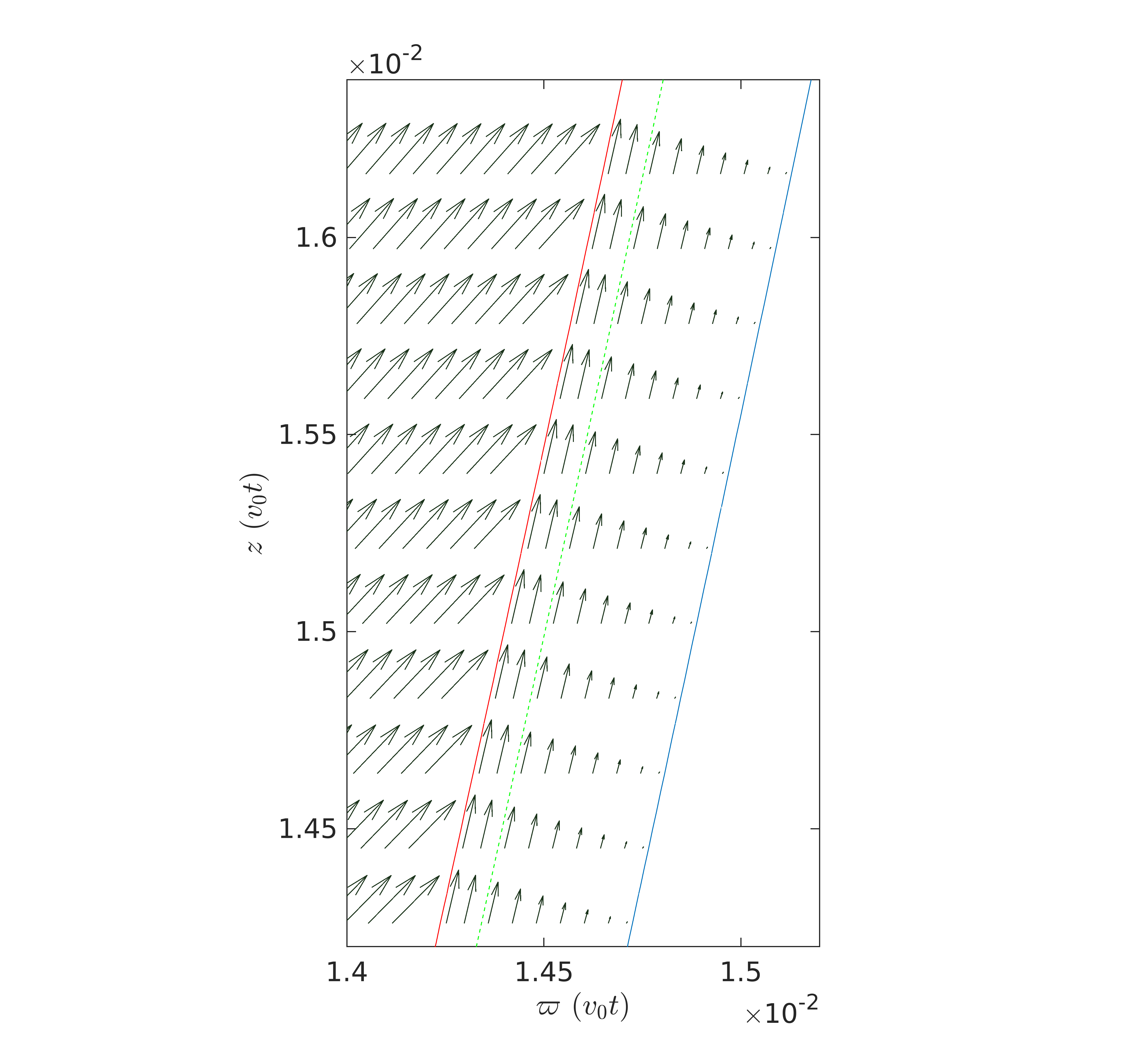}
\includegraphics[height=0.18\textheight]{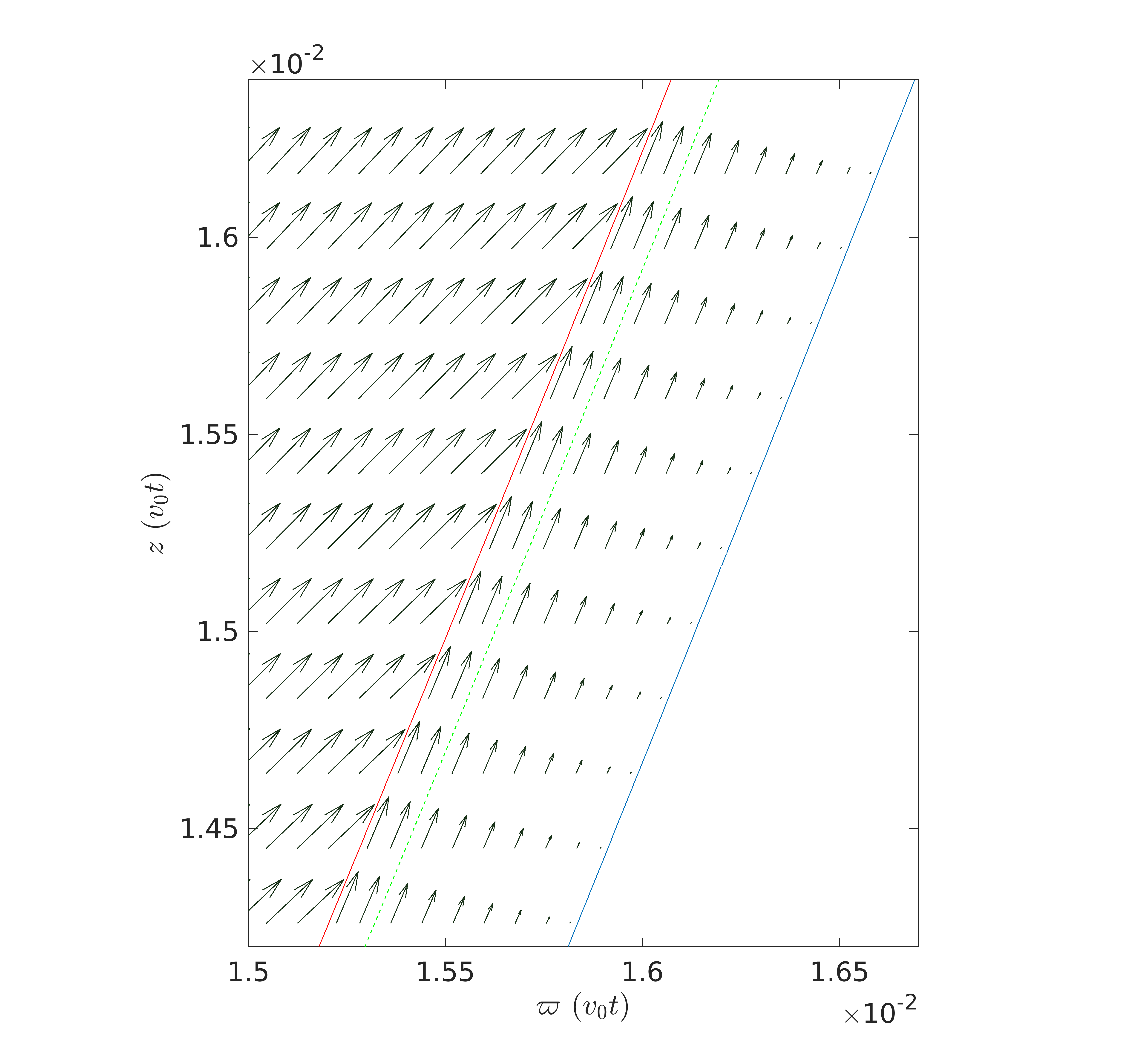}
\includegraphics[height=0.18\textheight]{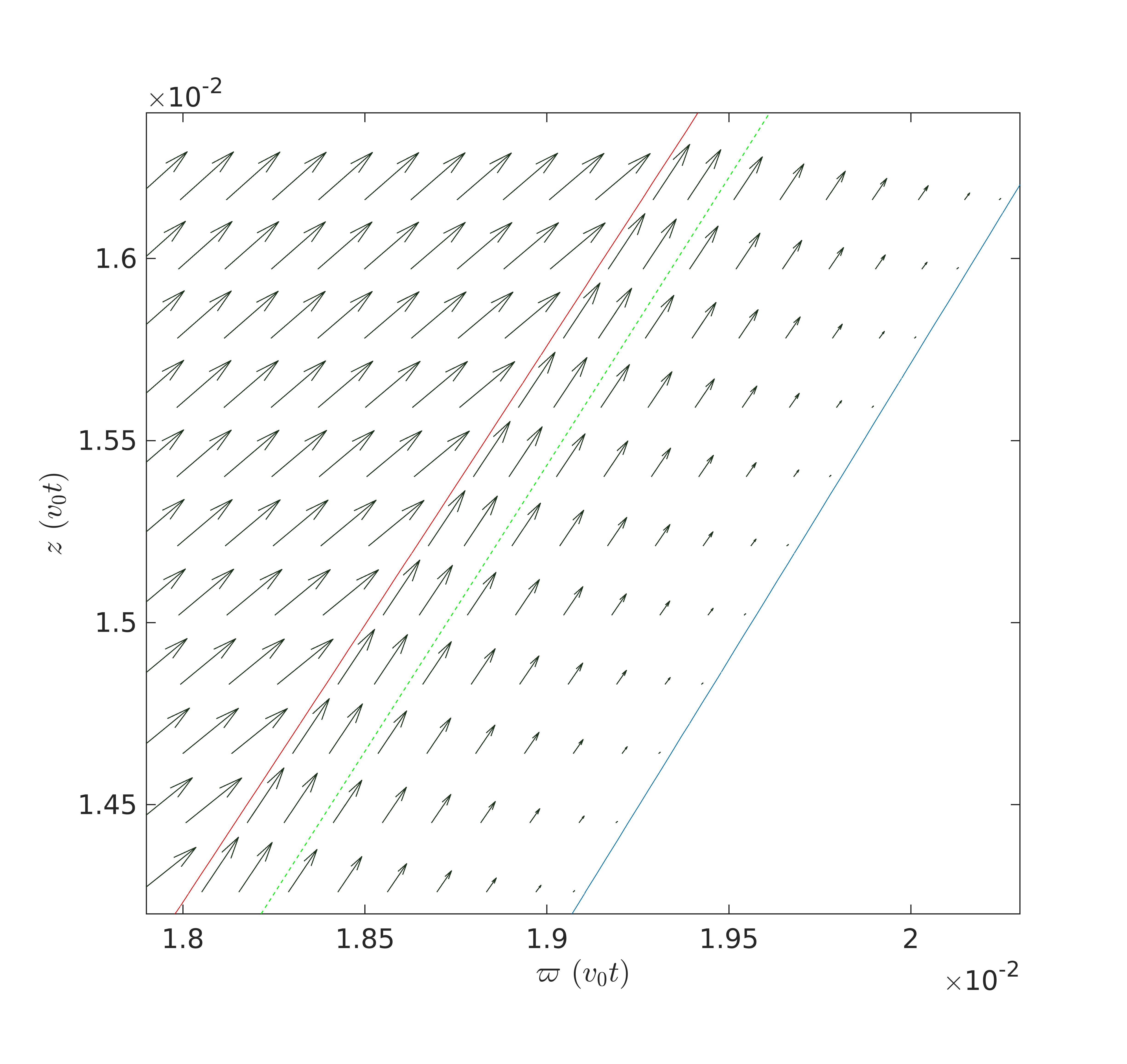}
\includegraphics[height=0.18\textheight]{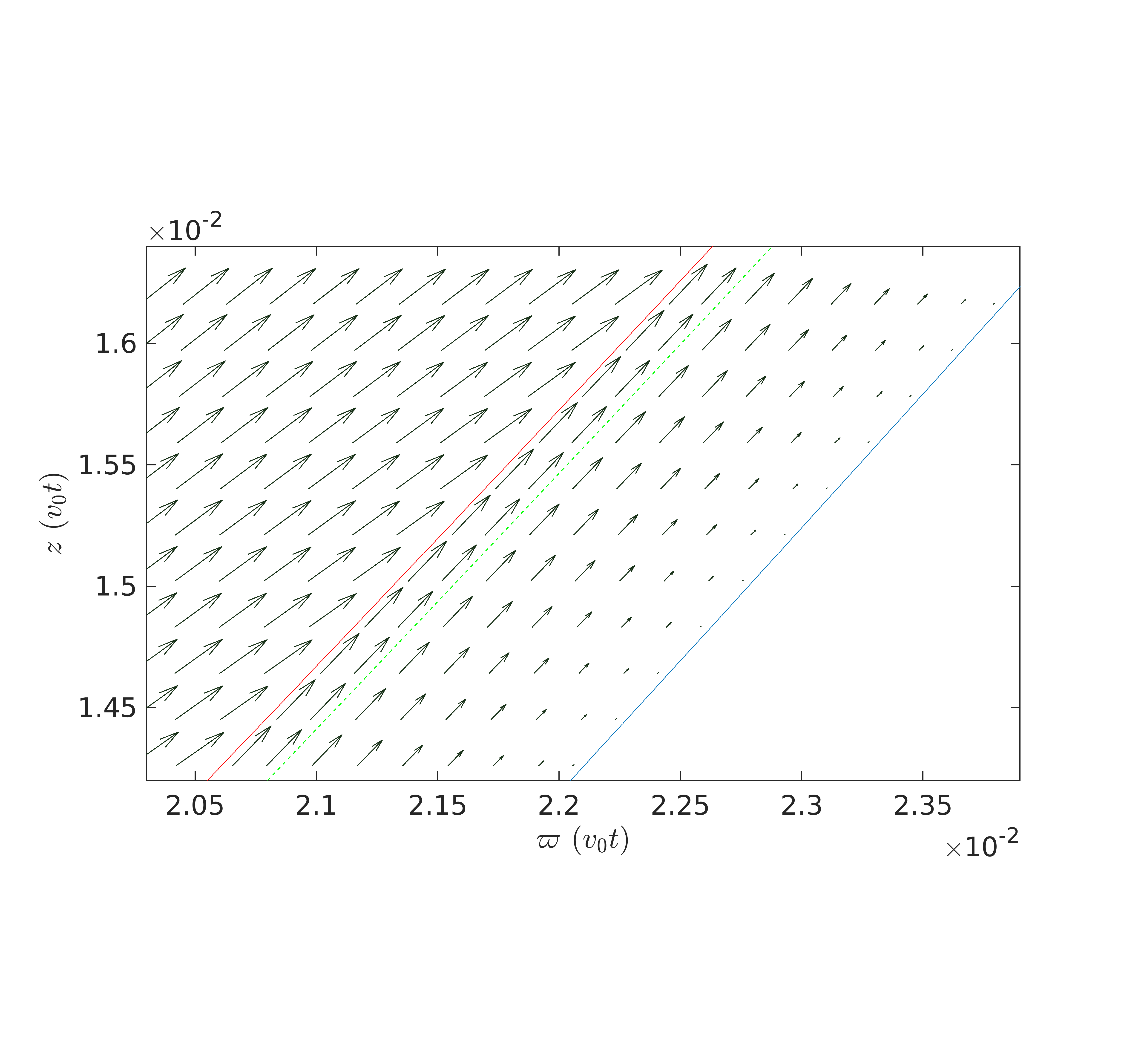}
\caption{Shear velocity panels, illustrating the results of Appendix \ref{sec:appendix:steps}. The panels are in deeply zoomed-in scales, able to show details of the velocity structure inside the shell as constructed with the models presented in Appendix \ref{sec:appendix:steps}. Columns, left to right: $n=1$, $2$, $4$, and $6$. Rows, top to bottom: top, middle, and base of the outflow. {\color{red} RS (red)}, {\color{green} CD (green)}, and {\color{blue} FS (blue)} drawn as solid lines. Velocity vector scales of each panel given by the free wind velocity $v_0$ in the left portions before the RS\@.
}
    \label{fig:shear}
\end{figure*}

Simple interpolation gives the constructed estimate for the shear velocity at $\Pv_g$:
\begin{equation}\label{eqn:step4_vP}
    \begin{split}
        f_\mathrm{RS} &\equiv \frac{d_\mathrm{FS}}{d_\mathrm{FS}+d_\mathrm{RS}} \quad\text{and}\quad
        \vv(\Pv_g) = f_\mathrm{RS} \vv^\mathrm{\,ff}_\mathrm{post}(\mathrm{RS}) + (1-f_\mathrm{RS}) \vv^\mathrm{\,ff}_\mathrm{post}(\mathrm{FS}),
    \end{split}
\end{equation}
thus completing the identification of velocity profile for any location in the post-shocked region. The resulting shear velocity field is illustrated in the panels of Figure \ref{fig:shear}.

In this step, we build upon the understanding of parallel oblique shock structures presented in Appendix C of \citetalias{shang2020} for the constructions of the analytic procedures here. There we had derived an analytical expression for the total shear across the post-shock region (without information of the structures inside), bounded by the RS and FS\@. 
Here we utilize the analytic formulations of the jump conditions and behavior of the oblique shocks, and construct analytically a whole new shear profile. We deploy methods of interpolation between two newly computed estimates of the locations of the RS and FS\@. 
Our new procedures are novel in giving analytical constructions for the shear velocity profile and for the thickness of the shells that possess 1D radial RS, CD, and FS structures, while extended to 2D in $\theta$. Because of the shift of focus, a new set of notations different from those in Appendix C of \citetalias{shang2020} for the basic schematics of the parallel oblique shocks has been adopted in this work. 

\newpage

\section{Small-Angle Obliqueness for the Momentum-Conserving Thin-Shell Curves}
\label{sec:appendix:chi_limit}

Under the momentum-conserving model, the shape of the outflow shell is controlled by two functions of $\theta$: the wind velocity $\vw(\theta)$ and the ratio $\delta(\theta)$ of wind to ambient densities. 
For this work, the wind velocity is adopted to be $\vw(\theta)=v_0$. The density ratio results from dividing a wind function $\rhow=\barrhow(\theta)/r^2$ over an ambient function $\rhoa=\barrhoa(\theta)/r^2$.
We consider the wind function to have the form $\rhow=D_0 (\sin\theta)^{-k}/r^2$, $k\geq0$. The ambient function of this work is the sum of two terms (Equation\ (\ref{eqn:alpharho})): a main term $\rho^\mathrm{T}(r,\theta) = (\lfrac{\aamb^{2}}{2\pi Gr^{2}}) R(\theta)$ (Equation\ (\ref{eqn:toroid_scaling})) representing the toroid solutions of \cite{li1996b}, parameterized by $n\geq0$, and a minor term $\rho^S(r)=D^S/r^2$ with $D^S\geq0$.

The momentum-conserving shape function corresponding to this $\delta$ is
$r(\theta)=v_0 t (1+\delta^{-1/2})^{-1}$. Obliqueness can be obtained from this shape function and the equation
\begin{equation}
\label{eqn:chi:f}
    \sin\chi=-r'/\sqrt{r'^2+r^2}=-(r'/r)/\sqrt{1+(r'/r)^2}\ ,
\end{equation}
where the ratio $r'/r\equiv(1/r) (dr/d\theta)$ needs to be obtained for a given $\theta$ value. The equations up to this point in the appendix are valid for all $\theta$. However, henceforth, in this appendix, we will focus on the small-angle limit $\theta\rightarrow0$.

For small $\theta$, the toroid solutions behave as $\rho^\mathrm{T} \propto (\sin\theta)^n$.
The $D^S>0$ cases are referred to as the tapered toroids, and the $D^S=0$ cases are the untapered toroids.
In the limit of small $\theta$ angles, the density ratio is
\begin{equation}
\label{eqn:delta_general}
\delta^{-1}=\rhoa/\rhow=\frac{\aamb^2}{2\pi G}\frac{a_0(n)}{D_0}(\sin\theta)^{n+k} + \frac{D^S}{D_0}(\sin\theta)^k\ ,
\end{equation}
where the ambient sound speed $\aamb$ sets the scale for the toroids, and $a_0(n)$ is a constant arising from their solutions.
Unless $D^S$ is set to zero, the dominant term (for small $\theta$) in the above expression has a power law of the form $(\sin\theta)^k$.
Using this value of $\delta$,
the ratio $r'/r$ for small values of $\theta$ is obtained as
\begin{equation}
\label{eqn:chi:r}
    r'/r=-\frac{\delta^{1/2}}{2}(1+\delta^{-1/2})^{-1}
    \cos\theta\left[
    (n+k) \frac{\aamb^2}{2\pi G}\frac{a_0(n)}{D_0}(\sin\theta)^{n+k-1} + k\frac{D^S}{D_0}(\sin\theta)^{k-1}
    \right]\ ,
\end{equation}
which can be used in combination with Equation (\ref{eqn:chi:f}) to obtain values of $\chi(\theta)$ valid for small $\theta$.
We need now to establish the leading form of these equations as $\theta\rightarrow0$, starting from Equation\ (\ref{eqn:delta_general}). 
For the untapered toroids with $D^S=0$, only the first term remains, and therefore it leads.
For $n=0$, even when tapered, both terms are of the same order in $\sin\theta$, and our choices of parameters are such that if $n=0$ we have that $a_0=1$ and
$(\aamb^2)/(2\pi G) \gg D^S$, making the tapered and untapered toroids effectively coincide in that the leading term is again the first one.
In the remaining scenario, for tapered toroids with $n>0$, for a sufficiently small value of $\theta$ it is the second term that dominates:
\begin{equation}
\label{eqn:delta_tapered}
\delta^{-1}=(D^S/D_0)(\sin\theta)^k \ .
\end{equation}
independent of $n$ for a sufficiently small $\theta$.
Therefore, the leading term always has the form
\begin{equation}
\label{eqn:delta_asymptotic}
\delta^{-1} = c (\sin\theta)^p \ , 
\end{equation}
where for $D^S=0$ we have $p=n+k$, and for $D^S>0$ we have $p=k$, and the coefficient $c$ is equal to either
$\frac{\aamb^2}{2\pi G}(a_0/D_0)$ (for the untapered $n>0$ toroids, and for $n=0$) or 
$(D^S/D_0)$ (for the tapered toroids with $n>0$).

\begin{figure*}
\epsscale{0.8}
\plotone{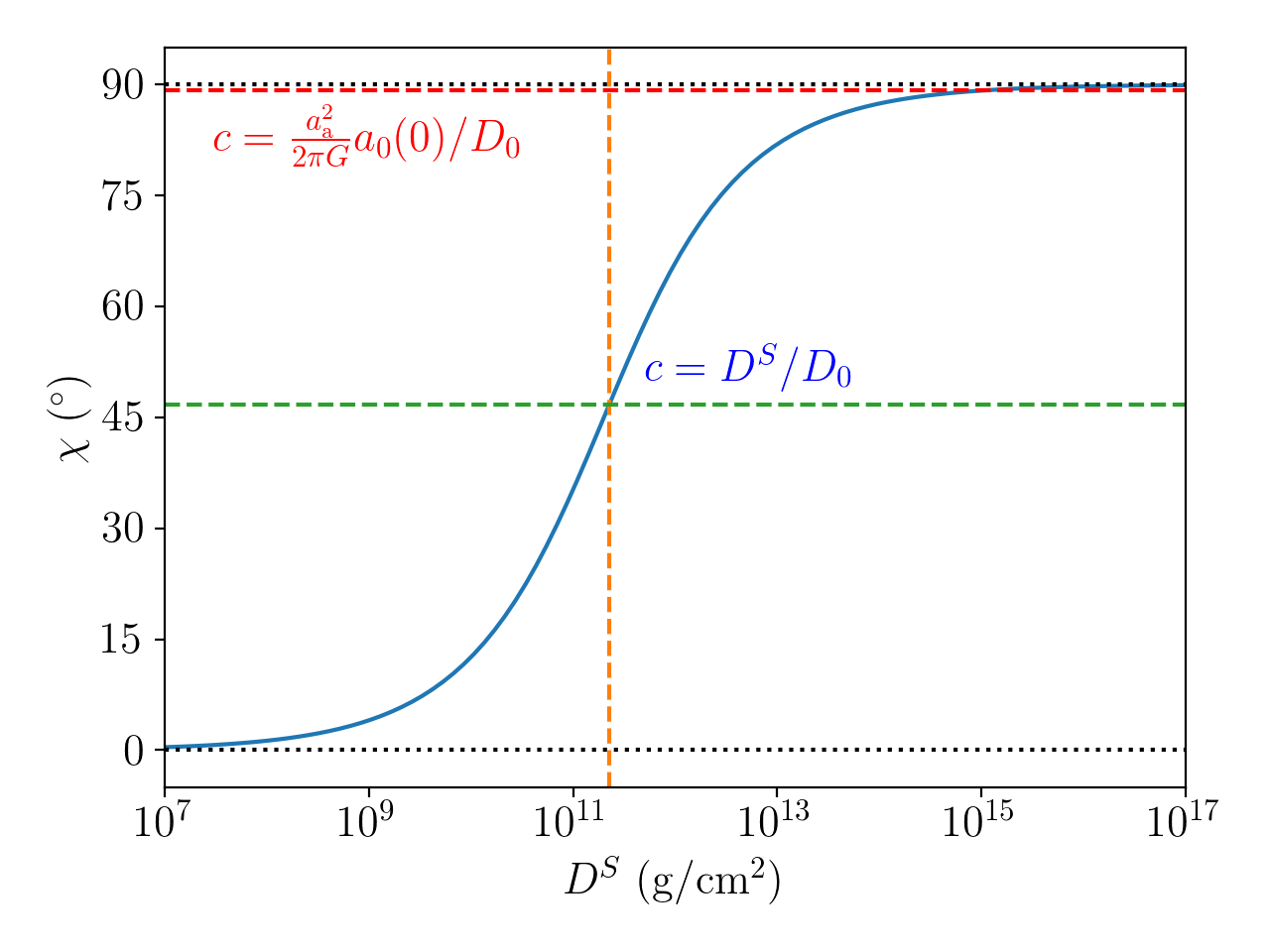}
\caption{
The blue curve shows the small-angle ($\theta\rightarrow0$) limit of obliqueness $\chi$ (ranging between $0\arcdeg$ to $90\arcdeg$, horizontal black dotted lines) as a function of the density factor $D^S$. This limit is computed using Equation (\ref{eqn:chi:small_theta_n0:orig}) with the tapered $n>0$ cases $c=D^S/D_0$ and $D_0=2.0025\times10^{11}$ g\,cm$^{-2}$. The $n>0$ tapered HD obliqueness approaches $\sim47\arcdeg$ (horizontal green dashed line) for $D^S=2.25\times10^{11}$ g\,cm$^{-2}$ (vertical orange dashed line). The horizontal red dashed line is the $\theta\rightarrow0$ limiting value of the untapered $n=0$ case ($\sim88\arcdeg$) with $c=(\lfrac{\aamb^2}{2\pi G})/D_0$. The tapered $\chi$ can be seen to approach the value of the untapered $n=0$ case as $D^S\approx10^{15}$ g\,cm$^{-2}$.}
\label{fig:chi_DScomparisons}
\epsscale{1.0}
\end{figure*}

The density ratio Equation\ (\ref{eqn:delta_asymptotic}) allows the computation of the radius $r(\theta)$ of the shell at a given $\theta$, with a derivative $r'(\theta)\equiv dr/d\theta$, whose ratio has the limiting behavior
$r'/r = -(1/2) c^{1/2} p (\sin\theta)^{p/2-1}$.
Using this limiting ratio in Equation\ (\ref{eqn:chi:arcsin}) yields the obliqueness for small angles as
\begin{equation}
\label{eqn:chi_p:orig}
\sin\chi = [1+(4/c) p^{-2} (\sin\theta)^{2-p} ]^{-1/2}\ .
\end{equation}
We now apply this result to our cases.
For the $n>0$ untapered toroids with our usual winds, we have that $p=n+k$ with $k=2$, and therefore the obliqueness is given by
\begin{equation}
\label{eqn:chi:small_theta:orig}
\sin\chi = [1+(4/c) (n+2)^{-2} (\sin\theta)^{-n} ]^{-1/2}\ ,
\end{equation}
an expression in which for $n>0$ the sine terms dominates the RHS of Equation\ (\ref{eqn:chi:small_theta:orig}), leading to
$\sin\chi \propto \theta^{n/2}$, which has a limiting value of zero at $\theta=0$ for the untapered toroids with $n>0$.
For the tapered toroids, and for the $n=0$ case, both terms are of the same order, leading to a nonzero limiting value of $\sin\chi$ for $\theta\rightarrow 0$, equal to
\begin{equation}
\label{eqn:chi:small_theta_n0:orig}
\sin\chi = [1+(1/c)]^{-1/2}\ ,
\end{equation}
depending on $c$, which is a function of problem parameters. This equation is illustrated in detail in Figure \ref{fig:chi_DScomparisons}.

\begin{figure*}
\plotone{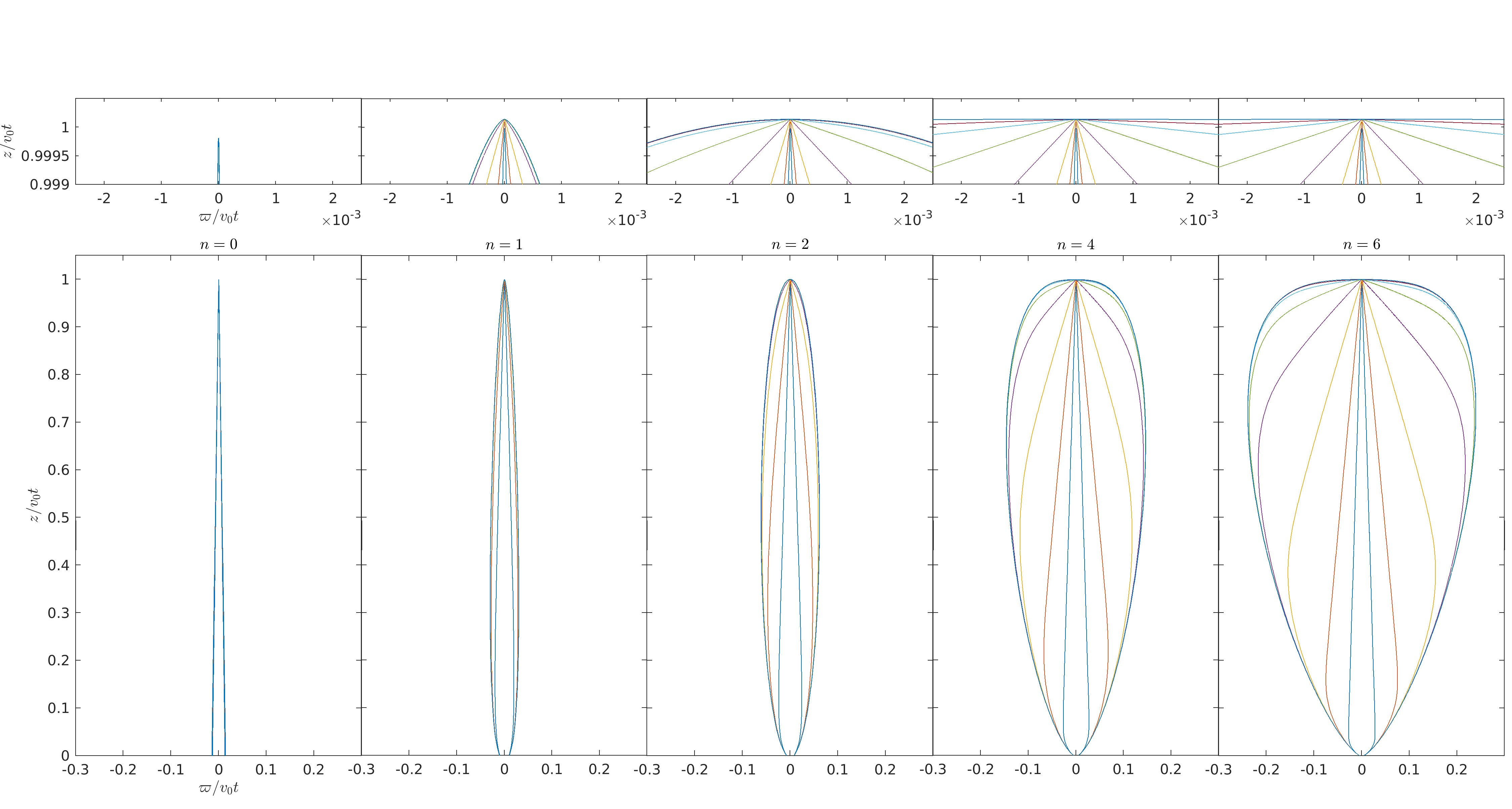}
\caption{
Shapes of the HD MC curves for $n=0$, $1$, $2$, $4$, and $6$ toroids, for variation of the density factor $D^S$ of $1000$, $100$, $10$, $1$, $0.1$, $0.01$, $0.001$, and $0$ relative to the current setup of $D^S=2.25\times10^{11}$ g\,cm$^{-2}$. The tips change from pointy to flat as $D^S$ decreases, except for $n=0$. The lower row shows the whole outflow shape, and the upper row shows the zoomed-in tips.}
\label{fig:MC_DScomparisons}
\end{figure*}

Figure \ref{fig:MC_DScomparisons} shows the outflow shapes at all angles, with its top row focusing on very small angles
to help in examining the $\theta\rightarrow 0$ limit of the inclination angle $\chi$. A variety of $D^S$ values are shown, including tapered cases ($D^S>0$ such as in our usual simulations) and untapered cases ($D^S=0$). Naturally due to their less-dense ambient medium, outflows of untapered cases can proceed faster, and they are the larger outflow curves in each panel of Figure \ref{fig:MC_DScomparisons}.
For any untapered $n>0$ case, it has been shown (from Equation (\ref{eqn:chi:small_theta:orig})) that $\lim_{\theta\rightarrow0}\chi=0$, and therefore the tangent to the outflow shape at that point $\theta=0\arcdeg$ is horizontal, parallel to the disk and perpendicular to the jet axis. The approach of the curves to their horizontal tangents is seen in the $n>0$ top panels of Figure \ref{fig:MC_DScomparisons}, most clearly for $n>1$, but nonetheless true for any $n>0$. However, approaching the horizontal tangent limit requires smaller values of $\theta\rightarrow0$ for $n=1$ than for, e.g., $n=2$, and it would require even smaller $\theta$ values if we want to consider $n=0.5$.
For the tapered cases $D^S>0$, and also for $n=0$, the tangent is oblique even at $\theta\rightarrow0$, following Equation (\ref{eqn:chi:small_theta_n0:orig}). Consistent with that, all of the curves with $D^S>0$ and in the $n=0$ column of Figure \ref{fig:MC_DScomparisons} have nonhorizontal tangents at all $\theta$, including $\theta=0\arcdeg$, for which the inclination is shown in Figure \ref{fig:chi_DScomparisons}.

We have adopted a wind with a density profile (Equation \ref{eqn:initcon}) $\rho\propto(\sin\theta)^{-k}$ with $k=2$, with a formal divergence at $\theta\rightarrow 0\arcdeg$, compatible with the simulations because of the implicit tapering provided by angular resolution, which keeps $\dot M$ finite and well controlled \citep{shang2006,wang2015}.
In addition to that, here we briefly discuss the $k=0$ cases.
In contrast to our usual $k=2$ cases, for a wind having an explicit tapering at narrow angles, or a spherically symmetric wind, there is a limiting behavior with $k=0$ for $\theta\rightarrow 0\arcdeg$. For the cases of $p=0$: the tapered toroids $D^S>0$ which have $p=k=0$, and the case $n=0$ for which $p=n+k=0$, both give a limit of zero axial obliqueness.
For the untapered toroids with $n>0$, we utilize the result of Equation (\ref{eqn:chi_p:orig}), for $D^S=0$, $p=n+k=n+0=n$. For a finite value of $c$ the limiting behaviors of Equation\ (\ref{eqn:chi_p:orig}) for $\theta\rightarrow 0$ depend on $n$ as follows:
\begin{equation}
\label{eqn:chi:small_theta_zerok:orig}
\chi = \arcsin\left\{[1+(4/c) n^{-2} (\sin\theta)^{2-n} ]^{-1/2}\right\}
=
\left\{
\begin{array}{ll}
90\arcdeg & \text{ for $k=0$ and untapered $0<n<2$}\\
\arcsin( (1+1/c)^{-1/2} ) & \text{ for $k=0$ and untapered $n=2$}\\
0\arcdeg & \text{ for $k=0$ and untapered $n>2$}\ .
\end{array}
\right.
\end{equation}
This expression has discontinuities at $n=0$ and $n=2$.
The $k=0$, $n=2$ case has similar algebraic behavior to the $k=2$, $n=0$ case, consistent with the definition $p=n+k$.

\section{Magnetic Forces and Pseudopulses}
\label{sec:appendix:mag_forces}

\begin{figure*}
    \centering
    \gridline{
\fig{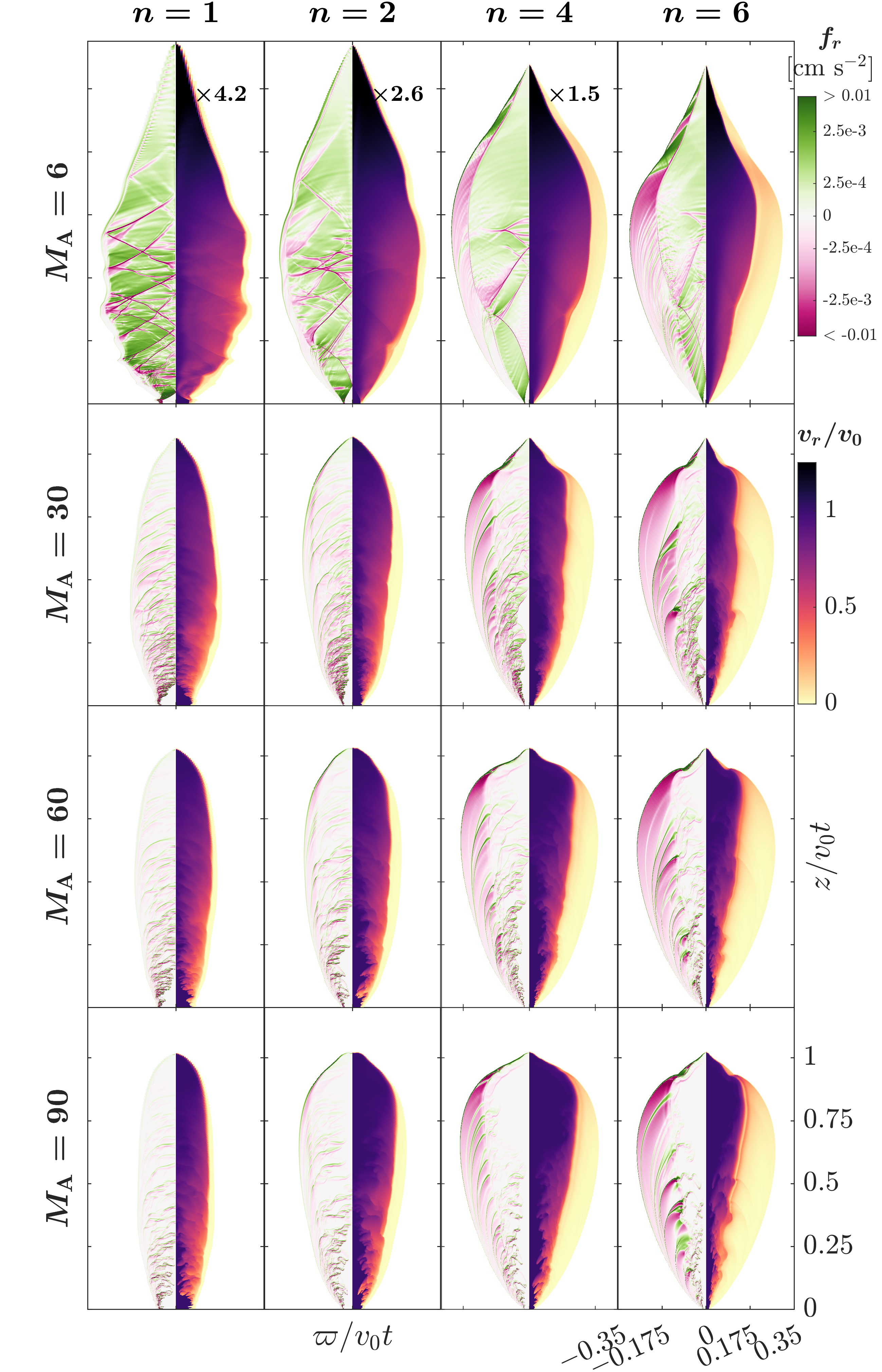}{0.33\textwidth}{(a) $f_r$, $v_r$ for $\alpha_b=1$}
\fig{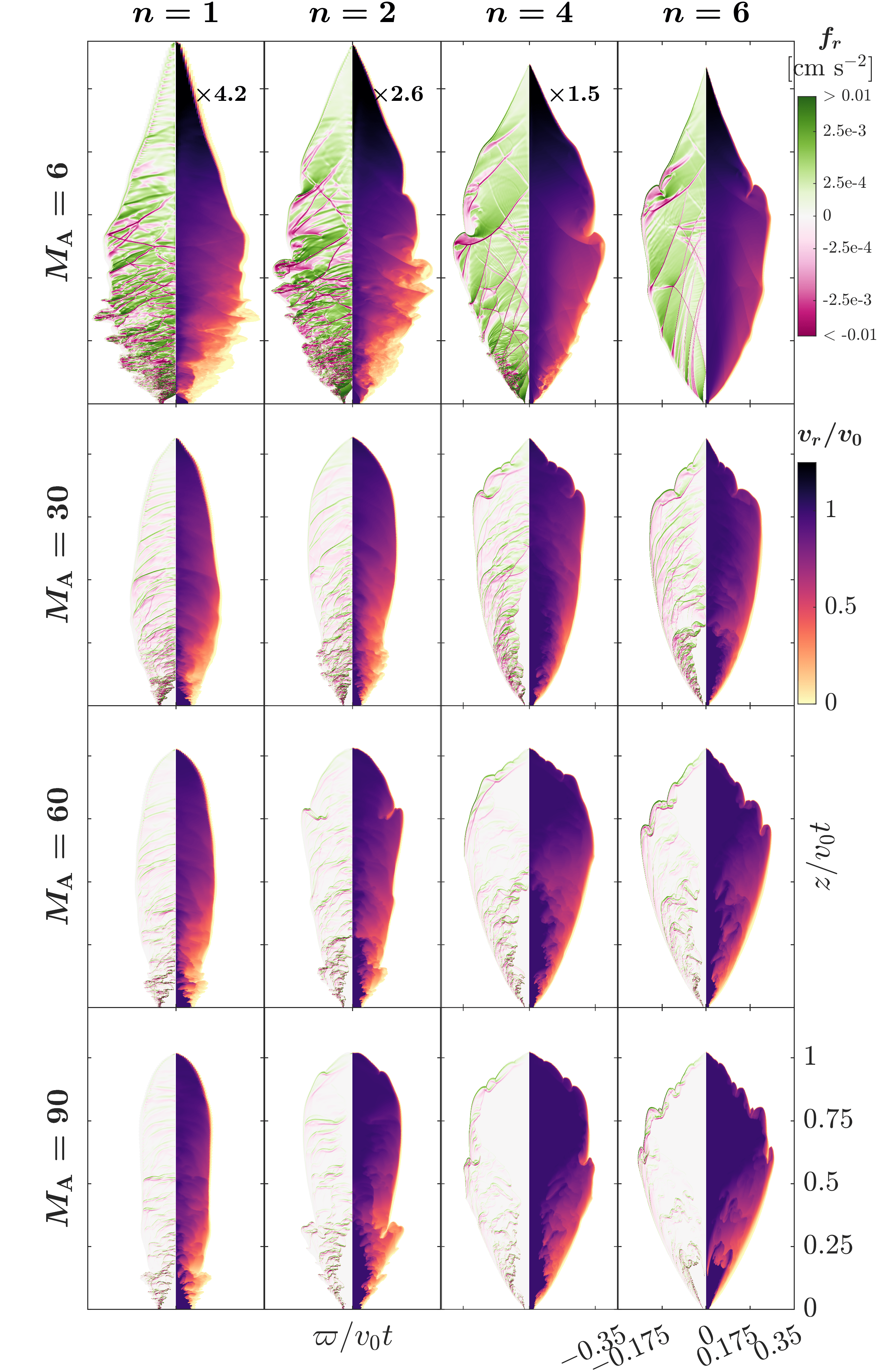}{0.33\textwidth}{(b) $f_r$, $v_r$ for $\alpha_b=0.1$}
\fig{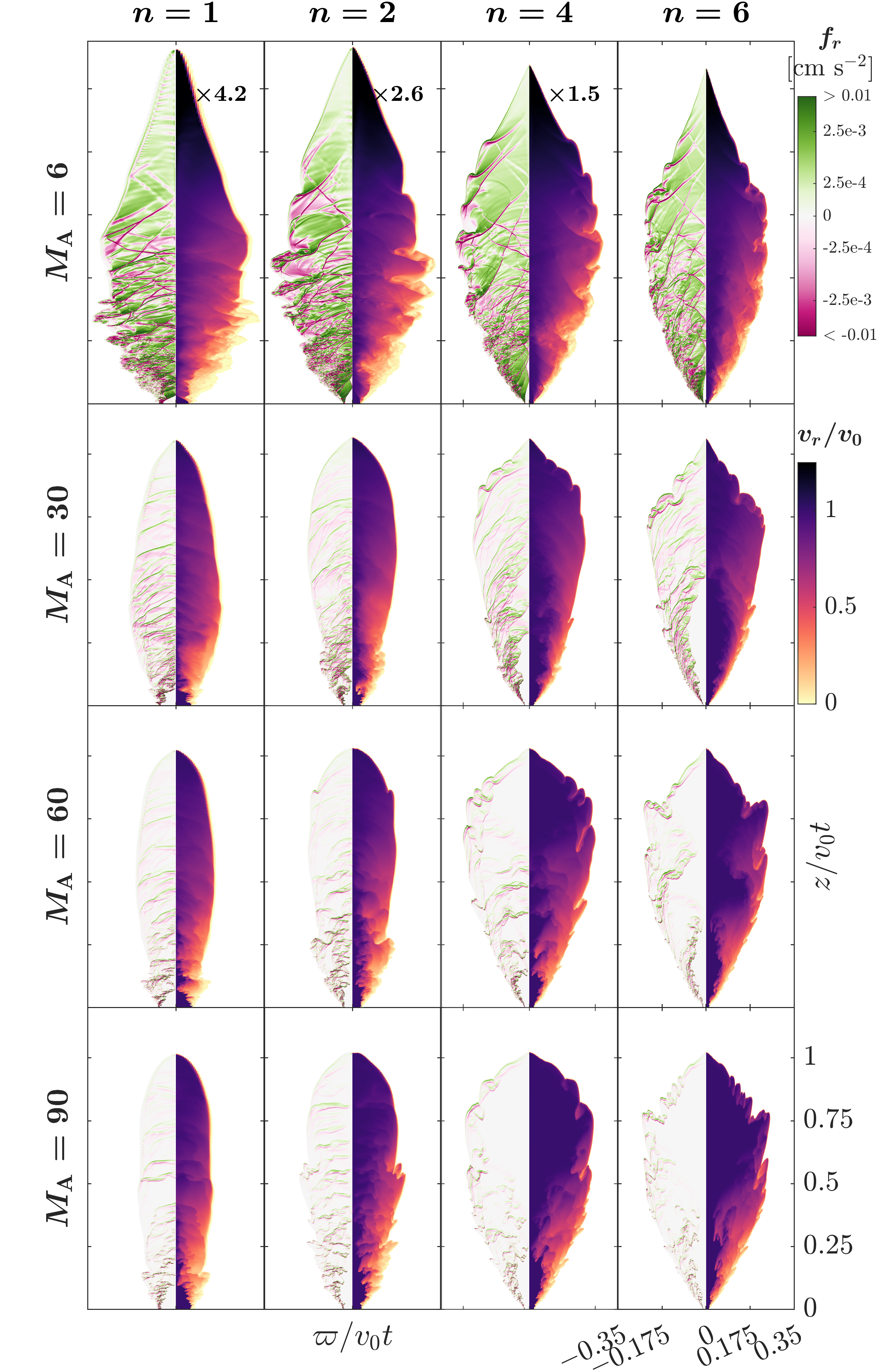}{0.33\textwidth}{(c) $f_r$, $v_r$ for $\alpha_b=0$}
}
    \gridline{
\fig{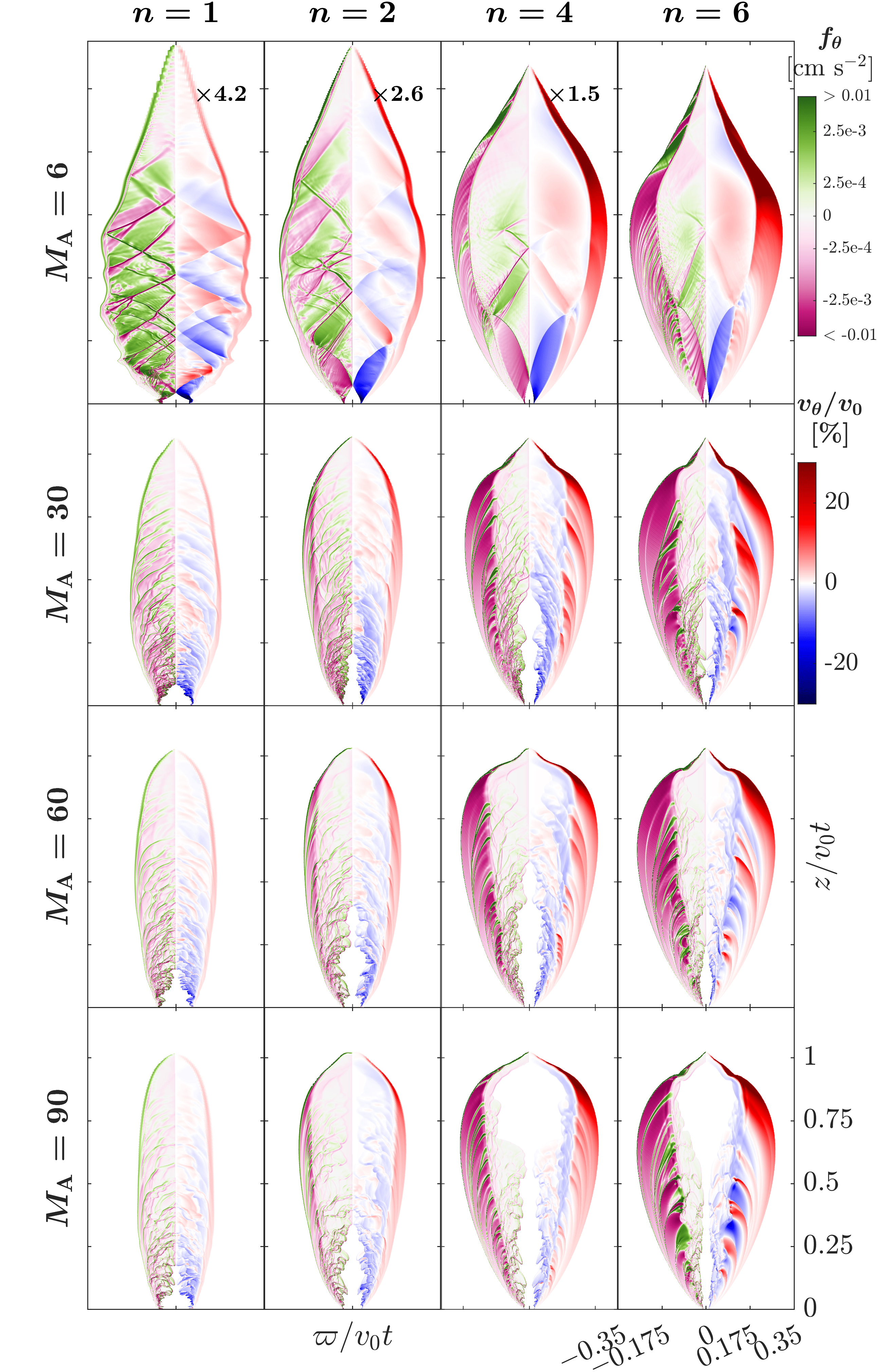}{0.33\textwidth}{(d) $f_\theta$, $v_\theta$ for $\alpha_b=1$}
\fig{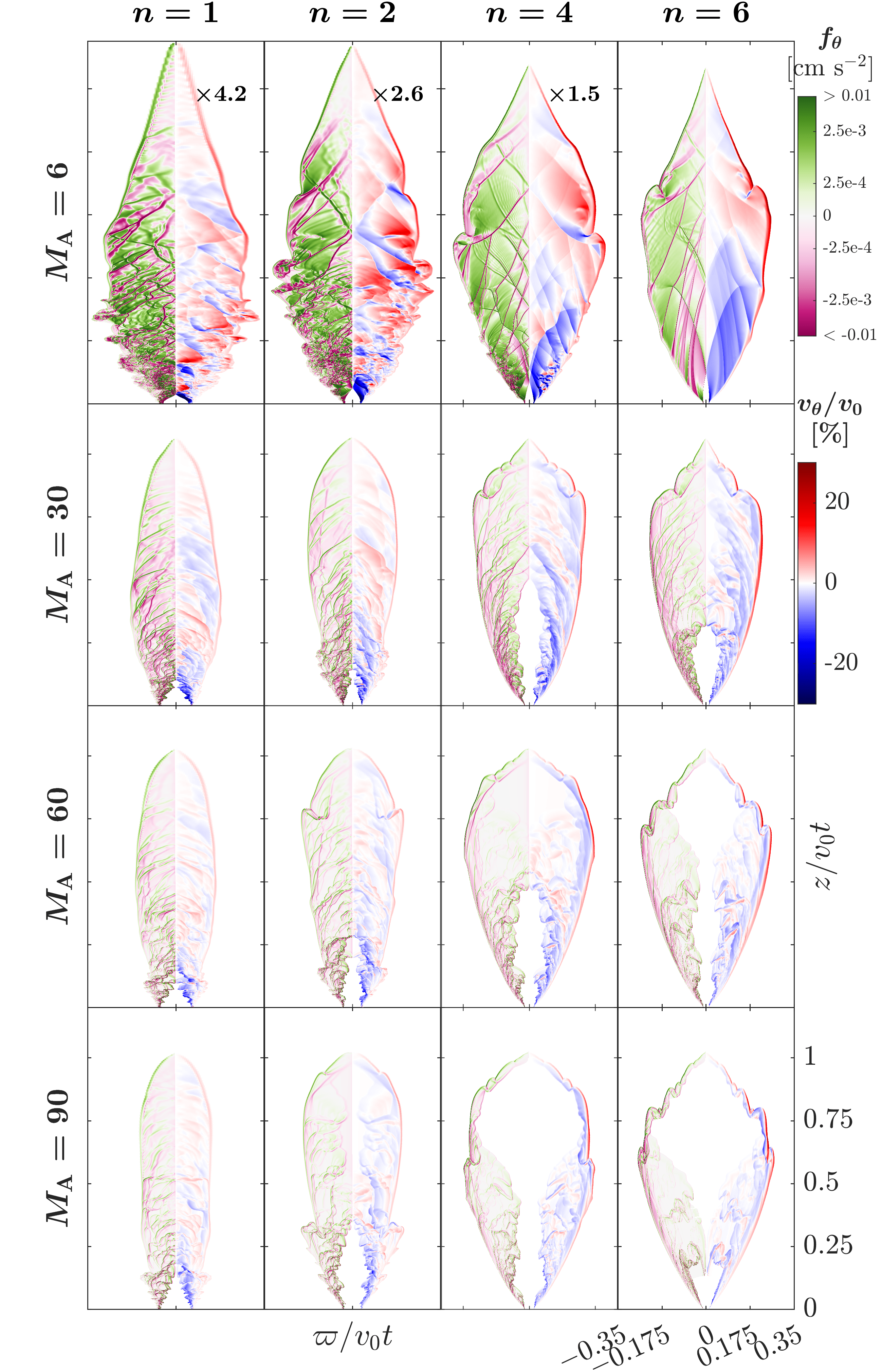}{0.33\textwidth}{(e) $f_\theta$, $v_\theta$ for $\alpha_b=0.1$}
\fig{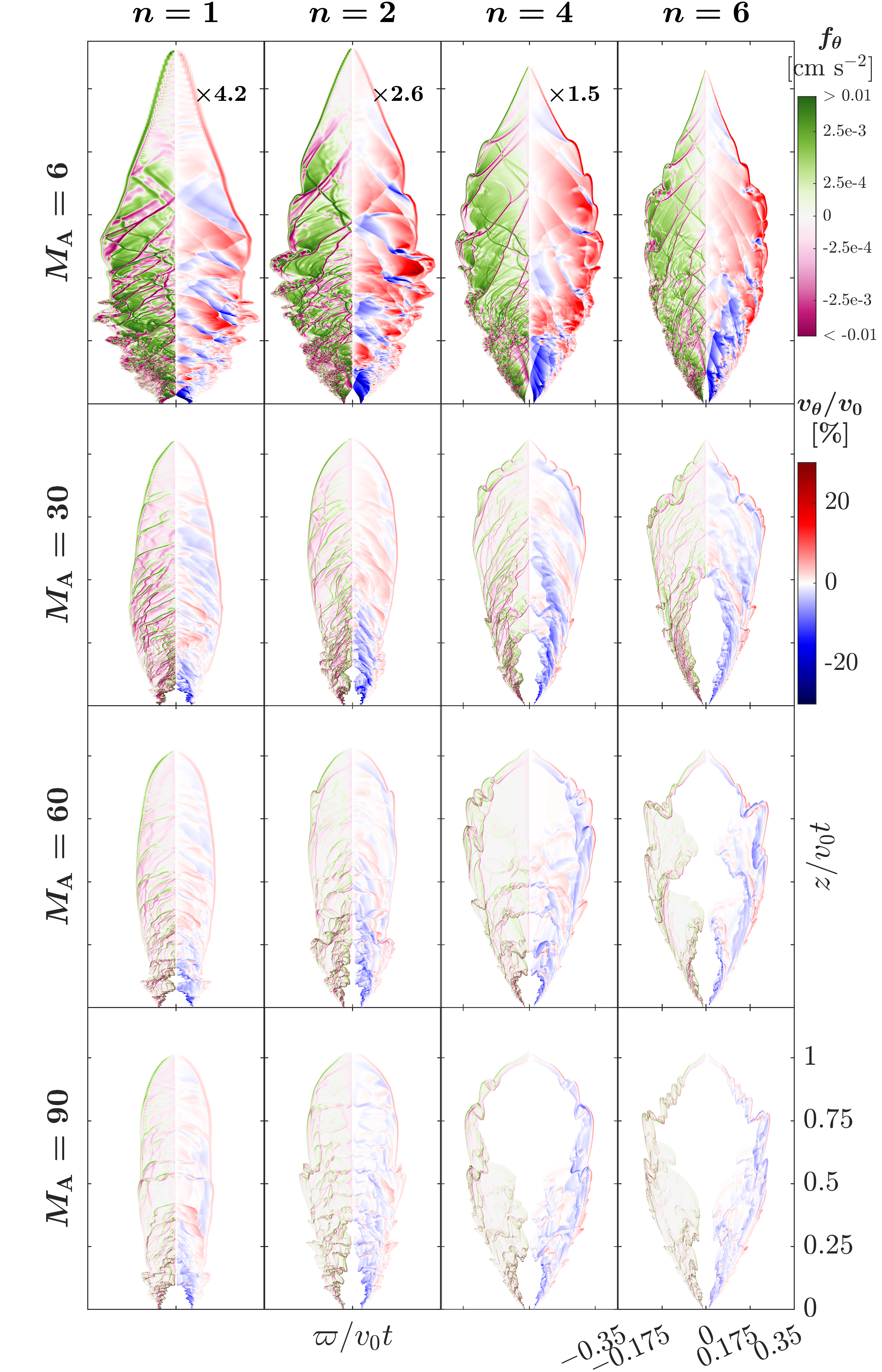}{0.33\textwidth}{(f) $f_\theta$, $v_\theta$ for $\alpha_b=0$}
}
\caption{
Top row: $r$-components of magnetic forces and of normalized $\vv/v_0$ velocities are respectively shown in the left and right half of each panel inside (a) $\alpha_b=1$, (b) $\alpha_b=0.1$, and (c) $\alpha_b=0$ cases. Bottom row: the respective $\theta$-components are shown for (d) $\alpha_b=1$, (e) $\alpha_b=0.1$, and (f) $\alpha_b=0$ cases. The $\varpi$ axis has been exaggerated to show detailed structures. The exaggeration factors are $4.2$, $2.6$, $1.5$, and $1.0$ for $n=1$, $2$, $4$, and $6$, respectively. Forces are shown in a symmetric logarithmic scale. Velocities are shown in linear scales (natural and percentage).
}
\label{fig:mag_forces}
\end{figure*}

The panels of Figure \ref{fig:mag_forces} are shown here to illustrate the effects and connections between the velocity and force components.

\section{Shapes of perpendicular PVDCD diagrams}
\label{sec:appendix:perpPVDCDshapes}

In this appendix, the diagram shapes sketched in Figure \ref{fig:transversePV_schematics} are listed and then compared to the PVDCD shapes of Figure \ref{fig:perpPV_vw100_1Bp-0Bp} computed using \Synline{}.
This continues (in more detail) the comparison of perpendicular PVDCDs, which was summarized with a few examples in Section\ \ref{subsec:perpPVDCDsketch}.

\begin{itemize}
\item
Intersecting ellipses or ovals $\,\ellipsesc\,$ as in Figure \ref{fig:transversePV_schematics}(c) or (e) are present in many cases, including Figure \ref{fig:perpPV_vw100_1Bp-0Bp}(c), (e), and (f) for $n=4$--$6$, all of Figure\  \ref{fig:perpPV_vw100_1Bp-0Bp}(a) and (b), and also Figure\ \ref{fig:perpPV_vw100_1Bp-0Bp}(f) ($n=2$,$\MA=6$).
\item
Rhombus shapes $\blacklozenge$ as in Figure \ref{fig:transversePV_schematics}(f) are also present in many cases: Figure \ref{fig:perpPV_vw100_1Bp-0Bp}(c) for $n=1$--$2$, nearly all of Figure \ref{fig:perpPV_vw100_1Bp-0Bp}(d), and Figure \ref{fig:perpPV_vw100_1Bp-0Bp}(f) ($n=1$,$\MA=6$). Panels in Figure \ref{fig:perpPV_vw100_1Bp-0Bp}(e) for ($n=1$--$2$,$\MA\geq60$) and ($n=2$,$\MA\geq30$) are slightly indented rhombuses, between Figure \ref{fig:transversePV_schematics}(f) and (e).
\item
Bulging oval shapes as in Figure \ref{fig:transversePV_schematics}(g) are also present (many of them close to the rhombus shape):
Figure \ref{fig:perpPV_vw100_1Bp-0Bp}(d) ($n=6$,$\MA=60$) and Figure \ref{fig:perpPV_vw100_1Bp-0Bp}(f) ($n=1$--$2$,$\MA\geq30$).
\item
In this set of simulations, the concentric ellipses $\ellipsesd$ as in Figure  \ref{fig:transversePV_schematics}(d) and (h) are present in many of the densest features (see below). The overall shape of Figure \ref{fig:perpPV_vw100_1Bp-0Bp}(e) ($n=1$--$2$,$\MA=6$) is also present in Figure \ref{fig:transversePV_schematics}(h), probably due to the CDIE cutoff level isolating the densest features in this case.
\end{itemize}
In Figure \ref{fig:perpPV_vw100_1Bp-0Bp}(e) ($n=1$,$\MA=30$), the overall shape is as that of Figure \ref{fig:transversePV_schematics}(e) due to a small high-velocity portion; otherwise, it would be as that of Figure \ref{fig:transversePV_schematics}(g).

Here we mention a few of the different shapes of the high CDIE features, which are the possible result in realistic observations, depending on excitation conditions and instrument sensitivity.
Figure \ref{fig:perpPV_vw100_1Bp-0Bp}(c) ($n=1$--$2$,$\MA=6$) contains high CDIE concentric ellipses as in Figure \ref{fig:transversePV_schematics}(h): the thicknesses $\Delta y$ and $\Delta v_x$ of these sets of ellipses are connected to the slit thickness $\Delta z$. Dense CDIE features shaped as in Figure \ref{fig:transversePV_schematics}(h) are also present in Figure \ref{fig:perpPV_vw100_1Bp-0Bp}(c) ($n=1$--$2$,$\MA=30$--$60$), Figure \ref{fig:perpPV_vw100_1Bp-0Bp}(d) ($n=1$), Figure \ref{fig:perpPV_vw100_1Bp-0Bp}(d) ($n=2$--$4$,$\MA=6$), Figure \ref{fig:perpPV_vw100_1Bp-0Bp}(e)--(f) ($n=1$,$\MA=6$), and other cases.
The high CDIEs of Figure \ref{fig:perpPV_vw100_1Bp-0Bp}(a) and (b) show shapes as in $\ellipsesdv$ in Figure \ref{fig:transversePV_schematics}(h) or (g), surrounding the dense dot at zero position--velocity coordinates due to the jet.

\end{document}